\documentclass[a4paper,11pt,british,twosides]{book}%

\usepackage{mathptmx}
\usepackage{indentfirst}
\usepackage{geometry}
\usepackage{amsfonts,amsmath,amssymb}
\usepackage{graphicx,wrapfig,color}
\usepackage{bbold}
\usepackage[Gray,squaren,thinqspace,thinspace]{SIunits}%
\usepackage[small,bf,hang]{caption}
\usepackage{slashed}
\usepackage{braket}
\usepackage{pslatex}
\usepackage[section]{placeins} 
\usepackage{float}
\usepackage{subfigure}
\usepackage[T1,T5]{fontenc}
\usepackage{textcomp}
\usepackage{cite}
\usepackage{comment}
\usepackage{dsfont}
\usepackage{epsfig}
\usepackage[dvips]{feynmp}
\usepackage{enumitem}
\usepackage{feyn}
\usepackage[pdftitle={Thesis: Aspects of the Gribov prolem in YM theories}]{hyperref}

\usepackage[normalem]{ulem}

\providecommand{\U}[1]{\protect\rule{.1in}{.1in}}

\usepackage{color}
\definecolor{darkgreen}{rgb}{0,0.35,0}
\definecolor{Rood}{rgb}{1, 0, 0}

\newcommand{\dslash}{\partial\!\!\!/}

\newcommand{\pslash}{p \hspace{-1.7mm} /}
\newcommand{\kslash}{k \hspace{-1.7mm} /}

\renewcommand{\d}{\ensuremath{\mathrm{d}}}

\newcommand{\ii}{\ensuremath{\mathrm{i}}}
\newcommand{\p}{\partial}
\newcommand{\Tr}{\ensuremath{\mathrm{Tr}}}
\newcommand{\tr}{\ensuremath{\mathrm{Tr}}}
\newcommand{\e}{\ensuremath{\mathrm{e}}}

\newcommand{\R}{\ensuremath{\mathrm{R}}}
\newcommand{\YM}{\ensuremath{\mathrm{YM}}}
\newcommand{\FP}{\ensuremath{\mathrm{FP}}}
\newcommand{\gf}{\ensuremath{\mathrm{gf}}}

\newcommand{\MSbar}{\overline{\mbox{MS}}}
\newcommand{\msbar}{\overline{\mbox{MS}}}

\newcommand{\lms}{\Lambda_{\overline{\mbox{\tiny{MS}}}}}

\newcommand{\omu}{\overline{\mu}}

\numberwithin{equation}{section}

\begin{document}


\begin{titlepage}
\pagestyle{empty}
\setlength{\topmargin}{0cm}

\large

\begin{figure}
\hspace{2.2cm}
\subfigure{\includegraphics[width=30mm]{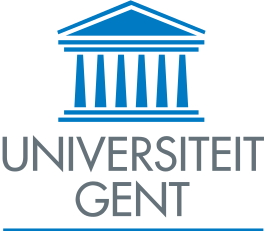}}
\hspace{4.4cm}
\subfigure{\includegraphics[width=27mm]{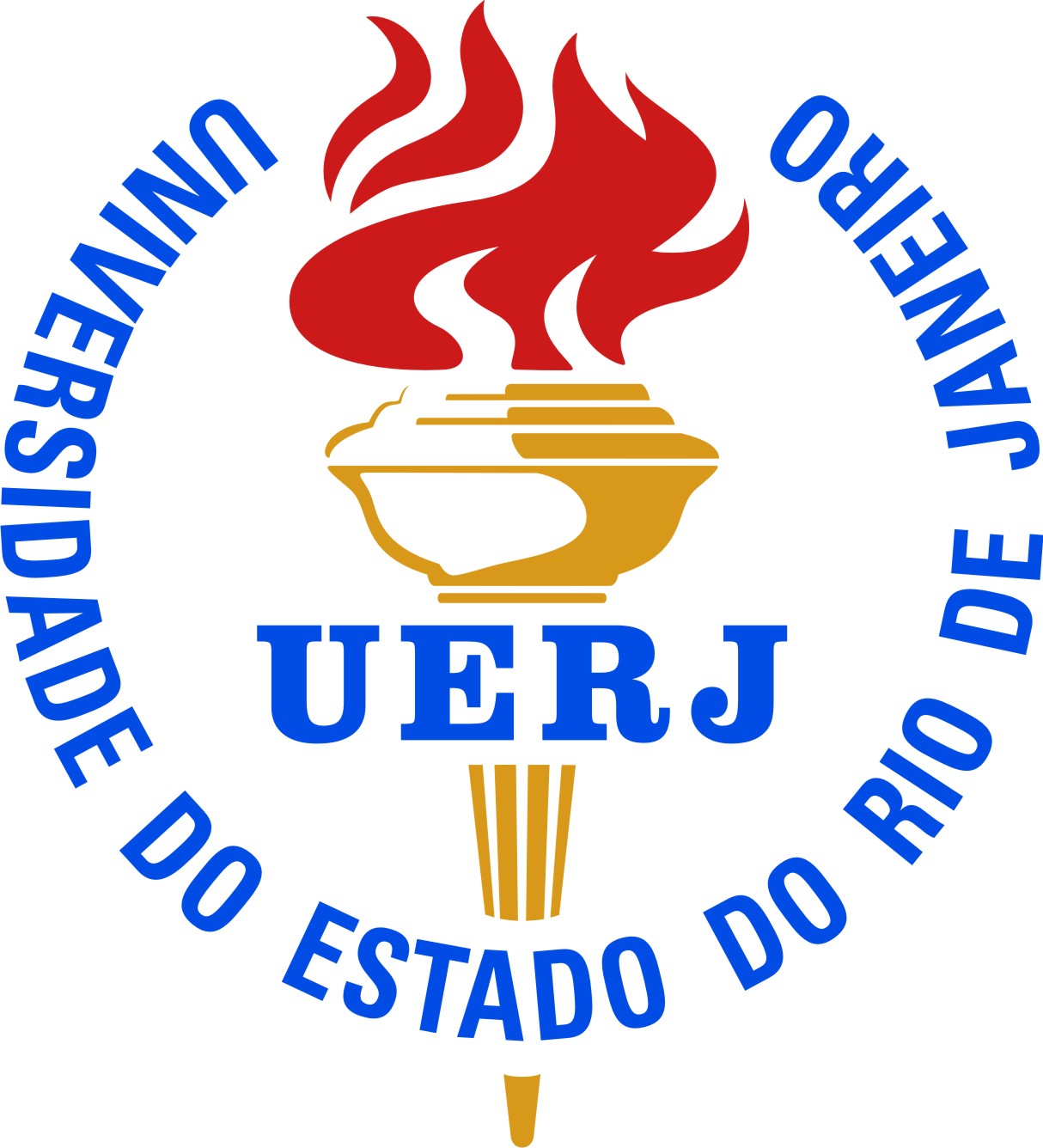}}
\end{figure}

\begin{minipage}[t]{0.45\textwidth}
\begin{center}
Faculty of Sciences \\
\end{center}
\end{minipage}
\begin{minipage}[t]{0.5\textwidth}  
\begin{center}
Science and Technology\\ Center \\
\end{center}
\end{minipage} 

\vspace{0.5cm}

\begin{minipage}[t]{0.45\textwidth}
\begin{center}
Department of Physics \\and Astronomy 
\end{center}
\end{minipage}
\begin{minipage}[t]{0.5\textwidth}  
\begin{center}
Armando Dias Tavares \\Physics Institute
\end{center}
\end{minipage}

\begin{center}

\vspace{1cm}

August, 2016 \\

\vspace*{2cm}

{\Huge\bf Aspects of the Gribov problem in Euclidean Yang-Mills theories} \\

\vspace*{1cm}

{\LARGE Igor F. \textsc{Justo}} 

\vspace*{2cm}

Promotors at Ghent University: Prof.\ Dr.\ David Dudal and Prof.\ Dr.\ Henri Verschelde \\ 
Promotors at Rio de Janeiro State University: Prof.\ Dr.\ Silvio Paolo Sorella and \\ Prof.\ Dr.\ Marcio Andr\'e L\'opes Capri

\vspace*{2cm}

Thesis submitted in fulfillment of the requirements for the degree of  \\ 
\textsc{Doctor in Sciences: Physics} \\ 
at Ghent University, and  \\
\textsc{Doctor in Sciences} \\ 
at Rio de Janeiro State University

\end{center}

\end{titlepage}

%
%
%
%
%
%


\newpage
\thispagestyle{empty}
\mbox{}
\newpage

\thispagestyle{empty}

\thispagestyle{empty}
    \null\vspace{\stretch {1}}
        \begin{flushright}
                To my wife Juliana V. Ford.
        \end{flushright}
\vspace{\stretch{2}}\null

\thispagestyle{empty}

\chapter*{Words of Thanks}
\addcontentsline{toc}{chapter}{Word of thanks}

It is always hard to thanks everyone that might helped me to reach this point of my career or
helped me to finish this thesis. Much probably some important person will be forgotten,
and for those (unfairly) forgotten people I promptly apologize.

At first place, I would like to thank my advisor Silvio~P.~Sorella. Thank you, Silvio. You did
inspire me to work in QFT and to always go as deep as possible in the world of non-perturbative
QCD. Thank you for the very long, fruitful and exciting meetings in your office (if it would
have a title, it would be ``The endless river''), and for the coffees after all. 

Dear Professor Dr. David~Dudal, thank you for everything you have made for me (and my wife) once
we have been in Belgium. You spared no efforts to make us feel at home: that was a marvelous
time, professionally and personally. Actually, it is really tough to separate the advisor
from the friend. I wish all the possible happiness to you and your family.

Thank you Prof. Marcio~Capri for have accepted be my co-advisor, for always be
accessible to discuss with me and for trying to teach me the art of Algebraic
Renormalization (yes, it is not a technique, it is an art) --- sorry, I will never reach your
high level of expertise. Thank you Prof. Marcelo Guimar\~aes, Prof. Bruno Mintz and Prof.
Leticia Palhares, which are not my officially advisors, but (lucky me!) were always there to
advise me and to have enlightening discussions with me. Thank you, each one of you, for making
pleasant our long meetings: not everyone is lucky enough to work with friends, as I was.

And talking about friends, thank you Luiz Gustavo, Gustavo Vicente, Ricardo Rodrigues (``o
Tol'') for turning easier, funnier and enjoyable those four years I did spend at UERJ. There
is no doubt that times of joy are fundamental for performing a good work. Equally, I would like
to thank Thomas Mertens. I know it was not easy for him to bear two Brazilian colleagues at
the same time in his office, at UGhent (it is not a fair thing to do with any Belgian). And
finally, to my great friend Diego R. Granado, thank you so much. We did start it together, we
will finish it together. Thank you.

I would like to thank CAPES for financially supporting me during these four years. It may not
seems to be, but physicists also need to eat.

\`A minha fam\'ilia, minha m\~ae, minha irm\~a e meu irm\~ao, agrade\c{c}o do fundo do meu
cora\c{c}\~ao, por voc\^es estarem sempre ao meu lado me dando suporte e muito carinho. Amo
todos voc\^es.

Por final, agrade\c{c}o \`a minha neguinha, Juliana V. Ford. Sem ela nada disso seria
poss\'ivel, ou mesmo faria sentido. Dedico a voc\^e esta tese.

Greetings,\\
\indent Igor F. Justo

\thispagestyle{empty}

\newpage

\thispagestyle{empty}
\

\pagenumbering{roman} 

\chapter*{English Summary}
\addcontentsline{toc}{chapter}{English Summary}

The content of the present thesis is based on the papers
\cite{Capri:2013oja,Capri:2013gha,Capri:2012ah,Capri:2012cr,Capri:2014jqa,Canfora:2015yia,Capri:2015mna}
and is devoted to the study of aspects of the Gribov problem in Euclidean Yang-Mills
theories coupled to matter fields. Here, we present some, mathematical and physical,
evidences that point to the existence of a possible interplay between the gauge sector
and the matter sector, in regimes of sufficiently low energy (known as the infrared regime). In
other words, we claim that an effect in the vector boson sector of Nature due to strong
interactions (\emph{cf.} the Standard Model), at low energies, may be reflected in
the matter sector, in the same regime. Specifically, we propose that the Gribov
horizon function of the gauge sector may be felt by the matter field, and that it would be
described by an effective non-local mass term attached to the matter field. Such a term seems
to be dynamically generated and accounts for non-perturbative aspects of the matter field.

To achieve our goal this study is divided in three parts: 
on the first part, comprehending chapter \ref{The Yang-Mills $+$ Higgs field
theory}, we present an analytical study of the Yang-Mills theory coupled to the scalar Higgs
field, in the framework of Gribov quantization scheme. In this framework the propagator of the
gauge field may be profoundly modified, depending on the values of the parameters of the
theory, presenting complex conjugate poles. Such scenario prevents us from
attaching any physical particle interpretation to gauge propagator, since it violates the
positivity principle of Osterwalder-Schrader. Our study, thus, concerns the analysis of
the gauge field propagator in the configuration space of parameters of the model. The scalar
field is accounted in its fundamental and adjoint representations, in a 3- and 4-dimensional
Euclidean space-time. At the end we try to make a parallel between our study, of gauge field
confinement, and the work by Fradkin-Shenker \cite{Fradkin:1978dv}, where the Wilson loop is
measured in a (mostly) equivalent scenario.

In the second part of this thesis we propose, and analyze, an effective model for the matter
sector that leads to a soft breaking of the BRST symmetry, by plugging in a non-local term to
this sector, equivalent to the Gribov-Zwanziger horizon of the gauge field.
We will show that such construction may be consistently implemented and that it leads to a
confinement interpretation of the matter field, according to the Gribov's conception of
confinement. By fitting our effective matter propagator, said to be of the Gribov-type, to the
most recent lattice data, we could verify that, indeed, the BRST symmetry is soft broken, by
measuring a local gauge invariant operator that is BRST exact. Furthermore, this new matter
field propagator is found to violate the positivity principle, according to the lattice
fit.  Besides, the UV safety of such effective model is studied: there we prove that such
confinement mechanism, which resembles Gribov's procedure, does not lead to new divergences
other than those from the original (non-effective) theory. This second part is comprehended in
chapters \ref{brstonmatter} and \ref{UVpropsofconfiningprop}; a proof of all order UV stability
is presented in the Appendex \ref{ARscalaraction}, and an example is developed in Appendex
\ref{algrenorm}.

\newpage
The third and last part of this thesis, concerning chapter \ref{Ploop1}, is devoted to the
analysis of the finite temperature theory of static quarks, within the Gribov-Zwanziger
framework. To probe for confinement phases transition in this model the Polyakov loop is
introduced by means of a background field framework, so that the quark phase transition can be
analysed by a single parameter of the theory. This gauge invariant order parameter, related to
the Polyakov loop, is accounted in different circumstances so that the interplay
between this one and the Gribov parameter could be probed. As an interesting outcome we could
see that the behavior of the Gribov parameter is clearly sensible to the quark phase
transition, while an unavoidable region of instability is present. A brief discussion/analysis
is made towards the refined-Gribov-Zwanziger approach to this model.


\chapter*{Resumo}
\addcontentsline{toc}{chapter}{Portuguese Summary}

A presente tese \'e baseada nos trabalhos
\cite{Capri:2013oja,Capri:2013gha,Capri:2012ah,Capri:2012cr,Capri:2014jqa,Canfora:2015yia,Capri:2015mna}
e destina-se ao estudo dos aspectos do problema de Gribov em teorias de Yang-Mills acoplada ao
campo de mat\'eria. Aqui apresentaremos algumas evid\^encias, matem\'aticas e f\'isicas, que
apontam para a exist\^encia de uma poss\'ivel influ\^encia entre os setores de calibre e de
mat\'eria, em regimes de baixa energia, conhecido como o regime infra-vermelho. Em outras
palavras, defenderemos que um certo efeito f\'isico no setor de bosons vetoriais da Natureza,
devido a intera\c{c}\~oes fortes na regi\~ao de baixas energias (cf. o Modelo Padr\~ao), pode
ser refletido no setor de mat\'eria, no mesmo regime de energia. 



Este trabalho est\'a dividido em tr\^es partes: na primeira parte, que compreende o cap\'itulo
\ref{The Yang-Mills $+$ Higgs field theory}, apresentaremos um estudo an\'itico das teorias
de Yang-Mills acopladas ao campo escalar de Higg, de acordo com o esquema de quantiza\c{c}\~ao
de Gribov. Neste esquema, proposto por Gribov para eliminar copias infinitesimais do campo de
calibre, o propagador do campo de calibre sofre profundas modifica\c{c}\~oes, apresentando
polos complexos conjugados, que o levam a violar a condi\c{c}\~ao de positividade, ou
condi\c{c}\~ao de realidade de Osterwalder-Schrader, n\~ao permitindo-nos interpret\'a-lo como
part\'icula f\'isica do espectro da teoria. Guardando as devidas diferen\c{c}as e
peculiaridades de cada modelo, faremos um paralelo entre o nosso modelo e os resultados obtidos
por Fradkin-Shenker no trabalho \cite{Fradkin:1978dv}.

Na segunda parte desta tese proporemos um modelo efetivo ao setor de mat\'eria no qual um termo
n\~ao-local ser\'a introduzido, provocando uma quebra suave da simetria BRST, equivalente \`a
fun\c{c}\~ao horizonte presente no formalismo de Gribov-Zwanziger no setor de calibre.
Mostraremos que a implementa\c{c}\~ao deste termo n\~ao-local, que provoca uma quebra suave da
simetria BRST, pode ser feita de maneira consistente. Como consequ\^encia, observamos que este
modelo se ajusta aos resultados mais recentes obtidos pelo m\'etodo da ``QCD na rede'', de
forma que os propagadores da mat\'eria violam a condi\c{c}\~ao de positividade. Al\'em disso,
mostraremos que este modelo efetivo de confinamento do campo de mat\'eria n\~ao acarreta em
novas diverg\^encias no regime ultravioleta, mas somente gera as diverg\^encias usuais do
modelo ``n\~ao-efetivo''. Esta segunda parte est\'a compreendida nos cap\'itulos
\ref{brstonmatter} e \ref{UVpropsofconfiningprop}. Algumas demonstra\c{c}\~oes est\~ao expostas
no Ap\^endice \ref{ARscalaraction} e \ref{algrenorm}.

A terceira e \'ultima parte desta tese, referente ao cap\'itulo \ref{Ploop1}, destina-se ao
ao estudo \`a temperatura finita do confinamento de \emph{quarks} est\'aticos, na teoria de calibre
n\~ao-Abeliana $SU(2)$, de acordo com o arcabou\c{c}o te\'orico de Gribov-Zwanziger. O estudo
ser\'a feito por meio do \emph{loop de Polyakov}, que ser\'a introduzido utilizando-se o
m\'etodo de campos-de-fundo, de forma que o confinamento dos \emph{quarks} est\'aticos ser\'a
analisado por apenas um \'unico par\^ametro do modelo. Estudaremos os efeitos deste
par\^ametro de ordem, associado ao \emph{loop de Polyakov}, sobre o par\^ametro de Gribov, e
para isso analisaremos o modelo proposto em v\'arias situa\c{c}\~oes distintas. Como
consequ\^encia, observamos que o par\^ametro de Gribov \'e claramente sens\'ivel \`a
transi\c{c}\~ao de fase dos \emph{quarks} est\'aticos, e que h\'a uma regi\~ao de instabilidade
nas visinhan\c{c}as da temperatura c\'itica. 

\chapter*{Nederlandse Samenvatting}
\addcontentsline{toc}{chapter}{Dutch Summary}

De inhoud van deze thesis is gebaseerd op de artikels
\cite{Capri:2013oja,Capri:2013gha,Capri:2012ah,Capri:2012cr,Capri:2014jqa,Canfora:2015yia,Capri:2015mna}
en gaat over de studie van bepaalde aspecten van het Gribov probleem in Euclidische Yang-Mills
theorie\"en qin de aanwezigheid van materievelden. In het bijzonder geven we enkele sterke
aanwijzingen, zowel wiskundig als fysisch, dat verder suggereert dat er een wederzijdse invloed
op de ijkveldynamica is ten gevolge van de materievelden, en omgekeerd, in het bijzonder in het
lage energie gebied (het zogenaamde infrarood regime). Meerbepaald claimen we dat er een effect
van de vectorbosonsector in de Natuur (cf. het Standaardmodel)  van de sterke/zwakke interactie
is op de materiesector, gekoppeld aan de sterke/zwakke interactie.

Deze thesis bevat 3 delen, volgend op 2 inleidende hoofdstukken. In een eerste deel, zijnde
Hoofdstuk 3, vindt de lezer een analytische studie van een Yang-Mills ijktheorie gekoppeld aan
een scalair Higgsveld, waarbij we rekening houden met het bestaan van Gribov ijkkopie\"en in de
kwantisatieprocedure. Dit laatste kwantisatiemechanisme, opgesteld door Gribov met als doel
ijkequivalente infinitesimale ijkkopie\"en te verwijderen, laat ons toe om een interessante
alternatieve visie op confinement te ontwikkelen, met name dat de deeltjespropagatoren complex
toegevoegde polen vertoont. Dit leidt tot wat men een schending van positiviteit noemt, hetgeen
betekent dat de deeltjes niet re\"eel waarneembaar kunnen zijn (de \"Osterwalder-Schrader
criteria zijn niet vervuld).
Rekening houden met de verschillen tussen en bijzonderheden van elk model, proberen we ook
telkens een parallel te trekken tussen onze analyse en de befaamde Fradkin-Shenker resultaten
over Yang-Mills-Higgs ijktheorie\"en.

In het tweede deel van de thesis stellen we een effectief model voor een zachte breking van de
BRST symmetrie voor, dit door een niet-lokale term in de materievelden aan de actie toe te
voegen. Deze speelt een rol gelijkaardig aan de Gribov-Zwanziger term die Gribov's
kwantisatieprocedure tot op alle ordes in de expansieparameter implementeert. We tonen o.a. aan
dat een consistente beschrijving van confined materie kan ge\"interpreteerd worden in termen van
een systematische zachte BRST breking. Dientengevolge is ons model in acceptabele overeenkomst
met recent rooster QCD data, met eveneens een materieveldpropagator die een schending van
positiviteit vertoont. Daarnaast is ons model ook veilig in het UV gebied, vermits we bewijzen
dat dergelijk confinement mechanisme ---analoog aan Gribov's originele analyse--- geen nieuwe
UV divergente termen kan genereren dan degene reeds aanwezig in de (niet-effectieve) theory.
Voor details verwijzen we hier naar de Hoofdstukken 4 en 5.

Het derde en tevens laatste deel van de thesis (Hoofdstuk 6) behandelt eindige
temperatuuraspecten van statische quarks, dit opnieuw gebruik maken van het Gribov-Zwanziger
kwantisatieformalisme. Met als bedoeling inzicht te verkrijgen in de confinement-deconfinement
fasetransitie, voeren we de Polyakovlus in via het achtergrondveldformalisme, uitgebreid met
Gribov-Zwanziger weliswaar. De statische quark fasetransitie kan zo onderzocht worden vermits
de Polyakovlus een ijkinvariante ordeparameter is. We bekijken de vacu\"umverwachtingswaarde in
verschillende omstandigheden, in het bijzonder om na te gaan hoe deze de dynamische Gribov
massaparameter be\"invloedt. We vinden het mooie resultaat dat deze massaschaal duidelijk
gevoelig is aan de fasetransitie, te zien aan het sterk veranderende gedrag rond de kritische
temperatuur. We vinden echter ook een gebied bij lage temperatuur dat thermodynamisch gezien
instabiel is. We eindigen met een korte discussie annex analyse wat een meer verfijnde
(Refined) Gribov-Zwanziger analyse hieraan zou kunnen verhelpen.


\tableofcontents


\newpage
\pagenumbering{arabic}

\chapter{Introduction}

Up to the present day confinement still is one of the most intriguing features of strong
interactions: why do quarks and gluons, being the fundamental excitations of fields of the
theoretical model that describes the strong interactions, widely known as QCD (Quantum
Chromodynamics), not appear in the spectrum of asymptotic physical particles? Or else, the
mechanism by which confinement happens in nature is another not answered question concerning strong coupling effects.

The strong interaction is one of the four fundamental interactions of Nature, next to the
gravitational, electromagnetic and weak interaction. Together, these four interactions reside
at the heart of the Standard Model (SM), which ought to theoretically describe all known
elementary particle physics processes, but neglecting quantized gravitational effects. For a
recent pedagogical review about the Standard Model (also referred to as a `Standard
Theory') take a look at \cite{'tHooft:2007zza}.

Despite the existence of some open questions surrounding the Standard Model (SM), such as too
many free parameters to fix, the not yet explained dark matter and
the quantization of gravity, its success in describing and foreseeing innumerable physical
process and particles, in scales of 10$^{-15}$cm and smaller than that, makes the SM the most
accepted theory to describe the physics of the fundamental particles, up to the present day \cite{Agashe:2014kda}. 

In particular, the present thesis is devoted to the study of effects of strong interactions,
more precisely those related to the interaction of fundamental colored particles (or color
sources), by making use of the QCD framework. Quantum chromodynamics (QCD) is a theoretical
model based on the theory of quantized relativistic fields, more widely known as Quantum Field
Theory, where particles are described by scalar fields, such as the Higgs one, and by
fermionic fields, accounting for the quarks, while the interaction between those particles
is mediated by gauge particles, which are described by vector fields
belonging to the adjoint representation of non-Abelian gauge groups such as the $SU(3)$.

Physically we may cite two typical characteristic features of strong interactions: confinement
and chiral symmetry breaking. Let us give a close look into these features.

\section{The confinement problem}

The modern understanding of confinement, in its physical sense, developed historically from a
more strict view of \emph{quark confinement} to a more general sense of \emph{color
confinement}. By color we mean a charge carried by particles described by QCD, which has
nothing to do with visual color, due to global gauge symmetry. The point is that quarks cannot
be found as free particles, but only \emph{confined} in hadrons (= composite state of quarks).
At the same time, gluons, or gauge particles, which are responsible for the mediation of
strong interactions, can also not be found as free particles in the physical spectrum of the
theory but, instead of that, only trapped inside \emph{glueballs} (= composite states of
gluons)\footnote{Up to the day of closing this thesis, there is no a definite particle to be
called {\it glueball}. However, the authors of \cite{Brunner:2015yha} claim that the resonance
``$f0(1710)$'' is the prefered candidate for a glueball. Further experimental results are
still expected to confirm (or not) the ``$f0(1710)$'' as the glueball.}. A hybrid composition of quarks and gluons is also possible, leading to the state of
quark-gluon plasma. Then, the modern concept of \emph{confinement} arises as only particles, or
composition of particles, carrying neutral color charge can be seen as asymptotic free
particles. Examples of hadrons are easy to find out, such as protons and neutrons that belong
to the set of composite states made up of three quarks called baryons.

Describing confinement is one of the tricky points of QCD. The fundamental particles of nature
are described in this framework as small excitations of quantum fields around the vacuum, with
these fields obeying specific rules of (local) gauge transformation of non-Abelian groups, such
as the $SU(3)$. Such gauge transformation leaves invariant the QCD action, so that the theory
is said to be gauge invariant.

Theoretically, it is widely believed that the phase transition ``confinement/deconfinement'' is
intimately related to the spontaneous breaking of a symmetry, or else, due to the existence (or
not) of a remnant symmetry. A {\it spontaneously broken symmetry} is meant to be a symmetry of
the system that does not leave invariant a physical state of the theory, such as the vacuum
state. In other words, suppose that $s$ stands for the (infinitesimal) transformation of a
certain symmetry of the theory ({\it i.e.} $sS=0$, where $S$ is the classical action of the
theory, but also $s\Gamma[\varphi]=0$, where $\Gamma[\varphi]$ is the quantum action of the
theory, as a function of the fields); and also suppose that $\varphi_{0}$ stands for the vacuum
configuration of that system ({\it i.e.} $\Gamma[\varphi_{0}]$ assumes its minimal value).
Therefore, the $s$ symmetry is said to be {\it spontaneously broken} if $s\varphi_{0} \neq
\varphi_{0}$. That is, in this case $\Gamma[\varphi_{0}] \neq \Gamma[s\varphi_{0}]$: the vacuum
is said to be degenerated in such cases, since there are two distinct vacuum configurations,
$\varphi_{0}$ and $s\varphi_{0}$ \cite{Weinberg:1996kr,Peskin:1995ev,Ryder:1985wq}.

A famous and simple example of phase transition due to spontaneous symmetry breaking can be
found in the Linear Sigma model, where a continuous symmetry group $O(N)$ is broken down to
$O(N-1)$; a more complex example is the Yang-Mills theory coupled to the Higgs field, known as
the Higgs mechanism, where the framework of spontaneous symmetry breaking is applied to the
theory of gauge fields. Times before the proposal of the Higgs mechanism, a similar procedure
had been applied by Ginzburg and Landau to the study of superconductors, although being a
classic (or statistical) model, where they plugged an external magnetic field into the model so
that the electromagnetic field could penetrate into the material only down to $m_{A}^{-1}$
depth; $m_{A}$ is the acquired mass by the electromagnetic field due to the spontaneous
symmetry breaking \cite{Weinberg:1996kr,Peskin:1995ev,Ryder:1985wq,Greensite:2011zz}. 

Two important theorems concerning symmetry breaking should be discussed in order to
better understand the link between the phase transition ``confinement/deconfinement'' and the
spontaneous symmetry breaking: one of them is the Goldstone theorem, while the other one is
the Elitzur's theorem.

\begin{itemize}
\item The Goldstone theorem:

\emph{There exists a spinless massless particle for every spontaneously broken continuous
symmetry,} \cite{Weinberg:1996kr,Peskin:1995ev,Ryder:1985wq,Greensite:2011zz}.

That is, in the case of the $SO(N)$ global transformations, for instance, there are $N(N-1)/2$
independent symmetries. It means that, in a theory with $\varphi^{i}$ real scalar fields, with
$i=1,2,\cdots,N$ and obeying an specific rule of transformation of the $SO(N)$ group, leaving invariant the theory, there are $N(N-1)/2$ independent transformations. The number of massless particles is,
\begin{eqnarray}
\frac{N(N-1)}{2} - \frac{(N-1)(N-2)}{2} ~=~ \frac{2N -2}{2} ~=~ N-1\,,
\end{eqnarray}
which is exactly the number of broken symmetries; $N(N - 1)/2$ comes form the original $SO(N)$
group symmetry, whilst $(N -1)(N-2)/2$ comes form the remaining $SO(N-1)$ group of symmetry
after the breaking.

Still in the instance of real scalar fields, let us consider a particular configuration of its
potential energy, which is given by
\begin{eqnarray}
V[\varphi^{i}] ~=~ - \frac12 \mu^{2} \pmb{\varphi}\cdot\pmb{\varphi} +
\frac{\lambda}{4}\left[ \pmb{\varphi}\cdot\pmb{\varphi} \right]^{2}\,,
\label{mexianhat}
\end{eqnarray}
called the \emph{Mexican hat} potential. In this case the vacuum expectation value (\emph{vev})
of the scalar field is not zero, but rather is degenerated $\langle  \varphi^{i} \rangle
=\nu\delta^{i0}$ around one chosen direction.
Supposing that this theoretical model has a \emph{global} $SO(N)$ symmetry, one can easily see
that, by performing a perturbation of the scalar field around its vacuum configuration, in the
chosen direction $\delta^{i0}$
\cite{Weinberg:1996kr,Peskin:1995ev,Ryder:1985wq,Greensite:2011zz},
\begin{eqnarray}
\pmb{\varphi}(x) ~=~ (\pi^{k}(x),\, \nu + \sigma(x))\,,
\end{eqnarray}
the potential becomes,
\begin{eqnarray}
V[\varphi^{i}] ~=~ - \frac12 (2\mu^{2}) {\sigma}^{2} - \frac{\lambda}{2}\pi^{2}\sigma^{2} -
\mu\sqrt{\lambda}\pi^{2}\sigma - \sqrt{\lambda}\mu \sigma^{3} - \frac{\lambda}{4}(\sigma^{2})^{2}
- \frac{\lambda}{4}(\pi^{2})^{2}
\,,
\end{eqnarray}
regarding that $\nu^{2} = \mu^2/\lambda$. Gathering the kinetic therm, the Lagrangian reads,
\begin{eqnarray}
{\cal L} ~=~  (\p_{\mu}\pi)^{2} + (\p_{\mu}\sigma)^{2} - V[\pi^{k},\,\sigma]\,.
\end{eqnarray}
So, it is not difficult to see that, at the end one gets one massive mode, the $\sigma$ one,
and $N-1$ massless and spinless modes, the $\pi^{k}(x)$ fields, corresponding to the foreseen
$N-1$ Goldstone bosons.

\item Elitizur's theorem:

\emph{It is not possible to spontaneously break a local (gauge) symmetry},
\cite{Elitzur:1975im}.

In other words, the vacuum expectation value (\emph{vev}) of a gauge non-invariant local
operator must vanish, and one should be careful when dealing with gauge theories coupled to
Higgs fields.

Let us provide again, but in other words, the concept of spontaneous symmetry breaking. A
symmetry is said to be \emph{spontaneously} broken in the sense that the vacuum configuration
is not symmetric under such (global) transformation. That is, if $\varphi_{0}$ is the vacuum
configuration of the scalar field and $\delta_{gl}\varphi_{0} \neq \varphi_{0}$, with
$\delta_{gl}$ standing for the variation of a global transformation, then we say that the global symmetry
$\delta_{gl} \Gamma[\varphi]=0$ is spontaneously broken, since $\delta_{gl}\Gamma[\varphi_{0}]
\neq 0$ (again $\Gamma[\varphi]$ stands for the quantum action of the theory)
\cite{Weinberg:1996kr,Peskin:1995ev,Ryder:1985wq}.

In the above example of the scalar field, where the theory is symmetric under global $SO(N)$
group transformations, the vacuum configuration before breaking the symmetry was the trivial
vacuum, $\varphi_{0} =0$. But after, when the potential acquires the \emph{Mexican hat}
potential form, the vacuum configuration is not symmetric anymore, but rather it is degenerated
out of $\varphi_{0} =0$.

Even though Elitzur's theorem forbids spontaneous symmetry breaking of local symmetries, it
seems that there exist in Nature exceptions to this rule: \emph{e.g.} the mechanism of mass
generation of the gauge field, such as in the Electroweak model, called {\it Higgs mechanism}.
Frequently we use to say that the \emph{gauge symmetry is spontaneously broken}, but what
exactly we mean by this expression? In order to answer, let us analyze one easy example, known
as the Georgi-Glashow model.

In the Georgi-Glashow model the Lagrangian reads,
\begin{eqnarray}
{\cal L} ~=~ -\frac{1}{4}F^{a}_{\mu\nu}F^{a}_{\mu\nu} + D_{\mu}\pmb{\varphi} D_{\mu}\pmb{\varphi} +
\frac{\lambda}{4}(\pmb{\varphi}\cdot\pmb{\varphi} - \nu^{2})^{2} \,,
\label{GGmodel}
\end{eqnarray}
where $D_{\mu} = \p_{\mu} -igA^{a}_{\mu}t^{a}$ stands for the covariant derivative, and
$F^{a}_{\mu\nu}$ stands for the field strength tensor:
\begin{eqnarray}
F^{a}_{\mu\nu} ~=~ \p_{\mu}A^{a}_{\nu} - \p_{\nu}A^{a}_{\mu} +
g^{2}f^{abc}A^{b}_{\mu}A^{c}_{\nu}\,.
\end{eqnarray}
Regarding
the fact that the Lagrangian \eqref{GGmodel} is invariant under the gauge transformation
$SU(N)$, let us {\it choose the direction of the broken symmetry} as being
\begin{eqnarray}
\langle \varphi^{i} \rangle = \nu\delta^{iN}\,,
\end{eqnarray}
with the scalar field in the fundamental representation. Expanding once again around the vacuum
configuration
\begin{eqnarray}
\pmb{\varphi} ~=~ (\tilde{\varphi}^{k},\,\nu + \sigma)
\label{unitgauge}
\end{eqnarray}
 and defining $\tilde{\varphi}$ rotated, so that it is perpendicular to $t^{a}\langle \varphi^{i} \rangle$,
\begin{eqnarray}
\tilde{\varphi}^{k}(t^{a})_{ki}\langle \varphi^{i}\rangle~=~ 0\,,
\label{unitycondit}
\end{eqnarray}
one ends up with a massive term for the gauge field of the form
$\frac{1}{2}A^{a}_{\mu}A^{b}_{\mu}\mu^{2}_{ab}$,
with \cite{Weinberg:1996kr,Peskin:1995ev,Ryder:1985wq}
\begin{eqnarray}
\mu^{2}_{ab} ~=~ (t^{a})_{iN}(t^{b})_{iN}\nu^{2}\,.
\label{gaugemass}
\end{eqnarray}

Let us now make some important remarks concerning this massive term. The first remark concerns
the importance of the condition \eqref{unitycondit}: this is the called \emph{unitary gauge
condition} and is nothing more than a rotation of the scalar field so as to end up with
$\tilde{\varphi}^{k}$ perpendicular to the broken directions $t^{a}\langle \varphi^{i}
\rangle$. The second remark concerns the \emph{spontaneous gauge symmetry breaking}: the
realization of the spontaneous symmetry breaking, in the sense that $\langle \varphi^{i}
\rangle \neq 0$, leads to the breaking of the (local) gauge symmetry, through the appearance of
the quadratic term $\frac{1}{2}A^{a}_{\mu}A^{b}_{\mu}\mu^{2}_{ab}$. However, it is clear that
this induced breaking in the gauge sector does not lead to the appearance of a vector Goldstone
boson. Instead of that, there exist $2N-1$ \emph{massive} vector bosons. Note that the massive
term \eqref{gaugemass} depends on the modes of the scalar fields associated to the broken
symmetries, $t^{a}\langle \varphi^{i} \rangle$, and since there exist  
\begin{eqnarray}
(N^{2}-1) - [(N-1)^{2} -1] ~=~ 2N -1 
\end{eqnarray} 
broken symmetries, then that is the number of massive vector bosons.

The Georgi-Glashow model is recovered for $N=2$ and, as mentioned before, with the Higgs field
in the fundamental representation the gauge group is said to be completely broken, yielding to
$3$ massive gauge bosons; another example is the electroweak gauge theory, where the
$SU(2)\times U(1)$ gauge symmetry is broken down to $U(1)$, providing mass to the $W^{\pm}$ and
$Z^{0}$ gauge bosons and to the matter field, leaving massless the photon and an (approximate)
massless pion \cite{Weinberg:1996kr,Peskin:1995ev,Ryder:1985wq,Greensite:2011zz}.

The point here is that, the breaking of a local gauge symmetry happens when the unitary gauge
is applied, \eqref{unitgauge} and \eqref{unitycondit}, in the \emph{Mexican hat} potential
configuration of the scalar field, which induces the breaking. However, in this local symmetry
breaking, there is no Goldstone boson associated. Otherwise, massive excitations of the vector
boson field appear, accounting for the ``missing'' Goldstone bosons degrees of freedom.
\end{itemize}

In order to probe for the existence of such global symmetry one should make use of gauge
invariant local operators that shall be sensible to the realization (or not) of that
symmetry. We call this operator an order parameter. Two well known order parameters are the
Wilson and the Polyakov loops. Both of them are related to the potential energy between two
color static sources. More precisely, the Wilson loop can be seen as a measure of the 
process of creation-interaction-annihilation of a pair of \emph{static} quark-antiquarks,
\cite{Wilson:1974sk}. Namely, its continuum expression is given by \cite{Greensite:2011zz}
\begin{eqnarray}
W ~=~ \left\langle P\,\exp\left[ig \oint_{C}\;dx^{\mu}A_{\mu}(x)\right]  \right\rangle ~\sim~
\e^{-V(R)T}\,,
\label{wilsonloop}
\end{eqnarray}
whence $T$ stands for the length in the time direction, and $V(R)$ is the quark-antiquark
potential, depending on their spacial distance $R$. In its discrete space-time version,
\emph{i.e.} on the lattice, the existence of such quark-antiquark creation (annihilation)
operator is evident, together with the creation operator of a gauge flux tube, mediating the
interaction between the quark-antiquark pair. We should emphasise the fact that the
\emph{vev} \eqref{wilsonloop} may be computed in a pure Yang-Mills theory, that is, in the
absence of any matter field, at all. Physically, this is a gauge theory in the presence of
heavy quark matter, which is theoretically achieved in the limit of infinite quark mass.

The Wilson loop is sensible to the existence of three possible phases, concerning the potential
between the static quarks \cite{Greensite:2011zz}:
\begin{itemize}
\item ``{\bf Yukawa, or massive} phase'', where the potential is given by
\begin{eqnarray}
V(R) ~=~ -g^{2}\frac{\e^{-mR}}{R} + 2V_{0}\,,
\label{perimeterlaw}
\end{eqnarray}
where $V_{0}$ stands for the self-energy of the system. The Wilson loop exhibit a
\emph{perimeter law} falloff, for a sequence of non self-intersecting loops,
\begin{eqnarray}
W  ~\sim~ \exp [-V_{0}P(\Gamma)]\,,
\end{eqnarray}
for situations where $R$ is large enough, in front of $1/m$; $P(C)$ is the perimeter of the
loop $\Gamma$ \footnote{At this point do not mistake $\Gamma$ for the quantum action.}.

\item ``{\bf Coulomb, or massless} phase'' phase, where the potential is proportional to the Coulomb
potential,
\begin{eqnarray}
V(R) ~=~ -\frac{g^{2}}{R} + 2V_{0}\,.
\end{eqnarray}
Also in this phase, if the self-energy contribution is considerably greater then the $1/R$ rule
potential, \emph{i.e.} $R\gg V^{-1}_{0}$, then the Wilson loop will also fall-off as 
\begin{eqnarray}
W  ~\sim~ \exp [-V_{0}P(\Gamma)]\,,
\end{eqnarray}
also when the loops do not intersect with themselves.

\item ``{\bf Disordered, or Magnetic disordered}'' phase, where the potential goes as
\begin{eqnarray}
V(R) ~=~ \sigma R + 2V_{0}\,.
\end{eqnarray}
In this case, the Wilson loop exhibit an \emph{area law} fall-off, for non self-intersecting
loops,
\begin{eqnarray}
W ~=~ \exp[ -\sigma RT - 2V_{0}T]\,.
\end{eqnarray}
Note that the potential energy grows as $\sim R$, that is, the bigger the distance between the
static quarks, the greater is the self-energy of the system. We use to attach a
\emph{confinement-like} interpretation to this scenario. On the other hand, in the first two
cases the potential behaves as $\sim V_{0}$, at the best, configuring a short-distance
interaction behaviour, and charged free particles can be found, such as the $W^{\pm}$ vector
bosons in the gauge + Higgs theory.

The expression ``magnetic disordered'', of the third regime of the Wilson loop, stems from the
fact that we are considering large enough loops $\Gamma$, so that its \emph{vev} equals the
product of vacuum expectation value of $n$ smaller loops ($\Gamma_{1}$, $\Gamma_{2}$, $\cdots$,
$\Gamma_{n}$) that lies inside the biggest one $\Gamma$ (\emph{cf.} \cite{Greensite:2011zz}).
Then we say that these $n$ loops are \emph{uncorrelated}, as much the same as the disordered
phase of the Ising model, whence the term was inspired.
\end{itemize}

The Polyakov loop \cite{Polyakov:1978vu} is, in its turn, sensible to a very important global
symmetry, intrinsic to gauge theories, called the \emph{center symmetry}. This symmetry group
is an intrinsic subgroup of a gauge group, let us say $G$, and is a set of elements that
commute with every elements of $G$. Said in other way, if $z_{n}$ belongs to the center
symmetry of a gauge group $G$, then $z_{n}$ commutes with every single element of $G$ --- so,
it is an element of the center of the group $G$. Therefore, considering the $SU(N)$ gauge
group, its associated center symmetry $Z_{N}$ is composed by elements of the following type,
\cite{Greensite:2011zz,Fukushima:2003fw,Fukushima:thesis}
\begin{eqnarray}
z_{n} ~=~ \exp\biggl( \frac{2i n\pi}{N} \biggr)\mathbb{1}\,,
\end{eqnarray}
whence $\mathbb{1}$ stands for the unity $N\times N$ matrix; $n = 0,\,1,\,2,\,\cdots,\,N-1$.
Since $Z_{N}$ is a subgroup of $SU(N)$, then it is straightforward to see that a pure gauge
theory is invariant under $Z_{N}$ transformations. However, that is not true in the Yang-Mills
+ matter field theory, with the matter field in any of its non-trivial group representation,
such as the fundamental one: the center symmetry is explicitly broken in this case. For the
matter field in the adjoint representation, which is an example of a trivial representation of
the gauge group, this center symmetry is still preserved, so that it may be spontaneously
broken afterwards (\emph{cf.} \cite{Greensite:2011zz} for further details) \footnote{Take
a look at the Chapter \ref{The Yang-Mills $+$ Higgs field theory} to see that in the theory of
Yang-Mills + Higgs field in its fundamental representation two distinct regimes, namely
the \emph{confinement-like} and the \emph{Higg-like}, coexiste in the same phase of the
theory, corresponding to the explicitly broken center symmetry (\emph{cf.} Fradkin \& Shenker
\cite{Fradkin:1978dv}).}.

The Polyakov loop may be defined in the continuum space-time as
\begin{eqnarray}
{\cal P} ~=~ \left\langle  P\,\exp\left[ ig\oint_{C}\, A_{0}dx^{0}  \right] \right\rangle ~\sim~
\e^{-FT}\,,
\label{polyakovloop}
\end{eqnarray}
and can be interpreted as an Euclidean space-time finite temperature torus circling around, and
accounts for the temporal component of the Wilson loop,
\cite{Fukushima:2003fw,Fukushima:thesis}. In equation \eqref{polyakovloop}, $F$ is the free
energy between the static quarks, and $T$ is the temperature.  As can be seen from
\eqref{polyakovloop}, the Polyakov loop, akin to the Wilson loop, is a measure of the
self-energy between static quarks. It is straightforward to see that if ${\cal P}\neq0$, then
the free energy between static quarks is finite, while that when ${\cal P}=0$, there is an
infinite free energy between them. So, one may classify as \emph{confined} and
\emph{deconfined} regimes situations of infinite binding energy between static quarks (${\cal
P}=0$) and situation of finite binding energy between static quarks (${\cal P}\neq0$),
respectively.

The sensibility of the Polyakov loop to the realization of the center symmetry can be seen from
its transformation under an element $z_{n}$ of $Z_{N}$, intrinsic to the gauge group $SU(N)$.
The {\it vev} of the Polyakov loop is understood to be computed in a pure Yang-Mills theory, or
coupled to an adjoint matter field. Namely, one has \cite{Fukushima:2003fw,Fukushima:thesis}
\begin{eqnarray}
{\cal P} ~\to~ z_{n} \left\langle  P\,\exp\left[ ig\oint_{C}\, A_{0}dx^{0}  \right]
\right\rangle ~=~ z_{n}{\cal P} \,.
\end{eqnarray}
Thus, the Polyakov loop is clearly not invariant under the center symmetry transformation, in
the specified situation. The Polyakov loop may, then, be seen as a suitable order parameter for
the center symmetry breaking.
Then, one has:
\begin{itemize}
\item[i.] The symmetric phase, where
\begin{eqnarray}
{\cal P} ~=~ 0\,,
\end{eqnarray}
which corresponds, as mentioned above, to the confined phase, of infinite energy between static
quarks;

\item[ii.] The broken phase, where
\begin{eqnarray}
{\cal P} ~\neq~ 0\,,
\end{eqnarray}
which corresponds to the deconfined phase, with finite energy between the static quarks
\footnote{The reader is pointed to
\cite{Fukushima:2002bk,Fukushima:thesis,Fukushima:2010bq,Bazavov:2009zn,Fukushima:2003fw,Greensite:2011zz}
for a detailed study on the Polyakov loop and the center symmetry}.
\end{itemize}

In both expressions \eqref{wilsonloop} and \eqref{polyakovloop} $P$ accounts for the \emph{path
ordering} of the (gauge field) operator $A_{\mu}$ as it appears in the closed path. At first
order in perturbation theory this path ordering operator is meaningless.

What happens if one has a gauge field theory coupled to a matter field in the fundamental
representation? In this case, the center symmetry is explicitly broken and no phase transition
occurs, \cite{Fradkin:1978dv,Greensite:2011zz}.

In the specific case of Yang-Mills theories coupled to scalar fields, in the adjoint
representation, whose potential energy is of the \emph{Mexican hat} type, the gauge symmetry is
said to be spontaneously broken after fixing the unitary gauge. However, it is not fully
broken, by leaving intact the global center symmetry (which does not happen in the fundamental
representation of the scalar field). Thus, phase transition may still be probed by means of the
Polyakov loop.

Another point of great interest is the order of the phase transition. In general, it can be
probed by measuring derivatives of the free energy with respect to thermodynamic order
parameters: divergences on the first derivative
would correspond to first order phase transitions; divergences on the second derivative
corresponds to a second order phase transition (we refer to \cite{LeBellac:1991cq} for a
pedagogical approach to this matter). Precisely, for pure gauge field theory it has been found
that for the $SU(2)$ gauge theory a second order phase transition takes place at a temperature
of $T_{c} = \unit{295}{\mega\electronvolt}$; and for $SU(N)$ gauge theories, with $N \ge 3$, a
first order phase transition is found to occur at $T_{c} = \unit{270}{\mega\electronvolt}$,
\cite{Fukushima:2002bk,Fukushima:thesis,Fukushima:2010bq,Bazavov:2009zn,Fukushima:2003fw,Banks:1979yr}.


But, what about the phase transition in the presence of dynamical quarks? In these cases things
get overcomplicated since the usual order parameters, \eqref{wilsonloop} and
\eqref{polyakovloop}, cannot be used anymore. Physically the scenario is that a threshold value
for the dynamical quark separation is reached, so that beyond this value the potential between
them goes flat, instead of growing linear with the separation length $R$, as happened in the
static quark scenario. It indicates a dynamical screening mechanism for the gauge field
known as the \emph{string breaking} effect. A possible interpretation is that the potential
energy between the quarks grows (linearly) up to a level high enough to create a pair of
quark-antiquarks. Theoretically the following happens: the traditional order parameter
\eqref{polyakovloop} work by measuring the existence of the center
symmetry, $Z(N)$, which is associated to the gauge symmetry. However, the presence of
dynamical quarks explicitly breaks the center symmetry, preventing the Polyakov
loop from measuring any phase transition (this is similar to what happens in the
Georgi-Glashow model with the Higgs field in the fundamental representation). Furthermore,
with dynamical quarks the Wilson loop is not sensible to the disordered phase anymore, since
due to the string breaking dynamical effect, at some point ({\it i.e.} at some distance $R$ from
the quarks) the potential energy between the quarks will not grows linear with $R$ but rather
becomes flat.

Something similar happens in the Yang-Mills + Higgs theory: for the scalar field in its
fundamental representation the global (center) symmetry is explicitly broken, so that no
Goldstone boson is present and no phase transition takes place; for the scalar field
in the adjoint representation the global (center) symmetry is not explicitly broken, and then a
phase transition is allowed to occur and to be measured (further details on subsection
\ref{FSresults} and in \cite{Greensite:2011zz}).

The present thesis is mainly devoted to add a small piece to this big puzzle called confinement
by attacking the problem with an alternative approach. Instead of searching for an order
parameter and analytically probing it --- which is far from being an easy task
--- we do apply the framework of Gribov (and Gribov-Zwanziger) to investigate the existence of
gauge field confinement. As it is introduced on chapter \ref{usefulbkground},
confinement in Gribov's framework is not (clearly) related to the breaking of the global
symmetry, neither with the potential energy between quarks. The way to probe for confinement,
however, relies on the alternative criterion firstly proposed by Gribov, where the existence of
complex conjugate poles in the gauge field propagator must indicate the gauge confinement. The
point is that in such cases the gauge field is deprived of any physical particle
interpretation. Inspired by what happens in the gauge sector and expecting a confinement
behaviour also for the quark sector, we will propose in this thesis an effective action for the
matter field where the matter propagator exhibit a similar non-physical interpretation. That is
not the first time that (gauge) confinement and the Gribov issue are linked to each other.
However, that {\it is} the first time that the Gribov issue is investigated in the presence of
matter fields, or else, that the quark sector presents a Gribov-type structure.

\section{Chiral symmetry breaking}

As mentioned before, the modern concept of fundamental physics is based on the
existence/break\-ing of symmetries. We have been discussing that confinement can be understood
as the symmetric, or magnetic disordered, phase of a gauge theory. However, nothing have been
said about the existence of \emph{approximate symmetries} in nature. One famous approximate
symmetry is the $SU(2)$ isotopic symmetry, related to the mass of quarks $u$ and $d$: from the
most recent Particle Data Group's (PDG) data \cite{Agashe:2014kda}, at the mass scale of
$\mu \approx \unit{2}{\giga\electronvolt}$ and in a
mass-independent subtraction scheme called $\msbar$, quark-$u$'s mass is about $1.8$ --
$\unit{3.0}{\mega\electronvolt}$, while quark-$d$'s mass is $4.5$ --
$\unit{5.3}{\mega\electronvolt}$, so the rate between their mass is $m_{u}/m_{d} = 0.35$ --
$0.58$. 

Therefore, it is clear that both of the quarks have masses of the same order of magnitude, 
allowing us to formulate a(n) (approximate) symmetric theory, where quarks $u$ and
$d$ belong to the same doublet,
\begin{eqnarray}
\psi ~=~ \left(
\begin{array}{ll}
u \\
d
\end{array}
\right)\,,
\end{eqnarray}
and with the corresponding action symmetric under $SU(2)$ gauge transformations,
\begin{eqnarray}
\left(
\begin{array}{ll}
u \\
d
\end{array}
\right)
~\to~ 
\exp\left\{ i\vartheta^{a} t^{a} \right\} 
\left(
\begin{array}{ll}
u \\
d
\end{array}
\right)\,.
\label{chiraltransf}
\end{eqnarray}
In the above equation $\vartheta$ is a real parameter; $t^{a}$ accounts for the three $SU(2)$
generators, \emph{i.e.} the Pauli matrices. We use to say that quarks $u$ and $d$
(approximately) belong to the same \emph{isospin}. The same approximation is not so good
regarding the quark $s$; his mass is about $90$ -- $\unit{100}{\mega\electronvolt}$, so it is
an order of magnitude grater than quarks $u$ and $d$ masses and such an approximation leads to
inaccurate results. For a realistic model, where the mass of quarks $u$, $d$ and $s$ are
considered as being different from each other, we say that the isotopic symmetry is explicitly
broken \cite{Weinberg:1996kr,Peskin:1995ev,Ryder:1985wq}. 

Besides the isotopic symmetry there may also be another approximate symmetry, if one notice that
the quarks $u$ and $d$ masses are relatively small enough to put it to zero, regarding that the
energy scale is $\mu \approx \unit{2}{\giga\electronvolt}$. One should also recall that the mass
of protons (composed of three quarks), $\sim \unit{938}{\mega\electronvolt}$, is much higher
than the mass of its constituents, as mentioned before (check for the most recent particle data
at \cite{Agashe:2014kda}). So, it is reasonable to make the quark massless approximation, and
one ends up with an $SU(2)\times SU(2)$ symmetric theory, known as the \emph{chiral symmetry},
$\chi$S. At this point, of massless quarks $u$ and $d$, the fermionic doublet $\psi$ must
transform as 
\begin{eqnarray}
\left(
\begin{array}{ll}
u \\
d
\end{array}
\right)
~\to~ 
\exp\left\{ i\vartheta^{a} t^{a} + i\gamma_{5}\theta^{a} t^{a} \right\} 
\left(
\begin{array}{ll}
u \\
d
\end{array}
\right)\,,
\end{eqnarray}
leaving invariant the action. In the equation above, \eqref{chiraltransf}, $\gamma_{5}$ is the
pseudo scalar Dirac matrix, defined in terms of the four Dirac matrices (take a look at Appendix
\ref{notations} for details),
\begin{eqnarray}
\gamma_{5} ~=~ i\gamma^{0}\gamma^{1}\gamma^{2}\gamma^{3} ~=~
\frac{i}{4!}\epsilon^{\mu\nu\rho\sigma}\gamma_{\mu}\gamma_{\nu}\gamma_{\rho}\gamma_{\sigma}\,,
\end{eqnarray}
and $\theta^{a}$ stands for a global real parameter related to the conserved axial-vector
current
\begin{eqnarray}
J_{\gamma_{5}}^{a\;\mu} ~=~ i \bar{\psi} \gamma^{\,\,\mu}\gamma_{5}t^{a}\psi\,.
\label{veccurrent}
\end{eqnarray}
Of course, the isotopic (approximate) symmetry does also have an associated conserved current,
which can be read as
\begin{eqnarray}
J^{a\; \mu} ~=~ i\bar{\psi}\gamma^{\,\,\mu}t^{a}\psi\,.
\end{eqnarray}

So, what we have seen is that for a massless $u$- and $d$-quark the theory enjoys an approximate
chiral symmetry, with conserved current given by \eqref{veccurrent} \footnote{For a detailed
analysis the reader is pointed to standard textbooks
\cite{Weinberg:1996kr,Peskin:1995ev,Ryder:1985wq} where this topic is fully covered.}.

Despite the fact that $u$- and $d$-quarks appear as (almost) massless particles in the QCD
action, related to an (approximate) chiral symmetry, the composite states of quarks, such as
protons and neutrons, are not found as (almost) massless particles in Nature. Instead of
that they are considerably heavier ($m_{p} = 938.272046 \pm 2.1\times
\unit{10^{-5}}{\mega\electronvolt}$) in contrast to quarks,
\cite{Roberts:1994dr,Agashe:2014kda}. Thus we are forced to ask if the approximate chiral
symmetry is indeed a reasonable approximation. If it is so, the chiral symmetry
$SU(2)\times SU(2)$ must be spontaneously broken down to the isotopic symmetry $SU(2)$, by
means of a dynamical process of mass generation for quarks, and with the rising of
massless Goldstone (we point the reader to \cite{'tHooft:1979bh,Banks:1979yr}
for a historical reference on this subject). Indeed, the pion meson ($\pi$) seems to
(approximately) fulfill these requirements, displaying the smallest mass of the known
particles, thus being identified with an (approximate) Goldstone boson and, then, pointing to
the effective existence of an spontaneous chiral symmetry breaking (S$\chi$SB),
\cite{Weinberg:1996kr,Peskin:1995ev,Ryder:1985wq}. To prove that nature really undergoes an
spontaneous chiral symmetry breaking is not an easy task, and as such has not been done, yet.
Despite this difficulty, it became clear that we do not need to fully comprehend the whole
process by which the chiral symmetry is broken to $SU(2)$, but rather that interesting process
of nature can be analyzed by just considering the existence of an approximate symmetry that is
spontaneously broken down to $SU(2)$,
\cite{'tHooft:1979bh,Banks:1979yr,Alexandru:2012sd,Weinberg:1996kr,Peskin:1995ev,Ryder:1985wq}.

In order to probe the breaking/restoration of the chiral symmetry one should measure the
existence, or not, of a non-zero \emph{chiral condensate} $\langle \bar{\psi}\psi \rangle$,
which can be done analytically, up to a certain accuracy (or energy level) in perturbation
theory. As effective theories we may cite some well known models, such as the Nambu-Jona-Lasino
model \cite{Nambu:1961tp,Volkov:2005kw,Palhares:2012fv,Fukushima:2003fw}; the MIT bag model
\cite{Palhares:2012fv,Canfora:2013zna,Bellac:2011kqa,Chodos:1974pn,DeTar:1979vb}; and also a quite new proposal by D. Dudal,
\emph{et al.} \cite{Dudal:2013vha} of introducing into the quark sector a nonlocal structure
similar to the Gribov-Zwanziger \emph{horizon} of the gluon sector, leading to a
renormalizable, confining and broken chiral symmetric theory. General properties of introducing
such a nonlocal term in the matter sector will be discussed on chapter \ref{brstonmatter},
while issues concerning UV divergences of such effective model will be treated on chapter
\ref{UVpropsofconfiningprop}.

%

\chapter{Backgrounds: Gribov, Gribov-Zwanziger and a bit of Refined Gribov-Zwanziger} 
\chaptermark{Backgrounds}
\label{usefulbkground}

The understanding of nonperturbative aspects of non-Abelian gauge theories is one of the main
challenging problems in quantum field theories. As an example, we may quote the transition
between the confined and the Higgs regimes in an Yang-Mills theory coupled to a scalar Higgs
field. See refs.\cite{Polyakov:1976fu,Cornwall:1998pt,Baulieu:2001vw} for analytical
investigations and
\cite{Fradkin:1978dv,Nadkarni:1989na,Hart:1996ac,Caudy:2007sf,Maas:2011yx,Maas:2010nc,
Greensite:2011zz} for results obtained through numerical lattice simulations.

Non-perturbative effects can be accounted perturbatively by considering ambiguities in the
gauge-fixing process, first noted by Gribov in \cite{Gribov:1977wm}. These ambiguities, also
referred to as Gribov copies, are unavoidably present in the Landau gauge, since the
(Hermitian) Faddeev-Popov operator admits the existence of zero modes. The widely accepted
mechanism to get rid of these ambiguities was firstly proposed by Gribov, in his famous work
\cite{Gribov:1977wm}, where the domain of integration of the gauge field should be restricted
to a closed region that satisfy specific requirements. As a consequence, the gauge propagator
does not belong to the physical spectrum of the theory anymore and the ghost propagator is
enhanced in the deep IR regime. This framework is an effective model of the Yang-Mills theory,
and as such is useful in analytical analysis of the gauge field theory.

This chapter is devoted to the introduction of the Gribov, Gribov-Zwanziger (GZ) and of the
refined version of the Gribov-Zwanziger (RGZ) frameworks, give the central role these
approaches play in this thesis. However, given the existence of a vast bibliography covering
this topic, hanging from scientific papers to pedagogical reviews
\cite{Gribov:1977wm,Sobreiro:2005ec,Vandersickel:2012tz,Dudal:2009bf}, this chapter is not meant to be one
more detailed pedagogical review, but rather it will provide the most important concepts and
equations that are useful to the comprehension of this thesis. For example, the main idea of
the Gribov mechanism to get rid of (infinitesimal) gauge copies and the consequent violation of
positivity by the gauge field propagator will be presented in the first section of this
chapter, while the horizon function, developed by Zwanziger, is presented in the section
\ref{gzintro} together with the refined version. At the end, the important concept of BRST
symmetry breaking will be introduced and a brief discussion will be developed.

\section{An introduction to Gribov's issue}
\label{introGribov}

This section is organized in a way to provide a brief introduction to Gribov's ambiguities by
following his seminal work \cite{Gribov:1977wm}. We do not intend to provide the final word on
this matter and as such we would refer to \cite{Vandersickel:2012tz} for a more complete and
pedagogical reference. The Faddeev-Popov quantization procedure is reproduced in the subsection
\ref{introductiontogribov1}, while the Gribov problem will be introduced and implemented on the
path integral formalism in the two subsequent subsections, \ref{Gribov's issue} and
\ref{Gribovimplementation}. At the end of this section we will present one of the most
important outcomes of the Gribov issue: a possible interpretation of gluon confinement, which
is encoded in the poles of the gluon propagators. We should say, to clarify matters, that our
present work concerns computations up to one-loop order in perturbation theory.

\subsection{Quantization of non-Abelian gauge field}
\label{introductiontogribov1}

The gauge-invariant action of a non-Abelian gauge field, or the Yang-Mills ($\YM$) action, is given by
\begin{eqnarray}
S_{\text{YM}} &=& \int \mathrm{d}^d x \frac{1}{4}  F_{\mu\nu}^a  F_{\mu\nu}^a \;,
\label{YMact}
\end{eqnarray}
with $
F^{a}_{\mu\nu} ~=~ \partial_{\mu}A^{a}_{\nu} - \partial_{\nu}A^{a}_{\mu} + gf^{abc}A^{b}_{\mu}A^{c}_{\nu}
$ 
being the field strength tensor. The action \eqref{YMact} enjoys the feature of being symmetric under gauge transformation, which is defined for the gauge field as
\begin{eqnarray}
A_{\mu}' ~=~ U^{\dagger}A_{\mu}U - \frac{\ii}{g} U^{\dagger}(\p_\mu U)\;,
\label{gaugetransf}
\end{eqnarray}
with $U(x) \in SU(N)$ and $N$ being the number of colors. A geometrical representation of this symmetry can be seen in Figure \ref{fig1}, where each orbit, representing equivalent gauge fields, is crossed by a gauge curve ${\cal F}$. It means that, the YM action is invariant under transformations that keep the transformed gauge field on the same gauge orbit.

As is well known, the gauge invariance of the action induces inconsistencies in the
quantization of the gauge field reflecting the existence of infinite physically equivalent
configurations. To get rid of those spurious configurations from the system one has to fix the
gauge, which is, in the geometrical view, to choose a convenient curve ${\cal F}$ that crosses
only once each gauge orbit. In the path integral formalism, the gauge-fixing procedure is
carried out by the Faddeev-Popov's procedure.

\begin{figure}[h]
\begin{center}
\includegraphics[width=8cm]{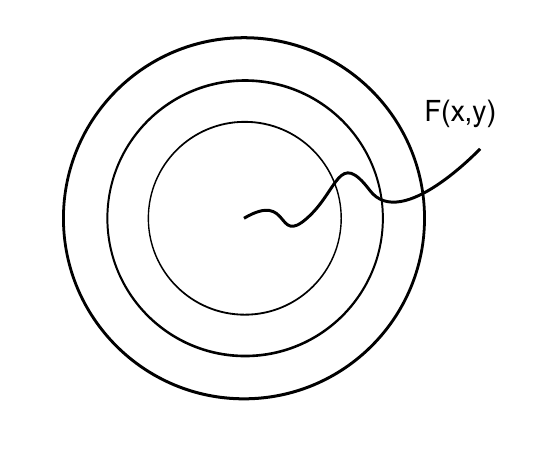}
\caption{Gauge orbits of a system with rotational symmetry in a plane and a function $\mathcal F$ which picks one representative from each gauge orbit.}
\label{fig1}
\end{center}
\end{figure}

The (Euclidean) gauge fixed partition function reads
\begin{eqnarray}
\label{genym}
Z_{\text{YM}}(J) ~=~ \int_{\cal F} [\d A]\;\; \e^{-S_{\text{YM}} + \int {\rm d} x J_\mu^a A_\mu^a}\;,
\label{YMaction0}
\end{eqnarray}
where $\int_{\cal F}$ denotes the path integral restricted to the curve $\cal F$, which can be recast in the form

\begin{eqnarray}
Z_{\YM}(J) ~=~ V\int [\d A] \;\; \Delta_{\cal F}(A) \delta({\cal F}(A')) \mathrm{e}^{-S_{\text{YM}} + \int {\rm d} x J_\mu^a A_\mu^a}\;.
\label{genfp}
\end{eqnarray}
The $V$ factor accounts for the (infinite) orbit's volume, while $\Delta_{\cal F}(A)$ stands for the Jacobian of infinitesimal gauge transformations 
\begin{eqnarray}
A'_{\mu} ~=~ A_{\mu} - D_{\mu}\theta(x) \,.
\label{inftgaugetransf}
\end{eqnarray}
The Jacobian is there since we are working with the gauge transformed integration measure. The
$\theta(x) ~=~ \theta^{a}(x)\tau^{a}$ stands for the infinitesimal gauge transformation
parameter, while $\tau^{a}$ are the $SU(N)$ generators.  $\delta({\cal F}(A))$ is the delta
function ensuring the gauge condition ${\cal F(A)}=0$. It is worthwhile to emphasize that the
Jacobian of a given transformation ({\it e.g.} the infinitesimal gauge transformation) is
defined as the absolute value of the determinant of the derivative --- with respect to the
transformation parameter ($\theta$) ---  of the transformed field. In our gauge-fixing case we
have
\begin{equation}
\Delta_{\mathcal F} (A) ~=~  |\det \mathcal  M_{ab} (x,y) | \qquad \text{with} \qquad  \mathcal
M_{ab} (x,y) ~=~ \left. \frac{\delta \mathcal F^a (A'_\mu (x))  }{\delta \theta^b(y)}
\right|_{\mathcal F(A') =0} \;.
\label{absvalue}
\end{equation}
The delta function can be written as
\begin{eqnarray}
\delta({\cal F}) ~\propto~ \exp\left\{-\frac{i}{2\xi}\int d^{d}x\,{\cal F}^{2}\right\} \,,
\end{eqnarray}
so that performing a sort of functional Fourier transformation, with the introduction of an
auxiliary field named ``Nakanishi-Lautrup field''. Firstly, let us take the example of a real
function, in order to clarify things:
\begin{eqnarray}
\hat{f}(b) ~=~ \int d^{d}x \;\e^{-ib\,x}f(x) ~=~ \int d^{d}x\;
\e^{-ib\,x}\e^{-\frac{i}{2\xi}x^{2}}\,,
\end{eqnarray}
which leads to the following Fourier transformed function,
\begin{eqnarray}
\hat{f}(b) ~\propto~ \e^{i\frac{\xi b^{2}}{2}} \,.
\end{eqnarray}
Thus, returning to the gauge-fixing, one ends up with an equivalent expression,
\begin{eqnarray}
\delta({\cal F}) ~\propto~ \int \left[\d b^{a} \right] \exp\left\{ i\int d^{d}x\, b^{a}{\cal
F}^{a} \right\}   \exp\left\{\frac{i\xi}{2}\int d^{d}x\,b^{a}b^{a}\right\}\,.
\label{deltaF}
\end{eqnarray}
Note that the Nakanishi-Lautrup field works as a Lagrange multiplier field, ensuring the gauge
fixing condition. The Landau gauge is recovered in the limit $\xi \to 0$.

In order to introduce the Gribov issue in its original form, the Landau gauge condition
will be chosen. Namely,
\begin{eqnarray}
\mathcal{F}^a (A'_\mu (x)) ~=~ \p_{\mu} A_{\mu }^{\prime a} (x) \;.
\label{notideal}
\end{eqnarray}
The $A'_{\mu}$ stands for the infinitesimal gauge transformation, given by the equation
\eqref{inftgaugetransf}.
After choosing the gauge condition, the Jacobian operator, named the Faddeev-Popov operator,
reads 
\begin{eqnarray}\label{mab}
 \mathcal M_{ab} (x,y) ~=~  \left.    - \p_\mu  D_\mu^{ab} \delta (y-x)    \right|_{ \mathcal
F(A) = 0 } \;,
\label{fpoperator}
\end{eqnarray}
while the delta function \eqref{deltaF} amounts to
\begin{eqnarray}
\delta({\cal F}) ~\propto~ \int \left[\d b^{a} \right] \exp\left\{ ib^{a} \p_{\mu}A^{a}_{\mu} +
\frac{i\xi}{2}b^{a} b^{a} \right\}\;.
\end{eqnarray}
In order to obtain the final expression of the gauge fixed Yang-Mills partition function, the
Jacobian must be rewritten as ``the exponential of something'', in order to be added into the
action. This will be achieved by introducing a couple of anticommuting {\it real} Grassmann
variables, named the ``Faddeev-Popov ghosts'' $(\bar{c}^{a},\,c^{a})$. The point is that, the
integration rule of a {\it Gaussian-like functional of Grassmann variables} is given by
\begin{eqnarray}
\int \left[ \d\bar{c} \right] \left[ \d c \right] \exp\left\{ \bar{c}^{a} M^{ab} c^{b} \right\}
~=~ \det[M^{ab}]\,.
\end{eqnarray}
Therefore, replacing $M^{ab}$ for the Faddeev-Popov operator ${\cal M}^{ab}$, one ends up with
the following generating function,
\begin{eqnarray}
Z[J] &=&   \int [\d A][\d c] [\d \overline c]         \exp \left[- S_\YM + \int \d
x \left(   \overline c^a   \p_\mu  D_\mu^{ab}  c^b -  \frac{1}{2 \xi} (\p_\mu A_\mu^a )^2
\right) + \int \d x J_\mu^a A_\mu^a  \right]\,.
\label{genfuncfp}
\end{eqnarray}

Let us emphasize two important (not so clear) assumptions made in the process to obtain
\eqref{genfuncfp}:
\begin{itemize}
\item The gauge condition ${\cal F}^{a}$ is said to pick up only one field configuration from each gauge orbit, representing the physical equivalent configurations related by gauge transformations;

\item The determinant of $\mathcal M_{ab} (x,y)$ is supposed to be always positive.
\end{itemize}
These assumptions were considered to be true during the quantization procedure developed by
Faddeev-Popov and described above. Gribov showed in his work \cite{Gribov:1977wm} that the
failure of these assumptions are closed related to the existence of zero-modes of the
Faddeev-Popov operator ${\cal M}^{ab}$. The problem surrounding the failure of these
assumptions defines the Gribov issue. In the next subsection the Gribov issue will be
described and subsequently the mechanism proposed by Gribov to fix these these quantization
inconsistencies will be presented. Important consequences of such mechanism will be discussed
in the subsection \eqref{A sign of confinement from gluon propagator}.

\subsection{Gribov's issue}
\label{Gribov's issue}

\begin{figure}[h]
\begin{center}
\includegraphics[width=8cm]{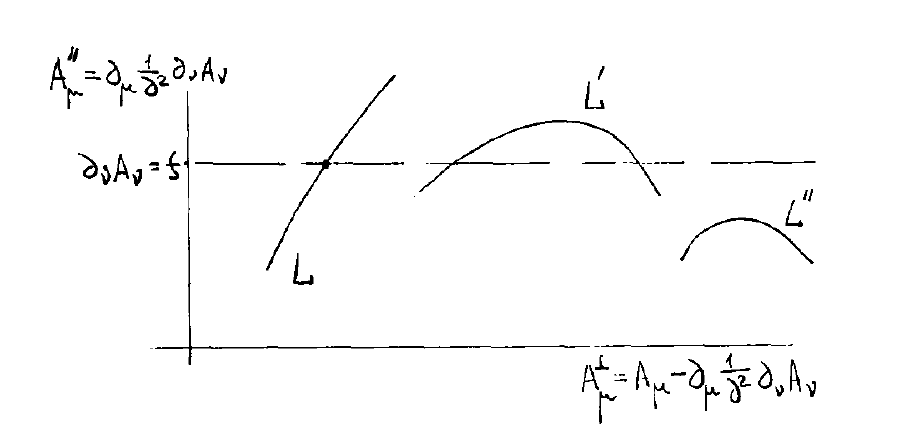}
\caption{The gauge condition curve can cross each orbit of equivalence once, more then once and
at no point at all. The horizontal axes denotes the transversal gluon propagator, while the
vertical axis represents the longitudinal one. This picture was taken from Gribov's
seminal paper \cite{Gribov:1977wm}.}
\label{2fig1}
\end{center}
\end{figure}

As stated before, one of the most important hypotheses required to derive the gauge fixed
Yang-Mills partition function is that: {\it once the gauge-fixing condition
is chosen, one should be able to find out only one gauge field configuration $A_{\mu}$ that
fulfils the gauge condition, ${\cal F}=0$, and that is related to another configuration
$A'_{\mu}$ through a gauge transformation}. This situation is represented in Figure \ref{2fig1}
by the curve $\bf L$ (where the Landau gauge is chosen). In other words, it means that in
principle one should {\it not} be able to find out two gauge-equivalent configurations, let us
say $A_{\mu}(x)$ and $A'_{\mu}(x)$, that satisfy, both of them, the gauge condition. Such
hypothetically forbidden situation is graphically depicted in Figure \ref{2fig1} by the curve
${\bf L}'$. Another forbidden configuration is the one described in Figure \ref{2fig1} by the
curve ${\bf L}''$. This curve describes the situation where no gauge-equivalent configuration
satisfy the gauge condition, at all.

While there is no examples of situations depicted by the curve $\bf L''$, the
situation described by the curve $\bf L'$ is quite typical in non-Abelian gauge theories and,
therefore, is worth analysing \cite{Gribov:1977wm,Sobreiro:2005ec,Vandersickel:2012tz,Dudal:2009bf}. To
this end, let us consider two gauge-equivalent configurations, $A'_{\mu}$ and $A_{\mu}$,
related by an {\it infinitesimal} gauge transformation \eqref{inftgaugetransf} and, both of
them, satisfying the Landau gauge condition. That is,
\begin{eqnarray}
A'_{\mu} ~=~ A_{\mu} - D_{\mu}\alpha(x)\;,  \qquad  \p_\mu A_\mu ~=~ 0 \quad \& \quad  \p_\mu A_\mu' ~=~ 0 \;.
\end{eqnarray}
Therefore, one should end up with
\begin{eqnarray}
- \p_\mu D_\mu \alpha &=& 0\;,
\label{FPeigenvlueqtion}
\end{eqnarray}
whence $D_{\mu}$ stands for the covariant derivative,
\begin{eqnarray}
D_{\mu} ~=~ \p_{\mu}\delta^{ab} - igA_{\mu}\,.
\end{eqnarray}
Such equation points to the existence of zero-modes, that are eigenstates of the FP operator
associated to null eigenvalues. Therefore, one may conclude that if there are (at least) two
infinitesimally gauge-equivalent fields satisfying the Landau gauge, defining gauge copies, the
Faddeev-Popov operator has zero-modes (eigenstates associated to zero eigenvalues). Note that
for the Abelian case the equation \eqref{FPeigenvlueqtion} reduces to the Laplace equation,
\begin{eqnarray}
\p^{2} \alpha &=& 0\;.
\label{abelianeingvalue}
\end{eqnarray}
Since it defines plane waves, which are not normalisable, we will not consider this case,
restricting ourselves to the analysis of fields that smoothly vanish at infinity. It becomes 
quite evident that a closer look at the space of eigenvalues of the FP operator is of great
importance for a better comprehension of the problem. Besides, let us regard that the FP
operator is Hermitian in the Landau gauge, which allows us to sort its eigenvalues in the real
axes. Therefore, let us start by considering a gauge configuration that is close enough to the
trivial vacuum. In this case the eigenvalue equation,
\begin{eqnarray}
- \p_\mu D_\mu \psi ~=~ \epsilon \psi\;,
\end{eqnarray}
reduces to
\begin{equation}
-\p_\mu^2 \psi = \epsilon \psi \;.
\end{equation}
Notice that this equation is solvable only for positive $\epsilon$, since in the momentum space
we have $p^{2} = \epsilon >0$. Then, for small enough field configurations there is no
zero-mode issues anymore. Otherwise, if $A_{\mu}$ turns out to be large, but still not too
large, one reaches the zero-mode solution $\epsilon = 0$, since a higher potential of the gauge
field tends to decrease the eigenvalue of the FP operator in the Landau gauge. Thus, for even
large amplitudes of $A_{\mu}$ the eigenvalue turns to be negative; if it keeps growing the
$\epsilon$ reaches zero again. Note that, in the Landau gauge, it is possible to identify a
threshold value for the squared norm of the gauge field, $\Vert A \Vert^{2}_{c}$, below
which the eigenvalue of the FP operator is positive, and for amplitudes whose squared norm
is greater than such critical value the eigenvalue is negative.

In his work \cite{Gribov:1977wm} Gribov showed that the domain of functional integration should
be restricted to the first region, named ``first Gribov region'' $\Omega$, where $\epsilon >0$,
in order to avoid gauge copies. It is, however, known that this region is actually not
completely free of copies. Besides, it has being analytically motivated that the existence of
those copies inside $\Omega$ do not influence physical results (see \cite{Dudal:2014rxa} and
references therein). Furthermore, until today there is no way to analytically implement the
restriction of the partition function to the region actually free of copies, known as the
Fundamental Modular Region, $\Lambda$.

The first Gribov region $\Omega$ enjoys some useful properties that have mathematical proof
still only in the Landau gauge, which justifies our interest in this gauge (see
\cite{Vandersickel:2012tz} and references therein) \footnote{{\it cf.}
\cite{Capri:2016aqq,Capri:2015nzw,Capri:2015pja,Capri:2015ixa} for recent developments on the
Gribov issue in the wider class of Linear Covariant Gauges.}. Namely,
\begin{itemize}
\item 
For every field configuration infinitesimally close to the border $\delta\Omega$  and belonging to the region immediately out side $\Omega$ (called $\Omega_{2}$), there exist a gauge-equivalent configuration belonging to $\Omega$ and infinitesimally close to the border $\delta\Omega$ as well \cite{Gribov:1977wm}. It was also proven that every gauge orbit intersects the first Gribov region $\Omega$ \cite{Capri:2005dy,Zwanziger:1982na}.

\item
The Gribov region is convex \cite{Zwanziger:2003cf}. This means that for two gluon fields $A_\mu^1$ and $A_\mu^2$ belonging to the Gribov region, also the gluon field $A_\mu = \alpha A_\mu^1 + \beta A_\mu^2$ with $\alpha, \beta \geq 0$ and $\alpha + \beta  =1$, is inside the Gribov region.

\item
One may also show that the Gribov region is bounded in every direction \cite{Zwanziger:2003cf}.
\end{itemize}

For more details concerning (proofs of) properties of Gribov regions see \cite{Vandersickel:2012tz} and references therein.

\subsection{Implementing the restriction to the first Gribov region $\Omega$}
\label{Gribovimplementation}
As was discussed before, our aim is to restrict the domain of the functional integration to the region where the $\FP$ operator has positive eigenvalues. Therefore, let us define the first Gribov region as the region where the $\FP$ operator is positive definite. Namely,
\begin{eqnarray}
\Omega ~\equiv ~ \{ A^a_{\mu}, \, \p_{\mu} A^a_{\mu} ~=~  0, \, \mathcal{M}^{ab}  >0  \} \;.
\label{defgribovregion}
\end{eqnarray}
Once again, $\mathcal{M}^{ab}$ stands for the Faddeev-Popov operator, defined innumerable times
and given once more,
\begin{eqnarray}\label{mab2}
\mathcal M^{ab}(x,y) ~=~ - \p_\mu  D_\mu^{ab} \delta(x-y)  ~=~ 
-\p_\mu \left( \p_\mu \delta^{ab} -
gf^{abc} A_\mu^c \right) \delta(x-y)\;.
\end{eqnarray}
The condition defines a positive definite operator, which means that only gauge field
configurations associated with positive eigenvalues of the FP operator will be considered.

One should notice that, since the $\FP$ operator is the inverse of the ghost propagator, thus
the ghost propagator plays a central role in the Gribov issue. Therefore, the propagator of the
ghost field is worthwhile to compute, which will be done up to the first loop order, by
following \cite{Gribov:1977wm}. The positive definite condition imposed on the FP operator can
be implemented into the partition function by means of the ghost propagator, which can be read
as
\begin{eqnarray}
\left\langle  c^{a}(k) \overline{c}^{b}(-k)  \right\rangle ~=~ \mathcal G (k^2, A)_{ab}\;.
\label{ghostequation0}
\end{eqnarray}
The effective computation of the equation \eqref{ghostequation0} shall be performed with the
partition function \eqref{genfuncfp} and treating the gauge field as an external classical
field. Namely, one gets,
\begin{eqnarray}
\mathcal G (k^2, A) &=&  \frac{1}{N^2 -1}\delta_{ab} \mathcal G (k^2, A)_{ab} ~=~
\frac{1}{k^2} + \frac{1}{V}\frac{1}{k^4} \frac{N g^2}{N^2 - 1} \int\frac{ \d^d q}{(2 \pi)^d}
A_\mu^\ell(-q) A_\nu^{\ell} (q)  \frac{(k+q)_\mu  k_\nu}{(k+q)^2} \nonumber\\
&=& \frac{1}{k^2}\left( 1 + \sigma(k, A) \right)\;,
\label{coloraway}
\end{eqnarray}
whereby
\begin{eqnarray}
\sigma(k,A) ~=~ \frac{1}{V}\frac{1}{k^2} \frac{Ng^2}{N^2 - 1} \int\frac{ \d^d q}{(2 \pi)^d}
A_\mu^\ell(-q) A_\nu^{\ell} (q) \frac{ (k+q)_\mu  k_\nu }{(k+q)^2}\;.
\label{1ghostformfacto}
\end{eqnarray}
Making use of the property
\begin{eqnarray}
A_{\mu }^{a}(q)A_{\nu }^{a}(-q) &=&
\left( \delta _{\mu \nu }-\frac{q_{\mu }q_{\nu }}{q^{2}}\right) \omega (A)(q)   
\nonumber \\
&\Rightarrow &\omega (A)(q)=\frac{1}{d-1}A_{\lambda }^{a}(q)A_{\lambda }^{a}(-q)\,,
\end{eqnarray}
which follows from the transversality of the gauge field, $q_\mu A^a_\mu(q)=0$, 
we will have
\begin{eqnarray}
\sigma(k,A) &=& \frac{1}{V}\frac{1}{k^2} \frac{Ng^2}{(N^2 - 1)(d-1)} 
\int\frac{ \d^d q}{(2 \pi)^d}\;
A_\mu^\ell(-q) A_\mu^{\ell}(q)
\left( \delta _{\mu \nu }-\frac{q_{\mu }q_{\nu }}{q^{2}}\right) 
\frac{ (k+q)_\mu  k_\nu }{(k+q)^2}
\;,
\nonumber \\
&=& 
\frac{1}{V}\frac{1}{k^2} \frac{Ng^2}{(N^2 - 1)(d-1)} 
\int\frac{ \d^d q}{(2 \pi)^d}\;
\frac{A_\mu^\ell(q) A_\mu^{\ell}(-q)}{(k+q)^2}
\left( (k+q)_\mu  k_\mu -\frac{q_{\mu }(k+q)_\mu  k_\nu q_{\nu}}{q^{2}} 
\right)
\,.
\nonumber \\
\label{1ghostformfacto3}
\end{eqnarray}
Such expression may be rewritten as follows,
\begin{eqnarray}
\sigma(k,A) &=& \frac{1}{V}\frac{1}{k^2} \frac{Ng^2}{(N^2 - 1)(d-1)} 
\int\frac{ \d^d q}{(2 \pi)^d}\;
\frac{A_\mu^\ell(q) A_\mu^{\ell}(-q)}{(k+q)^2}
\left(k^{2}  - \frac{q_{\mu}k_\mu k_\nu q_{\nu} }{q^{2}} 
\right) 
\,,
\nonumber \\
&=&
\frac{1}{V} \frac{Ng^2}{(N^2 - 1)(d-1)} 
\int\frac{ \d^d q}{(2 \pi)^d}\;
\frac{A_\mu^\ell(q) A_\mu^{\ell}(-q)}{(k+q)^2}
\frac{k_{\mu}k_{\nu}}{k^2}
\left(\delta_{\mu\nu} - \frac{q_{\mu}q_{\nu} }{q^{2}} 
\right)  \,.
\label{1ghostformfacto3}
\end{eqnarray}
Now, reminding the property
\begin{equation}
\int \frac{d^{d}p}{(2\pi )^{d}}
\mathcal{F}(p^2) \left( \delta _{\mu \nu }-\frac{p_{\mu }p_{\nu }}{p^{2}}\right)  
~=~ \left( \frac{d-1}{d} \right)  
\int \frac{d^{d}p}{(2\pi )^{d}}\;\mathcal{F}(p^2) \delta_{\mu\nu} \,,
\label{a}
\end{equation}
the ghost form factor becomes
\begin{eqnarray}
\sigma(k,A) &=&
\frac{1}{V} \frac{Ng^2}{d(N^2 - 1)} 
\int\frac{ \d^d q}{(2 \pi)^d}\;
\frac{A_\mu^\ell(q) A_\mu^{\ell}(-q)}{(k+q)^2}  \,.
\label{ghostfrmfact1}
\end{eqnarray}
The ghost propagator \eqref{coloraway} can be perturbatively approximated by
\begin{eqnarray}
\mathcal G (k^2, A) ~\approx~  \frac{1}{k^2}\frac{1}{( 1 - \sigma(k, A) )}
\label{ghstprop00}
\end{eqnarray}
whereby we can see that for $ \sigma(k, A) < 1$ the domain of integration is safely restricted
to $\Omega$, characterising the no-pole condition. It is not so dificult to see that 
$\sigma(k,A)$ is a
decreasing function of $k$, from \eqref{ghostfrmfact1}, which means that the largest value of
$\sigma$ is obtained at $k=0$. Therefore, if the condition $ \sigma(0, A) < 1$ is ensured, the
system is (or would be) completely safe of gauge copies for any non zero value of $k$. Note
that the only allowed pole is at $k^2=0$, which has the meaning of approaching the boundary of
the region $\Omega$. At the end, the ghost form may be computed by taking the limit 
$k^{2} \to 0$, which reads
\begin{eqnarray}
\sigma(0,A) ~=~
\frac{1}{V} \frac{1}{d} \frac{Ng^2}{N^2 - 1} \int\frac{ \d^d q}{(2 \pi)^4} A_\alpha^\ell(-q) A_\alpha^{\ell} (q) \frac{ 1 }{q^2}\;.
\label{nopole}
\end{eqnarray}
The partition function restricted to $\Omega$ then becomes,
\begin{eqnarray}\label{ZJ}
Z_{G} &=& \int_\Omega [\d A]   \exp \left[- S_\FP   \right] 
\nonumber\\
&=& \int [\d A][\d c] [\d \overline c]    V(\Omega)     \exp \left[- S_\YM  -   \int \d x \left(  \overline c^a   \p_\mu  D_\mu^{ab}  c^b -  \frac{1}{2 \xi} (\p_\mu A_\mu^a )^2  \right)  \right] \;,
\end{eqnarray}
with
\begin{eqnarray}
V(\Omega) &=& \theta (1 - \sigma(0,A))\;,
\label{nopolestepfunct}
\end{eqnarray}
where $\theta (1 - \sigma(0,A))$ is the Heaviside step function ensuring the no-pole
condition. Considering the transversality of the gauge field in the Landau gauge and by making
use of the integral representation of the Heaviside step function, one gets the following
expression of the partition function restricted to the first Gribov region $\Omega$,
\begin{equation}
Z_{G} ~=~  \mathcal N \int \frac{\d \beta}{2\pi \ii \beta} \int {\cal D} A    \e^{\beta (1 -
\sigma(0,A))}    \e^{ - S_\FP }
\;.
\label{fullgenertfunct}
\end{equation}
The final expression for the gauge fixed Yang-Mills action accounting for the Gribov 
copies reads
\begin{eqnarray}
S_{G} ~=~ S_{\YM} + S_{\gf} +  S_{\beta}\;,
\label{fullgribovact}
\end{eqnarray}
with $S_{\beta} ~=~ \beta\left( \sigma(0,A) - 1\right)$.

Roughly speaking, the Gribov restriction to the first Gribov region has already been
implemented into the partition function. However, notice that a new parameter has been
introduced, named the Gribov parameter $\beta$, which still deserves some analysis. As will
become clear, this new parameter is not a free parameter of the theory, but rather it is
dynamically determined by its gap equation, which amounts to ensure the no-pole condition. Akin
to the mass gap equation, the Gribov parameter gap equation will be derived from the vacuum
energy of the theory, computed up to the first loop order in perturbation theory. To that end,
only terms quadratic in the fields, from the Gribov action \eqref{fullgribovact}, must be taken
into account. Doing so, one should get
\begin{equation}
Z_{quad} ~=~ \iint \frac{\d\beta e^{\beta }}{2\pi i\beta } [\d A] \; 
\exp \left\{-\frac{1}{2}  \int \frac{\d^{d}q}{(2\pi )^{d}}  \;  A_{\mu }^{a}(q)\mathcal{P}_{\mu \nu }^{ab }A_{\nu }^{b }(-q)    \right\}  \;,
\label{Zq0}
\end{equation}
with
\begin{eqnarray}
\mathcal{P}_{\mu \nu }^{ab} &=&  \delta^{ab } \left(  q^{2}\delta _{\mu \nu }   +  \left( \frac{1}{\xi } -1 \right) q_{\mu }q_{\nu } +  \frac{2Ng^{2}\beta}{(N^{2}-1)Vd} \frac{\delta _{\mu \nu }}{q^{2}}   \right)   \;.
\label{P0}
\end{eqnarray}
It is straightforward to compute the functional integration on the gauge field, since a
Gaussian integration, leading to the functional determinant of $\mathcal{P}_{\mu \nu }^{ab}$:
\begin{equation}
Z_{quad}  ~=~ \int {\frac{\d \beta}{2\pi i} }
e^{({\beta} -\ln\beta)} \left[\det\mathcal{P}^{ab}_{\mu\nu}\right]^{-\frac{1}{2}} \;.
\label{Zq2f00}
\end{equation}
This functional determinant may be exponentiated by making use of the relation
\begin{equation}
\left[ \det \mathcal{P}_{\mu\nu}^{ab} \right]^{-\frac{1}{2}} ~=~ e^{-\frac{1}{2} \ln \det
\mathcal{P}_{\mu\nu}^{ab}} ~=~ e^{-\frac{1}{2}Tr \ln \mathcal{P}_{\mu\nu}^{ab}} \;.
\label{functdeterminant}
\end{equation}
Thus, after taking the trace of $\ln\mathcal{P}_{\mu \nu }^{ab}$, whose technicalities is
detailed in \cite{Vandersickel:2012tz}, one may finally get
\begin{eqnarray}
Z_{quad} ~=~ \int_{- \ii \infty + \epsilon}^{+ \ii \infty + \epsilon}\frac{\d \beta}{2\pi \ii} \e^{f(\beta)} \;,
\label{ven}
\end{eqnarray}
with
\begin{eqnarray}
f(\beta) ~=~ \beta - \ln \beta  -\frac{(d-1)(N^2 - 1)}{2}  V \int \frac{\d^d q}{(2\pi)^d} \ln \left( q^2 + \frac{\beta N g^2 }{N^2 - 1} \frac{2}{d V}\frac{1}{q^2} \right)\;.
\label{minusfreenergy}
\end{eqnarray}
The factors $(N^{2}-1)$ and $(d-1)$ in front of the integral came from the trace over the
$SU(N)$ gauge group indices and from the trace over the Euclidean space-time indices,
respectively\footnote{ A careful computation of the functional trace of $\ln
\mathcal{P}_{\mu\nu}^{ab}$ can be found in \cite{Vandersickel:2012tz}}. Note that the factor
$(d-1)$ is obtained only after the Landau gauge limit is taken.

In the thermodynamic limit (when $V\to \infty$) the saddle-point approximation becomes exact,
and the integral \eqref{ven} can be easily computed, resulting in
\begin{eqnarray}
\e^{-V{\cal E}_{v}} ~=~  \e^{f(\beta^{\ast})} \;.
\label{relatvaccener}
\end{eqnarray}
One should notice now that the vacuum energy has effectively been computed up to first loop
order, since 
\begin{eqnarray}
{\cal E}_{v} ~=~ -\frac{1}{V}f(\beta^{\ast})\,,
\end{eqnarray}
within the thermodynamic limit. The saddle-point approximation, that becomes exact within the
thermodynamic limit, states that the integral \eqref{ven} equals the integrated function
evaluated at its maximum value. Thus, the stared parameter $\beta^{\ast}$ accounts for the
value of $\beta$ that maximizes the integrated function, which amounts to computing the Gribov
parameter gap equation,
\begin{eqnarray}
\left.
\frac{\partial f}{\partial \beta}\right|_{\beta = \beta^{\ast}}=0\;,
\label{gapeq}
\end{eqnarray}
which lead us to
\begin{eqnarray}
\frac{d-1}{d}N g^2    \int \frac{\d^d q}{(2\pi)^d}  \;   \frac{1}{ \left( q^4 + \frac{2\beta^{\ast} N g^2}{(N^2 - 1)dV} \right) }  ~=~ 1   \;.
\label{finalgapeq2}
\end{eqnarray}
Note that the Gribov parameter $\beta$, introduced to get rid of gauge ambiguities by
restricting the path integral to the first Gribov region $\Omega$, is not in fact a free
parameter of the theory. Otherwise, it is dynamically determined by its gap equation
\eqref{finalgapeq2}. Besides, it has dimension of $[mass]^{4}$ and is proportional to the
space-time volume $V$. Consequently, in the thermodynamic limit the logarithmic term of
equation \eqref{minusfreenergy} becomes zero, leading to the equation \eqref{finalgapeq2}.

\subsection{The gauge propagator}
\label{A sign of confinement from gluon propagator}

In the present subsection we motivate that a possible sign of confinement could be read off
from the poles of the gluon propagator, putting this quantity at the centre of any further
discussion in the present work. At one-loop order only quadratic terms of the action
\eqref{fullgribovact} really matter, so that one can read off the two point function of the
gauge field from the inverse of the operator \eqref{P0}, setting $\xi \to 0$ at the very end of
the computation. Notice that the computation is performed within the thermodynamic limit, so
that the Gribov parameter must satisfy its gap equation. Namely,
\begin{equation}
\left\langle A_{\mu }^{a}(q)A_{\nu}^{b}(-q)\right\rangle ~=~ \frac{q^{2}}{q^{4}  +  \frac{2N
g^{2}\beta^{\ast} }{(N^{2}-1)dV} } \left( \delta _{\mu \nu }  -  \frac{q_{\mu }q_{\nu}}{q^{2}}
\right)\delta^{ab}  \;.
\label{Gribovprop0}
\end{equation}
Things become easier to analyze if we redefine the Gribov parameter as,
\begin{eqnarray}
\lambda^{4} = \frac{2\beta^{\ast} N g^2}{(N^2 - 1)dV}\;.
\end{eqnarray}
Consequently, the gauge propagator can be decomposed as,
\begin{eqnarray}
\left\langle A_{\mu }^{a}(q)A_{\nu}^{b}(-q)\right\rangle 
~=~ \frac{1}{2} \left( \frac{1}{q^{2}  +  i \lambda^{2} } +  \frac{1}{q^{2}  -  i \lambda^{2} }
\right)  \left( \delta _{\mu \nu }  -  \frac{q_{\mu }q_{\nu}}{q^{2}}  \right)\delta^{ab}\;.
\label{Gribovprop1}
\end{eqnarray}
Observe from \eqref{Gribovprop1} that the gluon propagator is suppressed in the infrared (IR)
regime, while displaying two complex conjugate poles, $m^{2}_{\pm} = \pm i\lambda^{2}$. That
feature does not allow us to attach the usual physical particle interpretation to the gluon
propagator, since such type of propagator is deprived of a spectral representation
\cite{Cucchieri:2007rg,Cucchieri:2008fc,Cucchieri:2011ig,Cucchieri:2004mf,Cucchieri:2014via}.
From the analytic point of view the gluon propagator \eqref{Gribovprop1} has not
a(n) (always) positive K\"{a}ll\'en-Lehmann spectral representation, which is necessary to
attach a probabilistic interpretation to the propagator\footnote{ See
\cite{Baulieu:2009ha,Sorella:2010it} and references therein for more details on the confinement
interpretation of gluons, $i$-particles and the existence of local composite operators, related
to these $i$-particles, displaying positive K\"{a}ll\'en-Lehmann spectral representation. For
lattice results pointing to the same confinement interpretation see
\cite{Cucchieri:2007rg,Cucchieri:2008fc,Cucchieri:2011ig,Cucchieri:2004mf,Cucchieri:2014via}.}.
These features lead us to interpret the gauge field as being confined.

As already mentioned in this thesis, our concept of confinement, throughout this work, will
always be concerned with the existence of a Gribov-kind of propagator. Particularly, not only
the gauge field will be susceptible to present such a Gribov-type propagator, but also the quark
field.

\section{A brief summary of the Gribov--Zwanziger framework}
\label{gzintro}

About ten years after Gribov's seminal paper has been published
\cite{Gribov:1977wm}, a generalization to the mechanism of getting rid of a leftover gauge
ambiguity after fixing the gauge was proposed by D. Zwanziger \cite{Zwanziger:1988jt}.
The main idea of his work is to take the trace of every positive eigenvalue of the
Faddeev-Popov operator,
\begin{eqnarray}
{\cal M} ~=~ - \p_{\mu}D_{\mu} ~=~ -\p_{\mu}\left(\p_{\mu} -igA_{\mu} \right)\,,
\end{eqnarray}
starting from the smallest eigenvalue. Regard that negative eigenvalues shall be avoided since
it is linked to the existence of gauge copies configurations --- and zero-modes ---, as was
presented in the previous section. Note that constant fields may also be eigenstates of the FP
operator in the Landau gauge related to zero eigenvalues. Since there is no gauge
configurations associated to {\it negative} eigenvalues  with constant eigenstates (the
constant fields), we will not consider such configurations.

Zwanziger did show that restricting the domain of integration of the gauge field to the first
Gribov region $\Omega$ is equivalent to take into account only gauge field configurations
that minimize the squared norm of the gauge field with respect to the gauge orbit
\cite{Zwanziger:1982na,Zwanziger:1988jt},
\begin{eqnarray}
 \left\Vert A\right\Vert^{2}_{min} ~=~ \min_{U\in SU(N)}\int d^{4}x \,
\left(A^{U}\right)^{2}\,.
\label{AAmin}
\end{eqnarray}
In other words, it means that the allegedly gauge physical configurations are those
that satisfy the (Landau) gauge condition and, furthermore, that minimizes the functional
\eqref{AAmin}. It should be clear that such minimized squared norm \eqref{AAmin} is, in 
fact, a gauge-invariant quantity, and that, at the same time, it is {\it nonlocal}
\footnote{The reader is pointed to a list of recent publications concerning the nonlocality of
such dimension 2 gauge field composite operator \eqref{AAmin},
\cite{Capri:2015pja,Capri:2015nzw,Capri:2016aqq,Fiorentini:2016rwx,Dudal:2006xd}.} (\emph{cf.}
\cite{Vandersickel:2012tz,Zwanziger:1982na,Zwanziger:1988jt,Zwanziger:1989mf,Zwanziger:1990tn,
Zwanziger:1992qr} and references therein).

For the sake of clarity, let us give once again the definition of the first Gribov region,
firstly introduced in Gribov's paper \cite{Gribov:1977wm} as
\begin{align}
\Omega \;= \; \{ A^a_{\mu}\;; \;\; \partial_\mu A^a_{\mu}=0\;; \;\; {\cal M}^{ab}=-(\partial^2
\delta^{ab} -g f^{abc}A^{c}_{\mu}\partial_{\mu})\; >0 \; \} \,.
\label{gr}
\end{align} 
Although we have already mentioned the important features of this region, let us state it
again, as a matter of completeness \cite{Vandersickel:2012tz,Zwanziger:1982na,Zwanziger:1988jt,Zwanziger:1989mf,Dell'Antonio:1989jn,Dell'Antonio:1991xt}:
\begin{itemize}
	\item[i.] $\Omega$  is convex and bounded in all direction in field space. Its
boundary, $\partial \Omega$, is the Gribov horizon, where the first vanishing eigenvalue of the
Faddeev-Popov operator shows up;
	\item[ii.] every gauge orbit crosses at least once the region $\Omega$. 
\end{itemize} 

In order to implement the restriction to this first Gribov's region, D. Zwanziger proposed an
all order procedure, by computing the FP operator's eigenvalue perturbatively, stating from the
lowest eigenvalue of the ``nonperturbative term'' of the FP operator,
\begin{eqnarray}
{\cal M}^{ab} ~=~ {\cal M}_{0}^{ab} + {\cal M}_{1}^{ab} ~=~ -\p^{2}\delta^{ab} +
gf^{abc}A^{c}_{\mu}\p_{\mu}\,.
\end{eqnarray}
Note that, it is straightforward to see that the lowest considered eigenvalue of the FP
operator must be always greater than zero. Thus, after perturbatively deriving the space of
positive eigenvalues, he took the trace over all of them obtaining, at the end, a positive
quantity, namely,
\begin{eqnarray}
dV(N^{2}-1) - H(A) ~>~ 0\,,
\label{positivitycondition}
\end{eqnarray}
where the functional $H(A)$ was identified with {\it horizon function} ({\it cf.}
\cite{Vandersickel:2012tz,Zwanziger:1982na,Zwanziger:1988jt,Zwanziger:1989mf}),
\begin{align}
H(A)  ~=~  {g^{2}}\int d^{4}x\;d^{4}y\; f^{abc}A_{\mu}^{b}(x)\left[ {\cal M}^{-1}\right]^{ad}
(x,y)f^{dec}A_{\mu}^{e}(y)   \;.
\label{hf2}
\end{align}
Therefore, the idea is to restrict the Yang-Mills path integral to the domain of integration of
the gauge field where the positivity condition \eqref{positivitycondition} is satisfied. It
amounts to make use of the following partition function, hereinafter called the
Gribov-Zwanziger partition function,
\begin{eqnarray}
Z_{GZ} ~=~ \int {\cal D} \Phi \theta (dV(N^{2}-1) - H(A))\e^{-S_{FP}}\,.
\label{gzpartition1}
\end{eqnarray}

The existence of such horizon function reflects the existence of a critical value for the
squared norm of the gauge field, $\left\Vert A \right\Vert_{c}$, beyond which the gauge
configuration corresponds to a negative eigenvalue of the FP operator.

The effect of the $\theta$-function into the Faddeev-Popov action will be derived in the
thermodynamic limit, where the $\theta$-function amounts to a $\delta$-function reflecting the
concept that in the limit $V\to \infty$ the volume of a $d$-dimensional sphere is directly
proportional to the surface of the border of this sphere. Thus, within the thermodynamic limit
the partition function \eqref{gzpartition1} can be rewritten as
\begin{eqnarray}
Z_{GZ} ~=~ \int {\cal D} \Phi \delta (dV(N^{2}-1) - H(A))\e^{-S_{FP}}\,.
\label{gzpartition2}
\end{eqnarray}
Finally, one  may use the same integral representation of the $\delta$-function, or even, may
use the equivalence between the microcanonical ensemble and the canonical ensemble in order to
obtain the GZ partition function,
\begin{equation}
 Z_{GZ} ~=~ \;
\int {\cal D}A\;{\cal D}c\;{\cal D}\bar{c} \; {\cal D} b \; e^{-\left[  S_{FP}+\gamma^4 H(A)
-V\gamma^4 4(N^2-1) \right]} \;, 
\label{zww2}
\end{equation}
The parameter $\gamma$ has the dimension of a mass
and is known as the Gribov parameter \footnote{Up to this point no relation exists between the
former Gribov parameter $\beta$ and the just derived $\gamma$ parameter. The authors of
\cite{Capri:2012wx} showed that the Gribov's mechanism amounts the Zwanziger's mechanism when
computed at first-order.}. It is not a free parameter of the theory; instead of
that, it is a dynamical quantity, being determined in a self-consistent way through a gap
equation called the \emph{horizon condition} \cite{Vandersickel:2012tz,Zwanziger:1982na,
Zwanziger:1988jt,Zwanziger:1989mf, Zwanziger:1990tn,Zwanziger:1992qr}, given by 
\begin{equation}
\left\langle H(A)   \right\rangle = 4V \left(  N^{2}-1\right) \;,   
\label{hc2}
\end{equation}
where the vacuum expectation value $\left\langle H(A)  \right\rangle$  has to be evaluated with
the measure defined in eq.\eqref{zww2}. The gap equation becomes exact due to the equivalence
between the microcanonical and canonical ensemble in the thermodynamic limit.

It is worth mentioning that most recently an all order proof has been published on the
equivalence between the Gribov's procedure and Zwanziger's one, \cite{Capri:2012wx}. Regarding
that the Gribov's approach relies on the perturbative expansion of the ghost propagator,
accounting up to the first non-null term of the expansion, and that the Zwanziger's one is an
all order computation of the FP operator's spectrum, the referred work computed the full
ghost propagator in perturbation theory, concluding at the end that both approaches are
equivalent at first order in perturbation theory.

\subsection{The local formulation of the Gribov-Zwanziger action}

Being able to construct a partition function for Yang-Mills theories that takes into account
the Gribov ambiguities, related to the gauge-fixing procedure, is a big achievement in the
direction of better comprehending the quantization procedure of non-Abelian fields. However,
equation \eqref{zww2} is not actually useful to compute physical quantities, not analytically
at least. The point is that one needs the action to be local in order to be able to compute
useful quantities, such as the two point function of the gauge field.

In this subsection we are going to present a localized version of the GZ action. Note, however,
that no details concerning its derivation will be provided, since such procedure has been
already extensively treated, \cite{Vandersickel:2012tz}. Rather, we will just mention the
mechanism with which one would obtain the same local expression.

Although the horizon function $H(A)$ is a nonlocal quantity, it can be recast
in a local form by means of the introduction of a set of auxiliary fields
$(\bar{\omega}_\mu^{ab}, \omega_\mu^{ab}, \bar{\varphi}_\mu^{ab},\varphi_\mu^{ab})$, where
$(\bar{\varphi}_\mu^{ab},\varphi_\mu^{ab})$ are a pair of bosonic fields, and
$(\bar{\omega}_\mu^{ab}, \omega_\mu^{ab})$ are a pair of anti-commuting fields. It turns out
that the Gribov-Zwanziger partition function $Z_{GZ}$, in equation \eqref{zww2}, can be
rewritten as \cite{Vandersickel:2012tz,Zwanziger:1988jt,Zwanziger:1989mf,Zwanziger:1992qr}
\begin{equation}
 Z_{GZ} ~=~ \;
\int {\cal D}\phi \; e^{-S_{GZ}} \;, \label{lzww1}
\end{equation}
with $\phi$ accounting for every single field of the theory,
$\{A,\,c,\,\bar{c},\,b,\,\omega,\,\bar{\omega},\,\varphi,\,\bar{\varphi}\}$. The Faddeev-Popov
action $S_{GZ}$ is then given by the local expression 
\begin{equation} 
S_{GZ} = S_{YM} + S_{gf} + S_0+S_\gamma  \;, 
\label{sgz2}
\end{equation}
with
\begin{equation}
S_0 =\int d^{4}x \left( {\bar \varphi}^{ac}_{\mu} (\partial_\nu D^{ab}_{\nu} )
\varphi^{bc}_{\mu} - {\bar \omega}^{ac}_{\mu}  (\partial_\nu D^{ab}_{\nu} ) \omega^{bc}_{\mu}
- gf^{amb} (\partial_\nu  {\bar \omega}^{ac}_{\mu} ) (D^{mp}_{\nu}c^p) \varphi^{bc}_{\mu}
\right) \;, 
\label{s0}
\end{equation}
and 
\begin{equation}
S_\gamma =\; \gamma^{2} \int d^{4}x \left( gf^{abc}A^{a}_{\mu}(\varphi^{bc}_{\mu} + {\bar \varphi}^{bc}_{\mu})\right)-4 \gamma^4V (N^2-1)\;. 
\label{hfl}
\end{equation} 
Let us now make some comments on the terms of the above action and about the mechanism one
should follow to obtain such an action. We are not going to provide a step-by-step construction
of the localized action \eqref{sgz2}. Otherwise, we will provide a backward construction.
Note that the term $gf^{amb} (\partial_\nu  {\bar \omega}^{ac}_{\mu} ) (D^{mp}_{\nu}c^p)
\varphi^{bc}_{\mu} $ has no physical meaning, in the sense that it is not possible to construct
any Feynman diagram with entering $\bar{c}$ and $\omega$ fields, so that the vertex with
$\bar{\omega}$ and $c$ would influence. This term is introduced into the action by means of a
shift in the $\omega$ field with the aim of writing the action $S_{0}$ as an exact BRST
quantity. Namely,
\begin{eqnarray}
S_{0} ~=~ s\;\int d^{4}x \,\left( \bar{\omega}^{ac}_{\mu} \p_{\nu}D_{\nu}  \varphi^{bc}_{\mu}
\right)\,.
\label{S0}
\end{eqnarray}
To check that the contribution $S_{0}$, given in equation \eqref{S0}, to the GZ action is
indeed BRST exact, consider the following BRST transformation rule of the fields,
\begin{eqnarray}
\label{brst0}
sA^{a}_{\mu} &=& - D^{ab}_{\mu}c^{b}\;,\nonumber \\
s c^{a} &=& \frac{1}{2}gf^{abc}c^{b}c^{c} \;, \nonumber \\
s{\bar c}^{a} &=& b^{a}\;, \qquad \; \; 
sb^{a} = 0 \;, \nonumber \\
s{\bar \omega}^{ab}_\mu & = & {\bar \varphi}^{ab}_\mu \;, \qquad  s {\bar \varphi}^{ab}_\mu =0\;, \nonumber \\
s { \varphi}^{ab}_\mu&  = & {\omega}^{ab}_\mu  \;, \qquad s {\omega}^{ab}_\mu = 0 \;.
\end{eqnarray}
Therefore, an equivalent shift may be performed in order to remove the referred term.
After that, one should ends up with the expression
\begin{eqnarray}
\int d^{4}x \left[ 
{\bar \varphi}^{ac}_{\mu} (\partial_\nu D^{ab}_{\nu} )
\varphi^{bc}_{\mu} - {\bar \omega}^{ac}_{\mu}  (\partial_\nu D^{ab}_{\nu} ) \omega^{bc}_{\mu}
+\gamma^{2} gf^{abc}A^{a}_{\mu}(\varphi^{bc}_{\mu} + {\bar \varphi}^{bc}_{\mu})
\right]\;,
\label{act9}
\end{eqnarray}
where the $\omega$ field must be regarded as being the shifted one. The functional integration
of the fermionic fields $(\bar{\omega},\omega)$ can easily be computed, leading to 
$\det\left[ \p_{\nu}D_{\nu} \right]$. In order to integrate out the fields
$(\bar{\varphi},\varphi)$ one has to define the sources $\bar{J}$ and $J$ as
\begin{eqnarray}
\bar{J}^{bc}_{\mu} ~=~ J^{bc}_{\mu} ~=~ \gamma^{2}gf^{abc}A^{a}_{\mu} 
\,,
\label{sourcesJJ}
\end{eqnarray}
so that the integral \eqref{act9} may be rewritten as
\begin{eqnarray}
\int d^{4}x \left[ 
{\bar \varphi}^{ac}_{\mu} (\partial_\nu D^{ab}_{\nu} ) \varphi^{bc}_{\mu} +
\bar{J}^{bc}_{\mu}\,\varphi^{bc}_{\mu} + J^{bc}_{\mu}\bar{\varphi}^{bc}_{\mu}
\right]\;.
\end{eqnarray}
The couple of fields $(\bar{\omega},\omega)$ were already integrated in the above expression.
Summing and subtracting the term 
$\bar{J}^{bc}_{\mu}(\partial_\nu D^{ab}_{\nu})^{-1}J^{ac}_{\mu}$ we can rewrite this integral
as following,
\begin{eqnarray}
\int d^{4}x \left\{ \left[ 
{\bar \varphi}^{ac}_{\mu}  + (\partial_\nu D^{ab}_{\nu} )^{-1}\bar{J}^{bc}_{\mu} \right]
(\partial_\nu D^{ab}_{\nu} )
\left[ \varphi^{bc}_{\mu} + (\partial_\nu D^{ab}_{\nu})^{-1} J^{ac}_{\mu} \right] -
\bar{J}^{bc}_{\mu}(\partial_\nu D^{ab}_{\nu})^{-1}J^{ac}_{\mu}
\right\}
\,.
\end{eqnarray}
Performing the shifts $ {\bar\varphi}^{ac}_{\mu}  + (\partial_\nu D^{ab}_{\nu}
)^{-1}\bar{J}^{bc}_{\mu} \to \bar{\varphi}^{\prime\, ac}_{\mu}$ and
$ \varphi^{bc}_{\mu} + (\partial_\nu D^{ab}_{\nu})^{-1} J^{ac}_{\mu} \to \varphi^{\prime\,
bc}_{\mu}$ one ends up with
\begin{eqnarray}
\int d^{4}x \left[ 
{\bar \varphi}^{\prime\, ac}_{\mu} (\partial_\nu D^{ab}_{\nu} ) \varphi^{\prime\, bc}_{\mu}  
- \bar{J}^{bc}_{\mu}(\partial_\nu D^{ab}_{\nu} )^{-1}J^{ac}_{\mu}
\right]\;.
\label{act10}
\end{eqnarray}
After all, one ends up with a Gaussian integration of the bosonic fields
$(\bar{\varphi}^{\prime},\varphi^{\prime})$, whose integral leads to $\det\left[ \p_{\nu}D_{\nu} \right]^{-1}$,
and with the horizon function in its nonlocal version. Therefore, in order to obtain the
localized version of the GZ action, starting from the GZ action of equation
\eqref{gzpartition2}, one should perform the process just described in the backward direction.

\subsubsection{The gap equation, or horizon condition:}

Back to the local formulation of the Gribov-Zwanziger action, the horizon condition \eqref{hc2}
takes the simpler form 
\begin{equation}
 \frac{\partial \mathcal{E}_v}{\partial\gamma^2}=0\;,   
\label{ggap}
\end{equation}
where $\mathcal{E}_{v}(\gamma)$ is the vacuum energy defined by,
\begin{equation}
 e^{-V\mathcal{E}_{v}}=\;Z_{GZ}\;  
\label{vce2} \;.
\end{equation}
The local action $S_{GZ}$ in eq.\eqref{sgz2} is known as the Gribov-Zwanziger action. It has
been shown to be renormalizable to all orders
\cite{Zwanziger:1988jt,Zwanziger:1989mf,Zwanziger:1992qr}.

\subsubsection{The gauge propagator:}

Finally, with the local version of the generating functional, the gluon and ghost propagator
could be computed. At first order in loop expansion, only quadratic terms in the fields of the
GZ action eq.\eqref{sgz2} have to be kept, while terms of great order will be ignored.
Therefore, performing the same step-by-step of the previous section, one would be able to
compute the gauge and ghost propagators, ending up with
\begin{eqnarray} 
\langle  A^a_\mu(k)  A^b_\nu(-k) \rangle  ~=~  \frac{k^2}{k^4 + 2Ng^2\gamma^4} \,\, \delta^{ab}
\left(\delta_{\mu\nu} - \frac{k_\mu k_\nu}{k^2}     \right)   \;, 
\label{glrgz1}
\end{eqnarray} 
for the gluon field, and 
\begin{eqnarray}
\left\langle  c^{a}(k) \overline{c}^{b}(-k)  \right\rangle ~=~ \mathcal G (k^2)_{ab}\;,
\end{eqnarray}
with
\begin{eqnarray}
\mathcal G (k^2) &=&
\frac{1}{k^2} + \frac{1}{V}\frac{1}{k^4} \frac{N g^2}{N^2 - 1} \int\frac{ \d^d q}{(2
\pi)^d} \frac{q^2}{q^4 + 2Ng^2\gamma^4} \left(\delta_{\mu\nu} - \frac{q_\mu q_\nu}{q^2}\right)
 \frac{(k-q)_\mu  k_\nu}{(k-q)^2} \nonumber\\
&=& \frac{1}{k^2}\left( 1 + \sigma(k) \right)\;,
\label{coloraway1}
\end{eqnarray}
for the ghost fields. Let us make, at this point, a brief analysis of \eqref{glrgz1} and
\eqref{coloraway1} in the IR regime. It is not difficult to see that in the deep IR regime the
gauge field propagator is strongly suppressed and tends to zero in the limit $k^2 \to 0$. As
can be checked in \eqref{Gribovprop0}, this behavior is shared by Gribov and Gribov-Zwanziger
approaches.

\subsubsection{The ghost propagator:}

Since we are performing a perturbative computation up to one-loop order, one must follow the
same step-by-step of the previous chapter in order to compute the ghost form factor. Therefore,
one should ends up with
\begin{eqnarray}
\sigma(k) ~=~ \frac{1}{V}\frac{1}{k^2} \frac{N g^2}{N^2 - 1} \int\frac{ \d^d q}{(2
\pi)^d} \frac{q^2}{q^4 + 2Ng^2\gamma^4} \left(\delta_{\mu\nu} - \frac{q_\mu q_\nu}{q^2}\right)
 \frac{(k-q)_\mu  k_\nu}{(k-q)^2}\,.
\end{eqnarray}
Note that the term linear in $q_{\mu}$ is zero, due to the transversal projector. Making use
of the identity $\int d^{d}q f(q)q_{\mu}q_{\nu}/q^{2} = 1/d\,\int d^{d}q f(q)\delta_{\mu\nu}$,
one ends up with,
\begin{eqnarray}
\sigma(k) ~=~ \frac{1}{V} \frac{N g^2 (d-1)}{d(N^2 - 1)} \int\frac{
\d^d q}{(2 \pi)^d} \frac{q^2}{q^4 + 2Ng^2\gamma^4} \frac{1}{(k-q)^2}\,.
\end{eqnarray}
Taking the limit $k \to 0$, we have
\begin{eqnarray}
\sigma(0) ~=~ \frac{1}{V} \frac{N g^2 (d-1)}{d(N^2 - 1)} \int\frac{
\d^d q}{(2 \pi)^d} \frac{1}{q^4 + 2Ng^2\gamma^4}\,,
\end{eqnarray}
which is a divergent integral and the ghost propagator is enhanced, just as in the Gribov
approach. However, before effectively solving the integral one has to fix the
Gribov parameter $\gamma^{2}$ dynamically, through the horizon condition \eqref{hc2} computed
with the quadratic generating functional.

\section{A brief introduction to the refined version of GZ}

Recently, a refinement of the Gribov-Zwanziger action has been worked out by the authors
\cite{Dudal:2007cw,Dudal:2008sp,Dudal:2011gd,Dudal:2008rm}, by taking into account the existence of certain
dimension two condensates\footnote{See \cite{Gracey:2010cg,Thelan:2014mza} for a recent
detailed investigation on  the structure of these condensates in color space.}.  The Refined
Gribov-Zwanziger (RGZ) action reads \cite{Dudal:2007cw,Dudal:2008sp,Dudal:2011gd,Dudal:2008rm}
\begin{equation}
S_{RGZ} = S_{GZ} + \int d^4x \left(  \frac{m^2}{2} A^a_\mu A^a_\mu  - \mu^2 \left( {\bar
\varphi}^{ab}_{\mu}  { \varphi}^{ab}_{\mu} -  {\bar \omega}^{ab}_{\mu}  { \omega}^{ab}_{\mu}
\right)   \right)  \;,  
\label{rgz}
\end{equation}
where $S_{GZ}$ stands for the Gribov-Zwanziger action,  eq.\eqref{sgz2}.  As much as the Gribov
parameter $\gamma^2$, the massive parameters $(m^2, \mu^2)$ have a dynamical origin, being
related to the existence of the dimension two condensates $\langle A^a_\mu A^a_\mu \rangle$ and
$\langle {\bar \varphi}^{ab}_{\mu}  { \varphi}^{ab}_{\mu} -  {\bar \omega}^{ab}_{\mu}  {
\omega}^{ab}_{\mu}  \rangle$, \cite{Dudal:2007cw,Dudal:2008sp,Dudal:2011gd,Dudal:2008rm}. 
The gluon propagator obtained from the RGZ action turns out to be suppressed in the infrared region, attaining a non-vanishing value at zero momentum, $k^2=0$, {\it i.e.}
\begin{eqnarray} 
\langle  A^a_\mu(k)  A^b_\nu(-k) \rangle  & = &  \delta^{ab}  \left(\delta_{\mu\nu} - \frac{k_\mu k_\nu}{k^2}     \right)   {\cal D}(k^2) \;, \label{glrgz} \\
{\cal D}(k^2) & = & \frac{k^2 +\mu^2}{k^4 + (\mu^2+m^2)k^2 + 2Ng^2\gamma^4 + \mu^2 m^2}  \;. \label{Dg}
\end{eqnarray} 
One should note that the gluon propagator obtained in the Gribov-Zwanziger approach differ from
the refined one by the terms proportional to $\mu^{2}$ and $m^{2}$. So, putting these
parameters to zero the GZ gluon propagator is recovered, with the well known suppressed
behavior in the IR regime, going to zero for $k^{2} \to 0$.
Also, unlike the case of the GZ action, the ghost propagator stemming from the Refined theory is not enhanced in the deep infrared:
\begin{equation}
{\cal G}^{ab}(k^2) = \langle  {\bar c}^{a} (k)  c^b(-k) \rangle \Big|_{k\sim 0} \; \sim \frac{\delta^{ab}}{k^2}   \;.\label{ghrgz} 
\end{equation}
The infrared behaviour of the  gluon and ghost propagators obtained from the RGZ  action turns
out to be in very good agreement with the most recent  numerical lattice simulations on large
lattices \cite{Cucchieri:2007rg,Cucchieri:2008fc,Cucchieri:2011ig}. Moreover, the numerical
estimates  \cite{Cucchieri:2011ig}  of the parameters $(m^2,\mu^2,\gamma^2)$ show that the RGZ
gluon propagator \eqref{glrgz} exhibits complex poles and violates  reflection positivity. This
kind of two-point function lacks the  K{\"a}ll{\'e}n-Lehmann spectral representation and cannot
be associated with the propagation of physical particles. Rather, it indicates that, in the
nonperturbative infrared region, gluons are not physical excitations of the spectrum of the
theory, {\it i.e.} they are confined.  It is worth mentioning here that the RGZ gluon
propagator has been employed in analytic calculation of the first glueball states
\cite{Dudal:2010cd,Dudal:2013wja}, yielding results which compare well with the available
numerical simulations as well as with other approaches, see \cite{Mathieu:2008me} for an
account on this topic. The RGZ gluon propagator has also been used in order to  study the
Casimir energy within the MIT bag model \cite{Canfora:2013zna}. The resulting energy has the
correct expected confining behaviour. Applications  of the RGZ theory at finite temperature can
be found in  \cite{Fukushima:2013xsa,Canfora:2013kma}.

\section{The BRST breaking}

One important aspect of both GZ and RGZ theories is that they exhibit a soft breaking of the
BRST symmetry. Indeed, it has been extensively studied that the breaking of the BRST symmetry
is intimately connected with the restriction of the domain of integration of the gauge field to
the region inside the Gribov horizon
\cite{Baulieu:2008fy,Zwanziger:2009je,Sorella:2009vt,Zwanziger:1993dh,vonSmekal:2008en,Dudal:2010hj,Dudal:2009xh,Sorella:2010it,Capri:2010hb,Dudal:2012sb,Reshetnyak:2013bga}.

In fact, considering either the GZ action \eqref{sgz2} or the RGZ action \eqref{rgz}, one
should be able to prove that the BRST variation of both of these actions is not zero, but
rather it equals an integrated polynomial of order smaller than $4$ ({\it i.e.} the space-time
dimension) and proportional to $\gamma^{2}$ \cite{Dudal:2007cw,Dudal:2008sp,Dudal:2011gd,Dudal:2008rm}.
Namely, 
\begin{equation}
s S_{GZ} ~=~ s S_{RGZ} ~=~ \gamma^2 \Delta  \;, \label{brstbrr}
\end{equation}
with
\begin{equation}
\Delta = \int d^{4}x \left( - gf^{abc} (D_\mu^{am}c^m) (\varphi^{bc}_{\mu} + {\bar \varphi}^{bc}_{\mu})   + g f^{abc}A^a_\mu \omega^{bc}_\mu            \right)  \;. \label{brstb1}
\end{equation}
To check the above statement, one has to consider the BRST variation rule of each field as the
one given in \eqref{brst0}.
%
%

Notice that the breaking term $\Delta$ is of dimension two in the fields and, as such, is said
to be a soft breaking. 
Equation \eqref{brstbrr} can be translated into a
set of softly broken Slavnov-Taylor  identities which ensure the all order renormalizability of
both GZ and RGZ actions. The presence of the soft breaking term $\Delta$ turns out to be
necessary in order to have a confining gluon propagator which attains a non-vanishing value at
zero momentum, eqs.\eqref{glrgz},\eqref{Dg}, in agreement with the lattice data
\cite{Cucchieri:2007rg,Cucchieri:2008fc,Cucchieri:2011ig}. It is worth underlining that this
property is deeply related to the soft breaking of the BRST symmetry. In fact, the
non-vanishing of the propagator at zero momentum relies on  the parameter $\mu^2$, which
reflects the existence of the   BRST-exact dimension-two condensate
\cite{Dudal:2007cw,Dudal:2008sp,Dudal:2011gd,Dudal:2008rm}.
Recently, the breaking of the BRST symmetry in the IR regime was firstly observed on the
lattice, as can be checked in \cite{Cucchieri:2014via}, by making use of the possibility of
fixing the (minimal) Landau gauge on the lattice. To that end, the authors investigated if the
so-called Bose-ghost propagator, at zero temperature, is zero or not. Such Bose-ghost
propagator can be read as
\begin{eqnarray}
{\cal Q}^{abcd}_{\;\;\mu\nu}(x,y) &=& \left\langle \omega^{ab}_{\mu}\bar{\omega}^{cd}_{\nu} +
\varphi^{ab}_{\mu} \bar{\varphi}^{cd}_{\nu}  \right\rangle 
\nonumber \\
&=&
\left\langle s \left( \varphi^{ab}_{\mu} \bar{\omega}^{cd}_{\nu} \right)  \right\rangle 
\;.
\label{boseghostprop}
\end{eqnarray}
Note that this is a BRST exact quantity and as
such shall be zero for a BRST symmetric theory. Otherwise, if the Bose-ghost propagator is not
zero, then it is an evidence that the BRST symmetry is broken. This Bose-ghost propagator has
been proposed as a carrier of long-range confining force in the minimal Landau gauge
\cite{Zwanziger:2009je}. In order to investigate the Bose-ghost propagator the authors of
\cite{Cucchieri:2014via} noticed that the quantity \eqref{boseghostprop} may be written as
\begin{eqnarray}
{\cal Q}^{abcd}_{\;\;\mu\nu}(x,y) ~=~  \left\langle {\cal R}^{ab}_{\;\;\mu}
{\cal R}^{cd}_{\;\;\nu}
\right\rangle \;,
\end{eqnarray}
where
\begin{eqnarray}
{\cal R}^{ab}_{\;\;\mu} ~=~ -g\int d^{4}z\; \left( {\cal M}^{-1}
\right)^{ab}f^{abc}A^{c}_{\mu}\;.
\end{eqnarray}
Such quantity may be accessed by taking the inverse of the FP operator for the gauge propagator
within the Gribov restriction. One must be careful to interpret these results: there is no
consistent proof of the equivalence of the minimal Landau gauge on the lattice and the usual
analytical Landau gauge, so far.


\chapter{The Yang-Mills $+$ Higgs field theory}
\label{The Yang-Mills $+$ Higgs field theory}

\section{Introduction}
\label{phasediag}

As mentioned in the Introduction, the confinement feature of QCD seems to be directly linked
to the existence of a (remnant) global symmetry: in the Linear Sigma model (LSM) that is the
$SO(N)$ symmetry, with $N$ standing for the number of flavors of the scalar field; in the
Yang-Mills theory coupled to a (static) matter field, such as the Higgs field, the center
symmetry $Z_{N}$ is the remaining global symmetry that has to be checked.

In this chapter we are going to discuss the specific model of Yang-Mills theory coupled to the
Higgs field. Specifically, the $SU(2)+$ Higgs and the Electroweak $SU(2)\times U(1)+$ Higgs
gauge theory will be analyzed. Here we adopt a perturbative analytical approach, by accounting
for non-perturbative effects through the quantization mechanism proposed by
Gribov\footnote{For details about the Gribov and Gribov-Zwanziger approaches the reader is
pointed to chapter \ref{usefulbkground} and advised to read references cited therein.}. Of
course, this is not the first time that such a model is studied. Instead of that, there
exist a vast bibliography concerning this topic, hanging from lattice works
\cite{Fradkin:1978dv,Lang:1981qg,Langguth:1985dr,Azcoiti:1987ua,Caudy:2007sf,Bonati:2009pf,Maas:2010nc,Maas:2012zf,Greensite:2011zz}
to the mean-field approach \cite{Horowitz:1983sr,Damgaard:1985nb,Baier:1986sa}.

Before effectively entering into details of our approach, we would like to briefly discuss the
work by Fradkin-Shenker on the lattice \cite{Fradkin:1978dv}, where the gauge field was
considered to be coupled to the Higgs field frozen in its state of vacuum configuration. For
such an end, in the next subsection we are going to present a summary of their work, with
details that may help us to understand differences between their discrete and our analytical
approach, leaving us in comfortable position to compare both results. Our results are shortly
exposed at the end of each section of this chapter. At the end of the chapter our conclusions
are exposed with a comparison between the referred lattice work of Fradkin-Shenker.

\subsection{Fradkin \& Shenker's results}
\label{FSresults}

By making use of a discrete space-time, called lattice, Fradkin \& Shenker reported a work on
the study of phase diagrams of gauge theories coupled to Higgs fields, \cite{Fradkin:1978dv}.
In order to properly address the feature of phase transition of gauge-Higgs theories, the
radial part of the scalar fields is considered to be frozen at its vacuum configuration state,
\begin{eqnarray}
{\phi}^{2} ~=~ \nu^2\,,
\label{untgauge}
\end{eqnarray}
by imposing the unitary gauge. The action describing this lattice Yang-Mills + Higgs fields
theory is given by
\begin{eqnarray}
S[\phi(\vec{r});U_{\mu}(\vec{r})] &=& \frac{K}{2} \sum_{(\vec{r},\,\mu\nu)} \Tr \bigg[
U_{\mu}(\vec{r}) U_{\nu}(\vec{r} + \hat{e}_{\mu}) U^{\dagger}_{\mu}(\vec{r} + \hat{e}_{\nu})
U^{\dagger}_{\nu}(\vec{r}) \bigg] + c.c. +
\nonumber \\
&&
\frac{\beta}{2} \sum_{\vec{r},\,\mu} \bigg[ \phi(\vec{r})\cdot D\left\{ U_{\mu}(\vec{r})
\right\} \cdot \phi^{\dagger}(\vec{r} + \hat{e}_{\mu}) + c.c. \bigg]\,,
\label{latticeaction}
\end{eqnarray}
making use of their notation, where $K$ stands for the inverse of the squared coupling
constant, $K = 1/g^{2}$, and $\beta$ stands for the squared fixed norm of the
scalar fields, $\beta = \nu^{2}$. The lattice is composed of sites, labeled by $\vec{r}$, and
of links, whose starting point is on the lattice site $\vec{r}$ with ending point on $\vec{r} +
\hat{e}_{\mu}$, with $\hat{e}_{\mu}$ denoting the fundamental direction vector. So a link is
labeled by $(\vec{r},\,\mu)$. 
In the action \eqref{latticeaction}, $U_{\mu}(\vec{r})$ denotes the gauge group element that
lives on the lattice link $(\vec{r},\,\mu)$, while $\phi(\vec{r})$ accounts for the scalar
fields living on the lattice site $\vec{r}$; $D\left\{U_{\mu} (\vec{r})\right\}$ accounts for
the group representation of the gauge link $U_{\mu}(\vec{r})$, and summation is taken over the
plaquette $(\vec{r},\,\mu\nu)$, which is defined as
\begin{eqnarray}
U_{\mu}(\vec{r}) U_{\nu}(\vec{r} + \hat{e}_{\mu}) U^{\dagger}_{\mu}(\vec{r} + \hat{e}_{\nu})
U^{\dagger}_{\nu}(\vec{r})\,.
\end{eqnarray}

In the example case of the (\emph{compact}) Electromagnetism quantum field theory, QED, the gauge
link is given by
\begin{eqnarray}
U_{\mu}(\vec{r}) ~=~ \exp\left[ i aeA_{\mu}(\vec{r})  \right]\,,
\end{eqnarray}
with $a$ denoting the lattice spacing --- the size of the link between two neighbor sites ---
and $e$ the electromagnetic coupling constant. The continuum limit is recovered for $a \to 0$
and the gauge sector of the action \eqref{latticeaction} goes as
\begin{eqnarray}
S[A] ~=~ \int d^{4}x \frac12 \Tr\bigg[ F_{\mu\nu}F_{\mu\nu} \bigg]\,,
\end{eqnarray}
with $F_{\mu\nu}$ standing for the electromagnetic field strength: $F_{\mu\nu} =
\p_{\mu}A_{\nu} - \p_{\nu}A_{\mu}$.

For the general case, the gauge group link $U_{\mu}(\vec{r})$ should transform under the $SU(N)$ gauge transformation as
\begin{eqnarray}
U_{\mu}(\vec{r}) ~\to~ G(\vec{r}) U_{\mu}(\vec{r}) G^{\dagger}(\vec{r})\,,
\label{gtlink}
\end{eqnarray}
while the scalar field transforms as 
\begin{eqnarray}
\phi(\vec{r}) ~\to~ D\left\{G(\vec{r})\right\}\phi(\vec{r})\,,
\label{gtscalar}
\end{eqnarray}
for $G(\vec{r}) \in SU(N)$. The lattice action \eqref{latticeaction} is left invariant under
such gauge transformation, \eqref{gtlink} and \eqref{gtscalar}. Particularly, the trace taken
over any gauge link $U_{\mu}(\vec{r})$ of the $SU(2)$ gauge group is real. In the general case
of $SU(N)$ gauge group the complex conjugate term (\emph{c.c.}) has to be added to the action,
so to end up with a real trace \cite{Fradkin:1978dv,Greensite:2011zz,Montvay:1994cy}.

When the unitary gauge is imposed, by choosing configurations of the scalar fields obeying
\footnote{The unitary gauge is not necessarily given by \eqref{untgauge}. Take a look at the
next subsection \ref{suitablegauge} for more details} \eqref{untgauge}, the gauge symmetry
is broken down to a \emph{local} group of symmetry, named the center symmetry $Z_{N}$. This is
a subgroup of the broken gauge group $SU(N)$ and whose elements commute with every element
of $SU(N)$. For instance, in the Georgi-Glashow model the
$SU(2)$ gauge group is spontaneously broken to the Abelian $U(1)$ group, after fixing the
unitary gauge, leaving the center symmetry $Z_{2}$ unbroken. As mentioned in the Introduction,
the confinement phase transition is to be understood as the ordered/disordered magnetic phase
transition related to the center symmetry $Z_{N}$.

Fradkin \& Shenker make use of the Wilson loop in order to probe for phase transition,
although such obsevable is not sensible to the center symmetry breaking at infinite volume. The
Wilson loop is, in fact, an order parameter (as discussed in the introduction) in the sense
that it is a gauge invariant quantity and is sensible to the existence of three different
phases, since it is a measure of the self-energy of static quarks. Namely, the Wilson loop is
defined on the lattice by
\begin{eqnarray}
W ~=~ \left\langle \Tr \left[ \prod_{(\vec{r},\,\mu)\in\Gamma} U_{\mu}(\vec{r}) \right]
\right\rangle\,,
\end{eqnarray}
where $\Gamma$ is the set of links forming closed loops.

Fradkin \& Shenker could find that for the Higgs fields in a trivial representation of the
$SU(N)$ gauge group, such as the adjoint representation, the gauge symmetry is broken
after fixing the unitary gauge, with the Higgs field frozen in its vacuum configuration.
However, the center symmetry $Z_{N}$ will always be left intact \cite{Fradkin:1978dv}. By
varying the parameters of the theory, $\beta=\nu^{2}$ and $K=1/g^{2}$, they could find three possible phases,
by probing the Wilson loop. Namely,
\begin{itemize}
\item[\emph{i}.] A Higgs-mechanism-type phase, with massive gauge bosons and a \emph{perimeter
law fall-off} for the Wilson loop. This region corresponds to large $\beta$ and $K$ values;

\item[\emph{ii}.] An intermediate phase, called \emph{free-charge} or \emph{Coulomb phase},
where the Wilson loop indicates a finite-energy between two static sources, and massless gauge
bosons;

\item[\emph{iii.}] A confined phase, where the Wilson loop develops an \emph{area law
fall-off}, the gauge bosons are massive with no free charges.
\end{itemize}
It should be emphasized that such spectrum concerns the Higgs fields in the adjoint
representation, see Figure \ref{fig1}.

\begin{figure}
\begin{center}
\includegraphics[width=5cm]{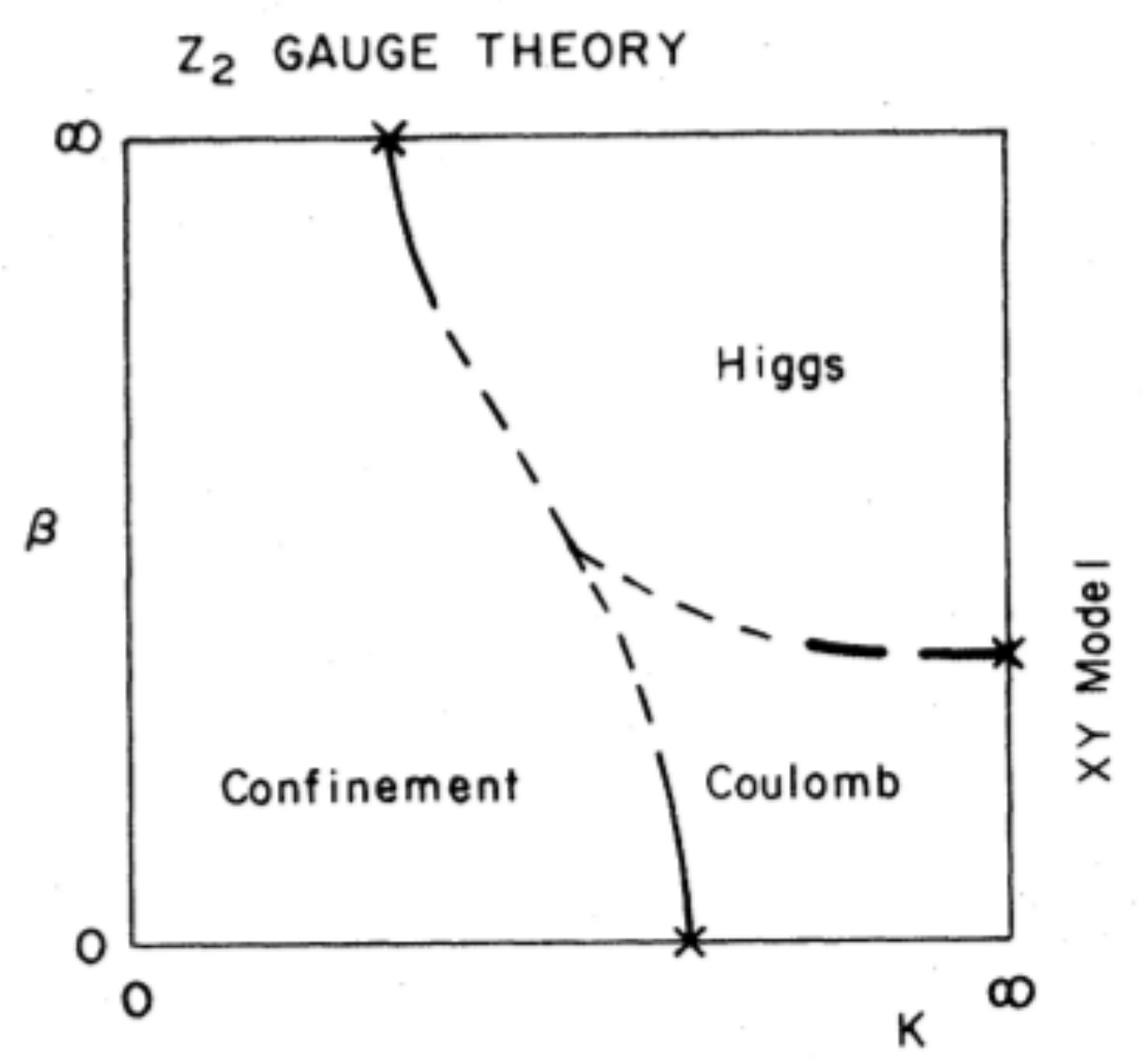}
\caption{Taken from Fradkin \& Shenker's work \cite{Fradkin:1978dv}. This phase diagram
corresponds to the non-Abelian gauge theory coupled to Higgs fields in the adjoint
representation. It also corresponds to the spectrum of the compact Abelian scenario.}
\label{fig1}
\end{center}
\end{figure}

For the Higgs filds in the fundamental representation, which is a non-trivial representation of
the gauge group, the situation is completely different. As they say, the unitary gauge
completely breaks the gauge symmetry, so that the center symmetry does not survive
\cite{Fradkin:1978dv}. In this case they found that the confinement-like regime and the
Higgs-like regime \emph{belong to the same phase of the theory}, on the whole configuration
space of the parameters. That is, there is no phase transition between the confinement- and
Higgs-like regimes. Furthermore, the transition between any two points of the confined regime
and of the Higgs-like regime are smoothly connected, which means that the \emph{vev} of local
composite operators develops continuously throughout the entire configuration space (some care
should be taken to the vicinity of $\beta =K=\infty$). By means of the Osterwalder-Seiler's
proof to the special case of fixed length of the Higgs field, they could prove the analyticity
of the whole configuration space. Those results were obtained for non-Abelian gauge fields
coupled to Higgs fileds in the fundamental representation. Things are considerably different
for the Abelian + Higgs gauge theory (for details, see \cite{Fradkin:1978dv}).

\subsection{A suitable gauge choice, but not the unitary one}
\label{suitablegauge}

Before properly starting to analyze the proposed model, let us state a few words on general features that are useful in this work. As mentioned before, the considered $SU(2)$ and $SU(2) \times U(1)$ Yang-Mills gauge theories are coupled to a scalar Higgs field. The Higgs field is considered in either the fundamental and the adjoint representation: in the $SU(2)$ case both the fundamental and the adjoint representation are analysed; while in the $SU(2)\times U(1)$ case only the the fundamental representation will be considered.

Usually the unitary gauge arises as a good choice when the Higgs mechanism is being treated, since in this gauge physical excitations are evident. However, instead of fixing the unitary gauge, we are going to choose the more general $\R_{\xi}$ gauge, whereby the unitary gauge is a special limit, $\xi \to \infty$, and the Landau gauge can be recovered when $\xi \to 0$. In the case of interest, the Landau gauge is imposed at the end of each computation. 

The $\R_{\xi}$ gauge condition reads,
\begin{eqnarray}
f^{a} ~=~ \p_{\mu}A^{a}_{\mu} -i\xi  \sum_{m,n} \varphi_{m}(x)(\tau^{a})_{mn} \nu_{n}   \;.
\label{Rxiguage}
\end{eqnarray}
Note that in eq. \eqref{Rxiguage} $i\varphi_{m}(x)(\tau^{a})_{mn} \nu_{n}$ is a possible form for the $b^{a}$ field, defined in \eqref{notideal}. The $\varphi$ field is defined as a small fluctuation around the vacuum configuration of the Higgs field $\Phi$,
\begin{eqnarray}
\Phi(x) ~=~ \varphi(x) + \nu  \;,
\end{eqnarray}
with vacuum expectation value $\langle \varphi \rangle = 0$. In order to apply the gauge condition \eqref{Rxiguage} we should follow the standard process described in many text books, \cite{Weinberg:1996kr,Peskin:1995ev,Srednicki:2007qs}. Rather than a Dirac delta function, as in eq. \eqref{genfp}, one should put a Gaussian function,
\begin{eqnarray}
\delta(f^{a}(A)) \to \exp \left( -\frac{1}{2\xi}\int \d^{d}x\; f^{a}f^{a} \right)   \;.
\label{higgsgauge}
\end{eqnarray}
In the limiting case of $\xi \to 0$ the Gaussian term \eqref{higgsgauge} oscillates very fast
around $f^{a} =0$ so that the Gaussian term \eqref{higgsgauge} behaves like a delta function,
ensuring the desired Landau gauge. On the other side, if $\xi \to \infty$ then we have the
unitary gauge. Needless to say, the limit $\xi \to 0$, recovering the Landau gauge, should be
applied at the very end of each computation\footnote{It is perhaps worthwhile pointing out here
that the Landau gauge is also a special case of the 't Hooft $R_\xi$ gauges, which have proven
their usefulness as being renormalizable and offering a way to get rid of the unwanted
propagator mixing between (massive) gauge bosons and associated Goldstone modes, $\sim A_\mu
\p_\mu \phi$. The latter term indeed vanishes upon using the gauge field transversality. The
upshot of specifically using the Landau gauge is that it allows to take into account potential
non-perturbative effects related to the gauge copy ambiguity.}.

In the present work we should deal with the Higgs field frozen at its vacuum configuration --- $i.e.$, $\Phi = \nu$. It is equivalent to replacing every Higgs field $\Phi$ in the action with its vacuum expectation value $\nu$. Since the $\R_{\xi}$ gauge fixing condition \eqref{Rxiguage} only depends on the vector gauge field $A_{\mu}$ and on the fluctuation of the scalar field $\varphi$ multiplied by the gauge parameter $\xi$, the Gribov procedure remains valid for the limit case $\xi \to 0$.


\section{$SU(2) +$Higgs field in the Fundamental representation}

In the present section nonperturbative effects of the $SU(2)+$Higgs model will be considered, by taking into account the existence of Gribov copies. The fundamental representation of the Higgs field, in $d=3$ and $d=4$, will be studied first. Subsequently, its adjoint representation, in $d=3$ and $d=4$, will be considered.

Working in Euclidean spacetime, the starting action of the current model reads
\begin{equation}
S=\int \d^{d}x\left(\frac{1}{4}F_{\mu \nu }^{a}F_{\mu \nu }^{a}+
(D_{\mu }^{ij}\Phi ^{j})^{\dagger}( D_{\mu }^{ik}\Phi ^{k})+\frac{\lambda }{2}\left(
\Phi ^{\dagger}\Phi-\nu ^{2}\right) ^{2} - \frac{(\partial _{\mu }A_{\mu}^{a})^{2}}{2\xi} +\bar{c}^{a}\partial _{\mu }D_{\mu }^{ab}c^{b}\right)  \;, 
\label{SYMhiggs}
\end{equation}
where the covariant derivative is given by
\begin{equation}
D_{\mu }^{ij}\Phi^{j} =\partial _{\mu }\Phi^{i} -ig \frac{(\tau^a)^{ij}}{2}A_{\mu }^{a}\Phi^{j} \;,
\end{equation}
and the vacuum expectation value of the scalar field is $\left\langle \Phi^{i} \right\rangle ~=~ \nu\delta^{2i} $, with $i=1,2$,
so that all components of the gauge field acquire the same mass, $m^2= \frac{g^2\nu^2}{2}$.

By following the procedure described in the subsection \ref{Gribov's issue}, the restriction to
the first Gribov region $\Omega$ relies on the computation of the ghost form factor, given by
equation \eqref{nopole}, and on the enforcement of the no-pole condition,
\eqref{ghstprop00}--\eqref{nopolestepfunct}. Since the presence of the scalar Higgs field does
not influence the procedure of quantizing the gauge field, due to the Landau gauge chosen with
the Higgs field frozen at its vacuum configuration\footnote{Take a look at the subsection
\ref{suitablegauge}.}, the non-local Gribov term, which is proportional to $\beta$, is not affected and one ends up with the action
\begin{eqnarray}
S\;' = S + \beta^* \sigma(0,A) -\beta^*   \;,
\label{effctactsu2}
\end{eqnarray}
with $S$ given by eq.\eqref{SYMhiggs}, the ghost form factor given by \eqref{nopole} and
$\beta^{\ast}$ stands for the Gribov parameter that solves the gap equation \eqref{gapeq},
where the function $f(\beta)$ is given in the following subsection.

\subsection{The gluon propagator and the gap equation}

In order to compute the gluon propagator up to one-loop order in perturbation theory let us
follow the steps described in the subsection \ref{A sign of confinement from gluon
propagator}. The condition of freezing the scalar field to its vacuum configuration is
equivalent to considering $\lambda$ large enough, so that the potential term of the scalar
field becomes a delta function of $\left(\Phi ^{\dagger}\Phi-\nu ^{2}\right)$: the quadratic
terms of the action \eqref{SYMhiggs} reads,
\begin{eqnarray}
S\,_{quad} ~=~ \int \d^{d}x\left( \frac{1}{4} { \left(  \partial_\mu A^a_\nu -\partial_\nu
A^a_\mu  \right)} ^2 - \frac{(\partial _{\mu }A_{\mu}^{a})^{2}}{2\xi}  +
\frac{g^{2}\nu^{2}}{4}A_{\mu }^{a}A_{\mu }^{a}  \right)
\;.
\label{quadf}
\end{eqnarray}

After implementing the Gribov's restriction of the gauge field configuration space
to the first Gribov region $\Omega$, and changing to the Fourier momentum space, one gets the following partition function

\begin{equation}
Z_{quad} ~=~ \int \frac{\d\beta  e^{\beta  }}{2\pi i\beta } [\d A] \; \exp \left\{  -\frac{1}{2} \int \frac{\d^{d}q}{(2\pi )^{d}}A_{\mu }^{a }(q)\mathcal{P}_{\mu \nu }^{ab }A_{\nu }^{b }(-q)   \right\}\;,  
\label{Zq1f}
\end{equation}
with
\begin{equation}
\mathcal{P}_{\mu \nu }^{ab } =\delta ^{ab }\left[  \delta _{\mu \nu }\left( q^{2}+\frac{\nu^{2}g^{2}}{2}\right) +\left( \frac{1}{\xi } -1  \right) q_{\mu }q_{\nu } +  \frac{4g^2\beta}{3dV} \frac{1}{q^2} \delta _{\mu \nu }  \right]   \;.  
\label{Pf}
\end{equation}

Computing the inverse of \eqref{Pf} and taking $\xi \to 0$ at the very end, so recovering the
Landau gauge, only the transversal component of will survive and the gauge propagators may be
identified as

\begin{equation}
\left\langle A_{\mu}^{a}(q)A_{\nu }^{b}(-q)\right\rangle ~=~  \delta^{ab}\frac{ q^2}{q^{4} + \frac{g^{2}\nu^{2}}{2} q^2
+  \frac{ 4g^2\beta^* }{3dV} }\left( \delta _{\mu\nu }-\frac{q_{\mu }q_{\nu }}{q^{2}}\right)   \;,
\label{propf0}
\end{equation}
whereby $\beta^{\ast}$ solves the gap equation, which is obtained by the means of the subsection \ref{Gribovimplementation}. After computing the Gaussian integral of the partition function and taking the trace over all indices, one ends up with the following partition function,
\begin{equation}  \label{Zf}
Z_{quad} ~=~ \int \frac{\d\beta}{2\pi i}\; e^{f(\beta)} ~=~ \e^{-V{\cal E}_{v}}   \;,
\end{equation}
whereby one reads the free-energy,
\begin{equation}
f(\beta)  ~=~  \beta - \ln \beta - \frac{3(d-1)V}{2} \int \frac{\d^d k}{(2\pi)^d} \; \ln\left( k^2 + \frac{g^2\nu^2}{2} +  \frac{4g^2\beta}{3dV} \frac{1}{k^2} \right)  \;, \label{ff}
\end{equation}
which is equivalent to \eqref{minusfreenergy}. In the thermodynamic limit the integral of equation \eqref{Zf} may be solved through the saddle-point approximation \eqref{gapeq} leading to the gap equation\footnote{We remind here that the derivative of the term ${\ln\beta}$  in expression \eqref{ff} will be neglected, for the derivation of the gap equation, eq.\eqref{gapf}, when taking the thermodynamic limit.}
\begin{equation}
\frac{2(d-1)}{d}g^2 \int \frac{\d^d q}{(2\pi)^d} \frac{1}{ q^{4} + \frac{g^{2}\nu^{2}}{2} q^2  +   \frac{2g^2\beta^{\ast}}{3dV} }  = 1  \;.
\label{gapf}
\end{equation}

In what follows the special case of $d=3$ and $d=4$ Euclidean space-time will be considered,
and the gap equation will be solved in both situations with a subsequent analysis of the gauge
field propagator, paying special attention to their pole: the applicability of Gribov's
confinement criterion (\emph{e.g.} the existence of complex conjugate poles) will be studied in
each space-time situation.

\subsection{The $d=3$ case}

Now, let us proceed with the solution  of the gap equation \eqref{gapf}. Since we are working in $d=3$, the gap equation contain a finite integral, easy to be computed, leading to
\begin{equation}
 \frac{4g^2}{3dV} \beta^{\ast} = \frac{1}{4} \left( \frac{g^2\nu^2}{2} -  \frac{g^4}{9\pi^2} \right)^2 \;. \label{solf}
\end{equation}
As done in the previous section, the analysis of the gluon propagator could be simplified by making explicit use of its poles. Namely,
\begin{equation}
\left\langle A_{\mu }^{a}(q)A_{\nu }^{b}(-q)\right\rangle
=  \frac{\delta^{ab}}{m^2_+-m^2_-}\left(  \frac{m^2_+}{q^2+m^2_+} -\frac{m^2_{-}}{q^2+m^2_-}  \right)
\left( \delta _{\mu\nu }-\frac{q_{\mu }q_{\nu }}{q^{2}}\right)  \;,  \label{decf}
\end{equation}
with
\begin{equation}
m^2_+ = \frac{1}{2} \left( \frac{g^2\nu^2}{2}  + \sqrt{\frac{g^6}{9\pi^2} \left(\nu^2-\frac{g^2}{9\pi^2}\right)}\; \right) \;,  \qquad m^2_- = \frac{1}{2} \left( \frac{g^2\nu^2}{2}  - \sqrt{\frac{g^6}{9\pi^2} \left(\nu^2-\frac{g^2}{9\pi^2}\right)} \;\right)
\label{massesf} \;.
\end{equation}
In this way, we may distinguish two regions in the $(\nu^2,g^2)$ plane:
\begin{itemize}
\item[\it i)] when $g^2 < 9\pi^2 \nu^2$ both masses $(m^2_+,m^2_-)$ are positive, as well as the residues. The gluon propagator, eq.\eqref{decf}, decomposes into two Yukawa modes. However, due to the relative minus sign in expression \eqref{decf} only the heaviest mode with mass $m^2_+$ represents a physical mode. We see thus that, for $g^2 < 9\pi^2 \nu^2$, all components of the gauge field exhibit a physical massive mode with mass $m^2_+$. This region is what can be called a Higgs phase. 

Let us also notice that, for the particular value $g^2=\frac{9\pi^2}{2}\nu^2$, corresponding to a vanishing Gribov parameter $\beta=0$, the unphysical Yukawa mode in expression  \eqref{decf} disappears, as $m^2_-~=~0$. As a consequence, the gluon propagator reduces to that of a single physical mode with mass $\frac{9\pi^2}{4}\nu^4$.
\item[\it ii)] when $g^2> 9 \pi^2 \nu^2$, the masses $(m^2_+,m^2_-)$ become complex. In this region, the gluon propagator, eq.\eqref{decf}, becomes of the Gribov type, displaying complex conjugate poles. All components of the gauge field become thus unphysical. This region corresponds to the confining phase.
\end{itemize}

\subsection{The $d=4$ case}
With quite the same process as for the $d=3$ case, let us analyse the poles of the gauge field propagator by solving the gap equation for $d=4$. To that end the following decomposition becomes useful
\begin{equation}
q^4 +  \frac{g^{2}\nu^{2}}{2} q^2
+ \frac{g^2}{3} \beta   ~=~ (q^2+m^2_+) (q^2+m^2_-) \;,  
\label{dec1}
\end{equation}
with
\begin{equation}
m^2_+ ~=~ \frac{1}{2} \left(\frac{g^2 \nu^2}{2} + \sqrt{\frac{g^4\nu^4}{4}  -\frac{4g^2}{3} \beta^*} \;  \right) \;,  \qquad    m^2_- ~=~ \frac{1}{2} \left(\frac{g^2 \nu^2}{2} -\sqrt{\frac{g^4\nu^4}{4}  -\frac{4g^2}{3} \beta^*} \;  \right)  \;.
\label{roots1}
\end{equation}
Making use of the $\MSbar$ renormalization scheme in $d=4-\varepsilon$ 
the gap equation \eqref{gapf} becomes
\begin{equation}
\left[1 + \frac{m^2_{-}}{m^2_+ -  m^2_-}\; \ln\left( \frac{m^2_-}{{\omu}^2} \right)  - \frac{m^2_+}{m^2_+ -  m^2_-}\; \ln\left( \frac{m^2_+}{{\omu}^2} \right)  \right] ~=~ \frac{32\pi^2}{3g^2}  \;. 
\label{gapf1}
\end{equation}
After a suitable manipulation we get a more concise expression for the gap equation
\begin{equation}
2 \sqrt{1-\zeta}\; \ln(a) = -  \left( 1 +  \sqrt{1-\zeta} \right) \; \ln\left( 1 +  \sqrt{1-\zeta} \right) +  \left( 1 -  \sqrt{1-\zeta} \right) \; \ln\left( 1 - \sqrt{1-\zeta} \right)  \;, 
\label{gapd1}
\end{equation}
where we have introduced the dimensionless variables
\begin{equation}
a ~=~ \frac{g^2 \nu^2}{4 \omu^2 e^{\left( 1 -\frac{32\pi^2}{3g^2} \right)}} \;, \qquad  \qquad \zeta ~=~ \frac{16}{3} \frac{\beta^*}{g^2\nu^4}  \geq0 \;, \label{vb1}
\end{equation}
with $0 \le \zeta < 1$ in order to have two real, positive, distinct roots $(m^2_+, m^2_-)$.
For $\zeta >1$, the roots $(m^2_+, m^2_-)$ become complex conjugate, and the gap equation takes the form
\begin{equation}
2 \sqrt{\zeta -1} \; \ln(a) =   -2 \; \arctan\left({\sqrt{\zeta-1}}\; \right)   - \sqrt{\zeta-1} \; \ln\;\zeta  \;. \label{d1}
\end{equation}
Moreover, it is worth noticing that both expressions \eqref{gapd1},\eqref{d1} involve only one function, {\it i.e.} they can be written as
\begin{equation}
2 \; \ln(a) = g(\zeta)  \;, \label{g}
\end{equation}
where for $g(\zeta)$ we might take
\begin{equation}
g(\zeta) =    \frac{1}{ \sqrt{1-\zeta}} \left(
- \left( 1 +  \sqrt{1-\zeta} \right) \; \ln\left( 1 +  \sqrt{1-\zeta} \right) +  \left( 1 -  \sqrt{1-\zeta} \right) \; \ln\left( 1 - \sqrt{1-\zeta} \right)  \right) \;,
\label{gex}
\end{equation}
which is a real function of the variable $\zeta \ge 0$. Expression \eqref{d1} is easily obtained from \eqref{gapd1} by rewriting it in the region $\zeta>1$.  In particular, it turns out that the function $g(\zeta)\leq -2\ln 2$ for all $\zeta \ge 0$, and strictly decreasing. As consequence, for each value of $a<\frac{1}{2}$, equation \eqref{g} has always a unique solution with $\zeta>0$.  Moreover, it is easy to check that $g(1)=-2$. Therefore,  we can distinguish ultimately three regions, namely
\begin{itemize}
\item[(a)]  when $a>\frac{1}{2}$,  eq.\eqref{g}  has no solution for $\zeta$. Since the gap equation \eqref{gapf} has been  obtained by applying the saddle point approximation in the thermodynamic limit, we are forced to set $\beta^{\ast}=0$.  This means that, when $a>\frac{1}{2}$, the dynamics of the system is such that the restriction to the Gribov region cannot be consistently implemented.  As a consequence, the standard Higgs mechanism takes place, yielding three components of the gauge field with mass $m^{2} ~=~ \frac{g^{2}\nu^{2}}{2}$. Note that, for sufficiently weak coupling $g^2$, $a$ will unavoidably be larger than $\frac{1}{2}$.

\item[(b)] when $\frac{1}{e}<a<\frac{1}{2}$, equation  \eqref{g} has a solution for  $0 \le \zeta <1$. In this region, the roots $(m^2_+, m^2_-)$  are real and the gluon propagator decomposes into the sum of two terms of the Yukawa type:
\begin{equation}
\left\langle A_{\mu }^{a }(q)A_{\nu }^{b }(-q)\right\rangle
=\frac{\delta^{ab}}{m^2_+-m^2_-}  \left(   \frac{m^2_+}{q^2+m^2_+} -   \frac{m^2_-}{q^2+m^2_-}   \right) 
 \left( \delta _{\mu
\nu }-\frac{q_{\mu }q_{\nu }}{q^{2}}\right)  \label{ffin} \;.
\end{equation}
Moreover, due to the relative minus sign in eq.\eqref{ffin} only the component proportional to
$m^{2}_{+}$ represents a physical mode.
\item[(c)]  for $a<\frac{1}{e}$, equation \eqref{g} has a solution for  $\zeta>1$. This
scenario will always be realized if $g^2$ gets sufficiently large, i.e.~at strong coupling. In
this region the roots  $(m^2_+, m^2_-)$  become complex conjugate and the gauge boson
propagator is of the Gribov type, displaying complex poles.  As usual, this can be interpreted
as the confining region.
\end{itemize}
In summary, we clearly notice that at sufficiently weak coupling, the standard Higgs mechanism will definitely take place, as $a>\frac{1}{2}$, whereas for sufficiently strong coupling, we always end up in a confining phase because then $a<\frac{1}{2}$.

Having obtained these results, it is instructive to go back where we originally started. For a fundamental Higgs, all gauge bosons acquire a mass that screens the propagator in the infrared. This effect, combined with a sufficiently small coupling constant, will lead to a severely suppressed ghost self energy, i.e.~the average of \eqref{nopole} (to be understood after renormalization, of course). If the latter quantity will a priori not exceed the value of 1 under certain conditions --- {\it i.e.}, satisfying the no-pole condition --- the theory is already well inside the Gribov region and there is no need to implement the restriction. Actually, the failure of the Gribov restriction for $a>\frac{1}{2}$ is exactly because it is simply not possible to enforce that $\sigma(0)=1$. Perturbation theory in the Higgs sector is \emph{in se} already consistent with the restriction within the 1st Gribov horizon. Let us verify this explicitly by taking the average of \eqref{nopole} with, as tree level input propagator, a transverse Yukawa gauge field with mass $m^2=\frac{g^2\nu^2}{2}$. Using that there are 3 transverse directions\footnote{We have been a bit sloppy in this paper with the use of dimensional regularization. In principle, there are $3-\epsilon$ transverse polarizations in $d=4-\epsilon$ dimensions. Positive powers in $\epsilon$ can (and will) combine with the divergences in $\epsilon^{-1}$ to change the finite terms. However, as already pointed out before, a careful renormalization analysis of the Gribov restriction is possible, see e.g.~\cite{Dudal:2010fq,Vandersickel:2012tz} and this will also reveal that the ``1'' in the Gribov gap equation will receive finite renormalizations, compatible with the finite renormalization in e.g.~$\sigma(0)$, basically absorbable  in the definition of $a$. } in $4d$, we have
\begin{eqnarray}
\sigma(0) =1-\frac{3g^2}{32\pi^2}\ln(2a)\;.
\end{eqnarray}
For $a>\frac{1}{2}$, the logarithm is positive and it is then evident that $\sigma(0)$ will not cross $1$, indicating that the theory already is well within the first Gribov horizon.

Another interesting remark is at place concerning the transition in terms of a varying value of $a$. If $a$ crosses $\frac{1}{e}$, the imaginary part of the complex conjugate roots becomes smoothly zero, leaving us with 2 coinciding real roots, which then split when $a$ grows. At $a=\frac{1}{2}$, one of the roots and its accompanying residue vanishes, to leave us with a single massive gauge boson. We thus observe that all these transitions are continuous, something which is in qualitative correspondence with the theoretical lattice predictions of the classic work \cite{Fradkin:1978dv} for a fundamental Higgs field that is ``frozen'' ($\lambda\to\infty$). Concerning the somewhat strange intermediate phase, {\it i.e.}  the one with a Yukawa propagator with a negative residue, eq.\eqref{ffin}, we can investigate in future work in more detail the asymptotic spectrum based on the BRST tools developed in \cite{Dudal:2012sb} when the local action formulation of the Gribov restriction is implemented. Recent works on the lattice confirm the existence of a cross-over region, where there is no line separating the ``phases'', as e.g. \cite{Maas:2013aia,Maas:2014pba} where the authors work in the non-aligned minimal Landau gauge and observe the transition between a QCD-like phase and a Higgs-like phase, in a region away from the cross-over region.

\subsection{The vacuum energy in the fundamental representation}

Let us look at the vacuum energy ${\cal E}_v$ of the system, which  can easily be read off from expression \eqref{Zf}, namely
\begin{equation}
{\cal E}_v = - \beta^* + \frac{9}{2} \int \frac{d^4k}{(2\pi)^4} \; \ln\left( k^2 + \frac{g^2\nu^2}{2} +\frac{\beta^*}{3} \frac{g^2}{k^2} \right)  \;, \label{ev}
\end{equation}
where $\beta^*$ is given by the gap equation \eqref{gapf}. Making use of the $\MSbar$ renormalization scheme, the vacuum energy may be written as:
\begin{itemize}
\item  for $a<\frac{1}{2}$, we have
\begin{eqnarray}
\frac{8}{9 g^4\nu^4}\; {\cal E}_v & = &  \frac{1}{32\pi^2} \left( 1 - \frac{32\pi^2}{3g^2} \right) - \frac{1}{2} \frac{\zeta}{32\pi^2} + \frac{1}{4}\frac{1}{32\pi^2} \left(  (4-2\zeta)\left( \ln( a) -\frac{3}{2} \right) \right)  \\ \nonumber & + &  \frac{1}{4}\frac{1}{32\pi^2} \left(   \left( 1+ \sqrt{1-\zeta} \right)^2 \ln  \left( 1+ \sqrt{1-\zeta} \right)
+ \left( 1- \sqrt{1-\zeta} \right)^2 \ln  \left( 1- \sqrt{1-\zeta} \right)
 \right)     \;, \label{v1}
\end{eqnarray}
where $\zeta$ is obtained through eqs.\eqref{g},\eqref{gex}. 

\item for $a>\frac{1}{2}$,
\begin{equation}
\frac{8}{9 g^4\nu^4}\; {\cal E}_v  =   \frac{1}{32\pi^2} \left( 1 - \frac{32\pi^2}{3g^2} \right)  + \frac{1}{32\pi^2} \left(  \left( \ln( a) -\frac{3}{2} \right) \right)   +   \frac{1}{32\pi^2} \ln2    \;. \label{v2}
\end{equation}
\end{itemize}
From these expressions we could check that the vacuum energy ${\cal E}_v(a)$ is a continuous  function of the variable $a$, as well as its first and second derivative, and that the third derivative develops a jump at $a=\frac{1}{2}$. We might be tempted to interpret this is indicating a third order phase transition at $a=\frac{1}{2}$. The latter value actually corresponds to a line in the $(g^2,\nu)$ plane according to the functional relation \eqref{vb1}. However, we should be cautious to blindly interpret this value. It is important to take a closer look at the validity of our results in the light of the made assumptions. More precisely, we implemented the restriction to the horizon in a first order approximation, which can only be meaningful if the effective coupling constant is sufficiently small, while simultaneously emerging logarithms should be controlled as well. In the absence of propagating matter, the expansion parameter is provided by $y\equiv\frac{g^2N}{16\pi^2}$ as in pure gauge theory. The size of the logarithmic terms in the vacuum energy (that ultimately defines the gap equations) are set by $m_+^2\ln\frac{m_+^2}{\omu^2}$ and $m_-^2\ln\frac{m_-^2}{\omu^2}$. A good choice for the renormalization scale would thus be $\omu^2\sim |m_+^2|$: for (positive) real masses, a fortiori we have $m_-^2<m_+^2$ and the second log will not get excessively large either because $m_-^2$ gets small and the pre-factor is thus small, or $m_-^2$ is of the order of $m_+^2$ and the log itself small. For complex conjugate masses, the size of the log is set by the (equal) modulus of $m_\pm^2$ and thus both small by our choice of scale.

Let us now consider the trustworthiness, if any, of the $a=\frac{1}{2}$ phase transition point.  For $a\sim\frac{1}{2}$, we already know that $\zeta\sim 0$, so a perfect choice is $\omu^2\sim m_+^2\sim\frac{g^2\nu^2}{2}$. Doing so, the $a$-equation corresponds to
\begin{equation}\label{aeq}
    \frac{1}{2}\sim e^{-1+\frac{4}{3y}}
\end{equation}
so that $y\sim 4$. Evidently, this number is thus far too big to associate any meaning to the ``phase transition'' at $a=\frac{1}{2}$. Notice that there is no problem for the $a$ small and $a$ large region. If $\nu^2$ is sufficiently large and we set $\omu^2\sim \frac{g^2\nu^2}{2}$ we have a small $y$, leading to a large $a$, i.e.~the weak coupling limit without Gribov parameter and normal Higgs-like physics. The logs are also well-tempered.    For a small $\nu^2$, the choice $\omu^2\sim\sqrt{g^2\theta^*}$ will lead to
\begin{equation}\label{aeq2}
a\sim (\textrm{small number})e^{-1+\frac{4}{3y}}
\end{equation}
so that a small $a$  can now be compatible with a small $y$, leading to a Gribov parameter dominating the Higgs induced mass, the ``small number'' corresponds to $\frac{g^2\nu^2}{\sqrt{g^2\theta^*}}$. Due to the choice of $\omu^2$, the logs are again under control in this case.

Within the current approximation, we are thus forced to conclude that only for sufficiently small or large values of the parameter $a$ we can probe the theory in a controllable fashion. Nevertheless, this is sufficient to ensure the existence of a Higgs-like phase at large Higgs condensate, and a confinement-like region for small Higgs condensate. The intermediate $a$-region is more difficult to interpret due to the occurrence  of large logs and/or effective coupling. Notice that this also might make the emergence of this double Yukawa phase at $a=\frac{1}{e}\approx 0.37$ not well established at this point.

\section{The $SU(2) +$Higgs field in the ajoint representation}
\label{Adjrep}

The Yang-Mills $+$ Higgs action with the scalar field in its adjoint representation may be
written as
\begin{eqnarray}
S ~=~ \int \,\d^{d}x \left[ \frac14 F^{a}_{\mu\nu}F^{a}_{\mu\nu} +
D^{ab}_{\mu}\Phi^{b}D^{ac}_{\mu}\Phi^{c} + \frac{\lambda}{2}\left( \Phi^{\dagger}\Phi - \nu^{2}
\right)^{2} - \frac{(\p A)^{2}}{\xi} + \bar{c}^{a}\p D^{ab}c^{b}
 \right]\;.
\end{eqnarray}
In the adjoint case the vacuum configuration that minimizes the energy is achieved by a
constant scalar field satisfying
\begin{equation}
\left\langle \Phi ^{a}\right\rangle ~=~ \nu \delta ^{a3}   \;,
\label{higgsv}
\end{equation}
leading to the standard Higgs mechanism. One should pay attention that the condition of 
degenerated vacuum, $\left\langle \Phi ^{a}\right\rangle \neq 0$, \eqref{higgsv} does not
automatically means that the \emph{unitary gauge} is being adopted. As has been emphasized
through out this chapter, the Higgs field is being considered to be frozen in its vacuum
configuration, which allows us to choose, under such hypothesis, the Landau gauge. Details
concerning this statement can be found in standard textbooks
\cite{Weinberg:1996kr,Peskin:1995ev,Ryder:1985wq} as well as in the section
\ref{suitablegauge}.

Just as in the fundamental case, fixing the scalar field in its vacuum configuration is
equivalent to consider a large enough value for the self-coupling $\lambda$, so that the
potential energy amounts to a delta function of $\left(\Phi ^{\dagger}\Phi-\nu ^{2}\right)$.
Thus, for the quadratic terms of the action, we have
\begin{equation}
S_{quad}=\int \d^{d}x\left( \frac{1}{4} { \left(  \partial_\mu A^a_\nu -\partial_\nu A^a_\mu  \right)} ^2 + b^a \partial_\mu A^a_\mu
+ \frac{g^{2}\nu ^{2}}{2}\left( A_{\mu }^{1}A_{\mu }^{1}+A_{\mu }^{2}A_{\mu
}^{2}\right)  \right)  \;. 
\label{quad}
\end{equation}

Following the standard procedure, before implementing the Gribov framework, one should notice
that the action \eqref{quad} has two independent sector, the \emph{diagonal} and the
\emph{off-diagonal} ones, corresponding respectively to the quadratic terms of $A_{\mu}^{3}$
and $A_{\mu}^{\alpha}$, with $\alpha=1,2$ (Greek letters should account for $1$ and $2$ in the
colour space). The existence of such split in the gauge sector reflects the breaking of the
gauge field, due to the gauge fixing after freezing the scalar field as 
\[
\left\langle \Phi^{a}\right\rangle ~=~ \nu \delta ^{a3}\,,
\]
leading to the existence of two massive vector modes, and a massless one. These massive vector
bosons and massless one may inferred from the following propagators,
\begin{equation}
\label{gluonoff}
\left\langle A_{\mu }^{\alpha }(p)A_{\nu }^{\beta }(-p)\right\rangle =\frac{\delta ^{\alpha \beta }}{p^{2}+m_{H}^{2}}\left( \delta _{\mu \nu }-%
\frac{p_{\mu }p_{\nu }}{p^{2}}\right) \;,
\end{equation}
from what, $m_{H}^{2} ~=~ g^{2}\nu ^{2}$ is the acquired mass after the symmetry breaking. The
massless mode amounts to the third component $A_\mu^3$, namely,
\begin{equation}
\left\langle A_{\mu }^{3}(p)A_{\nu }^{3}(-p)\right\rangle =\frac{1}{p^{2}}%
\left( \delta _{\mu \nu }-\frac{p_{\mu }p_{\nu }}{p^{2}}\right) \;. 
\label{zm}
\end{equation}

However, as was pointed out by Polyakov  \cite{Polyakov:1976fu}, the theory exhibits a different behaviour. The action \eqref{SYMhiggs} admits classical solitonic solutions, known as the 't Hooft-Polyakov monopoles\footnote{ These configurations are instantons in Euclidean space-time.} which play a relevant role in the dynamics of the model. In fact, it turns out that these configurations give rise to a monopole condensation at weak coupling, leading to a confinement of the third component $A^3_\mu$, rather than to a Higgs type behaviour, eq.\eqref{zm}, a feature also confirmed by lattice numerical simulations  \cite{Nadkarni:1989na,Hart:1996ac}.

Since our aim is that of analysing the nonperturbative dynamics of the Georgi-Glashow model by taking into account the Gribov copies, let's follow the procedure described in the subsection \ref{Gribovimplementation}. Due to the presence of the Higgs field in the adjoint representation, causing a breaking of the global gauge symmetry, the ghost two-point function has to be decomposed into two sectors, diagonal and off-diagonal:

\begin{equation}
\mathcal{G}^{ab}(k,A)=\left(
\begin{array}{cc}
\delta^{\alpha \beta}\mathcal{G}_{off}(k;A) & 0 \\
0 & \mathcal{G}_{diag}(k;A)
\end{array}
\right)
\end{equation}
where
\begin{eqnarray}
\mathcal{G}_{off}(k;A) 
&=&  \frac{1}{k^{2}}   \left( 1+\sigma _{off}(k;A)\right)       \approx \frac{1}{k^{2}}  \left( \frac{1}{1-\sigma
_{off}(k;A)}\right) 
\label{Goff} \;,  \\[5mm]
\mathcal{G}_{diag}(k;A) 
&=&  \frac{1}{k^{2}}  \left( 1+\sigma _{diag}(k;A)\right)  \approx \frac{1}{k^{2}}\left( \frac{1}{1-\sigma
_{diag}(k;A)}\right) \label{Gdiag} \;.
\end{eqnarray}

As we know, the quantities $\sigma_{off}(k;A), \; \sigma_{diag}(k;A)$ turn out to be  decreasing functions of the momentum $k$ and making use of the gauge field transversality, we have
\begin{eqnarray}
\sigma _{off}(0;A) &=&  \frac{g^{2}}{Vd}  \int \frac{\d^{d}q}{(2\pi )^{d}} \; \frac{\left( A_{\mu }^{3}(q)A_{\mu }^{3}(-q)+\frac{1}{2}A_{\mu }^{\alpha}(q)A_{\mu }^{\alpha }(-q)\right) }{q^{2}}  \;, 
\nonumber  \\
\sigma _{diag}(0;A) &=& \frac{g^{2}}{Vd}  \int {\frac{\d^{d}q}{(2\pi )^{d}} \; \frac{ \left( A_{\mu }^{\alpha }(q) A_{\mu }^{\alpha }(-q)\right) }{q^{2}}}    \;.
\label{sigma}
\end{eqnarray}
Once again, these expressions were obtained by taking the limit $k \rightarrow 0$ of eqs.\eqref{Goff},\eqref{Gdiag}, and by making use of the property
\begin{eqnarray}
A_{\mu }^{a}(q)A_{\nu }^{a}(-q) &=&\left( \delta _{\mu \nu }-\frac{q_{\mu }q_{\nu }%
}{q^{2}}\right) \omega (A)(q)   \nonumber \\
&\Rightarrow &\omega (A)(q)=\frac{1}{2}A_{\lambda }^{a}(q)A_{\lambda }^{a}(-q)
\end{eqnarray}
which follows from the transversality of the gauge field, $q_\mu A^a_\mu(q)=0$. Also, it is useful to remind that, for an arbitrary function $\mathcal{F}(p^2)$, we have
\begin{equation}
\int \frac{d^{3}p}{(2\pi )^{3}}\left( \delta _{\mu \nu }-\frac{%
p_{\mu }p_{\nu }}{p^{2}}\right) \mathcal{F}(p^2)=\mathcal{A}\;\delta _{\mu \nu } \,.
\end{equation}%
Therefore, the no-pole condition for the ghost function $\mathcal{G}^{ab}(k,A)$ is implemented by imposing that \cite{Gribov:1977wm,Vandersickel:2012tz,Sobreiro:2005ec}
\begin{eqnarray}
\sigma _{off}(0;A) &\leq &1\;, \nonumber \\
\sigma _{diag}(0;A) &\leq &1   \label{np} \;.
\end{eqnarray}



After that two different parameters are needed in order to implement the no-pole condition in
the action, so restricting the path integral to the first Gribov region. Thus, we are led to
the following action accounting for the Gribov ambiguities,
\begin{eqnarray}
S\;' = S + \beta^* \left( \sigma_{off}(0,A) - 1 \right) + \omega^* \left( \sigma_{diag}(0,A) - 1 \right)   \;.
\label{effctactadjoit1}
\end{eqnarray}
In the action \eqref{effctactadjoit1} $\beta^*$ and $\omega^*$ are given dynamically through its own gap equation.

\subsection{The gluon propagator and the gap equation}

In order to obtain the partition function associated to the action \eqref{effctactadjoit1}, the
first step is to consider the standard Yang-Mills partition function within the first Gribov
region, $\Omega$. Namely, this restricted partition function reads
\cite{Gribov:1977wm,Sobreiro:2005ec,Vandersickel:2012tz},
\begin{equation}
Z=\int {[DA_{\mu }]\delta (\partial A)(\det\mathcal{M})\theta (1-\sigma
_{diag}(A))\theta (1-\sigma _{off}(A))e^{-S_{YM}}}\,.
\end{equation}
Since we are interested in the study of the gluon propagators, we shall consider the quadratic
approximation for the partition function, namely  
\begin{eqnarray}
Z_{quad} &=&\int \frac{d\beta }{2\pi i\beta }\frac{d\omega }{2\pi i\omega }%
DA_{\mu }e^{\beta (1-\sigma _{diag}(0,A))}e^{\omega \left( 1-\sigma
_{off}(0,A)\right) }  \nonumber  \label{Zq} \\
&\times &e^{-\frac{1}{4}\int d^{d}x(\partial _{\mu }A_{\nu }^{a}-\partial
_{\nu }A_{\mu }^{a})^{2}-\frac{1}{2\xi }\int {d^{d}x(\partial _{\mu }A_{\mu
}^{a})^{2}-}\frac{{g^{2}\nu ^{2}}}{2}\int {d^{d}xA_{\mu }^{\alpha }A_{\mu
}^{\alpha }}} \;,
\end{eqnarray}
where use has been made of the integral representation
\begin{equation}
\theta(x) = \int_{-i \infty +\epsilon}^{i\infty +\epsilon} \frac{d\beta}{2\pi i \beta} \; e^{\beta x}  \;. \label{step}
\end{equation}
The partition function accounting only for quadratic terms of the action
\eqref{effctactadjoit1} can be written as
\begin{equation}
Z_{quad} ~=~ \int \frac{\d\beta e^{\beta }}{2\pi i\beta }\frac{\d\omega e^{\omega }} {2\pi i\omega }[\d A^{\alpha }][\d A^{3}] \; e^{-\frac{1}{2}  \int \frac{\d^{d}q}{(2\pi )^{d}}  \;  A_{\mu }^{\alpha }(q)\mathcal{P}_{\mu \nu }^{\alpha
\beta }A_{\nu }^{\beta }(-q)-\frac{1}{2}  \int \frac{\d^{d}q}{(2\pi )^{d}} \;  A_{\mu }^{3}(q)\mathcal{Q}_{\mu \nu }A_{\nu }^{3}(-q)},  
\label{Zq1}
\end{equation}
with
\begin{eqnarray}
\mathcal{P}_{\mu \nu }^{\alpha \beta } ~=~  \delta^{\alpha \beta } \left(\delta _{\mu \nu }
\left( q^{2}+\nu^{2}g^{2}\right) +\left( \frac{1}{\xi } -1 \right) q_{\mu }q_{\nu } +
\frac{2g^{2}}{Vd}\left( \beta +\frac{\omega }{2} \right) \frac{1}{q^{2}}\delta _{\mu \nu
}\right)  \label{P}  \;,
\end{eqnarray}
and
\begin{eqnarray}
\mathcal{Q}_{\mu \nu } ~=~ \delta_{\mu \nu }\left( q^{2} - \frac{2\omega
g^{2}}{Vd}\frac{1}{q^{2}}\right) +\left( \frac{1}{\xi }-1\right) q_{\mu }q_{\nu } \;.
\label{Q}
\end{eqnarray}
The parameter $\xi$ stands for the usual gauge fixing parameter, to be put to zero at the end in order to recover the Landau gauge. Evaluating  the inverse of the expressions \eqref{Q} and taking the limit $\xi\rightarrow 0$, the gluon propagators become
\begin{eqnarray}
\left\langle A_{\mu }^{3}(q)A_{\nu }^{3}(-q)\right\rangle &=&\frac{q^{2}}{q^{4}+  \frac{2\omega g^{2}}{Vd} }  \left( \delta _{\mu \nu }-\frac{q_{\mu }q_{\nu}}{q^{2}}\right)  \label{Pdiag} \;, 
\\
\left\langle A_{\mu}^{\alpha}(q) A_{\nu}^{\beta }(-q)\right\rangle  &=&  \delta^{\alpha\beta} 
\frac{q^{2}}{q^{2} \left( q^{2}+g^{2}\nu^{2}\right) +  \frac{2g^{2}}{Vd} \left(  \beta + \frac{\omega}{2}\right)}  
\left(\delta_{\mu\nu} - \frac{q_{\mu}q_{\nu}}{q^{2}}\right)  \;.
\label{NPoff} 
\end{eqnarray}
The off-diagonal sector of the gluon propagator can be put in a more convenient form, where its
poles are explicitly written,
\begin{eqnarray}
\left\langle A_{\mu}^{\alpha}(q) A_{\nu}^{\beta }(-q)\right\rangle 
&=&  \frac{\delta ^{\alpha \beta }}{m^2_+-m^2_-}   \left(  \frac{m^2_+}{q^2+m^2_+} -\frac{m^2_{-}}{q^2+m^2_-}  \right)
\left( \delta _{\mu\nu }-\frac{q_{\mu }q_{\nu }}{q^{2}}\right)  \;,  
\label{NPoff_f1}
\end{eqnarray}
with
\begin{equation}
m^2_+ = \frac{g^2\nu^2 + \sqrt{g^4 \nu^4 - 4 \tau}}{2}  \;,  \qquad m^2_- = \frac{g^2\nu^2 - \sqrt{g^4 \nu^4 - 4 \tau}}{2} 
\;, \qquad  \tau = \frac{2g^2}{Vd} \left( \beta +\frac{\omega}{2} \right)  \;.\label{masses}
\end{equation}

Since the Gribov parameters $(\beta, \omega)$ are fixed dynamically through the gap equation,
now we should integrate out the gauge field from equation \eqref{Zq1} and make use of the
saddle-point approximation, in the thermodynamic limit, which will gives us two gap equations,
enabling us to express $\beta$ and $\omega$ in terms of the parameters of the starting model,
{\it i.e.} the gauge coupling constant $g$ and the {\it vev} of the Higgs field $\nu$. That is,
firstly, we integrate out the gauge fields, obtaining
\begin{equation}
Z_{quad}=\int{\frac{d\beta}{2\pi i\beta}\frac{d\omega}{2\pi i\omega}}%
e^{\beta}e^{\omega}\left(\det\mathcal{Q}_{\mu\nu}\right)^{-\frac{1}{2}%
}\left(\det\mathcal{P}^{\alpha\beta}_{\mu\nu}\right)^{-\frac{1}{2}} \;.
\label{Zq2}
\end{equation}
By making use of the following property of functional determinants,
\begin{equation}
\left(\det \mathcal{A}_{\mu\nu}^{ab}\right)^{-\frac{1}{2}}=e^{-\frac{1}{2}%
\ln \det \mathcal{A}_{\mu\nu}^{ab}}=e^{-\frac{1}{2}Tr \ln \mathcal{A}%
_{\mu\nu}^{ab}} \;,
\end{equation}
for those determinants in expression (\ref{Zq2}), one gets
\begin{eqnarray}
\left( \det \mathcal{Q}_{\mu \nu }\right) ^{-\frac{1}{2}} &=&\exp \left[
-\int {\frac{d^{d}q}{(2\pi )^{d}}\ln \left( q^{2}+\frac{2\omega
g^{2}}{Vd}\frac{1}{q^{2}}\right) }\right] \;, \nonumber \\
\left( \det \mathcal{P}_{\mu \nu }^{\alpha \beta }\right) ^{-\frac{1}{2}}
&=&\exp \left[ -2\int {\frac{d^{d}q}{(2\pi )^{d}}\ln \left(
(q^{2}+g^{2}\nu^{2}) + \frac{g^{2}}{Vd} \left( 2\beta + \omega \right)
\frac{1}{q^{2}}\right) }\right] \;.
\end{eqnarray}
At the end, we have
\begin{equation}  \label{Zq3}
Z_{quad}=\int{\frac{\d\beta}{2\pi i}\frac{\d\omega}{2\pi i}}e^{f(\omega,\beta)} \;,
\end{equation}
with
\begin{eqnarray}\label{Zq4}
f(\omega ,\beta ) &=& \beta +\omega -\ln \beta -\ln \omega - \frac{(d-1)V}{2}  \int
{\frac{\d^{d}q}{(2\pi )^{d}}  \; \ln \left( q^{2}+\frac{2\omega
g^{2}}{Vd}\frac{1}{q^{2}}\right)
} \nonumber  \\
&-&  \frac{2(d-1)V}{2}  \int {\frac{\d^{d}q}{(2\pi)^{d}}  \; \ln \left((q^{2}+g^{2}\nu^{2}) + \frac{g^{2}}{Vd} \left( 2\beta + \omega \right)
\frac{1}{q^{2}}\right) } \;.
\end{eqnarray}
Since in the thermodynamic limit, as mentioned in the section \ref{Gribovimplementation}, the
integral \eqref{Zq3} can be solved through the saddle point approximation,
\begin{equation}
\frac{\partial f}{\partial \beta^*}=\frac{\partial f}{\partial \omega^*}=0 \;,
\end{equation}
 leading to
\cite{Gribov:1977wm,Sobreiro:2005ec,Vandersickel:2012tz}
\begin{equation}
Z_{quad}\approx e^{f(\beta^*,\omega^*)} \;,
\end{equation}
one gets the following two gap equations
\begin{eqnarray}
\frac{4(d-1)g^{2}}{2d} \int \frac{\d^{d}q}{(2\pi )^{d}} \frac{1}{q^{4}+\frac{2\omega^{\ast}g^{2}}{d}}  &=&1 \;, 
\label{gap1} \\
\frac{4(d-1)g^{2}}{2d}  \int \frac{\d^{d}q}{(2\pi )^{d}}\left( \frac{1}{q^{2}(q^{2}+g^{2}\nu^{2})+g^{2}\left( \frac{2\beta ^{\ast}}{d}+\frac{\omega^{\ast}}{d}\right) }\right) &=&1 \;.
\label{gap2}
\end{eqnarray}
Therefore, $\beta^*$ and $\omega^*$ can be expressed in terms of the parameters $\nu,g$. To solve the gap equations the denominator of eq.\eqref{gap2} can be decomposed into its poles, which is similar to \eqref{NPoff}--\eqref{masses}.

Let us assume the particular cases of $d=3$ and $d=4$ Euclidean space-times. In the light of
the gap equation in each situation, we will analyse what happens to the diagonal and
off-diagonal propagators.




\subsection{The $d=3$ case}
\label{3dsu2}




In the three-dimensional case both gap equations cause not many difficulties to be solved, as there are no divergences to be treated. Namely, the first gap equation, eq.\eqref{gap1}, leads to the following result,
\begin{eqnarray}
\omega^*(g)= \frac{3}{2^{11}\pi^4} \;g^6        \;,  \label{omega}
\end{eqnarray}
while the second one, given by eq.\eqref{gap2}, leads to
\begin{equation}
\tau =\beta^* \frac{2g^{2}}{3}+\omega^* \frac{g^{2}}{3}  = \left[ \frac{1}{2}g^{2}\nu^{2}-\frac{g^{4}}{32\pi^2 }\right] ^{2}  \;. \label{feq}
\end{equation}

Now we can look at the gluon propagators, \eqref{Pdiag} and \eqref{NPoff}, and analyse the different regions in the $(g,\nu)$ plane. Let us start by the diagonal component $A^3_\mu$. Namely, we have
\begin{equation}
\left\langle A_{\mu }^{3}(q)A_{\nu }^{3}(-q)\right\rangle =\frac{q^{2}}{%
q^{4}+\frac{2\omega^* g^{2}}{3}}\left( \delta _{\mu \nu }-\frac{q_{\mu }q_{\nu
}}{q^{2}}\right)  \label{Pdiagf} \;.
\end{equation}
One observes that expression \eqref{Pdiagf} turns out to be independent from the $vev$ $\nu$ of the Higgs field, while displaying two complex conjugate poles. This gauge component is thus of the Gribov type. In other words, the mode $A^3_\mu$ is always confined, for all values of the parameters $g, \nu$. Concerning now the off-diagonal gluon propagator \eqref{NPoff}, after decomposing it into two Yukawa modes  \eqref{NPoff_f1}, we could find the following regions in the $(g,\nu)$ plane:

\begin{itemize}
\item[\it i)] 
when $g^2 < 32 \pi^2 \nu^2$, corresponding to $\tau < \frac{g^2\nu^2}{4}$, both masses $m^2_+, m^2_-$ are real, positive and  different from each other. Moreover, due to the presence of the relative minus sign in expression \eqref{NPoff_f1}, only the heaviest mode with mass $m^2_+$ represents a physical excitation --- {\it i.e.}, despite the existence of two real positive poles, $m^2_{+}$ and $m^{2}_{-}$, only the contribution related to the $m^{2}_{+}$ pole has physical meaning.

It is also worth observing that, for the particular value $g=16 \pi^2 \nu^2$, corresponding to $\tau=0$, the unphysical mode in the decomposition \eqref{NPoff_f1} disappears. Thus, for that particular value of the gauge coupling, the off-diagonal propagator reduces to a single physical Yukawa mode with mass $16\pi^2\nu^4$.
\item[\it ii)] when $g^2>32\pi^2\nu^2$, corresponding to $\tau> \frac{g^2\nu^2}{4}$, all masses become complex and the off-diagonal propagator becomes of the Gribov type with two complex conjugate poles. This region, called Gribov region since all modes are of Gribov type, corresponds to a phase in which all gauge modes are said to be confined.
\end{itemize}
In summary, when the Higgs field is in the adjoint representation we could find two distinct regions. For $g^2<32\pi^2\nu^2$ the $A_3$ mode is confined while the off-diagonal propagator displays a physical Yukawa mode with mass $m^2_+$. This phase is referred to as the $U(1)$ symmetric phase \cite{Nadkarni:1989na,Hart:1996ac}. When $g^2>32\pi^2\nu^2$ all propagators are of the Gribov type, displaying complex conjugate poles leading to a confinement interpretation. According to \cite{Nadkarni:1989na,Hart:1996ac} this regime is referred to as the $SU(2)$ confined phase. 

Since our results were obtained in a semi-classical approximation ({\it i.e.}, lowest order in the loop expansion), let us comment on the validity of such approximation. In general, the perturbation theory is reliable when the \emph{effective coupling constant} is sufficiently small. The effective coupling depends, in $3d$, on the factor $\frac{g^2}{(4\pi)^{3/2}}$. However, since $g^2$ has mass dimension $1$ the effective coupling is not complete yet. In the presence of a mass scale $M$, the perturbative series --- for e.g.~the gap equation --- will organize itself automatically in a series in $G^2/M$. Let us analyse, for example, the case where $g^2 < 32 \pi^2 \nu^2$, the called the ``Higgs phase''. In this case the effective coupling will be sufficiently small when $\frac{g^2}{\nu^2(4\pi)^{3/2}}$ is small compared\footnote{The Higgs mass $\nu^2$ is then the only mass scale entering the game.} to $1$. Such condition is not at odds with the retrieved condition $g^2 < 32 \pi^2 \nu^2$. Next, assuming the coupling $g^2$ to get large compared to $\nu^2$, thereby entering the confinement phase with $cc$ masses, $g^2$ dominates everything, leading to a Gribov mass scale $\tau\propto g^8$, and an appropriate power of the latter will secure a small effective expansion parameter consistent with the condition $g^2 > 32 \pi^2 \nu^2$. We thus find that at sufficiently small and large values of $\frac{g^2}{\nu^2}$ our approximation and results are trustworthy.

\subsection{The $d=4$ case}

Let us start by considering the second gap equation, eq.\eqref{gap2}. Performing the decomposition described in eq.\eqref{masses} the referred gap equation becomes of the same form as the one obtained in the fundamental $d=4$ case, eq.\eqref{gapf1}. The difference between the fundamental and adjoint $d=4$ cases appears in the definition of the mass parameter $m^{2}_{\pm}$ (see eq.\eqref{roots1} and eq.\eqref{masses}, respectively). Namely,
\begin{equation}
\left(1 + \frac{m^2_-}{m^2_+ -  m^2_-}\; \ln\left( \frac{m^2_-}{{\omu}^2} \right)  - \frac{m^2_+}{m^2_+ -  m^2_-}\; \ln\left( \frac{m^2_+}{{\omu}^2} \right)  \right) = \frac{32\pi^2}{3g^2}  \;. \label{gap2bb}
\end{equation}
Introducing now the dimensionless variables\footnote{We introduced the renormalization group invariant scale $\lms$. }
\begin{eqnarray}
b ~=~  \frac{g^2\nu^2}{2 {\bar \mu}^2\; e^{\left(1-\frac{32\pi^2}{3g^2}\right)} } =\frac{1}{2\; e^{\left( 1 -\frac{272 \pi^2}{21 g^2} \right)}}  \;\frac{g^2\nu^2}{\lms^2  } \;, \qquad \text{and} \qquad 
 \xi  ~=~  \frac{4\tau}{g^4\nu^4}  \geq 0\;,
\label{vb}
\end{eqnarray}
with $0 \le \xi < 1 \;$. Proceeding as in the fundamental $d=4$ case, eq. \eqref{gap2bb} can be recast in the following form
\begin{equation}
2 \sqrt{1-\xi}\; \ln(b) = -  \left( 1 +  \sqrt{1-\xi} \right) \; \ln\left( 1 +  \sqrt{1-\xi} \right) +  \left( 1 -  \sqrt{1-\xi} \right) \; \ln\left( 1 - \sqrt{1-\xi} \right)  \;. \label{gapdd}
\end{equation}
or compactly,
\begin{equation}
2\ln b=g(\xi)\;. \label{gapddbis}
\end{equation}
Also here, in the adjoint $d=4$ case, eq.\eqref{gapddbis} remains valid also for complex conjugate roots, viz.~$\xi>1$. We are then led to the following cases.

\subsubsection{When $b<\frac{1}{2}$}
Using the properties of $g(\xi)$, it turns out that eq.\eqref{gapdd} admits a unique solution for $\xi$, which can be explicitly constructed with a numerical approach. More precisely, when the mass scale $g^2\nu^2$ is sufficiently smaller  than $\lms^2$, {\it i.e.}
\begin{equation}
g^2 \nu^2 < 2 \; e^{\left( 1 -\frac{272 \pi^2}{21 g^2} \right)} \; \lms^2   \;, \label{mh1}
\end{equation}
we have what can be called the $U(1)$ confined phase. In fact, in this regime the gap equation \eqref{gap1} leads to a non-null $\omega^*$, so that the diagonal component of the gauge field is said to be of the Gribov type, {\it i.e.} with confinement interpretation.

On the other side, the second gap equation \eqref{gapdd} splits this region in the two subregions:
\begin{itemize}
\item[(i)]  when $\frac{1}{e}<b<\frac{1}{2}$ equation  \eqref{gapdd} has a single solution with $0 \le \xi <1$. In this region, the roots $(m^2_+, m^2_-)$  are thus real and  the off-diagonal propagator decomposes into the sum of two Yukawa propagators.

However, due to the relative minus sign in eq.\eqref{NPoff_f1},  only the component with $m^{2}_+ $ pole can be associated to a physical mode, analogously as in the fundamental case. Due to the confinement of the third component $A^3_\mu$, this phase is recognized as the $U(1)$ confining phase. It is worth observing that it is also present in the $3d$ case, with terminology coined in \cite{Nadkarni:1989na}, see also \cite{Capri:2012cr}.

\item[(ii)]  for $b<\frac{1}{e}$, equation \eqref{gapdd} has a solution for  $\xi>1$. In this
region the roots  $(m^2_+, m^2_-)$  become complex conjugate and the off-diagonal gluon
propagator is of the Gribov type, displaying complex poles.  In this region all gauge fields
display a propagator of the Gribov type. This is recognized as the $SU(2)$ {\it confined-like}
regime.
\end{itemize}
Similarly, the above regions are continuously connected when $b$ varies. In particular, for $b \stackrel{<}{\to} \frac{1}{2}$, we obtain $\xi=0$ as solution.

\subsubsection{The case $b>\frac{1}{2}$}
Let us consider now the case in which $b>\frac{1}{2}$. Here, there is no solution of the equation \eqref{gapdd} for the parameter $\xi$, as it follows by observing that the left hand side of eq.\eqref{gapdd} is always positive, while the right hand side is always negative. This has a deep physical consequence. It means that for a Higgs mass $m_{Higgs}^2 = g^2\nu^2$ sufficiently larger than $\lms^2$, {\it i.e.}
\begin{equation}
g^2 \nu^2 > 2 \; e^{\left( 1 -\frac{272 \pi^2}{21 g^2} \right)} \; \lms^2   \;, \label{mh}
\end{equation}
the gap equation \eqref{gap2} is inconsistent. It is then important to realize that this is actually the gap equation obtained by acting with $\frac{\p}{\p \beta}$ on the vacuum energy  ${\cal E}_v= -  f(\omega ,\beta )$.
So, we are forced to set $\beta^*=0$, and confront the remaining $\omega$-equation, viz.~eq.\eqref{gap1},
which can be transformed into
\begin{eqnarray}
4 \; \ln(b) & = &  \frac{1}{\sqrt{1-\xi}} \left[ -\left( 1 +  \sqrt{1-\xi} \right) \; \ln\left( 1 +  \sqrt{1-\xi} \right) +  \left( 1 -  \sqrt{1-\xi} \right) \; \ln\left( 1 - \sqrt{1-\xi} \right) \right.  \nonumber \\
 &\; \;&-\left. \sqrt{1-\xi}\; \ln\xi - \sqrt{1-\xi} \ln 2\right]\equiv h(\xi)
 \label{gapddn}
\end{eqnarray}
after a little algebra, where $\xi=\frac{\omega^*}{g^4\nu^4}$. The behaviour of $h(\xi)$ for $\xi\geq0$ is more complicated than that of $g(\xi)$. Because of the $-\ln\xi$ contribution, $h(\xi)$ becomes more and more positive when $\xi$ approaches zero. In fact, $h(\xi)$ strictly decreases from $+\infty$ to $-\infty$ for $\xi$ ranging from 0 to $+\infty$.

It is interesting to consider first the limiting case $b \stackrel{>}{\to} \frac{1}{2}$, yielding $\xi\approx 1.0612$.   So, there is a discontinuous jump in $\xi$ (i.e.~the Gribov parameter for fixed $v$) when the parameter $b$ crosses the boundary value $\frac{1}{2}$.

We were able to separate the $b>\frac{1}{2}$ region as follows:
\begin{itemize}
\item[(a)] For $\frac{1}{2}<b<\frac{1}{\sqrt{\sqrt{2}e}}\approx 0.51$, we have a unique solution $\xi>1$, i.e.~we are in the confining region again, with all gauge bosons displaying a Gribov type of propagator with complex conjugate poles.
\item[(b)] For $\frac{1}{\sqrt{\sqrt{2}e}}< b<\infty$, we have a unique solution $\xi<1$, indicating again a combination of two Yukawa modes for the off-diagonal gauge bosons. The ``photon'' is still of the Gribov type, thus confined.
\end{itemize}
Completely analogous as in the fundamental case, it can be checked by addressing the averages of the expressions \eqref{sigma} that for $b>\frac{1}{2}$ and $\omega$ obeying the gap equation with $\beta=0$, we are already within the Gribov horizon, making the introduction of the second Gribov parameter $\beta$ obsolete.

It is obvious that the transitions in the adjoint case are far more intricate than in the earlier studied fundamental case. First of all, we notice that the ``photon'' (diagonal gauge boson) is confined according to its Gribov propagator. There is never a Coulomb phase for $b<\infty$. The latter finding can be understood again from the viewpoint of the ghost self-energy. If the diagonal gluon would remain Coulomb (massless), the off-diagonal ghost self-energy, cfr.~eq.\eqref{sigma}, will contain an untamed infrared contribution from this massless photon\footnote{The ``photon'' indeed keeps it coupling to the charged (= off-diagonal) ghosts, as can be read off directly from the Faddeev-Popov term $c^a \p_\mu D_\mu^{ab}c^b$.}, leading to an off-diagonal ghost self-energy that will cross the value 1 at a momentum $k^2>0$, indicative of trespassing the first Gribov horizon. This crossing will not be prevented at any finite value of the Higgs condensate $\nu$, thus we are forced to impose at any time a nonvanishing Gribov parameter $\omega$. Treating the gauge copy problem for the adjoint Higgs sector will screen (rather confine) the a priori massless ``photon''.

An interesting limiting case is that of infinite Higgs condensate, also considered in the lattice study of \cite{Brower:1982yn}. Assuming $\nu\to\infty$, we have $b\to\infty$ according to its definition \eqref{vb}. Expanding the gap equation \eqref{gapddn} around $\xi=0^+$, we find the limiting equation $b^4=\frac{1}{\xi}$, or equivalently $\omega^*\propto \lms^8/g^4\nu^4$. Said otherwise, we find that also the second Gribov parameter vanishes in the limit of infinite Higgs condensate. As a consequence, the photon becomes truly massless in this limit. This result provides ---in our opinion--- a kind of continuum version of the existence of the Coulomb phase in the same limit as in the lattice version of the model probed in \cite{Brower:1982yn}. It is instructive to link this back to the off-diagonal no pole function, see Eq.~\eqref{sigma}, as we have argued in the proceeding paragraph that the massless photon leads to $\sigma_{off}(0)>1$ upon taking averages. However, there is an intricate combination of the limits $\nu\to\infty$, $\omega^*\to 0$ preventing such a problem here. Indeed, we find in these limits, again using dimensional regularization in the $\MSbar$ scheme, that
\begin{align}\label{dv}
  \sigma_{off}(0) &= \frac{3g^2}4 \left(\int \frac{d^4q}{(2\pi)^4} \frac1{q^4+\frac{\omega^\ast g^2}2} + \int \frac{d^4q}{(2\pi)^4} \frac1{q^2(q^2+g^2\nu^2)+\frac{\omega^\ast g^2}4}\right) \nonumber\\
	&= -\frac{3g^2}{128\pi^2} \left(\tfrac12 \ln\frac{\omega^\ast g^2}{2\overline\mu^4} + \ln\frac{g^2\nu^2}{\overline\mu^2}-2\right) \nonumber \\
	&= -\frac{3g^2}{128\pi^2} \left(\tfrac12 \ln\frac{\omega^\ast g^6\nu^4}{2\overline\mu^8}-2\right) \stackrel{b^4=\xi^{-1}}{\longrightarrow} -\frac{3g^2}{64\pi^2} \ln 8g^2 + \frac12.
\end{align}
The latter quantity is always smaller than $1$ for $g^2$ positive, meaning that we did not cross the Gribov horizon. This observation confirm in an explicit way the intuitive reasoning also found in section 3.4 of \cite{Lenz:2000zt}, at least in the limit $\nu\to\infty$. The subtle point in the above analysis is that it is not allowed to naively throw away the 2nd integral in the first line of \eqref{dv} for $\nu\to\infty$. There is a logarithmic $\ln\nu$ ($\nu\to\infty$) divergence that conspires with the $\ln \omega^*$ ($\omega^*\to0$) divergence of the 1st integral to yield the final reported result. This displays that, as usual, certain care is needed when taking infinite mass limits in Feynman integrals.

\subsection{The vacuum energy in the adjoint representation}
As done in the case of the fundamental representation, let us work out the expression of the vacuum energy ${\cal E}_v$, for which we have the one loop  integral representation given by eq.\eqref{Zq4} multiplied by $-1$.
Making use of the $\MSbar$ renormalization scheme in $d = 4 - \varepsilon$ the vacuum energy becomes

\begin{eqnarray}\label{ddvac4}
\frac{{\cal E}_v}{g^4\nu^4}&=& -\frac{1}{g^2}-\frac{3\xi'}{128\pi^2}\left(\ln(2b^2\xi')-1\right)+\frac{3(4-2\xi)}{128\pi^2}\left(\ln b-\frac{1}{2}\right)\nonumber\\
&&+\frac{3}{128\pi^2}\left(\left(1-\sqrt{1-\xi}\right)^2\ln\left(1-\sqrt{1-\xi}\right)+\left(1+\sqrt{1-\xi}\right)^2\ln\left(1+\sqrt{1-\xi}\right)\right)\;,
\end{eqnarray}
where $b$ was introduced via its definition \eqref{vb}, while
\begin{equation}\label{ddvac5}
    \xi'=\frac{4\tau'}{g^4\nu^4}\,,\qquad \xi=\frac{4\tau}{g^4\nu^4} 
\qquad \text{and} \qquad 
\tau'=\frac{g^2\omega^*}{4}\,,\qquad \tau=g^2\left(\frac{\beta^*}{2}+\frac{\omega^*}{4}\right)    \;.
\end{equation}

\begin{figure}[h!]
\center
\includegraphics[width=8cm]{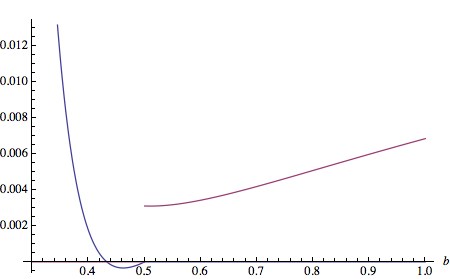}
\caption{Plot of the vacuum energy in the adjoint representation as a function of the parameter $b$. The discontinuity at $b=\frac{1}{2}$ is evident. }
\label{BS-d4}
\end{figure}

Since we found scenarios completely different for $b<1/2$ and $b>1/2$ with the scalar Higgs
field in its adjoint representation, it becomes of great importance analysing the plot of the
vacuum energy as a function of $b$. From Figure \ref{BS-d4} one can easily find out a clear jump for $b=1/2$, which can be seen as a reflection of the discontinuity of the parameter  $\xi$.

Investigating the functional \eqref{ddvac4} in terms of $\xi$ and $\xi'$, it is numerically (graphically) rapidly established there is always a solution to the gap equations $\frac{\p \mathcal{E}_v }{\p \xi}=\frac{\p \mathcal{E}_v }{\p \xi'}=0$ for $b<\frac{1}{2}$, but the solution $\xi^*$ is pushed towards the boundary $\xi=0$ if $b$ approaches $\frac{1}{2}$, to subsequently disappear for $b>\frac{1}{2}$ \footnote{The gap solutions correspond to a local maximum, as identified by analysing the Hessian matrix of 2nd derivatives.}. In that case, we are forced to return on our steps as in the fundamental case and conclude that $\beta=0$, leaving us with a single variable $\xi=\xi'$ and a new vacuum functional to extremize. There is, a priori, no reason for these 2 intrinsically different vacuum functionals be smoothly joined at $b=\frac{1}{2}$. This situation is clearly different from what happens when a potential has e.g.~2 different local minima with different energy, where at a first order transition the two minima both become global minima, thereafter changing their role of local vs.~global. Evidently, the vacuum energy does not jump since it is by definition equal at the transition.

Nevertheless, a completely analogous analysis as for the fundamental case will learn that $b=\frac{1}{2}$ is beyond the range of validity of our approximation\footnote{A little more care is needed as the appearance of two Gribov scales complicate the log structure. However, for small $b$ the Gribov masses will dominate over the Higgs condensate and we can take  $\omu$ of the order of the Gribov masses to control the logs and get a small coupling. For large $b$, we have $\beta^*=0$ and a small $\omega^*$: the first log will be kept small by its pre-factor and the other logs can be managed by taking $\omu$ of the order of the Higgs condensate.}. The small and large $b$ results can again be shown to be valid, so at large $b$ ($\sim$ large Higgs condensate) we have a mixture of off-diagonal Yukawa and confined diagonal modes and at small $b$ ($\sim$ small Higgs condensate) we are in a confined phase. In any case we have that the diagonal gauge boson is \emph{not} Coulomb-like, its infrared behaviour is suppressed as it feels the presence of the Gribov horizon.

\section{$SU(2)\times U(1)+$Higgs field in the fundamental representation}
\label{The Electroweak theory}

From now on in this work only the fundamental case of the Higgs field will be treated, for reasons relying on the physical relevance of the fundamental representation of this field. As a first step, we are going to present, as in the previous sections, general results for $d$-dimension. Afterwards, the $3$ and $4$-dimensional cases will be considered in the subsections \ref{d=3} and \ref{d=4}. The starting action of the $SU(2) \times U(1)+$Higgs field reads
\begin{eqnarray}
S ~=~ \int \d^{d}x  \;  \bigg(\frac{1}{4}  F_{\mu \nu }^{a} F_{\mu \nu }^{a}  +  \frac{1}{4} B_{\mu\nu} B_{\mu\nu} +{\bar c}^a \partial_\mu D^{ab}_\mu c^b - \frac{(\partial_\mu A^a_\mu)^{2}}{2\xi} 
 + {\bar c}\partial^2 c  - \frac{(\partial_\mu B_\mu)^{2}}{2\xi}  +
\nonumber \\
+
(D_{\mu }^{ij}\Phi^{j})^{\dagger}( D_{\mu }^{ik}\Phi^{k})+\frac{\lambda }{2}\left(\Phi^{\dagger}\Phi - \nu^{2}\right)^{2}   \bigg)  \;,
\label{Sf}
\end{eqnarray}
where the covariant derivative is defined by
\begin{equation}
D_{\mu }^{ij}\Phi^{j} =\partial _{\mu }\Phi^{i} - \frac{ig'}{2}B_{\mu}\Phi^{i} -   ig \frac{(\tau^a)^{ij}}{2}A_{\mu }^{a}\Phi^{j}  \;.
\end{equation}
and the vacuum expectation value (\textit{vev}) of the Higgs field is $\langle \Phi^{i} \rangle ~=~ \nu\delta^{2i}$.
The indices $i,j=1,2$ refer to the fundamental representation of $SU(2)$ and $\tau^a, a=1,2,3$, are the Pauli matrices. The coupling constants $g$ and $g'$ refer to the groups $SU(2)$ and $U(1)$, respectively. The field strengths $F^a_{\mu\nu}$ and $B_{\mu\nu}$ are given by
\begin{equation}
F^a_{\mu\nu} = \partial_\mu A^a_\nu -\partial_\nu A^a_\mu + g \varepsilon^{abc} A^b_\mu A^c_\nu \;, \qquad B_{\mu \nu} = \partial_\mu B_\nu -\partial_\nu B_\mu  \;.
\label{fs}
\end{equation}

In order to obtain the boson propagators only quadratic terms of the starting action are required and, due to the new covariant derivative, this quadratic action is not diagonal any more. To diagonalize this action one could introduce a set of new fields, related to the standard ones by
\begin{subequations} \begin{gather}
W^+_\mu = \frac{1}{\sqrt{2}} \left( A^1_\mu + iA^2_\mu \right) \;, \qquad W^-_\mu = \frac{1}{\sqrt{2}} \left( A^1_\mu - iA^2_\mu \right)  \;,
\label{ws} \\
Z_\mu =\frac{1}{\sqrt{g^2+g'^2} } \left(  -g A^3_\mu + g' B_\mu \right) \qquad \text{and}\qquad A_\mu =\frac{1}{\sqrt{g^2+g'^2} } \left(  g' A^3_\mu + gB_\mu \right) \;.
\label{za}
\end{gather} 
\end{subequations}
The inverse relation can be easily obtained.
With this new set of fields the quadratic part of the action reads,
\begin{eqnarray}
S_{quad} &=&  \int d^3 x   \left( \frac{1}{2} (\partial_\mu W^+_\nu - \partial_\nu W^+_\mu)(\partial_\mu W^-_\nu - \partial_\nu W^-_\mu)  + \frac{g^2\nu^2}{2}W^+_\mu W^-_\mu   \right) 
\nonumber \\
&+& \int d^3x  \left(  \frac{1}{4} (\partial_\mu Z_\nu - \partial_\nu Z_\mu)^2  + \frac{(g^2+g'^2)\nu^2}{4}Z_\mu Z _\mu  +    \frac{1}{4} (\partial_\mu A_\nu - \partial_\nu A_\mu)^2  \right)  \;,
\label{qd}
\end{eqnarray}
from which we can read off the masses of the fields $W^+$, $W^-$, and $Z$:
\begin{equation}
m^2_W = \frac{g^2\nu^2}{2} \;, \qquad m^2_Z =  \frac{(g^2+g'^2)\nu^2}{2}  \;. \label{ms}
\end{equation}

The restriction to the Gribov region $\Omega$ still is needed and the procedure here becomes quite similar to what was carried out in the section \ref{Adjrep}. Due to the breaking of the global gauge invariance, caused by the Higgs field (through the covariant derivatives), the ghost sector can be split up in two different sectors, diagonal and off-diagonal. Namely, the ghost propagator reads,

\begin{equation}
\mathcal{G}^{ab}(k;A) ~=~ \left(
  \begin{array}{ll}
   \delta^{\alpha \beta} \mathcal{G}_{off}(k;A) & \,\,\,\,\,\,\,\,0 \\
   \,\,\,\;\;\;\;\;\;0 & \mathcal{G}_{diag}(k;A)
  \end{array}
\right).
\label{gh prop offdiag}
\end{equation}
By expliciting the ghost form factor we have
\begin{eqnarray}
\mathcal{G}_{off}(k;A) 
~\simeq~  \frac{1}{k^{2}} \left( \frac{1}{1 - \sigma_{off}(k;A)} \right) \;,
\label{gh off}
\end{eqnarray}
and
\begin{eqnarray}
\mathcal{G}_{diag}(k;A) 
~\simeq ~ \frac{1}{k^{2}} \left( \frac{1}{1 - \sigma_{diag}(k;A)} \right)\;,
\label{gh diag}
\end{eqnarray}
where
\begin{subequations} \begin{equation}
\sigma_{off}(0;A) ~=~ \frac{g^{2}}{dV} \int\!\! \frac{\d^{d}p}{(2\pi)^{d}} \;  \frac{1}{p^{2}} \left( \frac{1}{2} A^{\alpha}_{\mu}(p)A^{\alpha}_{\mu}(-p) + A^{3}_{\mu}(p)A^{3}_{\mu}(-p)\right)  \;,
\label{sigma off}
\end{equation}
and
\begin{equation}
\sigma_{diag}(0;A) ~=~ \frac{g^{2}}{dV} \int\!\! \frac{\d^{d}p}{(2\pi)^{d}} \; \frac{1}{p^{2}} A^{\alpha}_{\mu}(p)A^{\alpha}_{\mu}(-p)\;.
\label{sigma diag}
\end{equation} \end{subequations}
In order to obtain expressions  \eqref{sigma off} and \eqref{sigma diag}, where $V$ denotes the (infinte) space-time volume, the transversality of the gluon field and the property that $\sigma(k;A)_{off}$ and $\sigma(k;A)_{diga}$ are decreasing functions of $k$ were used\footnote{For more details concerning the ghost computation see \cite{Capri:2013oja,Capri:2013gha,Capri:2012ah,Vandersickel:2012tz}}. From equations \eqref{gh off} and \eqref{sigma off} one can easily read off the two no-pole conditions. Namely,
\begin{subequations} \begin{equation}
\sigma_{off}(0;A) < 1  \;,
\label{sigmaoffnopole}
\end{equation}
and
\begin{equation}
\sigma_{diag}(0;A) < 1\;.
\label{sigmadiagnopole}
\end{equation} \end{subequations}

At the end, the partition function restricted to the first Gribov region $\Omega$ reads,
\begin{eqnarray}
Z &=&  \int \frac{\d \omega}{2\pi i \omega}\frac{\d \beta}{2\pi i \beta} [\d A] [\d B]  \; \; e^{\omega (1-\sigma_{off})} \, e^{\beta (1-\sigma_{diag})} e^{-S}\;.
\label{ptionfucnt22}
\end{eqnarray}

\subsection{The gluon propagator and the gap equation}

The perturbative computation at the semi-classical level requires only quadratic terms of the full action, defined in eq.\eqref{ptionfucnt22} (with $S$ given by eq.\eqref{Sf}), yielding a Gaussian integral over the fields. Inserting external fields to obtain the boson propagators, one gets, after taking the limit $\xi \to 0$, the following propagators,

\begin{subequations} \label{propsaandb} \begin{gather}
\langle  A^{\alpha}_\mu(p) A^{\beta}_\nu(-p) \rangle ~=~ \frac{p^2}{p^4 + \frac{\nu^2g^2}{2} p^2 + \frac{2g^2\beta}{dV}} \; \delta^{\alpha \beta} \left( \delta_{\mu\nu} - \frac{p_\mu p_\nu}{p^2} \right)  \;,   
\label{aalpha} \\
\langle  A^{3}_\mu(p) A^{3}_\nu(-p) \rangle ~=~ \frac{p^2 \left(p^2 +\frac{\nu^2}{2} g'^{2}\right)}{p^6 + \frac{\nu^2}{2} p^4 \left(g^2 +g'^2 \right)  + \frac{2\omega g^2}{dV} \left( p^2 + \frac{\nu^2 g'^2}{2} \right)} \;  \left( \delta_{\mu\nu} - \frac{p_\mu p_\nu}{p^2} \right)  \;,   \label{a3a3}
\\ %
\langle  B_\mu(p) B_\nu(-p) \rangle ~=~ \frac{ \left(p^4 +\frac{\nu^2}{2} g^{2} p^2+\frac{2\omega g^2}{dV}  \right)}{p^6 + \frac{\nu^2}{2} p^4 \left(g^2 +g'^2 \right)  + \frac{2\omega g^2}{dV} \left( p^2 +  \frac{\nu^2 g'^2}{2} \right)} \;  \left( \delta_{\mu\nu} - \frac{p_\mu p_\nu}{p^2} \right)  \;,   \label{bb}
\\
\langle  A^3_\mu(p) B_\nu(-p) \rangle ~=~  \frac{ \frac{\nu^2}{2} g g'  p^2}{p^6 + \frac{\nu^2}{2} p^4 \left(g^2 +g'^2 \right)  + \frac{2\omega g^2}{dV} \left( p^2 + \frac{\nu^2 g'^2}{2} \right)} \;  \left( \delta_{\mu\nu} - \frac{p_\mu p_\nu}{p^2} \right)  \;.   \label{ba3}
\end{gather} \end{subequations}
Moving to the fields $W^{+}_\mu, W^{-}_\mu, Z_\mu, A_\mu$, one obtains 
\begin{subequations} \label{propszandgamma} \begin{gather}
\langle  W^{+}_\mu(p) W^{-}_\nu(-p) \rangle ~=~ \frac{p^2}{p^4 + \frac{\nu^2g^2}{2} p^2 + \frac{2g^2\beta}{dV} } \;  \left( \delta_{\mu\nu} - \frac{p_\mu p_\nu}{p^2} \right)  \;,   \label{ww}
\\
\langle  Z_\mu(p) Z_\nu(-p) \rangle ~=~ \frac{\left( p^4 +\frac{2\omega}{dV} \frac{g^2 g'^2}{g^2+g'^2}  \right)}{p^6 + \frac{\nu^2}{2} p^4 \left(g^2 +g'^2 \right)  + \frac{2\omega g^2}{dV} \left( p^2 + \frac{\nu^2 g'^2}{2} \right) } \;  \left( \delta_{\mu\nu} - \frac{p_\mu p_\nu}{p^2} \right)  \;,   \label{zz}
\\
\langle  A_\mu(p) A_\nu(-p) \rangle ~=~ \frac{\left( p^4 +\frac{\nu^2}{2} p^2 (g^2+g'^2) +\frac{2\omega}{dV} \frac{g^4}{g^2+g'^2}\right)}{p^6 + \frac{\nu^2}{2} p^4 \left(g^2 +g'^2 \right)  + \frac{2\omega g^2}{dV} \left( p^2 + \frac{\nu^2 g'^2}{2} \right) } \;  \left( \delta_{\mu\nu} - \frac{p_\mu p_\nu}{p^2} \right)  \;,   \label{aa}
\\
\langle  A_\mu(p) Z_\nu(-p) \rangle ~=~ \frac{\frac{2\omega}{dV} \frac{g^3 g'}{g^2+g'^2} }{p^6 + \frac{\nu^2}{2} p^4 \left(g^2 +g'^2 \right)  + \frac{2\omega g^2}{dV} \left( p^2 + \frac{\nu^2 g'^2}{2} \right)} \;  \left( \delta_{\mu\nu} - \frac{p_\mu p_\nu}{p^2} \right)  \;.   \label{az}
\end{gather} \end{subequations}
As expected, all propagators get deeply modified in the IR by the presence of the Gribov parameters $\beta$ and $\omega$. Notice, in particular, that due to the parameter $\omega$ a mixing between the fields $A_\mu$ and $Z_\mu$ arises, eq.\eqref{az}. As such, the original photon and the boson $Z$ loose their distinct particle interpretation.  Moreover, it is straightforward to check that in the limit $\beta \rightarrow 0$ and $\omega \rightarrow 0$, the standards propagators are recovered.

Let us now proceed by deriving the gap equations which will enable us to (dynamically) fix the Gribov parameters, $\beta$ and $\omega$, as function of $g$, $g'$ and $\nu^2$. Thus, performing the path integral of eq.\eqref{ptionfucnt22}, in the semi-classical level, we get
\begin{eqnarray}
f(\omega, \beta) ~=~ \frac{\omega}{2} + \beta - \frac{2(d-1)}{2} \int \frac{\d^{d}p}{(2\pi)^{d}} \; \log \left(p^{2} + \frac{\nu^{2}}{2}g^{2} + \frac{2g^{2}\beta}{dV} \frac{1}{p^{2}}\right) -
\nonumber \\
- \frac{(d-1)}{2} \int \frac{\d^{d}p}{(2\pi)^{d}} \; \log \lambda_{+}(p, \omega)\, \lambda_{-}(p, \omega)\;. 
\label{f eq}
\end{eqnarray}
In eq.\eqref{f eq}, $f(\omega,\beta)$ is defined according to eq.\eqref{Zq3} and
\begin{equation}
\lambda_{\pm} ~=~ \frac{\left( p^{4} + \frac{\nu^{2}}{4} p^{2}(g^{2} + g'^{2}) + \frac{g^{2}\omega}{dV} \right) \pm \sqrt{\left[ \frac{\nu^{2}}{4}(g^{2} + g'^{2})p^{2} + \frac{g^{2}\omega}{dV}\right]^{2} - \frac{\omega}{3}\nu^{2}g^{2}\,g'^{2}p^{2}}}{p^{2}}  \;. 
\label{ev1}
\end{equation}
Making use of the thermodynamic limit, where the saddle point approximation takes place, we have the two gap equations given by\footnote{For more details see \cite{Capri:2013oja,Capri:2013gha,Capri:2012ah}.}

\begin{equation}
\frac{4(d-1)}{2d}g^{2} \int \frac{\d^{d}p}{(2\pi)^{d}} \;  \frac{1}{p^{4}+\frac{g^{2}\nu^{2}}{2}p^{2} + \frac{2g^{2}\beta^{\ast}}{dV} } ~=~ 1   \;,
\label{beta gap eq}
\end{equation} 
and
\begin{equation}
\frac{2(d-1)}{d}g^{2}\int\!\! \frac{\d^{d}p}{(2\pi)^{d}} \; \frac{p^{2} + \frac{\nu^{2}}{2}g'^{2}}{p^{6} + \frac{\nu^{2}}{2}(g^{2} + g'^{2})p^{4} + \frac{2\omega^{\ast}g^{2}}{dV}p^{2} + \frac{\nu^{2}g^{2}\,g'^{2}\omega^{\ast}}{dV} } ~=~ 1 \;.
\label{omega gap eq}
\end{equation}

Given the difficulties in solving the gap equations \eqref{beta gap eq} and \eqref{omega gap eq}, we propose an alternative approach to probe the gluon propagators in the parameter space $\nu$, $g$ and $g'$. Instead of explicitly solve the gap equations, let us search for the necessity to implement the Gribov restriction. For that we mean to compute $\langle \sigma_{off}(0) \rangle $ and $\langle \sigma_{diag}(0) \rangle$ with the gauge field propagators unchanged by the Gribov terms, {\it i.e.}, before applying the Gribov restriction. Therefore, if $\langle \sigma_{off}(0;A) \rangle  < 1$ and $\langle \sigma_{diag}(0;A) \rangle < 1$ already in this case (without Gribov restrictions), then we would say that there is no need to restrict the domain of integration to $\Omega$. In that case we have, immediately, $\beta^* = \omega^* =0$ and the standard Higgs procedure takes place. Namely, the expression of each ghost form factor is
\begin{eqnarray}
\langle \sigma_{off}(0) \rangle & = &  \frac{(d-1)g^{2}}{d}  \int\!\!  \frac{\d^d p}{(2\pi)^d} \frac{1}{p^{2}}\left(\frac{1}{p^{2} + \frac{\nu^{2}}{2}g^{2}} + \frac{1}{p^{2} + \frac{\nu^{2}}{2}(g^{2}+g'^{2})} \right)  \;.
\label{sgoff1}
\end{eqnarray}
and
\begin{equation}
\langle \sigma_{diag}(0) \rangle ~=~ \frac{2(d-1)g^{2}}{d} \int\!\!\frac{\d^{d}p}{(2\pi)^{d}}\frac{1}{p^{2}}\left(\frac{1}{p^{2} + \frac{\nu^{2}}{2}g^{2}}\right)  \;.
\label{sgdiag1}
\end{equation}





\subsection{The $d=3$ case} 
\label{d=3}

\begin{figure}\begin{center}
\includegraphics[width=.25\textwidth]{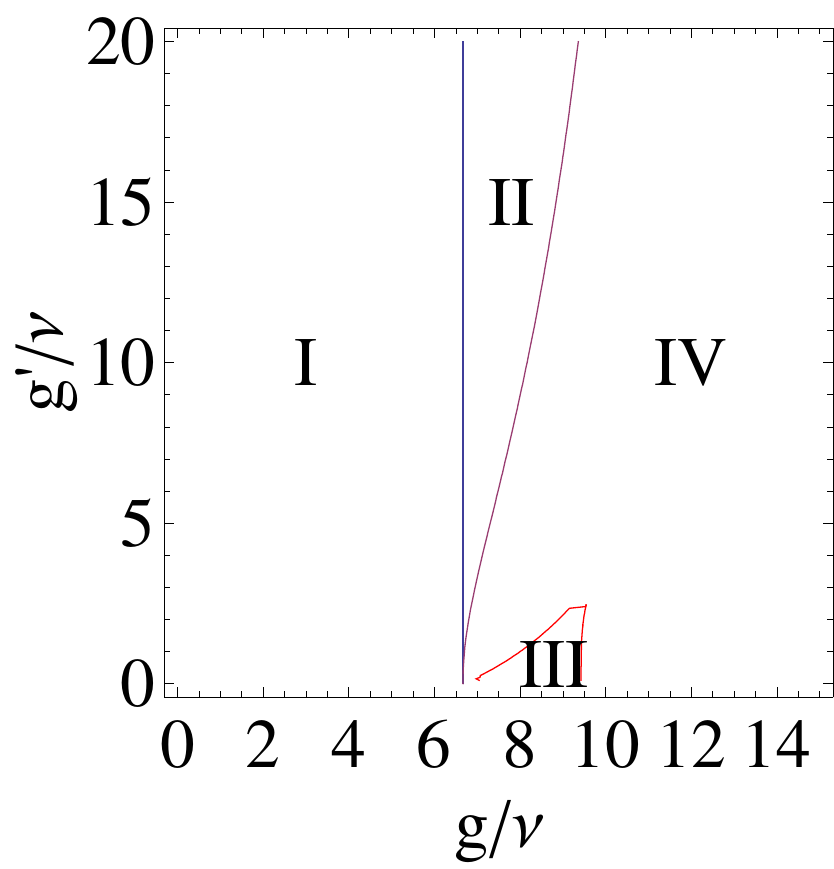}
\caption{There appear to be four regions in phase space. The region I is defined by condition \eqref{secondcond} and is characterized by ordinary Yang--Mills--Higgs behaviour (massive $W$ and $Z$ bosons, massless photon). The region II is defined by \eqref{firstcond} while excluding all points of region I --- this region only has electrically neutral excitations, as the $W$ bosons are confined (see Section \ref{sect6}); the massive $Z$ and the massless photon are unmodified from ordinary Yang--Mills--Higgs behaviour. Region III has confined $W$ bosons, while both photon and $Z$ particles are massive due to influence from the Gribov horizon; furthermore there is a negative-norm state. In region IV all $SU(2)$ bosons are confined and only a massive photon is left. Mark that the tip of region III is hard to deal with numerically --- the discontinuity shown in the diagram is probably an artefact due to this difficulty.  Details are collected in Section \ref{sect7}. \label{regionsdiag1}}
\end{center}\end{figure}

In the three-dimensional case things become easier since there is no divergences to treat. Therefore, computing the ghost form factors \eqref{sgoff1} and \eqref{sgdiag1} we led to the following conditions

\begin{subequations} 
\label{conds} 
\begin{eqnarray}
(1+\cos(\theta_{W}))\frac{g}{\nu} &<& 3\sqrt{2}\pi \label{firstcond} \\
2\frac{g}{\nu} &<& 3\sqrt{2}\pi \label{secondcond} \;,
\end{eqnarray} 
\end{subequations}
where $\theta(W)$ stands for the Weinberg angle,
\begin{equation}
\cos(\theta_{W}) = \frac{g}{\sqrt{g^{2}+g'^{2}}}\;.
\label{thW}
\end{equation}
These two conditions make phase space fall apart in three regions, as depicted in \figurename\ \ref{regionsdiag1}.
\begin{itemize}
	\item If $g/\nu<3\pi/\sqrt2$, neither Gribov parameter is necessary to make the integration cut off at the Gribov horizon. In this regime the theory is unmodified from the usual perturbative electroweak theory.
	\item In the intermediate case $3\pi/\sqrt2<g/\nu<3\sqrt2\pi/(1+\cos\theta_W)$ only one of the two Gribov parameters,  $\beta$, is necessary. The off-diagonal ($W$) gauge bosons will see their propagators modified due to the presence of a non-zero $\beta$, while the $Z$ boson and the photon $A$ remain untouched.
	\item In the third phase, when $g/\nu>3\sqrt2\pi/(1+\cos\theta_W)$, both Gribov parameters are needed, and all propagators are influenced by them. The off-diagonal gauge bosons are confined. The behaviour of the diagonal gauge bosons depends on the values of the couplings, and the third phase falls apart into two parts, as detailed in section \ref{sect7}.
\end{itemize}
Note that here in the $3$-dimensional $SU(2)\times U(1)+$Higgs case, as well as in the $3d$ $SU(2)+$Higgs treated in section \ref{3dsu2}, an effective coupling constant becomes of utmost importance when discussing the trustworthiness of the our semi-classical results.

\subsubsection{The off-diagonal ($W$) gauge bosons} 
\label{sect6}
Let us first look at the behaviour of the off-diagonal bosons under the influence of the Gribov horizon. The propagator \eqref{ww}  only contains the $\beta$ Gribov parameter, meaning that $\omega$ need not be considered here.

In the regime $g/\nu<3\pi/\sqrt2$ (region I in \figurename\ \ref{regionsdiag1}) the parameter $\beta$ is not necessary, due to the ghost form factor $\langle\sigma_{diag}(0)\rangle$ always being smaller than one. In this case, the off-diagonal boson propagator is simply of massive type, with mass parameter $\frac{\nu^{2}}{2}g^{2}$.

In the case that $g/\nu>3\pi/\sqrt2$ (regions II, III, and IV in \figurename\ \ref{regionsdiag1}), the relevant ghost form factor is not automatically smaller than one any more, and the Gribov parameter $\beta$ becomes necessary. The value of $\beta^{\ast}$ is determined from the gap equations \eqref{beta gap eq}. After rewriting the integrand in partial fractions, the integral in the equation becomes of standard type, and we readily find the solution
\begin{equation}
  \beta^{\ast} = \frac{3g^2}{32} \left(\frac{g^2}{2\pi^2}-\nu^2\right)^2 \;.
\end{equation}
Mark that, in order to find this result, we had to take the square of both sides of the equation twice. One can easily verify that, in the region $g/\nu>3\pi/\sqrt2$ which concerns us, no spurious solutions were introduced when doing so.

Replacing this value of $\beta^{\ast}$ in the off-diagonal propagator \eqref{ww} one can immediately check that it
clearly displays two complex conjugate poles. As such, the off-diagonal propagator  cannot describe a physical excitation of the physical spectrum, being adequate for a confining phase. This means that the off-diagonal components of the gauge field are confined in the region $g/\nu>3\pi/\sqrt2$.

\subsubsection{The diagonal $SU(2)$ boson and the photon field} \label{sect7}
The other two gauge bosons --- the $A^3_\mu$ and the $B_\mu$ --- have their propagators given by \eqref{a3a3}, \eqref{bb}, and \eqref{ba3} or equivalently --- the $Z_\mu$ and the $A_\mu$ --- by \eqref{zz}, \eqref{aa} and \eqref{az}. Here, $\omega$ is the only one of the two Gribov parameters present.

In the regime $g/\nu<3\sqrt2\pi/(1+\cos\theta_W)$ (regions I and II) this $\omega$ is not necessary to restrict the region of integration to within the first Gribov horizon. Due to this, the propagators are unmodified in comparison to the perturbative case.

In the region $g/\nu>3\sqrt2\pi/(1+\cos\theta_W)$ (regions III and IV) the Gribov parameter $\omega$ does become necessary, and it has to be computed by solving its gap equation, eq. \eqref{omega gap eq}. Due to its complexity it seems impossible to do so analytically. Therefore we turn to numerical methods. Using Mathematica the gap equation can be straightforwardly solved for a list of values of the couplings. Then we determine the values where the propagators have poles. 

The denominators of the propagators are a polynomial which is of third order in $p^2$. There are two cases: there is a small region in parameter space where the polynomial has three real roots, and for all other values of the couplings there are one real and two complex conjugate roots. In \figurename\ \ref{regionsdiag1} these zones are labelled III and IV respectively. Let us analyze each region separately.

\subsubsection{Three real roots (region III)}


Region III is defined by the polynomial in the denominators of \eqref{a3a3}, \eqref{bb}, and \eqref{ba3} having three real roots. This region is sketched in \figurename\ \ref{regionsdiag1}. (Mark that the tip of the region is distorted due to the difficulty in accessing this part numerically.) 


The residues of related to these poles were computed numerically. Only the two of the three roots have positive residue and can correspond to physical states. Those are the one with highest and the one with lowest mass squared. The third of the roots, the one of intermediate value, has negative residue and thus belongs to some negative-norm state, which cannot be physical.

All three states have non-zero mass for non-zero values of the electromagnetic coupling $g'$, with the lightest of the states becoming massless in the limit $g'\to0$. In this limit we recover the behaviour found in this regime in the pure $SU(2)$ case \cite{Capri:2012cr} (the $Z$-boson field having one physical and one negative-norm pole in the propagator) with a massless fermion decoupled from the non-Abelian sector.

\subsubsection{One real root (region IV)}
\begin{figure} \begin{center}
\includegraphics[width=.5\textwidth]{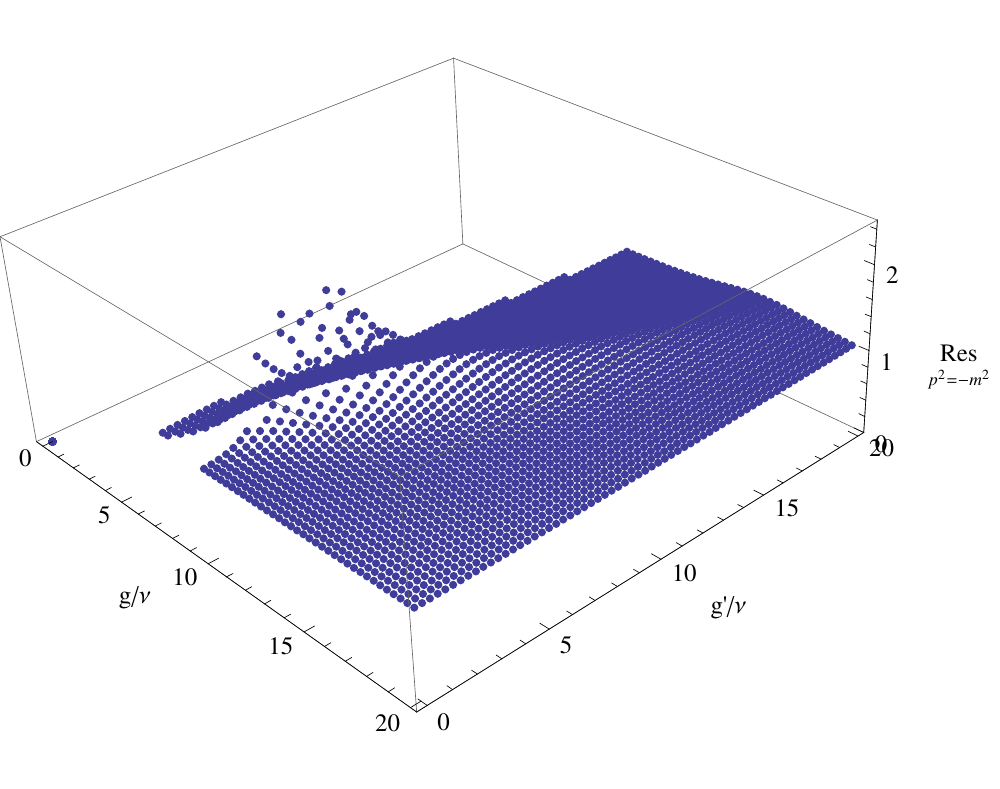}
\caption{The residue of the pole of the photon propagator. It turns out to be positive for all values of the couplings within the region IV. \label{resrealmass}}
\end{center} \end{figure}

In the remaining part of the parameter space, there is only one state with real mass-squared.
The two other roots of the polynomial in the denominators of \eqref{a3a3}, \eqref{bb}, and
\eqref{ba3} have non-zero imaginary part and are complex conjugate to each other. In order to
determine whether the pole coming from the real root corresponds to a physical particle
excitation, we computed its residue, which can be read off in the partial fraction
decomposition (the result is plotted in \figurename\ \ref{resrealmass}). It turns out the
residue is always positive, meaning that this excitation has positive norm and can thus be
interpreted as a physical, massive contributions. The poles coming from the complex roots
cannot, of course, correspond to such physical contributions.


In the limit $g'\to0$ we once more recover the corresponding results already found in the pure $SU(2)$ case \cite{Capri:2012cr} (two complex conjugate poles in the propagator of the non-Abelian boson field) plus a massless photon not influenced by the non-Abelian sector.

We shall emphasise here the complexity of the found ``phase spectrum'' in the $3d$ case. For the most part of the $(g'/\nu, g/\nu)$ plane we found the diagonal component of the bosonic field displaying a mix of physical and non-physical contributions, regarding the regions III and IV. The off-diagonal component was found to have physical meaning only in the region I. The transition between those regions was found to be continuous with respect to the effective perturbative parameter $\sim g/\nu$.

\subsection{The $d=4$ case}
\label{d=4}

\begin{figure}\begin{center}
\parbox{.5\textwidth}{\includegraphics[width=.5\textwidth]{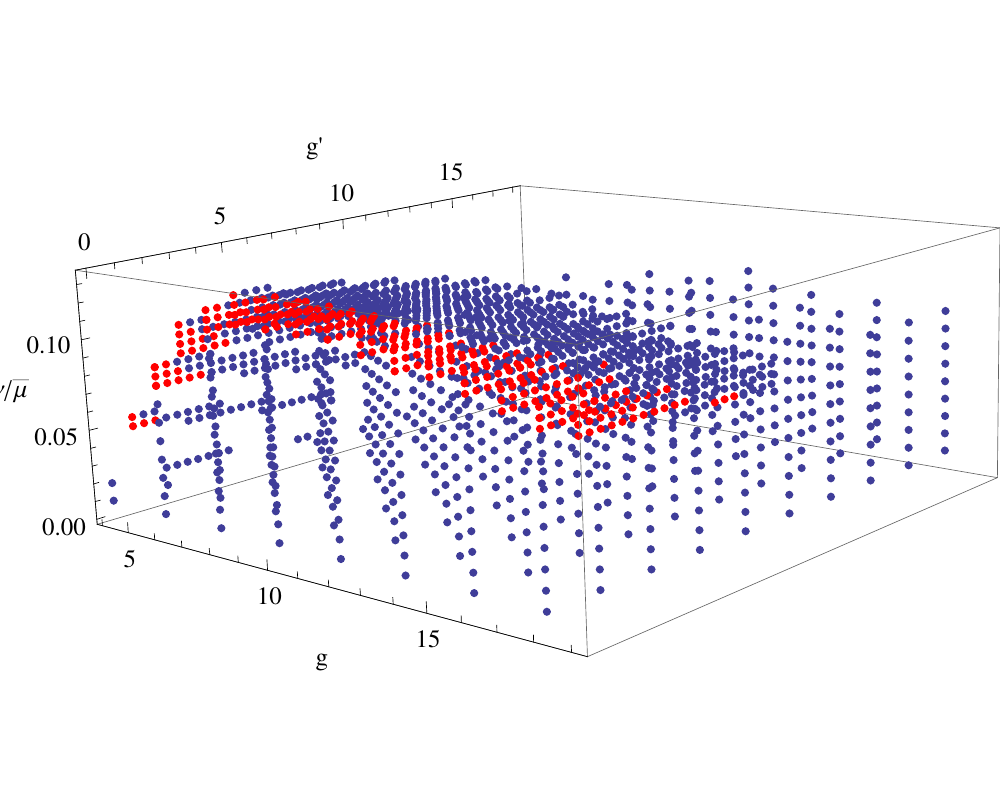}} \quad \parbox{.4\textwidth}{\includegraphics[width=.4\textwidth]{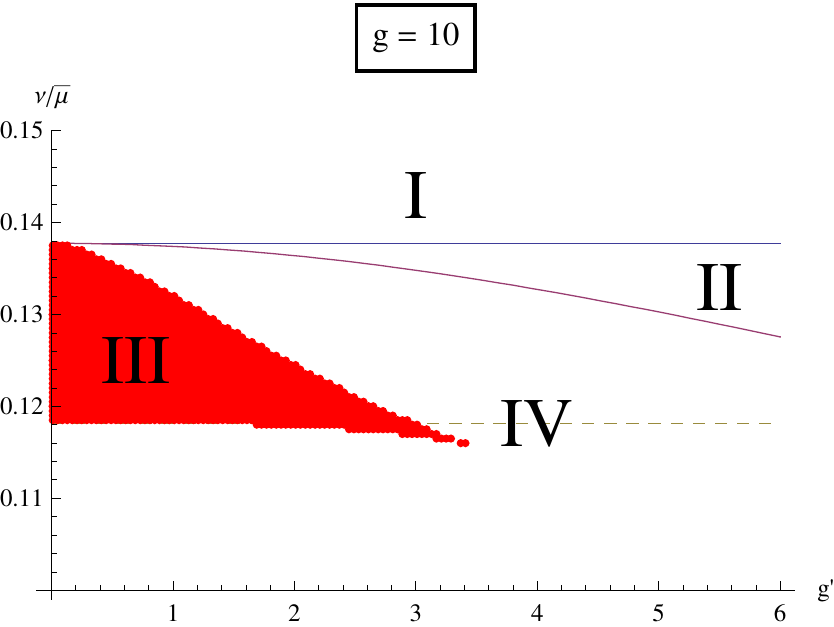}}
\caption{Left is a plot of the region $a'<1/2$ (the region $a'>1/2$ covers all points with higher $\nu$). In red are points where the polynomial in the denominator of \eqref{a3a3} - \eqref{ba3} has three real roots, and in blue are the points where it has one real and two complex conjugate roots. At the right is a slice of the phase diagram for $g=10$. The region $a>1/2$ and $a'>1/2$ is labelled I, the region $a<1/2$ and $a'>1/2$ is II, and the region $a<1/2$ and $a'<1/2$ is split into the regions III (polynomial in the denominator of \eqref{a3a3} - \eqref{ba3} has three real roots, red dots in the diagram at the left) and IV (one real and two complex conjugate roots, blue dots in the diagram at the left). The dashed line separates the different regimes for off-diagonal SU(2) bosons (two real massive poles above the line, two complex conjugate poles below). \label{regionsdiag}}
\end{center}\end{figure}

In the $4$-dimensional case the diagonal and off-diagonal ghost form factors read, using the standard $\MSbar$ renormalization procedure,
\begin{equation}
\label{constas} 
\langle \sigma_\text{off}(0) \rangle = 1 - \frac{3g^{2}}{32\pi^{2}}\ln\frac{2a}{\cos(\theta_{W})}
\,, \qquad
\langle \sigma_\text{diag}(0) \rangle = 1-\frac{3g^{2}}{32\pi^{2}}\ln(2a)\;,
\end{equation}
where
\begin{equation}
\label{consta} a = \frac{\nu^{2}g^{2}}{4\bar{\mu}^{2}\,e^{1-\frac{32 \pi^{2}}{3g^{2}}}}\,, \qquad a' = \frac{\nu^{2}(g^{2}+g'^{2})}{4\bar{\mu}^{2}\,e^{1-\frac{32 \pi^{2}}{3g^{2}}}} = a \frac{g^{2}+g'^{2}}{g^{2}} = \frac{a}{\cos^{2}(\theta_{W})}
\end{equation}
and $\theta_{W}$ stands for the Weinberg angle. With expression \eqref{constas} we are able to identify three possible regions, depicted in \figurename\ \ref{regionsdiag}:
\begin{itemize}
\item Region I, where $\langle \sigma_\text{diag}(0)\rangle < 1$ and $\langle \sigma_\text{off}(0)\rangle < 1$, meaning $2a > 1$. In this case the Gribov parameters are both zero so that we have the massive $W^{\pm}$ and $Z$, and a massless photon. That region can be identified with the ``Higgs phase''.
\item Region II, where $\langle \sigma_\text{diag}(0)\rangle > 1$ and $\langle \sigma_\text{off}(0)\rangle < 1$, or equivalently $\cos \theta_{W} < 2a < 1$. In this region we have $\omega = 0$ while $\beta \neq 0$, leading to a modified $W^{\pm}$ propagator, and a free photon and a massive $Z$ boson.
\item The remaining parts of parameter space, where $\langle \sigma_\text{diag}(0)\rangle > 1$ and $\langle \sigma_\text{off}(0)\rangle > 1$, or $0 < 2a < \cos\theta_{W}$. In this regime we have both $\beta \neq 0$ and $\omega \neq 0$, which modifies the $W^{\pm}$, $Z$ and photon propagators. Furthermore this region will fall apart in two separate regions III and IV due to different behaviour of the propagators (see \figurename\ \ref{regionsdiag}).
\end{itemize}

\subsubsection{The off-diagonal gauge bosons} \label{sect4}
Let us first look at the behaviour of the off-diagonal bosons under the influence of the Gribov horizon. The propagator \eqref{ww} only contains the $\beta$ Gribov parameter, meaning $\omega$ does not need be considered here.

As found in the previous section, this $\beta$ is not necessary in the regime $a>1/2$, due to the ghost form factor $\langle\sigma_\text{diag}(0)\rangle$ always being smaller than one. In this case, the off-diagonal boson propagator is simply of the massive type.

In the case that $a<1/2$, the relevant ghost form factor is not automatically smaller than one anymore, and the Gribov parameter $\beta$ becomes necessary. The value of $\beta$ is given by the gap equations \eqref{beta gap eq}, which has exactly the same form as in the case without electromagnetic sector. Therefore the results will also be analogous. As the analysis is quite involved, we just quote the results here.

For $1/e<a<1/2$ the off-diagonal boson field has two real massive poles in its two-point function. One of these has a negative residue, however. This means we find one physical massive excitation, and one unphysical mode in this regime. When $a<1/e$ the two poles acquire a non-zero imaginary part and there are no poles with real mass-squared left. In this region the off-diagonal boson propagator is of Gribov type, and the $W$ boson is completely removed from the spectrum. More details can be found in \cite{Capri:2012ah}.

\subsubsection{The diagonal SU(2) boson and the photon} \label{zandgamma}
The two other gauge bosons --- the diagonal SU(2) boson and the photon, $Z_\mu$ and the $A_\mu$ --- have their propagators given by \eqref{zz}, \eqref{aa} and \eqref{az}. Here, $\omega$ is the only of the two Gribov parameters present.

In the regime $a'>1/2$, $\omega$ is not necessary to restrict the region of integration to $\Omega$. Due to this, the propagators are unmodified in comparison to the perturbative case.

In the region $a'<1/2$ the Gribov parameter $\omega$ does become necessary, and it has to be computed by solving its gap equation. Due to its complexity it seems impossible to compute analytically. Therefore we turn to numerical methods. 
Once the parameter $\omega$ has been (numerically) determined, we look at the propagators to investigate the nature of the spectrum.

As the model under consideration depends on three dimensionless parameters ($g$, $g'$ and $\nu/\bar\mu$), it is not possible to plot the parameter dependence of these masses in a visually comprehensible way. Therefore we limit ourselves to discussing the behaviour we observed.

In region III, when there are three real poles in the full two-point function, it turns out that only the two of the three roots we identified have a positive residue and can correspond to physical states, being the one with highest and the one with lowest mass squared. The third one, the root of intermediate value, has negative residue and thus belongs to some negative-norm state, which cannot be physical. All three states have non-zero mass for non-zero values of the electromagnetic coupling $g'$, with the lightest of the states becoming massless in the limit $g'\to0$. In this limit we recover the behaviour found in this regime in the pure $SU(2)$ case \cite{Capri:2012ah} (the $Z$-boson field having one physical and one negative-norm pole in the propagator) with a massless boson decoupled from the non-Abelian sector.

In region IV there is only one state with real mass squared --- the other two having complex mass squared, conjugate to each other --- and from the partial fraction decomposition follows that it has positive residue. This means that, in this region, the diagonal-plus-photon sector contains one physical massive state (becoming massless in the limit $g'\to0$), and two states that can, at best, be interpreted as confined.

\section{Discussions about the results}
\label{discussion}

In this chapter we presented results achieved during the study of Yang-Mills models, in the
Landau gauge, coupled to a Higgs field, taking into account non-perturbative effects. More
specifically, the $SU(2)$ and $SU(2)\times U(1)$ models were analysed in $3$- and
$4$-dimensional Euclidean space-time, while the Higgs field was considered in its fundamental
and adjoint representation. The non-perturbative effects were taken into account by considering
the existence of Gribov ambiguities, or Gribov copies, in the (Landau) gauge fixing process. In
order to get rid of those ambiguities we followed the procedure developed by Gribov in his
seminal work \cite{Gribov:1977wm}, which consists in restricting the configuration space of the
gauge field into the first Gribov region $\Omega$. As found by Gribov, that restriction of the
path integral domain leads to a modification of the gluon propagator in a way that it is not
possible to attach any physical particle interpretation to it. The gauge field
propagator develops, after the Gribov restriction, two complex conjugate poles, preventing
any physical particle interpretation, since it presents positivity violation, which is a
reality condition of Osterwalder-Schrader. This may be interpreted as a sign of confinement of
the gauge field. In the same sense we could observe similar modifications of the gluon
propagator in the Yang-Mills + Higgs models. In general, the poles of the gauge field
propagator are functions of the parameters in each Yang-Mills model (including the Gribov
parameter $\gamma$ and the Higgs self mass parameter $\nu$), so that we could identify regions
in the parameter space where the gauge propagator has contributions coming from Yukawa type
modes (with real poles) and/or contributions from Gribov-like modes (presenting {\it cc}
poles). Regions where only Yukawa contribution exists are called {\it Higgs-like regimes}. On
the other hand, regions where there is only Gribov type contributions are named {\it confined
regimes} or {\it confined-like} regions. Note that contributions with negative residue, despite
being of the Yukawa type, have no physical particle interpretation as well.

In general we could find that the Higgs-like regime corresponds to the region of weak coupling,
{\it i.e.} small $g$ and sufficiently large $\nu$, reached in the UV regime. In that region of
the parameter space (coupling constant and Higgs vacuum expectation value), where we expect
perturbation theory to works, we do recover the standard perturbative Yang-Mills-Higgs
propagators. This is an important observation, since it means that the Gribov ambiguity does
not spoils the physical vector boson interpretation of the gauge sector where it is relevant.
On the other side, the confined-like regime corresponds to the strong coupling region in the
parameter space, characterized by large values of $g$ and sufficiently small $\nu$, but still
keeping logarithmic divergences under control. For higher values of the non-Abelian coupling
constant the Gribov horizon lets its influence be felt and the propagators become modified. In
general, for the $SU(2)$ and for the $SU(2)\times U(1)$ cases we could find an intermediate
region where contributions from physical modes (with real poles) mix with contributions from
non-physical modes (with {\it cc} poles or negative residues). 

It is very important to emphasise that the whole analysis strongly depends on the group
representation of the Higgs field, just as in Fradkin \& Shenker's work \cite{Fradkin:1978dv}.

For the fundamental representation of the Higgs field, either in the $SU(2)$ or in the
$SU(2)\times U(1)$ model, we could find that the two detected regimes, Higgs- and
confined-like, may be continuously connected, in the sense that the parameters of the theory
are allowed to continuously vary from one region to another, without leading to any
discontinuity or singularity of the vacuum energy or the two-point Green function. However, we
have to be careful when talking about \emph{analyticity region} in our perturbative approach,
since we could not rigorously prove such property, since there exist a region of the parameter space where our perturbative approximation is not reliable.

Something quite different happens in the adjoint representation of the scalar field. There we
could explicitly show (\emph{cf.} the section \ref{Adjrep}) the existence of an specific
configuration in the parameter space where the vacuum energy develops a jump discontinuity.
However, at the very point of the jump our approximation is beyond its range of
validity, and we cannot make any statement for sure about analyticity.

In the adjoint representation of the Higgs field the scenario looks quite different. Besides
the confined-like regime, in which the gluon propagator is of the Gribov-type, our results
indicated the existence of what can be called a $U(1)$ confined-like regime for finite values
of the Higgs condensate. This is a regime in which the third component $A^3_\mu$ of the gauge
field  displays a propagator of the Gribov-type, while the remaining off-diagonal components
$A^{\alpha}_\mu$, $\alpha=1,2$, exhibit a propagator of the Yukawa type. Interestingly,
something similar to this has been already detected on lattice studies of the three-dimensional
Georgi-Glashow model \cite{Nadkarni:1989na,Hart:1996ac}. A second result is the absence of a
Coulomb regime for finite Higgs condensate. For an infinite value of the latter, we were able
to clearly reveal the existence of a massless photon, in agreement with the lattice suggestion,
\emph{e.g.} the \cite{Brower:1982yn}.

One should keep in mind that in our analytic non-perturbative model, where Gribov ambiguities
where taken into account, we cannot properly speak about physical phases, at all. The non-local
Gribov term leads to a soft breaking of the BRST symmetry, forbidding us from defining in the
usual sense physical states. At the same time, there is no obvious gauge invariant order
parameter available in our model, which would be useful for probing phase transition.

Keeping safe the due difference between Fradkin \& Shenker's approach to the Yang-Mills + Higgs
theory and ours perturbative approach, it is fair enough to acknowledge the matching of a
couple of remarkable results. Fradkin \& Shenker clearly say that in the fundamental
representation of the Higgs field there is no phase transition to occur, being the theory in
the symmetric (or ordered) phase in the (almost) entire parameter space, despite of the
particular case of null Higgs coupling constant, $\nu =0$, where the theory is found to be in
the disordered phase  \cite{Fradkin:1978dv}. Besides, they did show that there exist two
different regimes in the configuration space, called \emph{confinement-like regime} and
\emph{Higgs-like regime}, and that any point of these regions of the configuration space are
smoothly connected to each other. In other words, the system is allowed to smoothly hang from
one point in the \emph{confined-like regime} to another point in the \emph{Higgs-like regime}.
It should be emphasized that in their case a phase transition is properly defined, since they
work on the lattice, measuring the gauge invariant Wilson loop order parameter. They could also
prove the existence of the analyticity region.

%


\chapter{Quark confinement and BRST soft breaking}
\chaptermark{Quark confinement and BRST breaking}
\label{brstonmatter}

Being restricted to the matter sector of Nature, the directly observed particles are called
hadrons, which are colorless composition of quarks. For instance, protons are elements of the
baryon set of particles (composed of three quarks) whose mass is about
$\unit{938}{\mega\electronvolt}$, while quark's mass is of order $\sim
\unit{1}{\mega\electronvolt}$. So, where does come from the huge difference of mass? How does
it happen? Nature gives us clues of an spontaneous symmetry breaking (of an approximate
symmetry), known as \emph{chiral symmetry breaking}, with a mass term being dynamically
generated.

Besides chiral symmetry breaking, there exists the confinement scenario, where quarks cannot 
be asymptotically detected as free particles or, in other words, they do not belong to the
physical particle spectrum of Nature. In a general sense, the chiral phase transition is not
directly related to the deconfinement transition; it is known that for the two-flavors and
three colors scenario the chiral phase transition takes place at a critical temperature of
about $\unit{170}{\mega\electronvolt}$, while the deconfinement second order phase transition
occur at a temperature around $\unit{270}{\mega\electronvolt}$,
\cite{Fukushima:2002bk,Fukushima:thesis,Fukushima:2010bq,Bazavov:2009zn,Fukushima:2003fw,Banks:1979yr}.

In this chapter we propose, and analyze, an effective model to the matter sector, by means of
introducing a non-local mass term to the matter field, leading to a soft breaking of the
BRST symmetry, in analogy to what happens in the gauge sector. In a sense, this non-local mass
term would represent a generalization of the \emph{horizon function} of the GZ scheme of the
gauge field applied to the matter field. We provide a general analysis of this procedure by
specializing the matter sector to the scalar field and to the quark field. A comparison is made
with the most recent lattice data of the matter field and we could find a clear agreement
between them. The matter field, in this scenario, is deprived of an asymptotic physical
particle interpretation, since its propagator displays positivity violation, so not satisfying
every reality condition of Osterwalder-Schrader, just the same as the gauge field. The ${\cal
N}=1$ supersymmetric case is presented at the of the chapter as an example.

The content of the second chapter, concerning the Gribov and Gribov-Zwanziger mechanism of
quantizing the gauge field, is useful to the comprehension of the present one. More precisely,
the fate of BRST symmetry breaking due to the non-local horizon function must be kept in mind.
Therefore, to the benefit of the reader, some recurrent expressions will be rewritten here,
preventing going back and forth to the second chapter from being repeated overmuch. The first
one is the GZ action, which reads
\begin{equation} 
S_{GZ} = S_{YM} + S_{gf} + S_0+S_\gamma  \;, 
\label{sgz5}
\end{equation}
with
\begin{equation}
S_0 =\int d^{4}x \left( {\bar \varphi}^{ac}_{\mu} (\partial_\nu D^{ab}_{\nu} )
\varphi^{bc}_{\mu} - {\bar \omega}^{ac}_{\mu}  (\partial_\nu D^{ab}_{\nu} ) \omega^{bc}_{\mu}
- gf^{amb} (\partial_\nu  {\bar \omega}^{ac}_{\mu} ) (D^{mp}_{\nu}c^p) \varphi^{bc}_{\mu}
\right) \;, \label{s05}
\end{equation}
and 
\begin{equation}
S_\gamma =\; \gamma^{2} \int d^{4}x \left( gf^{abc}A^{a}_{\mu}(\varphi^{bc}_{\mu} + {\bar
\varphi}^{bc}_{\mu})\right)-4 \gamma^4V (N^2-1)\,.
\label{hfl5}
\end{equation} 
The RGZ action can written as
\begin{equation}
S_{RGZ} = S_{GZ} + \int d^4x \left(  \frac{m^2}{2} A^a_\mu A^a_\mu  - \mu^2 \left( {\bar
\varphi}^{ab}_{\mu}  { \varphi}^{ab}_{\mu} -  {\bar \omega}^{ab}_{\mu}  { \omega}^{ab}_{\mu}
\right)   \right)  \;.
\label{rgz5}
\end{equation}
The soft breaking of the BRST symmetry can be directly seen from the aplication of the BRST
transformation $s$ on the (R)GZ action, given by the BRST transformation of each field that is
given in \eqref{brst0}. At the
end one gets
\begin{equation}
s S_{GZ} = \gamma^2 \Delta  \;, 
\label{brstbrr5}
\end{equation}
with
\begin{equation}
\Delta = \int d^{4}x \left( - gf^{abc} (D_\mu^{am}c^m) (\varphi^{bc}_{\mu} + {\bar
\varphi}^{bc}_{\mu})   + g f^{abc}A^a_\mu \omega^{bc}_\mu            \right)  \;.
\label{brstb15}
\end{equation}
Finally, the gluon and ghost propagators read as,
\begin{eqnarray} 
\langle  A^a_\mu(k)  A^b_\nu(-k) \rangle  & = &  \delta^{ab}  \left(\delta_{\mu\nu} -
\frac{k_\mu k_\nu}{k^2}     \right)   {\cal D}(k^2) \;, 
\label{glrgz5} \\
{\cal D}(k^2) & = & \frac{k^2 +\mu^2}{k^4 + (\mu^2+m^2)k^2 + 2Ng^2\gamma^4 + \mu^2 m^2}  \;.
\label{Dg5}
\end{eqnarray} 
and
\begin{equation}
{\cal G}^{ab}(k^2) = \langle  {\bar c}^{a} (k)  c^b(-k) \rangle \Big|_{k\sim 0} \; \sim
\frac{\delta^{ab}}{k^2}   \,.
\label{ghrgz5} 
\end{equation}

Moreover, despite the soft breaking, eq.\eqref{brstbrr5}, a set of BRST  invariant composite operators whose correlation functions exhibit the K{\"a}ll{\'e}n-Lehmann spectral representation with positive spectral densities can be consistently introduced \cite{Baulieu:2009ha}. 

Although a satisfactory understanding of the physical meaning of the soft breaking of the BRST
symmetry in presence of the Gribov horizon and of its relationship with confinement is still
lacking, it is worth  underlining here that the first concrete numerical lattice evidence  of
the existence of such breaking has been provided by the authors of \cite{Cucchieri:2014via},
where the Bose-ghost propagator 
\begin{eqnarray}
{\cal Q}^{abcd}_{\;\;\mu\nu} ~=~ \langle \omega^{ab}_{\mu} \bar{\omega}^{cd}_{\nu} +
\varphi^{ab}_{\mu}
\bar{\varphi}^{cd}_{\nu}   \rangle
\label{ss1}
\end{eqnarray}
has being numerically computed on the lattice formulation, since it can be written as
\begin{eqnarray}
{\cal Q}^{abcd}_{\;\;\mu\nu} ~=~ \langle s\varphi^{ab}_{\mu}\bar{\omega}^{cd}_{\nu} \rangle\,,
\label{ss1}
\end{eqnarray}
which is evidently a BRST exact correlation function. So, if it is non-zero, it is a signal of
the (soft) BRST symmetry breaking. As $(\bar{\omega},\omega)$ and $(\bar{\varphi},\varphi)$
are localizing auxiliary fields of the GZ framework, thus there must be a non-local version of
the Boson-ghost propagator, and indeed there is. Evaluating \eqref{ss1} is equivalent to
measuring
\begin{eqnarray} 
\langle {\cal R}^{ab}_{\;\;\mu}(x)  {\cal R}^{cd}_{\;\;\nu}(y) \rangle    \;, 
\label{rr} 
\end{eqnarray}
with
\begin{eqnarray}
{\cal R}^{ac}_{\;\;\mu}(x) & = &  \int d^4z\;  ({\cal M}^{-1})^{ad} (x,z) \; g f^{dec}
A^{e}_\mu(z)  \;, 
\label{ra} 
\end{eqnarray} 
where $\cal M$ accounts for the inverse of the Faddeev-Popov operator.
The relation of the correlation function \eqref{rr} with the breaking of the BRST symmetry can
be understood by observing that, within the local formulation of the Gribov-Zwanziger
framework, expression \eqref{rr} corresponds exactly to the Bose-ghost propagator \eqref{ss1}.
In fact, integrating out the auxiliary fields $(\bar{\omega}_\mu^{ab}, \omega_\mu^{ab}, \bar{\varphi}_\mu^{ab},\varphi_\mu^{ab})$ in expression 
\begin{equation}
\int [{\cal D} {\Phi}] \; \left( \omega^{ab}_\mu(x) {\bar \omega}^{cd}_\nu(y) + \varphi^{ab}_\mu(x) {\bar \varphi}^{cd}_\nu(y) \right)   \; e^{-S_{GZ}} \;,  \label{loce}
\end{equation}
one ends up with
\begin{equation} 
\frac{ \int [{\cal D} {\Phi}] \;   \left( s \left( \varphi^{ab}_\mu(x) {\bar
\omega}^{cd}_\nu(y)  \right)   \right) \; e^{-S_{GZ}} }{ \int [{\cal D \phi}]    \;
e^{-S_{GZ}}}  =\gamma^4 \; \frac{  \int {\cal D}A\; \delta(\partial A) \; \left( det{\cal M}
\right) \; {\cal R}^{ab}_{\;\;\mu}(x)  {\cal R}^{cd}_{\;\;\nu}(y)  \; e^{-(S_{YM}+\gamma^4 H(A)
)} } {\int {\cal D}A\; \delta(\partial A) \; \left( det{\cal M} \right)   \;
e^{-(S_{YM}+\gamma^4 H(A) )} } \;. 
\label{brstbra}
\end{equation}
This equation shows  that the investigation of the correlation function \eqref{rr} with a
cutoff at the Gribov horizon is directly related to the existence of the BRST breaking. This is
precisely what has been done in \cite{Cucchieri:2014via}, where the correlator \eqref{rr} has
been shown to be non-vanishing, see Fig.1 of \cite{Cucchieri:2014via}. Moreover, from
\cite{Cucchieri:2014via}, it turns out that in the deep infrared the Fourier transform of the
correlation function \eqref{rr} is deeply enhanced, see Fig.2 of \cite{Cucchieri:2014via},
behaving as $\frac{1}{k^4}$, namely 
\begin{equation} 
\langle \tilde{\cal R}^{ab}_{\;\;\mu}(k)  \tilde {\cal R}^{cd}_{\;\;\nu}(-k)  \rangle  \sim
\frac{1}{k^4} \;.  
\label{enhanc}
\end{equation}  
As observed in \cite{Cucchieri:2014via}, this behaviour can be  understood by making use of the
analysis \cite{Zwanziger:2010iz}, {\it i.e.} of the cluster decomposition 
\begin{equation}
 \langle {\tilde {\cal R}} ^{ab}_{\;\;\mu}(k)  {\tilde {\cal R}}^{cd}_{\;\;\nu}(-k)  \rangle   \sim  g^2 {\cal G}^2(k^2) {\cal D}(k^2) \;, \label{clust} 
\end{equation} 
where ${\cal D}(k^2)$ and ${\cal G}(k^2)$ correspond to the   gluon and ghost propagators,
eqs.\eqref{Dg5},\eqref{ghrgz5}. A non-enhanced ghost propagator, {\it i.e.}  ${\cal G}(k^2)
\Big|_{k\sim 0} \sim \frac{1}{k^2}$, and an infrared finite gluon propagator, {\it i.e.} ${\cal
D}(0) \neq 0$, nicely yield the behaviour of eq.\eqref{enhanc}. 

Thus, we are going to show in this chapter that the quantity ${\cal R}$, eq.\eqref{ra}, and
the correlation function  $ \langle {\cal R}(x)  {\cal R}(y)  \rangle $, eq.\eqref{rr}, can be
consistently generalized to the case of matter fields, {\it e.g.} for the quarks and scalar
fields.

\section{A horizon-like term to the matter field: the $\langle {\tilde {\cal R}}(k) {\tilde
{\cal R}}(-k) \rangle$ in the light of lattice data}
\sectionmark{A horizon-like term to the matter field}

Let $F^{i}$ denote a generic matter field in a  given
representation of $SU(N)$, specified by the generators $(T^a)^{ij}$, $a=1,..,(N^2-1)$, and let
${\cal R}^{ai}(x)$ stand for the quantity
\begin{equation}
{\cal R}^{ai}(x)  =  g \int d^4z\;  [{\cal M}^{-1}]^{ab} (x,z)   \;(T^b)^{ij} \;F^{j}(z)   \label{rmatter}  \;, 
\end{equation}
which is a convolution of the inverse Faddeev-Popov operator with a given colored matter field,
being clearly the matter counterpart of the operator ${\cal R}^{ab}_{\;\;\mu}$ in the pure
gauge case. We shall be able to prove that, in analogy with the case of the gauge field
$A^a_\mu$, a non-trivial correlation function 
\begin{equation} 
 \langle {\cal R}^{ai}(x)  {\cal R}^{bj}(y)  \rangle    \;, \label{rcm} 
\end{equation} 
can be obtained from a local and renormalizable action  which is constructed by adding to the
starting conventional matter action a non-local term which shares great similarity with the
horizon function $H(A)$, eq.\eqref{hf2}, namely 
\begin{equation}
{g^{2}}   \int d^{4}x\;d^{4}y\; F^{i}(x) (T^a)^{ij} \left[ {\cal M}^{-1}\right]^{ab} (x,y)
(T^b)^{jk} F^{k} (y) \;.    
\label{hmatter}
\end{equation} 
The introduction of such \emph{horizon-like} functional into the sector of the matter field of
the action has the physical meaning of a non-local mass term, due to the inverse of the FP
operator, which would account for non-perturbative features of the matter sector. Therefore,
the proposed non-local effective action would looks like,
\begin{eqnarray}
S_{non-loc} &=& \iint d^4x\,d^{4}y\; \Biggl\{
\frac{1}{4}F^a_{\mu\nu}(x) F^a_{\mu\nu}(y)
+ b^{a}(x)\partial_{\mu}A^{a}_{\mu}(y)
+ \bar{c}^{a}(x)\partial_{\mu}D^{ab}_{\mu}(x,y)c^{b}(y)
\nonumber \\
&+&
\text{T}[F^{i}](x,y)
- U[F^{i}](x,y)
+ {g^{2}}\gamma^{4} f^{abc}A_{\mu}^{b}(x)\left[ {\cal M}^{-1}\right]^{ad}
(x,y)f^{dec}A_{\mu}^{e}(y) 
\nonumber \\
&+&
{g^{2}}\sigma^{4}  F^{i}(x) (T^a)^{ij} \left[ {\cal M}^{-1}\right]^{ab} (x,y)
(T^b)^{jk} F^{k} (y)
-\gamma^{4}4(N^{2}-1)
\Biggr\}\,,
\label{mnlocalact}
\end{eqnarray}
whence $\text{T}[F^{i}](x,y)$ accounts for the kinetic term of the matter sector, and
$U[F^{i}](x,y)$ stands for the potential term. The by-hand introduced parameter $\sigma^{2}$
has dimension of $\pmb{[mass]}^{2}$, just as the GZ parameter $\gamma^{2}$, although being a
free parameter of the theory.

As it happens in the case of the Gribov-Zwanziger theory, the non-local action
\eqref{mnlocalact} can be cast in a local form by means of the introduction of suitable
auxiliary fields. The resulting local action enjoys a large set of Ward identities which
guarantee its renormalizabilty (take a look at the next chapter). The introduction of the term \eqref{hmatter} deeply modifies
the infrared behavior (IR) of the correlation functions of the matter fields, while keeping
safe the well known UV perturbative results. One of the most interesting outcomes of this
procedure is that the matter's propagators are of the confining type, displaying positivity
violation, while being in good agreement with the available lattice data, as in the case of the
scalar matter fields, \cite{Maas:2011yx,Maas:2010nc}, and of quarks
\cite{Furui:2006ks,Parappilly:2005ei}.

Moreover, relying on the numerical data for the two-point correlation functions of quark and
scalar fields, the \emph{vev} \eqref{rcm} turns out to be non-vanishing and, interestingly
enough, it appears to behave exactly as the Boson-ghost propagator \eqref{enhanc} of the gauge
sector in the  deep IR, \emph{i.e.}
\begin{equation} 
 \langle {\tilde {\cal R}} ^{ai}(k)  {\tilde {\cal R}}^{bj}(-k)  \rangle  ~\sim~ \frac{1}{k^4}
\;. 
\label{menhanc}
\end{equation} 
Furthermore, just as in the case of the gauge sector, expression \eqref{rcm} signals the
existence of a (soft) BRST breaking in the matter field sector of the theory.

In the next section we shall show how the correlation function $\langle {\cal R}^{ai}(x) {\cal R}^{bj}(y)  \rangle $ can be obtained from a local and renormalizable action exhibiting a soft breaking of the BRST invariance in the matter sector.

\section{The local version of the proposed model and the analysis of $\langle {\tilde
{\cal R}}(k) {\tilde {\cal R}}(-k) \rangle$}
\sectionmark{The local version of the proposed model}
\label{localhorizoninmatter}

Useful quantities in QFT can only be obtained through a local (and renormalizable) action, such
as the $n$-point correlation functions, \emph{vev} of composite operators and the vacuum
energy. Therefore, since we have proposed an effective non-local action for the matter field,
in order to describe non-perturbative features of matter, it is very important to check, and
prove, that the proposed action can be recast in a local form. To achieve this goal, a couple
of auxiliary fields must be introduced, just as in the gauge sector. Furthermore, after
properly localizing the action, the propagator of the matter field will be derived in a Refined
theory, where dynamical condensates of the auxiliary fields are taken into account. Naturally,
the existence --- energetically favorable --- of such condensates is also checked. This
procedure will be developed in both example cases, for the scalar field and for the quark
field.


\subsection{The scalar field in the adjoint representation } 

We start by considering the following non-local action  
\begin{eqnarray}
\label{acs}
S^{\phi} &=& \int d^4x\; \left(
 \frac{1}{2}(D^{ab}_{\mu}\phi^{b})^{2} + \frac{m^2_{\phi}}{2} \phi^a \phi^a 
+ \frac{\lambda}{4!}(\phi^{a}\phi^{a})^{2} \right)   + 
\nonumber \\
&+&
{g^{2}} \sigma^4  \int d^{4}x\;d^{4}y\;
f^{abc}\phi^{b}(x)\left[ {\cal M}^{-1}\right]^{ad} (x,y)f^{dec}\phi^{e}(y) \;, 
\end{eqnarray}
where, once again, $\sigma$ is a massive parameter which, to some extent, plays a role akin to that of the
Gribov parameter $\gamma^2$ of the Gribov-Zwanziger action. eq.\eqref{sgz2}. It should be
noticed that, despite of any mathematical similarity with the Gribov-Zwanziger's parameter,
$\gamma^{2}$, $\sigma^{2}$ has no dynamical origin, nor geometrical interpretation, until now.
However, they indeed share algebraic similarities, so that
most of the tools already known from GZ framework can be used here for the matter sector.

Following, then, the same procedure adopted in the case of the Gribov-Zwanziger action, it is
not difficult to show that the non-local action \eqref{acs} can be cast in a local form. This
is achieved by introducing  a set of auxiliary fields $(\tilde{\eta}^{ab},\eta^{ab})$,
$(\tilde{\theta}^{ab},\theta^{ab})$, where $(\tilde{\eta}^{ab},\eta^{ab})$ are commuting fields
while  $(\tilde{\theta}^{ab},\theta^{ab})$ are anti-commuting. For the local version of
\eqref{acs} one gets 
\begin{equation}
S_{loc}^{\phi}   =     S_{0}^{\phi} + S_{\sigma}   \;, \label{lphi} 
\end{equation}
with 
\begin{eqnarray}
\label{act0}
S_{0}^{\phi}  ~=~  \int d^4x\; \bigg(  \frac{1}{2}  (D^{ab}_{\mu}\phi^{b})^{2}  + \frac{m^2_{\phi}}{2} \phi^a \phi^a
+  \frac{\lambda}{4!}(\phi^{a}\phi^{a})^{2}
+ \tilde{\eta}^{ac}(\partial_{\mu} D^{ab}_{\mu})\eta^{bc} -
\nonumber \\
-
\tilde{\theta}^{ac}(\partial_{\mu} D^{ab}_{\mu})\theta^{bc}      -
gf^{abc}(\partial_{\mu}\tilde{\theta}^{ae})(D^{bd}_{\mu}c^{d})\eta^{ce}  \bigg)  \;  
\end{eqnarray}
and
\begin{equation}
S_{\sigma}  =  \sigma^{2}g  \int d^4x   \; f^{abc}\phi^{a}(\eta^{bc} + \tilde{\eta}^{bc}) \;.     \label{ss}
\end{equation}
As in the case of the Gribov-Zwanziger action, the auxiliary fields $(\tilde{\eta}^{ab},\eta^{ab})$, $(\tilde{\theta}^{ab},\theta^{ab})$ appear quadratically, so that they can be easily integrated out, giving back precisely the non-local starting expression \eqref{acs}. Moreover, in full analogy with the Gribov-Zwanziger case, the local action $S_{loc}^{\phi}$ exhibits a soft breaking of the BRST symmetry. In fact, making use of eqs.\eqref{brst1} and of 
\begin{eqnarray}
&&
s\phi^{a}=-gf^{abc}\phi^{b}c^{c} \;,    \nonumber \\
&&
s\tilde{\theta}^{ab} = \tilde{\eta}^{ab}\;, \qquad s\tilde{\eta}^{ab} =0\;, \nonumber \\
&&
s\eta^{ab}=\theta^{ab}\;, \qquad s\theta^{ab}=0\;, 
\end{eqnarray}
it follows that 
\begin{equation}
s  S_{loc}^{\phi} = \sigma^2 \Delta^{\phi}   \;, \label{bs}
\end{equation}
 where 
\begin{equation}
 \Delta^{\phi}  = g \int d^4x\; f^{abc} \left( -g f^{amn} \phi^{m} c^n (\eta^{bc} + \tilde{\eta}^{bc}) + \phi^a \theta^{bc}     \right)   \;. 
\label{dphi}
\end{equation}
Being of dimension two in the fields (smaller than the space-time dimension $4$, in general),
the breaking term  $\Delta^{\phi} $ \eqref{dphi} is in fact a soft breaking. 

Now the local action \eqref{lphi} is added to the Gribov-Zwanziger action \eqref{sgz2},
obtaining 
\begin{eqnarray}
\label{actlc}
S_{ loc} &=& \int d^4x\; \Biggl\{
\frac{1}{4}F^a_{\mu\nu} F^a_{\mu\nu}
+ b^{a}\partial_{\mu}A^{a}_{\mu}
+ \bar{c}^{a}\partial_{\mu}D^{ab}_{\mu}c^{b}
+ \frac{1}{2}(D^{ab}_{\mu}\phi^{b})^{2}  + \frac{m^2_{\phi}}{2} \phi^a \phi^a
+ \frac{\lambda}{4!}(\phi^{a}\phi^{a})^{2}
\nonumber \\
&&
+ \varphi^{ac}_{\nu}\partial_{\mu}D^{ab}_{\mu}\bar{\varphi}^{bc}_{\nu}
- \omega^{ac}_{\nu}\partial_{\mu}D^{ab}_{\mu}\bar{\omega}^{ac}_{\nu}
+ \gamma^{2}gf^{abc}A^{a}_{\mu}(\varphi^{bc}_{\mu} + \bar{\varphi}^{bc}_{\mu})
- gf^{abc}(\partial_{\mu}\bar{\omega}^{ae}_{\nu})(D^{bd}_{\mu}c^{d})\varphi^{ce}_\nu
\nonumber \\
&&
- \gamma^{4}4(N^{2}-1)
+ \tilde{\eta}^{ac}(\partial_{\mu} D^{ab}_{\mu})\eta^{bc}
- \tilde{\theta}^{ac}(\partial_{\mu} D^{ab}_{\mu})\theta^{bc}
+ \sigma^{2}gf^{abc}\phi^{a}(\eta^{bc} + \tilde{\eta}^{bc})
\nonumber \\
&&
- gf^{abc}(\partial_{\mu}\tilde{\theta}^{ae})(D^{bd}_{\mu}c^{d})\eta^{ce}
\Biggr\}
\;.
\end{eqnarray}
As it happens in the case of the Gribov-Zwanziger action, the local action  $S_{ loc}$ can be
proven to be renormalizable to all orders. This important property follows from the existence
of a large set of Ward identities which can be derived in the matter scalar sector and which
restrict very much the possible allowed counterterms.  For the sake of completeness, the
Appendix \ref{ARscalaraction} has been devoted to the detailed algebraic proof of the
renormalizability of the local action \eqref{actlc}. 

As in the case of the Gribov-Zwanziger action, expression
\eqref{actlc} is well suited to investigate the correlation function 
\begin{equation} 
 \langle {\cal R}^{ab}(x)  {\cal R}^{cd}(y)  \rangle    \;, \label{cphi} 
\end{equation} 
\begin{equation}
{\cal R}^{ab}(x)  =  g \int d^4z\;  ({\cal M}^{-1})^{ac} (x,z)   \; f^{cdb} \phi^{d}(z)   \label{rmsc}  \;, 
\end{equation}
and its relation with the soft BRST breaking in the scalar field sector, eq.\eqref{bs}. In fact, repeating the same reasoning of eqs.\eqref{ss}, \eqref{loce},\eqref{brstbra}, one is led to consider the exact BRST correlation function in the matter scalar field sector
\begin{equation} 
\langle \; s ( \eta^{ab}(x) {\tilde \theta}^{cd}(y)  \; ) \rangle_{S_{ loc}}  = \langle   \theta^{ab}(x) {\tilde \theta}^{cd}(y) + \eta^{ab}(x) {\tilde \eta}^{cd}(y)      \rangle_{S_{ loc}} \;.
\end{equation}
Integrating out the auxiliary fields $(\tilde{\theta}^{ab}, \theta^{ab}, \tilde{\eta}^{ab},\eta^{ab})$ in expression 
\begin{equation}
\int [{\cal D} {\Phi}] \; \left( \theta^{ab}(x) {\tilde \theta}^{cd}(y) + \eta^{ab}(x) {\tilde \eta}^{cd}(y) \right)   \; e^{-S_{loc}} \;,  \label{locphi}
\end{equation}
gives
\begin{equation} 
\frac{ \int [{\cal D} {\Phi}] \;   \left( s \left( \eta^{ab}(x) {\tilde \theta}^{cd}(y)
\right)   \right) \; e^{-S_{loc}} }{ \int [{\cal D} {\Phi}]    \; e^{-S_{loc}}}  =\sigma^4 \;
\frac{  \int {\cal D}A {\cal D}{\phi}\; \delta(\partial A) \; \left( det{\cal M} \right) \;
{\cal R}^{ab}(x)  {\cal R}^{cd}(y)  \; e^{-(S_{YM}+\gamma^4 H(A) + S^{\phi})} } {\int {\cal D}A
{\cal D}{\phi} \; \delta(\partial A) \; \left( det{\cal M} \right)   \; e^{-(S_{YM}+\gamma^4
H(A) + S^{\phi})}  } \;,\label{brstphi}
\end{equation}
showing that, in analogy with the case of the gauge field,  the correlation function \eqref{cphi}  with a cutoff at the Gribov horizon is directly related to the existence of the BRST breaking in the matter sector. 

We can now have a look at the two-point correlation function of the scalar field. Nevertheless, before that, an additional effect has to be taken into account. In very strict analogy with the case of the Refined Gribov-Zwanziger action, eq.\eqref{rgz}, the soft breaking of the BRST symmetry occurring in the scalar matter sector, eq.\eqref{bs}, implies the existence of a non-vanishing BRST exact dimension two condensate, namely 
\begin{equation}
\langle s ( \tilde{\theta}^{ab}(x)  {\eta}^{ab}(x) ) \rangle = \langle ( \tilde{\eta}^{ab}(x)  {\eta}^{ab}(x)  -  \tilde{\theta}^{ab}(x)  {\theta}^{ab}(x) ) \rangle \neq 0 \;.  \label{condphi}
\end{equation}
In order to show that expression \eqref{condphi} in non-vanishing, we couple the operator $( \tilde{\eta}^{ab}(x)  {\eta}^{ab}(x)  -  \tilde{\theta}^{ab}(x)  {\theta}^{ab}(x) ) $ to the local action $S_{ loc}$, eq.\eqref{actlc}, by means of a constant external source $J$, 
\begin{equation}
S_{ loc} - J \int d^4x\;  ( \tilde{\eta}^{ab}(x)  {\eta}^{ab}(x)  -  \tilde{\theta}^{ab}(x)  {\theta}^{ab}(x) )  \;, \label{cj}
\end{equation}
and we evaluate the vacuum energy $\mathcal{E}(J)$ in the presence of $J$, namely 
\begin{equation}
 e^{-V\mathcal{E}(J)}= \int {\cal D}{\Phi} \; e^{ -\left( S_{ loc} - J \int d^4x\;  ( \tilde{\eta}^{ab}(x)  {\eta}^{ab}(x)  -  \tilde{\theta}^{ab}(x)  {\theta}^{ab}(x) ) \right) }   \;. \label{ej}
\end{equation}
Thus, the condensate $\langle ( \tilde{\eta}^{ab}(x)  {\eta}^{ab}(x)  -  \tilde{\theta}^{ab}(x)  {\theta}^{ab}(x) ) \rangle$  is obtained by differentiating $\mathcal{E}(J)$ with respect to $J$ and setting $J=0$ at the end, {\it i.e.}
\begin{equation}
\frac{\partial \mathcal{E}(J)}{\partial J} \Big|_{J=0} = - \langle ( \tilde{\eta}^{ab}(x)  {\eta}^{ab}(x)  -  \tilde{\theta}^{ab}(x)  {\theta}^{ab}(x) ) \rangle    \;. \label{vj}
\end{equation}
Employing dimensional regularisation, to the first order, we have 
\begin{equation}
\mathcal{E}(J) = \frac{(N^2-1)}{2} \int \frac{ d^dk}{(2\pi)^d} \; \log\left( k^2 +m^2_{\phi} +\frac{2N\sigma^4 g^2}{k^2+J} \right)  \; + \;{\hat {\cal E}}   \;, \label{fo}
\end{equation}
where ${\hat {\cal E}} $ stands for the part of the vacuum energy which is independent from $J$. Differentiating eq.\eqref{fo} with respect to $J$ and setting $J=0$, we get 
\begin{equation}
 \langle ( \tilde{\eta}^{ab}(x)  {\eta}^{ab}(x)  -  \tilde{\theta}^{ab}(x)  {\theta}^{ab}(x) ) \rangle = (N^2-1) N \sigma^4 g^2  \int \frac{ d^dk}{(2\pi)^d} \frac{1}{k^2} 
 \frac{1}{k^4 + m^2_\phi \;k^2 +  2 N \sigma^4 g^2}   \neq 0 \;. \label{vcondphi}
\end{equation}
Notice that the integral in the right hand side of eq.\eqref{vcondphi} is ultraviolet convergent in $d=4$. Expression  \eqref{vcondphi} shows that, as long as the parameter $\sigma$ in non-vanishing, the condensate $\langle ( \tilde{\eta}^{ab}(x)  {\eta}^{ab}(x)  -  \tilde{\theta}^{ab}(x)  {\theta}^{ab}(x) ) \rangle$ is dynamically generated. The effect of the condensate 
 \eqref{condphi}  can be taken into account by adding to the action $S_{ loc}$ the novel term 
\begin{equation}
\mu^2_\phi \int d^4x \; s ( \tilde{\theta}^{ab} {\eta}^{ab} )  = \mu^2_\phi \int d^4x\;  ( \tilde{\eta}^{ab}  {\eta}^{ab}  -  \tilde{\theta}^{ab}  {\theta}^{ab} )   \;,  \label{accondphi}
\end{equation}
giving rise to the  Refined action 
\begin{equation}
{\tilde S}_{Ref} =  S_{ loc} +  \int d^4x \left(  \frac{m^2}{2} A^a_\mu A^a_\mu  - \mu^2 \left( {\bar \varphi}^{ab}_{\mu}  { \varphi}^{ab}_{\mu} -  {\bar \omega}^{ab}_{\mu}  { \omega}^{ab}_{\mu} \right)   \right)    - \mu^2_\phi \int d^4x\;  \left( \tilde{\eta}^{ab}  {\eta}^{ab}  -  \tilde{\theta}^{ab}  {\theta}^{ab} \right)  \;. \label{refphi}
\end{equation}
Finally, for the propagator of the scalar field, we get 
\begin{equation} 
\langle  \phi^a(k)  \phi^b(-k) \rangle   =   \delta^{ab}   \frac{k^2 +\mu^2_{\phi}}{k^4 +
(\mu^2_\phi+m^2_{\phi})k^2 + 2Ng^2\sigma^4 + \mu^2_\phi m^2_\phi}  \;. 
\label{phiprop}
\end{equation} 

In the subsection \ref{latticefit} we are going to fit this perturbative propagator to the
correspondent lattice data, so that the free parameters of the theory can be estimated.

\subsection{The quark field} 
In this subsection we generalise the previous construction to the case of quark fields. The  starting non-local action \eqref{acs} is now given by 
\begin{eqnarray}
\label{apsi}
S^{\psi} &=& \int d^4x\; \left( {\bar \psi}^{i} \gamma_\mu D_{\mu}^{ij} \psi^{j} - m_{\psi}  {\bar \psi}^{i} \psi^{i}  \right) 
 \nonumber \\
 &-&
 M^3 g^2   \int d^{4}x\;d^{4}y\;   {\bar \psi}^{i}(x)  (T^a)^{ij} \left[ {\cal M}^{-1}\right]^{ab} (x,y)  (T^b)^{jk} \psi^{k}(y) \;, 
\end{eqnarray}
where the massive parameter $M$ is the analogue of the parameter $\sigma$ of the scalar field and 
\begin{equation}
D^{ij}_\mu = \delta^{ij} \partial_\mu - i g  (T^a)^{ij} A^a_\mu \;, \label{covf}
\end{equation}
is the covariant derivative in the fundamental representation, specified by the generators $(T^a)^{ij}$. As in the previous case, the non-local action \eqref{apsi} can be cast in local form through the introduction of a suitable set of auxiliary fields: $({\bar \lambda}^{ai}, {\lambda}^{ai})$ and $({\bar \xi}^{ai}, {\xi}^{ai})$. The fields $({\bar \lambda}^{ai}, {\lambda}^{ai})$ are Dirac spinors  with two color indices $(a,i)$ belonging, respectively,  to the adjoint and to the  fundamental representation. Similarly, $({\bar \xi}^{ai}, {\xi}^{ai})$ are a pair of spinor fields with ghost number $(-1,1)$. The spinors  $({\bar \lambda}^{ai}, {\lambda}^{ai})$ are anti-commuting, while $({\bar \xi}^{ai}, {\xi}^{ai})$ are commuting. For the local version of the action, we get 
\begin{equation}
S^{\psi}_{loc} = S_0 + S_M \;, \label{locpsi}
\end{equation}
where 
\begin{eqnarray}
S_0  =  \int d^4x\; \left( {\bar \psi}^{i} \gamma_\mu D_{\mu}^{ij} \psi^{j} - m_{\psi}  {\bar \psi}^{i} \psi^{i}   +  {\bar \lambda}^{ai}( -\partial_\mu D^{ab}_\mu) \lambda^{bi} 
+ {\bar \xi}^{ai}( -\partial_\mu D^{ab}_{\mu} ) \xi^{bi}  \right.
\nonumber \\
\left. -(\partial_\mu {\bar \xi}^{ai}) g f^{acb} (D^{cm}_\mu c^m) \lambda^{bi}  \right)  \;, \label{szpsi}
\end{eqnarray}
and 
\begin{equation}
S_M = g M^{3/2} \int d^4x \; \left(   {\bar \lambda}^{ai} (T^a)^{ij} \psi^{j} +  {\bar \psi}^{i} (T^a)^{ij} \lambda^{aj}    \right)    \;. \label{sM}
\end{equation}
The non-local action $S^{\psi}$ is easily recovered by integrating out the auxiliary fields $({\bar \lambda}^{ai}, {\lambda}^{ai})$ and $({\bar \xi}^{ai}, {\xi}^{ai})$. As in the case of the scalar field, the term $S_M$ induces a soft breaking of the BRST symmetry. In fact, from 
\begin{eqnarray}
s \psi^{i} & = & -ig c^a (T^a)^{ij} \psi^{j} \;, \nonumber \\
s {\bar \psi}^{i} & = & -ig {\bar \psi}^{j} c^a (T^a)^{ji} \;, \nonumber \\
s {\bar \xi}^{ai} & = & {\bar \lambda}^{ai} \;, \qquad s {\bar \lambda}^{ai} = 0\;, \nonumber \\
s {\lambda}^{ai} & = & {\xi}^{ai} \;, \qquad s{\xi}^{ai}=0 \;,  \label{spsi}
\end{eqnarray}
one easily checks that 
\begin{equation}
s S^{\psi}_{loc} = s  S_M = M^{3/2}  \Delta^M   \;, \label{sbM} 
\end{equation}
where
\begin{equation}
\Delta^M = \int d^4x \; \left(  ig^2  {\bar \lambda}^{ai} (T^a)^{ij} c^b (T^b)^{jk}\psi^{k} -ig^2 {\bar \psi}^{k} c^b (T^b)^{ki}(T^a)^{ij} \lambda^{aj}  
- g {\bar \psi}^{i} (T^a)^{ij} \xi^{aj}   \right)   \;. \label{dm}
\end{equation}
Again, being of dimension $5/2$ in the fields, $\Delta^M$ is a soft breaking. In the present case, for the quantity  \eqref{rmatter} we have 
\begin{eqnarray}
{\cal R}^{ai}_{\;\alpha} (x)  & = &  g \int d^4z\;  ({\cal M}^{-1})^{ab} (x,z)   \;(T^b)^{ij} \psi^{j}_{\alpha} (z)     \;, \nonumber \\
{\bar {\cal R}}^{bj}_{\;\beta} (x)  & = &  g \int d^4z\;  ({\cal M}^{-1})^{bc } (x,z)  {\bar \psi}^{k}_{\beta}(z)  \;(T^c)^{kj}     \;, \label{rpsi}
\end{eqnarray}
where we have explicitated  the Dirac indices $\alpha,\beta=1,2,3,4$. 

As in the case of the scalar field, the action $S^{\psi}_{loc} $ can be added to the Gribov-Zwanziger action. The resulting action, $(S_{GZ} + S^{\psi}_{loc})$, turns out to be renormalizable. Although we shall not give here the details of the proof of the renormalizability of the action $(S_{GZ} + S^{\psi}_{loc})$, it is worth mentioning that it can be given by following the framework already outlined in \cite{Baulieu:2009xr}, where a similar non-local spinor action has been considered. 

Proceeding now as in the case of the scalar field, one finds 
\begin{eqnarray} 
	&&
	\frac{ \int [{\cal D} {\Phi}] \;   \left[ s \left( {\bar \xi}^{ai}_{\alpha}(x) {\lambda}^{bj}_\beta(y)  \right)   \right] \; e^{-(S_{GZ}+S^{\psi}_{loc})} }{ \int [{\cal D} {\Phi}]    \; e^{-(S_{GZ}+S^{\psi}_{loc})}}  =
	\nonumber \\
	&=&
	M^3 \; \frac{  \int {\cal D}A {\cal D}{\psi} {\cal D}{\bar \psi} \; \delta(\partial A)  \left( det{\cal M} \right) {\cal R}^{ai}_{\;\alpha}(x)  {\bar {\cal R}}^{bj}_{\;\beta}(y)  \; e^{-(S_{YM}+\gamma^4 H(A) + S^{\psi})} } {\int {\cal D}A {\cal D}{\psi} {\cal D}{\bar \psi} \; \delta(\partial A) \; \left( det{\cal M} \right)   \; e^{-(S_{YM}+\gamma^4 H(A) + S^{\psi})}  } \;,\label{brstpsi}
\end{eqnarray}
showing that  the correlation function $\langle {\cal R}^{ai}_{\;\alpha}(x)  {\bar {\cal R}}^{bj}_{\;\beta}(y) \rangle$  with a cutoff at the Gribov horizon is  related to the existence of the BRST breaking, eq.\eqref{sbM}. 

Let us end this section by discussing the two-point correlation function of the quark field. As before, an additional effect has to be taken into account. Also here,  the soft breaking of the BRST symmetry, eq.\eqref{sbM}, implies the existence of a non-vanishing BRST exact dimension two condensate, namely 
\begin{equation}
\langle s ( {\bar {\xi}}^{ai}(x)  {\lambda}^{ai}(x) ) \rangle = \langle ( {\bar \lambda}^{ai}(x)  {\lambda}^{ai}(x)  + {\bar  \xi}^{ai}(x)  {\xi}^{ai}(x) ) \rangle \neq 0 \;,  \label{condpsi}
\end{equation}
whose effect can be taken into account by adding to the action $S^{\psi}_{loc}$ the term 
\begin{equation}
\mu^2_\psi \int d^4x \; s   ( {\bar {\xi}}^{ai}(x)  {\lambda}^{ai}(x) )   = \mu^2_\psi \int d^4x\;  ( {\bar \lambda}^{ai}(x)  {\lambda}^{ai}(x)  + {\bar  \xi}^{ai}(x)  {\xi}^{ai}(x) )  \;.  \label{accondpsi}
\end{equation}
Therefore, including the dimension two condensates, we end up with the Refined action 
\begin{equation}
	{\tilde S}_{Ref}^{\psi} =  S_{RGZ} + S^{\psi}_{loc} +  \mu^2_\psi \int d^4x\;  \left[ {\bar \lambda}^{ai}(x)  {\lambda}^{ai}(x)  + {\bar  \xi}^{ai}(x)  {\xi}^{ai}(x) \right]   \;. \label{refpsi}
\end{equation}
Finally, for the propagator of the quark field, we get 
\begin{equation} 
\langle  \psi^{i}(k)  {\bar \psi}^{j}(-k) \rangle   =   \delta^{ij} \;  \frac{-ik_\mu
\gamma_\mu + {\cal A}(k^2)}{k^2 + {\cal A}^2(k^2)}  \;, \label{psiprop}
\end{equation} 
 where 
\begin{equation}
{\cal A}(k^2) = m_{\psi} + \frac{g^2 M^3 C_F}{k^2+\mu^2_\psi} \;, \label{A}
\end{equation}
and 
\begin{equation}
 (T^a)^{ij} (T^a)^{jk} = \delta^{ik} C_F \;, \qquad C_F= \frac{N^2-1}{2N}  \;.    \label{norm}
\end{equation}
%

In the following section \ref{latticefit} we are going to fit our perturbative Refined matter
propagators \eqref{phiprop} and \eqref{psiprop} with the most recent curves of scalar and quark
propagators from the lattice data. As will be shown, for the fitted parameters of the theory
both, the scalar and quark, propagators exhibit positivity violation, so that they do not
belong to the spectrum of the asymptotically free physical particles of the theory; they are
said to be confined.

\section{Analysis of $\langle {\tilde {\cal R}}(k) {\tilde {\cal R}}(-k) \rangle$ in the light
of the available lattice data}
\sectionmark{Analysis of $\langle {\tilde {\cal R}}(k) {\tilde {\cal R}}(-k) \rangle$}
\label{latticefit}

In the present section we present a discussion of the correlation function \eqref{rcm}  in
the case of  quark and scalar fields, relying on the available lattice data for the quark and
scalar propagators. This will be done by working out in detail the case of a scalar field in
the adjoint representation. We shall also discuss how $\langle {\cal R}^{ai}(x)  {\cal
R}^{bj}(y)  \rangle $ encodes information on the soft  breaking of the BRST symmetry. In the
same section we generalize the previous construction  to the case of quark fields. The final Appendix collects  the details of the algebraic proof of
the renormalizability of the local action obtained by the addition of the term  \eqref{hmatter}
in the case of a scalar matter field in the adjoint representation.

Let us, then, investigate the correlation function $\langle {\tilde {\cal R}}(k) {\tilde {\cal R}}(-k) \rangle$, that signals soft BRST breaking in the matter sector, in  light of available lattice data for gauge-interacting matter propagators in the Landau gauge.

As in the pure gauge case, one may rely on the general cluster decomposition property in order
to obtain the leading behavior in the deep infrared region. The point is that, in one side we
have the highly non-local operator
\begin{eqnarray}
{\cal R}^{ai}(x)  {\cal R}^{bj}(y) ~=~ g^{2} \iint d^{4}z\,d^{4}z'\,\, [{\cal
M}^{-1}]^{ad} (x,z)   \;(T^d)^{il} \;F^{j}(z) \;(T^e)^{jl} \;F^{j}(z')  \,[{\cal M}^{-1}]^{be}
(y,z')\,,
\end{eqnarray}
whose non-locality stems from the squared inverse of the FP operator. In the other side we have
\begin{eqnarray}
\langle c^{a}(y)\bar{c}^{b}(x) \rangle ~=~ [{\cal M}^{-1}]^{ab} (x,y)\,,
\end{eqnarray}
so that the non-local operator may be rewritten as,
\begin{eqnarray}
{\cal R}^{ai}(x)  {\cal R}^{bj}(y) ~=~ \iint d^{4}z\,d^{4}z' 
\langle c^{a}(z)\bar{c}^{d}(x) \rangle  \langle c^{e}(z')\bar{c}^{b}(y) \rangle \;(T^d)^{il}
\;F^{j}(z) \;(T^e)^{jl} \;F^{j}(z')  \,.
\label{RRop}
\end{eqnarray}
Therefore, the \emph{vev} of this operator can be written as
\begin{eqnarray}
\langle {{\cal R}}^{ai}(x) {{\cal R}}^{bj}(y)\rangle 
~=~ 
{g^{2}}   \int d^{4}z\;d^{4}z'\; \langle \bar{c}^{a}(x) c^{a'}(z) F^{i'}(z) (T^{a'})^{i'i}
\bar{c}^b(y) c^{b'}(z') (T^{b'})^{j'j} F^{j'} (z') 
\rangle
\end{eqnarray}
whence the cluster decomposition principle applies to the ghost and matter propagators,
yielding to
\begin{eqnarray}
\langle {{\cal R}}^{ai}(x) {{\cal R}}^{bj}(y)\rangle 
&=& g^2  (T^{a})^{i'i} (T^{b})^{i'j}\int d^4k {\rm e}^{ik(x-y)}{\cal G}(k^{2})D(k^2)+
\nonumber\\
&+&
{g^{2}}   \int d^{4}z\;d^{4}z'\; \langle \bar{c}^{a}(x) c^{a'}(z) F^{i'}(z) (T^{a'})^{i'i}
\bar{c}^b(y) c^{b'}(z') (T^{b'})^{j'j} F^{j'} (z') 
\rangle_{1PI}
\;.
\label{<RR>}
\end{eqnarray}
The cluster decomposition principle could be seen as a reflection of the non-locality of the
operator ${{\cal R}}^{ai}(x) {{\cal R}}^{bj}(y)$. Notice that the operator \eqref{RRop} depends
on two non-local quantities,$\langle c^{a}(z)\bar{c}^{d}(x) \rangle$ and $\langle
c^{e}(z')\bar{c}^{b}(y) \rangle$, measured in two unrelated points $x$ and $y$.

On equation \eqref{<RR>} ${\cal G}(k^2)$ is the ghost propagator, while $D(k^2)$ now stands for the propagator of
the associated matter field. The one-particle-irreducible (1PI) contribution above (the second
term) becomes subleading in the IR limit, since in this case the points $x$ and $y$ are largely
separated and
the cluster decomposition applies. This can also be seen diagrammatically: since the external legs
are ghosts, these corrections will involve at least two ghost-gluon vertices, that carry a
derivative coupling.
In fact, as a consequence of the transversality  of the gluon propagator, factorization of the
external momentum takes place, implying the subleading character of the 1PI contributions.

Therefore, in the limit $k\to 0$, the (full) ghost and matter propagators alone dictate the momentum-dependence of the correlation function $\langle {\tilde {\cal R}}(k) {\tilde {\cal R}}(-k) \rangle$, i.e.
\begin{eqnarray}
\langle \tilde{\cal R}^{ai}(k) \tilde{\cal R}^{bj}(-k)
\rangle
&\sim& g^2 {\cal G}^2(k)D(k^2)
\;.
\end{eqnarray}
Having in mind the non-enhanced ghost propagator, ${\cal G}(k^2)\sim 1/k^2$ (as observed in high-precision pure gauge simulations in the Landau gauge \cite{Cucchieri:2007rg,Cucchieri:2008fc,Cucchieri:2011ig}), it is straightforward to conclude that a finite zero-momentum value for the matter propagators is a sufficient condition for a $\sim 1/k^4$  behavior of the correlation function $\langle {\tilde {\cal R}}(k) {\tilde {\cal R}}(-k) \rangle$ in the deep IR.

As we shall see in the following subsections, both scalar and fermion propagators display, when coupled to non-Abelian gauge fields, a shape compatible with a finite zero-momentum value in the currently available lattice data. We expect thus a $\sim 1/k^4$ behavior of  the correlation function $\langle {\tilde {\cal R}}(k) {\tilde {\cal R}}(-k) \rangle$ in the matter sector, being in this sense a universal property associated with the Faddeev-Popov operator -- when coupled to any colored field --  in confining Yang-Mills theories that can be easily probed in the future via direct lattice measurements.

Moreover, fits of the lattice data are presented for adjoint scalars in
subsection \ref{scalars} and for fermions in subsection \ref{quarks}. This analysis shows that the propagators for gauge-interacting scalars and fermions are compatible not only with a finite zero-momentum limit, but also with a complete analytical form that can be extracted from an implementation of soft BRST breaking in the matter sector to be presented below, in Sect.3.

\subsection{The scalar field in the adjoint representation }
\label{scalars}

In this subsection we consider real scalar fields coupled to a confining Yang-Mills theory:
\begin{eqnarray}
{\cal L} ~=~ \frac{1}{4} F_{\mu\nu}^aF_{\mu\nu}^a +\frac{1}{2} [D_{\mu}^{ab}\phi^b]^2 + \frac{m_{\phi}^2}{2}\phi^a\phi^a +\frac{\lambda}{4!} [\phi^a\phi^a]^2  + {\cal L}_{GF}\;,
\end{eqnarray}
where  ${\cal L}_{GF}$ is the Landau gauge fixing term and $\phi$ is a real scalar field in the adjoint representation of $SU(N)$ and there is no Higgs mechanism, namely $\langle\phi\rangle = 0$.  

We are interested in analyzing the infrared non-perturbative regime, focussing especially on
the adjoint scalar propagator. We resort to the lattice implementation of this system:
currently available in the quenched approximation with the specific setup described in
\cite{Maas:2010nc}. Preliminary and unpublished data points for larger lattice sizes (with
lattice cutoff $a^{-1}=4.94 $ GeV and $N=30$ lattice sites) \cite{axel} are displayed in Fig. 1
for different values of the bare scalar mass ($m_{bare}=0,\,1,\,10$ GeV). It should be noticed
that this data is unrenormalized in the lattice sense. The renormalization procedure that fixes
the data to a known renormalization scheme and the resulting points will be discussed below.

\begin{figure}[h!]
\center
\includegraphics[width=9cm]{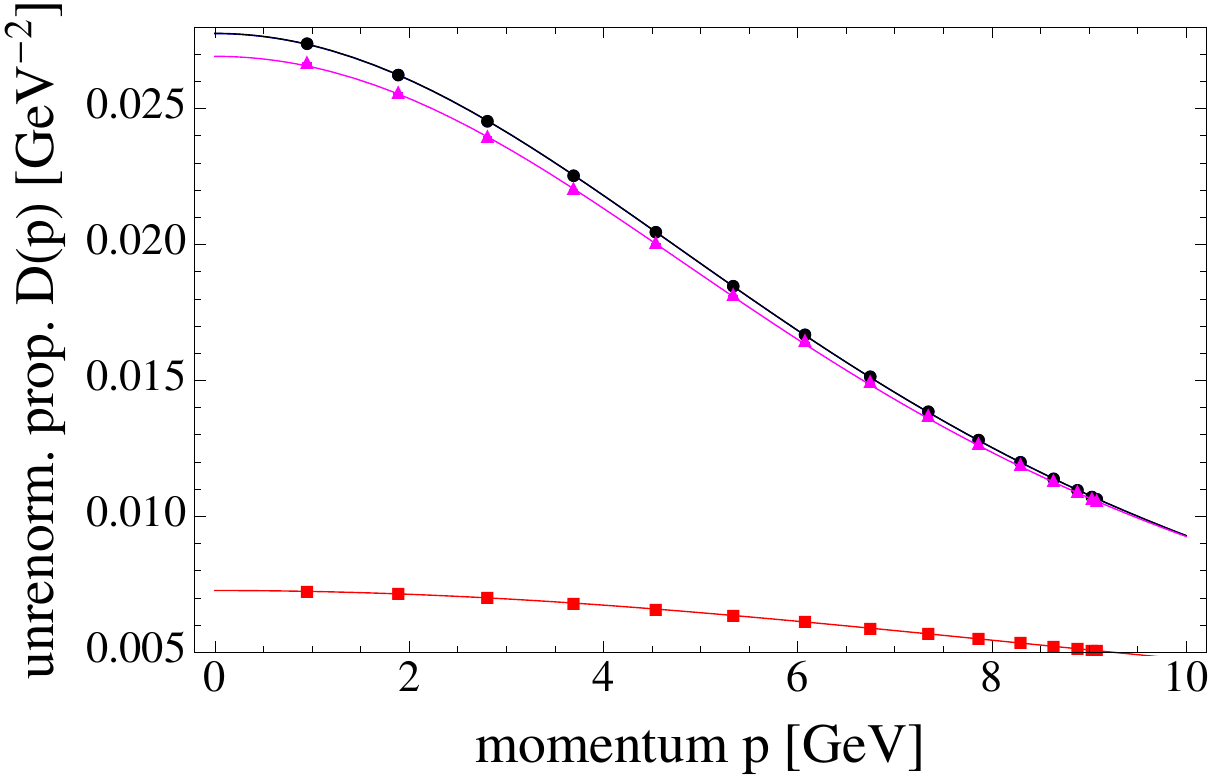}
\caption{Unrenormalized propagator for different bare masses of the scalar field:
$m_{bare}=0$(top, black), 1 and 10 GeV (bottom, red). The points are preliminary and
unpublished lattice data from quenched simulations \cite{axel} (for lattice cutoff $a^{-1}=4.54
$ GeV, $N=30$ and $\beta=2.698$; cf. also \cite{Maas:2010nc} for more details on the lattice
setup and measurements) and the curves are the corresponding fits, whose parameter values can
be found in Table \ref{table:par}.}
\label{scalarlattfig}
\end{figure}

First of all, the data tends to show a finite zero-momentum value for the scalar propagator,
irrespective of its bare mass. This indicates   -- together with the well-stablished
non-enhanced ghost propagator -- that the correlation function $\langle \tilde{R} \tilde
R\rangle_k$ is indeed non-vanishing in the IR limit, presenting the power-law enhancement
$\sim1/k^4$ that we have anticipated above.

The curves in Fig. \ref{scalarlattfig} further show that the data is compatible with fits of a
propagator of the same type as the one we found in \eqref{phiprop},
\begin{eqnarray}
D(p) ~=~ Z\,
\frac{p^2+\mu_{\phi}^2}{p^4+p^2(m_{\phi}^2+\mu_{\phi}^2)+\sigma^4+m_{\phi}^2\mu_{\phi}^2}
\,,\label{RGZfit}
\end{eqnarray}
where $Z,\mu_{\phi},m_{\phi},\sigma$ are the fit parameters, whose values are presented in
Table \ref{table:par}. In this case we may extrapolate the fits in order to obtain the specific values at
zero momentum: $D(p=0)\approx 0.028,\, 0.027,\, 0.0073$ GeV$^{-2}$, for the bare mass $m_{bare}
= \unit{0,1,10}{\giga\electronvolt}$, respectively, so that the non-trivial IR limit is clear.
Moreover, the $\sigma$ parameter -- which is directly related to the non-vanishing of the
\emph{vev} of an exact BRST local operator, $\langle s(\eta^{ab}(x)\tilde{\theta}^{cd}(x))
\rangle \neq 0$ \eqref{brstphi} -- seems to be non-vanishing.  It is also interesting to point
out that the obtained fits correspond to a combination of two complex-conjugate poles for all
values of bare scalar mass, indicating the absence of a K\"all\'en-Lehmann spectral
representation for this two-point function and the presence of positivity violation. In this
sense the adjoint scalar propagators  consistently represent confined degrees of freedom, that
do not exhibit a physical propagating pole.
    
    \begin{table}[ht]
 \caption{Fit parameters for the unrenormalized propagator in powers of GeV.}
 \vspace{0.3cm}
  \centering
   \begin{tabular}{c ||c| c| c| c||c}
    $m_{bare}$ & $\mu_{\phi}^2$ & $m_{\phi}^2$& $\sigma^4$ & $Z$&$\chi^2/\textrm{dof}$\\
    \hline\hline 
    0 & 120
    & 0 & 4913 & 1.137 &0.31\\
      \hline 
    1 & 46 & 34  & 644 & 1.28 & 1.84\\
      \hline 
    10 & 88
    & 158 & 1267 & 1.26 & 0.10
     \end{tabular}
     \label{table:par} 
    \end{table}
    
An important issue to be addressed is the possibility of scheme dependence of those findings. To check for this, we have also analyzed the scalar propagators after renormalization in another scheme.
As usual, renormalization is implemented through the inclusion of mass $\delta m_{\phi}$ and wave-function renormalization $\delta Z $ counterterms:
\begin{eqnarray}
D_{ren}^{-1}(p)&=& D^{-1}(p) +\delta m_{\phi}^2 +\delta Z (p^2+m_{bare}^2)
\,,
\end{eqnarray}
where the counterterms are obtained by imposing the following renormalization conditions (for $\Lambda=2$ GeV):
\begin{enumerate}
	\item[i)] $\partial_{p^2}D_{ren}^{-1}(p=\Lambda)=1 $;
	\item[ii)] $D_{ren}^{-1}(p=\Lambda)=\Lambda^2+m_{bare}^2$.
\end{enumerate}
The fit functions were used to compute the counterterms and the renormalized points are obtained from the original lattice data by adding the same counterterms\footnote{Direct renormalization of lattice data was avoided, since we did not have access to the measurement of $\partial_{p^2}D$ and the number of data points available was not sufficient for a reliable numerical derivative to be computed.}. Results are shown in Fig. 2 and Table 2.

\begin{figure}[h!]
\center
\includegraphics[width=10cm]{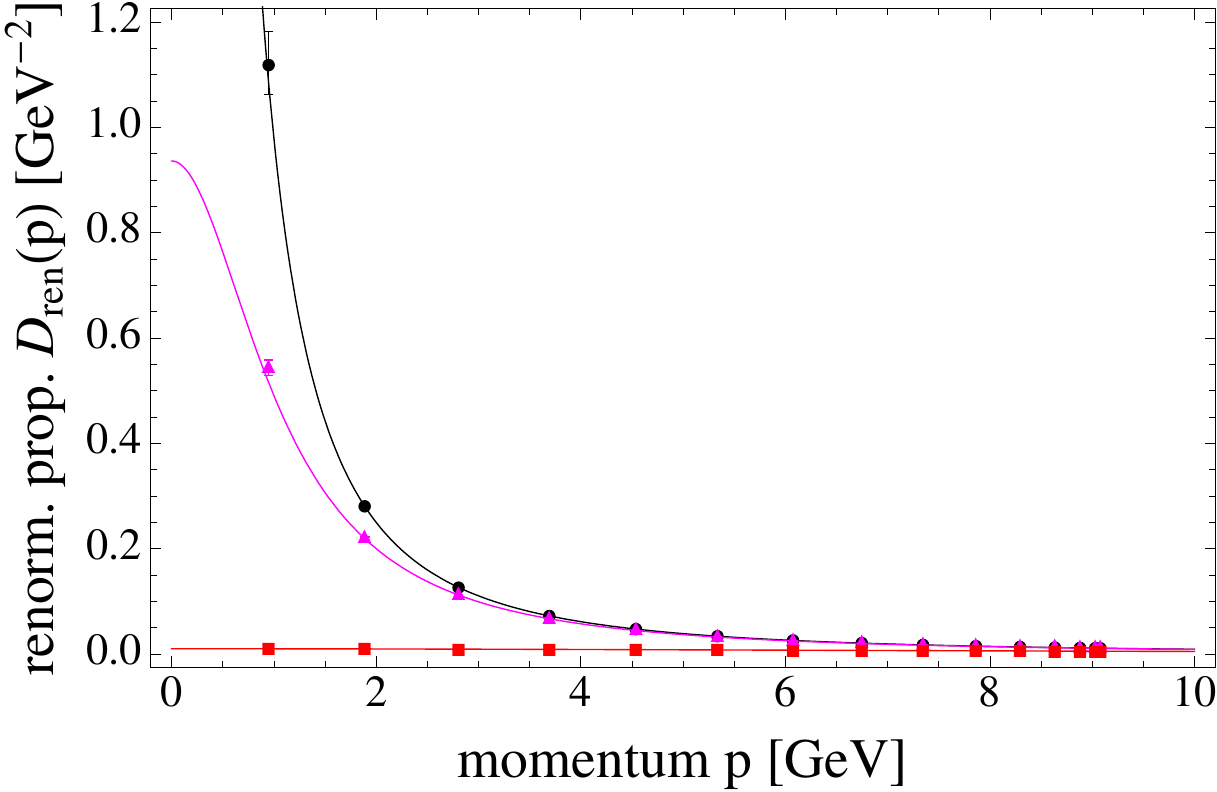}
\caption{Renormalized propagator for different bare masses of the scalar field: $m_{bare}=0$(top, black), 1 and 10 GeV (bottom, red). The points are obtained from the unrenormalized lattice data  \cite{axel,Maas:2010nc} displayed in Fig.1.}
\end{figure}

The renormalized propagator may be rewritten in the form \eqref{RGZfit}, with redefined parameters $m_{\phi}',\sigma',Z'$:
\begin{eqnarray}
D_{ren}(p)&=&
Z'\, \frac{p^2+\mu_{\phi}^2}{p^4+p^2(m_{\phi}'^2+\mu_{\phi}^2)+\sigma'^4+m_{\phi}'^2\mu_{\phi}^2}
\end{eqnarray}    
    
        \begin{table}[ht]
 \caption{Counterterms, redefined fit parameters and zero-momentum values of the renormalized propagator in powers of GeV.}
 \vspace{0.3cm}
  \centering
   \begin{tabular}{c ||c| c|| c| c|c||c}
    $m_{bare}$ & $\delta m_{\phi}^2$ & $\delta Z$& $m_{\phi}'^2$ & $\sigma'^4$ & $Z'$& $D_{ren}(p=0)$\\
   \hline\hline 
    0 & -35.98    & 0.40 & -28.09 & 3374.32 & 0.781&26.7\\
      \hline 
    1 & -36.49 & 0.416 & -8.18  & 420.84 & 0.834 &0.94\\
      \hline 
    10 &  -69.69    & 0.322 & 79.19 & 902.23 & 0.894 &0.01
     \end{tabular}
     \label{table:CT} 
    \end{table}

All the interesting qualitative properties observed in the unrenormalized data remain valid, namely: $(i)$ finite IR limit, $(ii)$ compatibility with 4-parameter fits of the same form, with non-trivial $\sigma$ values, $(iii)$ the fit parameters yield complex-conjugate poles, so that the renormalized propagator is still compatible with positivity violation and confinement.

We underline that the present analysis for the scalar fields is meant to be a preliminary study
of the propagator. As such, the results are still at the qualitative level. A more quantitative
analysis would require further simulations with improved statistics and even larger lattices.

\subsection{The quark field}
\label{quarks}

Now we consider the case of gauge-interacting fermionic fields coupled to a confining Yang-Mills theory. Of course, the case of QCD is the emblematic example. 
We will verify that the same qualitative properties shown above for scalar fields can also be found in this case, indicating that the IR enhancement of the correlation function $\langle\tilde {\cal R}\tilde {\cal R}\rangle\sim 1/k^4$ seems to be universally present in the confined matter sector.

The fermionic propagator is decomposed as usual,
\begin{eqnarray}
{\cal S}(p)=Z(p^2)\frac{-ip_{\mu}\gamma_{\mu}+{\cal A}(p^2)}{p^2+{\cal A}(p^2)}\,,
\end{eqnarray}
and our interest resides solely on the mass function ${\cal A}(p^2)$, whose lattice data will be analyzed here.

\begin{figure}[h!]
   \centering
       \includegraphics[width=9cm]{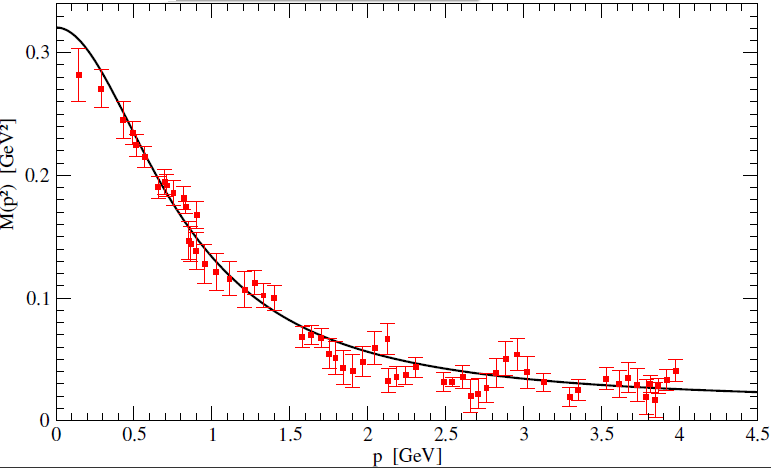}
               \caption{Lattice quark mass function \cite{Parappilly:2005ei} with its fit ${\cal A}(p^2)$. Figure extracted from \cite{Dudal:2013vha}; fit obtained by O. Oliveira \cite{Orlando}.}
\end{figure}

As already discussed and shown in \cite{Dudal:2013vha}, the data of \cite{Parappilly:2005ei} for the mass function of the propagator of degenerate up ($u$) and down ($d$) quarks with current mass $\mu=0.014~\text{GeV}$ can be fitted excellently with
\begin{equation}\label{fit}
{\cal A}(p^2)=\frac{M^3}{p^2+m^2}+\mu~\text{with}~M^3=0.1960(84)~\text{GeV}^3\,, m^2=0.639(46)~\text{GeV}^2 \quad (\chi^2/\text{d.o.f.}~=~1.18)\,.
\end{equation}
as can be seen in Fig. 3.
 
The quark propagator presents clearly a finite IR limit. This is, in fact,  well-known in QCD
as dynamical mass generation and is intimately related to chiral symmetry breaking.
Interestingly enough, this is also a sufficient condition -- supposing a non-enhanced ghost
propagator -- for the soft BRST breaking in the quark sector through the IR enhancement of the
correlation function $\langle \tilde{\cal R}\tilde{\cal R}\rangle$. Again, we predict a $\sim
1/k^4$ IR scaling for this observable, now in the quark sector. This suggests a close relation
between soft BRST breaking and chiral symmetry breaking, and may provide an interesting
underlying connection between confinement and chiral symmetry breaking.

\section{Discussions about the results\label{conc4}}

One of the striking features of the (R)GZ formulation of non-perturbative Euclidean continuum
Yang-Mills theories is the appearance of the soft breaking of the BRST symmetry, which seems to
be deeply related to gluon confinement. Recently, direct lattice investigations
\cite{Cucchieri:2014via} have confirmed the existence of such a breaking, through the analysis
of the Boson-ghost correlation function:
\begin{eqnarray}
\langle \tilde {\cal R}^{ab}_{\;\;\mu}(k)\tilde {\cal
R}^{cd}_{\;\;\nu}(-k)\rangle&\stackrel{k\to 0}{\sim}&\frac{1}{k^4} \,,
\label{RRgluon}
\end{eqnarray}
with
\begin{eqnarray}
{\cal R}^{ac}_{\;\;\mu}(x)&=&g \int d^4z ({\cal M}^{-1})^{ad}(x,z) f^{dec}A^e_{\mu}(z)
\,.
\end{eqnarray}
As pointed in \cite{Cucchieri:2014via}, the non-vanishing of such correlator signals the
breaking of the BRST invariance, since it is related with the \emph{vev} of an exact BRST local
operator,
\begin{eqnarray}
\langle s(\varphi^{ab}_{\mu}(x)\bar{\omega}^{cd}_{\nu}(y))  \rangle_{GZ} ~=~ \gamma^{4} \langle
{\cal R}^{ab}_{\;\;\mu}(x) {\cal R}^{cd}_{\;\;\nu}(y)\rangle_{GZ}\,.
\end{eqnarray}
Interestingly enough, the behavior \eqref{RRgluon}, in the gauge sector, is in quite good agreement with the RGZ framework.

Inspired by the gauge sector, we proposed an effective model for the matter sector, where a
similar structure to that one of the gauge sector can be consistently implemented. It must be
clear that we had no geometrical motivation, or any ambiguity issues in the matter field
quantization procedure, that would have led us to this effective model. However, it seems to be
reasonable that, in some sense, non-perturbative features of the gauge field play any role in
the non-perturbative feature of the matter field \footnote{More on the interplay between
non-perturbative features of the gauge sector and quark confinement will be treated in the
chapter \ref{Ploop1}.}. Therefore, within the framework of (R)GZ quantization procedure, the
fate of restricting the space of configuration of the gauge field to the first Gribov region,
$\Omega$, may be reflected in the matter sector.

Two main interesting cases were considered in this chapter, the adjoint scalar and the quark
fields. In these cases we could show that it is possible to construct an analogous operator ${\cal
R}^{ai}_{\;\;F}$ for matter field,
\begin{eqnarray}
{\cal R}^{ai}_{\;\;F}(x)  &=&  g \int d^4z\;  ({\cal M}^{-1})^{ab} (x,z)   \;(T^b)^{ij}
\;F^{j}(z)  
\,,
\end{eqnarray}
so that the correlation function $\langle{\cal R}_{\;\;F}{\cal R}_{\;\;F}\rangle$ is
non-vanishing and, from the available lattice data, seems to behave like the Boson-ghost
propagator in the IR regime, \eqref{RRgluon}, namely
\begin{eqnarray}
\langle \tilde {\cal R}^{ai}_{\;\;F}(k)\tilde {\cal R}^{bj}_{\;\;F}(-k)\rangle&\stackrel{k\to 0}{\sim}&\frac{1}{k^4} 
\,.
\end{eqnarray}
Again, the non-vanishing of $\langle{\cal R}_{\;\;F}{\cal R}_{\;\;F}\rangle$ indicates the soft
breaking of the BRST symmetry in the matter sector, since the \emph{vev} of ${\cal
R}_{\;\;F}{\cal R}_{\;\;F}$ can be written in terms the \emph{vev} of a BRST exact local
operator of the localizing fields,
\begin{eqnarray}
\langle s(\eta^{ab}(x)\bar{\theta}^{cd}(y))  \rangle ~=~ \gamma^{4} \langle
{\cal R}^{ab}(x) {\cal R}^{cd}(y)\rangle\,.
\end{eqnarray}
In this sense, the correlation function $\langle{\cal R}_{\;\;F}{\cal R}_{\;\;F}\rangle$ could
be regarded as a direct signature for BRST breaking, being accessible both analytically as well
as through numerical lattice simulations. 

Concerning the analytic side, we have been able to construct a local and renormalizable action
including matter fields which accommodates the non-trivial correlation functions $\langle{\cal
R}_{\;\;F}{\cal R}_{\;\;F}\rangle$. Our analysis further suggests that the inverse of the
Faddeev-Popov operator ${\cal M}^{-1}$, whose existence is guaranteed by the restriction to
the first Gribov region $\Omega$ of the gauge field, couples in a universal way to any coloured
field $G^i$ ({\it e.g.} gluon and matter fields),
\begin{eqnarray}
{\cal R}^{ai}_{\;\;G}(x)  &=&  g \int d^4z\;  ({\cal M}^{-1})^{ab} (x,z)   \;(T^b)^{ij}
\;G^{j}(z) \,,
 \label{RG}
\end{eqnarray}
giving rise to a non-vanishing correlation function 
\begin{eqnarray}
\langle \tilde {\cal R}_{\;\;G}(k)\tilde {\cal R}_{\;\;G}(-k)\rangle&\stackrel{k\to
0}{\sim}&\frac{1}{k^4} \,.
\label{RRG}
\end{eqnarray}

The construction carried out here was restricted to the Landau gauge, although something
similar could be developed in other gauges, \emph{e.g.} the Maximal Abelian Gauge
\cite{Capri:2015pxa}, or even in the wider class of Linear Covariant Gauges, in a framework that
lives invariant the action under a non-perturbative version of the BRST symmetry
\cite{Capri:2015ixa,Capri:2015nzw,Capri:2016aqq}.

\chapter{The UV safety of any Gribov-like confined theory}
\chaptermark{The UV safety of any Gribov-like theory}
\label{UVpropsofconfiningprop}

In the present Chapter we are going to continue the analysis started in the previous Chapter,
concerning general effective models presenting non-local terms in the action \emph{\`a la}
Gribov. Precisely, here we present some interesting observations about the UV behavior of such
models, guided by the already known UV safety of Yang-Mills models within the Gribov horizon,
and prove to all orders the renormalizability of such general models.

The feature that we want to explore is the fact that both the GZ and the RGZ
tree-level propagators hold the key for the good UV behavior of the theory. More precisely,
notice that the gluon propagator ${\cal D}(k^2)$ \eqref{Dg} can be rewritten as a sum of its UV
perturbative term plus an effective non-perturbative contribution,
 \begin{eqnarray} 
{\cal D}(k^2) & = & \frac{k^2 +\mu^2}{k^4 + (\mu^2+m^2)k^2 + 2Ng^2\gamma^4 + \mu^2 m^2}  \;.\nonumber\\ 
& = &  \frac{1}{k^2 +  m^2} - \frac{2Ng^2\gamma^4}{\left(k^2 +  m^2\right)\left(k^2 +  M_{+}^2\right)\left(k^2 +  M_{-}^2\right)}
\label{Dg2}
\end{eqnarray}
where
\begin{eqnarray} 
M^2_{\pm}= \frac{\mu^2 + m^2}{2} \pm \frac 12 \sqrt{\left(\mu^2 + m^2 \right)^2 -
8Ng^2\gamma^4} \,.
\label{masses1}
\end{eqnarray}
The first term in \eqref{Dg2} represents the usual propagator of a massive vector boson. The
second term is the contribution coming from the restriction to the Gribov region. Notice the
negative sign that points to an unphysical contribution that violates positivity requirements.
The important feature we want to emphasize is the subleading contribution of the second term in
the $UV$: it presents a $\sim 1/k^4$ suppression with respect to the standard first term, which
will always produce a UV convergent loop contribution in dimension 4. The renormalization of
the RGZ and GZ (which corresponds to $\mu = m =0$) follows from this important property and, as
already mentioned, it is well known that $\gamma$ does not renormalize independently and thus
cannot be considered as an independent dynamically generated scale.

One is thus led to conjecture that this is a general property of theories displaying such
confining propagators, with $\gamma$ standing for a general mass scale associated with
confinement of the fundamental fields; $\gamma$ must be understood as a scale determined by
other dynamically generated scales of the theory. More precisely, the second term in
\eqref{Dg2} cannot generate any new $UV$ divergences in the theory and therefore cannot change
the renormalization properties of the theory, which must be the same as with $\gamma = 0$. In a
diagrammatic approach, only positive powers of propagators appear, so that it is clear that the
highly-suppressed Gribov contribution (cf. \eqref{Dg2}, e.g.) will not influence the deep UV
behavior of the theory. Furthermore, it follows that if the theory with $\gamma = 0$ does not
generate a mass scale , then, since there can be no divergences proportional to $\gamma$, no
mass scale will be generated in the  $\gamma \neq 0$ theory. This in turn means that it is not
possible to assign a dynamical meaning to the $\gamma$ parameter in this case, i.e., the only
possible solution is to have  $\gamma = 0$ in these cases.

In the following sections we will study a variety of examples that support these claims. In section \ref{confscalar} we discuss the case of an interacting scalar field theory displaying a confining propagator. In section 
\ref{confscalarferm} we consider the inclusion of confined fermions interacting with the confined scalars through a Yukawa term. In section \ref{symN1} we discuss the case of Super Yang-Mills with ${\cal N} =1$ supersymmetries
and show to all orders via the algebraic renormalization approach that the adoption of Gribov-type propagators does not produce any new UV divergences, with the renormalization of the IR parameters being completely defined by the UV renormalization of the parameters of the original theory. 

\section{The confined scalar field}
\label{confscalar}

Let us begin with the theory of a scalar field $\phi$, whose action is given by \eqref{acs}
with a decoupled gauge field. In this situation we have
\begin{eqnarray} 
S ~=~ \int d^4 x \Biggl[ \frac 12  \left(\p{\phi}\cdot \p{\phi} \right)^{2} + 
\frac12 m_{\phi}^2 \; {\phi}\cdot{\phi}
 + \frac{\lambda}{4}  ({\phi}\cdot{\phi})^{2} + 
\frac{\sigma^4}{2}  \left(  \frac{{\phi}\cdot{\phi}}{-\p^2}\right)  \Biggr] \,,
\label{scalar}
\end{eqnarray}
The parameter $m_{\phi}^2$ is the mass of the scalar field in the deconfined ($\sigma \to 0$)
theory and $\lambda$ is the quartic coupling constant. Here,  $\sigma$ is the confining
parameter, or infrared parameter, that shall play a similar role for the scalars as the Gribov
mass does for the confined gluons, as detailed in the previous Chapter. Our claim in this case
is that for the case that the parameter $\sigma$ is non-zero the deep UV behavior of the theory
is not affected, at all.

One should notice that the model considered here follows the construction of the previous
Chapter, \ref{brstonmatter}. Since the action \eqref{scalar} is equivalent to the action
\eqref{acs} defined in the previous Chapter, for a decoupled gauge field, and we are
interested in computing the corresponding propagators, we have just to follow the same
procedure as developed therein: introduce a couple of auxiliary fields in
order to localize the action; change to the Fourier space, obtaining
\begin{eqnarray}
S^{\,quad} ~=~ \int \frac{d^{4}k}{(2\pi)^{4}}\, 
\Bigg\{ 
\frac 12 \phi \left( k^2 + m_{\phi}^2 \right) \phi ~ - ~
\tilde{\eta} k^{2} \eta ~+~ \tilde{\theta} k^{2} \theta ~+~
\sigma^{2}\phi(\eta + \tilde{\eta}
\Bigg\}\,;
\end{eqnarray}
and integrating out the auxiliary fields. One should end up, afterwards, with
\begin{eqnarray}
S^{\,quad} ~=~ \int \frac{d^{4}k}{(2\pi)^{4}}\, 
\Bigg\{ 
\frac 12 \phi \Biggl[ \frac{ k^{4} +m_{\phi}^{2}k^{2} - \sigma^{4}}{k^{2}} \Biggr] \phi
\Bigg\}\,.
\label{scquadact}
\end{eqnarray}
Finally, from the functional generator, one can identify the inverse of the momentum dependent
factor of the quadratic term $\phi^{2}$ of equation \eqref{scquadact} as the \emph{tree-level
confining propagator} of the scalar field:
\begin{eqnarray} 
{\cal D}(k^2) & = & \frac{k^2}{k^4 +m_{\phi}^2k^2 + \sigma^4}  \;.\nonumber\\ 
& = &  \frac{1}{k^2 +  m_{\phi}^2} - \frac{\sigma^4 }{\left(k^2 +  m_{\phi}^2\right)\left(k^2 +
M_{+}^2\right)\left(k^2 +  M_{-}^2\right)}
\nonumber\\ 
& = &  \frac{1}{k^2 +  m_{\phi}^2} -
\sigma^4 \Delta(k^2)
\label{scalarprop}
\end{eqnarray}
where we have isolated the confining contribution to the scalar propagator, $\sigma^4\Delta$,
with
\begin{eqnarray} 
\Delta(k^2)&=& \frac{1}{\left(k^2 +  m_{\phi}^2\right)\left(k^2 +  M_{+}^2\right)\left(k^2 +
M_{-}^2\right)}\label{delta}\,,
\end{eqnarray}
which is highly suppressed in the UV: $\Delta\sim 1/k^6$. The mass parameters $M^2_{\pm}$ are
written in terms of $\sigma$ and $m_{\phi}$:
\begin{eqnarray} 
M^2_{\pm}&=& \frac{m_{\phi}^2}{2} \pm \frac 12 \sqrt{m_{\phi}^4 - 4\sigma^4} \,.
\label{scalarmasses}
\end{eqnarray}
Note that $M^2_{\pm}$ may become complex for large enough $\sigma/m_{\phi}$. The complexity of
these IR mass parameters is closely related to positivity violation and, then, with the absence
of a physical particle interpretation for these excitations, leading to the Gribov-kind
confinement interpretation.

It is not difficult to see that there are no new $UV$ divergences associated with the
non-local contribution (\emph{i.e.} proportional to $\sigma^{4}$) to the action \eqref{scalar}
by looking at the diagrams of primitive divergences of the
theory.

In fact, the one-loop scalar selfenergy is
\begin{eqnarray}
\Diagram{ 
&& c & \\
&&& \\
 f& & f &f
} \propto \int d^4 p{\cal D}(p) = \int d^4 p \frac{1}{p^2 +  m_{\phi}^2}  + \sigma^4 \int d^4p
\Delta (p^2)\nonumber\\
 = \int d^4 p \frac{1}{p^2 +  m_{\phi}^2}  + \sigma^4 (\text{UV finite})\,.
\end{eqnarray}
The correction to the quartic coupling at one loop reads:
\begin{eqnarray}
\Diagram{ 
fd \;\;\;\;\;\;\; & !{fl}{k-p} !{flu}{p}  & \;\;\;\;\;\;\; fu \\
fu \;\;\;\;\;\;\; &&  \;\;\;\;\;\;\; fd
} \propto \int d^4 p {\cal D}(k-p) {\cal D}(p) = \int d^4 p \frac{1}{p^2 +  m_{\phi}^2}
\frac{1}{(k-p)^2 +  m_{\phi}^2} +\nonumber\\
&&\hspace{-10cm}+ \,\sigma^4 \int d^4p \Delta (p^2) \frac{1}{(k-p)^2 +  m_{\phi}^2} 
 + \sigma^4 \int d^4p  \frac{1}{p^2 +  m_{\phi}^2}\Delta ((k-p)^2) \nonumber \\
 &&\hspace{-10cm}+\, \sigma^8  \int d^4p \Delta (p^2) \Delta ((k-p)^2) \nonumber\\
= \int d^4 p \frac{1}{p^2 +  m_{\phi}^2} \frac{1}{(k-p)^2 +  m_{\phi}^2} + {\cal O}(\sigma^4,
\sigma^8) (\text{UV finite }) 
\end{eqnarray}

As a representative example at two-loop order, we may look at the scalar selfenergy sunset
diagram:
\begin{eqnarray}
\Diagram{ 
& !{fl}{k-p-q} !{flu}{p}  &\\
&& \\
f f & !{f}{q} f
} &\propto& \int d^4 p \int d^4 q {\cal D}(k-p-q) {\cal D}(q){\cal D}(p)\nonumber\\
&=&  \int d^4 p \int d^4 q \frac{1}{p^2 +  m_{\phi}^2} \frac{1}{q^2 +
m_{\phi}^2}\frac{1}{(k-p-q)^2 +  m_{\phi}^2} + \nonumber\\
 && + \,\sigma^4 \int d^4 p \int d^4 q  \Delta (p^2)  \frac{1}{q^2 +
m_{\phi}^2}\frac{1}{(k-p-q)^2 +  m_{\phi}^2} +\nonumber\\
&& + \,\sigma^4     \int d^4 p \int d^4 q \frac{1}{p^2 +  m_{\phi}^2} \Delta (q^2)
\frac{1}{(k-p-q)^2 +  m_{\phi}^2}+\nonumber\\
 && + \,\sigma^4      \int d^4 p \int d^4 q \frac{1}{p^2 +  m_{\phi}^2} \frac{1}{q^2 +
m_{\phi}^2} \Delta ((k-p-q)^2) +{\cal O}(\sigma^ 8)\nonumber\\
&=&  \int d^4 p \frac{1}{p^2 +  m_{\phi}^2} \frac{1}{q^2 +  m_{\phi}^2}\frac{1}{(k-p-q)^2 +
m_{\phi}^2} +
\nonumber\\
&&
\phantom{\int d^4 p}
\ \ \ \ \ \ \ \ \ \ \ \ \ \ \  + {\cal O}(\sigma^4, \sigma^8, \sigma^{12}) (\text{UV
finite})\,.
\end{eqnarray}

In all  examples above, the appearance of a general form for the contributions of the confining
scale with increasingly UV convergent momentum integrals is clear.
It is straightforward to realize then that this pattern will spread throughout all orders of
the diagrammatic expansion, so that we are led to infer that  contributions proportional to
$\sigma$  cannot give rise to new primitive divergences, besides the ones coming from the
standard theory (that one with $\sigma =0$).

\section{The confined fermion and scalar fields interacting}
\sectionmark{The fermion and scalar fields interacting}
\label{confscalarferm}

The same reasoning can be applied when Dirac fermions are added to the theory, with an Yukawa
coupling and a fermionic Gribov-type term rendering the fermionic excitations also confined.  

We consider here the theory in the absence of scalar condensates. In this case, the full action
reads 
\begin{eqnarray} 
S &=& \int d^4 x \Biggl[ \frac 12  {\phi}\left(  -\p^{2} +  m_{\phi}^2 \right){\phi} + 
 + \frac{\lambda}{4}  ({\phi}\cdot {\phi})^{2} + 
\frac{\sigma^4}{2}  \left(  \frac{{\phi}\cdot{\phi}}{-\p^2}\right)  +
\nonumber \\
&&
\phantom{\int d^4 x }
\bar{\psi} \left(\dslash + m_{\psi}\right) \psi + g\phi\bar{\psi} \psi + \frac 12 \phi \left(  \frac{\gamma^4}{-\partial^2}\right)
\phi  + \bar{\psi}\left(\frac{M^3}{-\partial^2}\right)\psi \Biggr]\,,
\label{scalarferm}
\end{eqnarray}
where $m_{\psi}$ is the mass of the original fermion field (i.e. for $M\to 0$) and $g$ is the
Yukawa coupling. In the fermionic sector the IR mass scale analogous to the Gribov parameter is
$M$, such as in the previous Chapter. 

Analogously to the previous purely scalar case, the \emph{tree-level propagators} of both the
scalar and fermion fields are obtained through the insertion of auxiliary fields,
different doublets to each sector, and the subsequent integration of such fields in the Fourier
space. After all, it is not difficult to see that there are no UV divergences associated to
the non-local terms proportional to $\sigma$ and $M$. The scalar excitations display the same
confining propagator as the one
derived in the last section,  \eqref{scalarprop}, while for fermion field we have
\begin{eqnarray} 
{\cal S}(k^2) & = & \frac{i\kslash + m_{\psi} + \frac{M^3}{k^2}}{k^2 + (m_{\psi} +
\frac{M^3}{k^2})^2}  \;.\nonumber\\ 
& = &  \frac{i\kslash + M }{k^2 + m_{\psi}^2} + M^3 \frac{(k^2+m_{\psi}^2)k^2 -(i\kslash +
m_{\psi})(2Mk^2 + M^3)  }{(k^6 + (m_{\psi} k^2 + M^3)^2)(k^2 + m_{\psi}^2)}
\nonumber\\ 
& = &  \frac{i\kslash + m_{\psi} }{k^2 + m_{\psi}^2} + M^3 \Sigma(k^2)\,,
\label{fermprop}
\end{eqnarray}
Again, the isolated confining contribution to the propagator is highly suppressed in the UV
with respect to the standard massive Dirac term ($\sim 1/k$):
\begin{eqnarray} 
\Sigma(k)& = &   \frac{(k^2+m_{\psi}^2)k^2 -(i\kslash + m_{\psi})(2m_{\psi}k^2 + M^3)  }{(k^6 +
(m_{\psi} k^2 + M^3)^2)(k^2 + m_{\psi}^2)}\sim 1/k^4
 \,,
\label{sigma1}
\end{eqnarray}
and we anticipate that the primitive divergences of the theory with confined propagators will
be exactly the ones coming from terms of the original (local) theory, since any contribution
proportional to $\sigma$ or $M$ will be strongly suppressed in the UV regime.

At one loop order, besides the diagrams already analyzed in the previous section, new diagrams contributing to primitive divergences appear, due to the presence of fermion lines (dashed ones):
\begin{figure}[h!]
   \centering
       \includegraphics[width=0.8\linewidth]{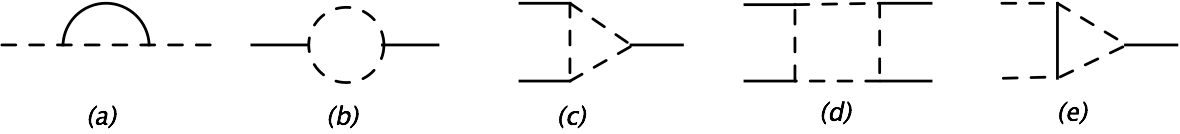}
               \caption{One-loop diagrams containing fermion (dashed) lines for the fermion and scalar selfenergies and cubic, quartic and Yukawa couplings, respectively. \label{fermiondiags}}
\end{figure}

It should be noticed that the Yukawa coupling breaks the discrete symmetry $\phi\to -\phi$
originally present in the scalar sector, generating at the quantum level a cubic scalar
interaction. This means that the renormalizable version of this theory requires a counterterm
for the cubic scalar interaction, even if the physical value of this coupling is set to zero.
In the case of a pseudoscalar Yukawa coupling (i.e. $g\phi\bar\psi\psi \to
g\phi\bar{\psi} \gamma^{5} \psi $), parity symmetry guarantees that the cubic terms vanish
identically. We emphasize, however, that our statement concerning the UV properties of
Gribov-type confining propagators remains valid in any case, as will be made explicit below via
the whole set of primitive divergences at one loop order. 

In order to investigate the influence of the confining propagators in the UV regime, we may
isolate the free fermion and scalar propagators from the confining contributions, namely
$\Sigma(k)\stackrel{UV}{\sim} 1/k^4$ and $\Delta(k^2)\stackrel{UV}{\sim} 1/k^6$, being both
highly suppressed in the UV. Writing down explicitly the momentum integrals in the
corresponding expressions for the one-loop diagrams in Fig. \ref{fermiondiags}, we have,
respectively:
\begin{enumerate}[label=(\alph*)]
\item the one-loop fermion self energy:
\begin{eqnarray}
 \int d^4 p {\cal D}(k-p) {\cal S}(p)
 &=&   \int d^4 p  \frac{1}{(k-p)^2 +  m_{\phi}^2} \frac{i\pslash +m_{\psi}}{p^2 +  m_{\psi}^2}
+ \sigma^4  \int d^4 p {\Delta}((k-p)^2) \frac{i\pslash +m_{\psi}}{p^2 +  m_{\psi}^2}
\nonumber\\&&
+ M^3 \int d^4 p  \frac{1}{(k-p)^2 +  m_{\phi}^2} \Sigma(p)
+ \sigma^4 M^3 \int d^4 p {\Delta}((k-p)^2)\Sigma(p) \nonumber
 \\
& =& \int d^4 p \frac{i\pslash +m_{\psi}}{p^2 +  m_{\psi}^2} \frac{1}{(k-p)^2 +  m_{\phi}^2} +
{\cal O}(\sigma^4, M^3, \sigma^4 M^3) (\text{UV finite}) 
\end{eqnarray}

\item  the fermion loop contributing to the scalar self energy:
\begin{equation}
 \int d^4 p\, {\rm Tr}[{\cal S}(p){\cal S}(k-p) ] ~=~ \int d^4 p {\rm Tr}\Big[\frac{i\pslash
+m_{\psi}}{p^2 +  m_{\psi}^2} \frac{i(\kslash -\pslash)+ m_{\psi}}{(k-p)^2 +  m_{\psi}^2}
\Big]+ {\cal O}( M^3, M^6) (\text{UV finite}) 
\end{equation}

\item the triangular diagram contributing to the scalar cubic interaction:
\begin{eqnarray}
 \int d^4 p\, {\rm Tr}[{\cal S}(p){\cal S}(p-k){\cal S}(p-k-k') ]
 &=& \int d^4 p {\rm Tr}\Big[ \frac{i\pslash +m_{\psi}}{p^2 +  m_{\psi}^2}
\frac{i(\pslash-\kslash) +m_{\psi}}{(p-k)^2 +  m_{\psi}^2} \frac{i(\pslash -\kslash -\kslash')+
m_{\psi}}{(p-k-k')^2 +  m_{\psi}^2}\Big] +\nonumber \\
 &&+ {\cal O}( M^3, M^6, M^{9}) (\text{UV finite}) 
\end{eqnarray}

\item the fermion loop correction to the $\phi^4$ vertex:
\begin{eqnarray}
 \int d^4 p\,{\rm Tr}\Big[{\cal S}(p){\cal S}(p-k){\cal S}(p-k-k'){\cal S}(p-k-k'-k'')  \Big]
 &=& \nonumber\\&&\hspace{-10cm}=
  \int d^4 p {\rm Tr}\Big[ \frac{i\pslash +m_{\psi}}{p^2 +  m_{\psi}^2}
\frac{i(\pslash-\kslash) +m_{\psi}}{(p-k)^2 +  m_{\psi}^2} \frac{i(\pslash -\kslash -\kslash')+
m_{\psi}}{(p-k-k')^2 +  m_{\psi}^2}
  \frac{i(\pslash -\kslash -\kslash'-\kslash'')+ m_{\psi}}{(p-k-k'-k'')^2 +  m_{\psi}^2}\Big]
+\nonumber \\
 &&
\hspace{-6cm}+ {\cal O}( M^3, M^6, M^{9},M^{12}) (\text{UV finite}) 
\end{eqnarray}

\item the modification of the Yukawa coupling:
\begin{eqnarray}
 \int d^4 p\,\Big[{\cal S}(p){\cal D}(p-k){\cal S}(p-k-k') \Big]
 &=& 
  \int d^4 p {\rm Tr}\Big[ \frac{i\pslash +m_{\psi}}{p^2 +  m_{\psi}^2}\frac{1}{(p-k)^2 +
m_{\phi}^2} \frac{i(\pslash-\kslash-\kslash') +m_{\psi}}{(p-k-k')^2 +  m_{\psi}^2} 
 \Big] +\nonumber \\
 &&+ {\cal O}( \sigma^4,M^3, M^6, \sigma^4 M^3, \sigma^4 M^6) (\text{UV finite}) 
\end{eqnarray}

\end{enumerate}

As already occurred for the confining scalar theory in the previous section, the highly
suppressed UV behavior of the confining pieces $\Sigma(k)\stackrel{UV}{\sim} 1/k^4$ and
$\Delta(k^2)\stackrel{UV}{\sim} 1/k^6$ enforces the convergence of all terms proportional to
the new massive parameters introduced ($\sigma$ and $M$).
The divergent integrals in all diagrams above are exactly the ones coming from the original
action, i.e. the one obtained in the limit $\sigma \to 0$ and $M \to 0$. 
In the theory including the confining quadratic non-local terms, the absence
of new primitive divergences then guarantees that the parameters $\sigma$ and $M$ can be
consistently related to dynamically generated scales and do not affect the UV regime of the
theory.

Realizing that any diagrammatic expression at higher loops will involve higher powers of the
propagators, it becomes straightforward to envision the generalization of our claim in the full
diagrammatic expansion of this general Yukawa theory. Therefore, given the renormalizability of
the original theory, one concludes that the resulting action with confining, Gribov-type
propagators is renormalizable and the IR confining parameters in both fermionic and bosonic
sectors do not display an independent renormalization, being thus consistent with dynamically
generated mass scales.

\section{$\mathcal N=1$ Super Yang--Mills within the Gribov--Zwanziger approach}
\sectionmark{$\mathcal N=1$ Super Yang--Mills within the GZ approach}
\label{symN1}

Let us now investigate a more intricate theory with confining propagators, including gauge interactions as well as Majorana fermions. We consider here Yang-Mills theory in $D=4$ spacetime dimensions with $\mathcal N=1$ supersymmetry in the presence of the Gribov horizon. We shall use this (most complicated) example to prove, to all-orders in the loop expansion, our claim concerning the good UV behavior of Gribov-type propagators. The IR parameters introduced will be shown to have renormalization parameters that are completely determined by the renormalization of the original theory.

This theory has already been put forward and investigated in \cite{Capri:2014xea}. There, the extension of the Gribov-Zwanziger framework to $\mathcal  N = 1$ 
Super-Yang-Mills (SYM) theories quantized in the Wess-Zumino gauge by imposing the Landau gauge condition was presented. The resulting 
effective action is
\begin{equation} 
S_{SGZ}^{N=1} = S_{SYM}^{N=1}  + Q \int d^4x \left( {\bar c}^a \partial_\mu A^a_\mu + {\bar \omega}^{ac}_{\mu}  (-\partial_\nu D^{ab}_{\nu} ) \varphi^{bc}_{\mu}  \right) +  S_{\gamma} + S_{\tilde{G}} \;, \label{sgzn1}
\end{equation} 
where $Q$ is the full transformation accounting for the supersymmetryc transformation and the
BSRT transformation, and is defined in the appendix so that the action \eqref{sgzn1} results in
\eqref{fnlact}; $S_{\gamma}$ is the horizon term in its local form, eq.\eqref{hfl}, namely 
\begin{equation}
S_\gamma =\; \gamma^{2} \int d^{4}x \left( gf^{abc}A^{a}_{\mu}(\varphi^{bc}_{\mu} + {\bar \varphi}^{bc}_{\mu})\right)-4 \gamma^4V (N^2-1)\;. \label{hfln1}
\end{equation} 
and the term $S_{\tilde{G}}$ is given by 
\begin{equation}
S_{\tilde{G}} = - \frac{1}{2}M^3\int d^{4}x \left( \bar{\lambda}^{a\alpha}\frac{\delta_{\alpha\beta}}{\partial^{2}}\lambda^{a\beta}\right) \;, \label{sslambda}
 \end{equation} 
which also has a new massive constant $M$. This quantum action takes into account the existence of Gribov copies in the path-integral quantization of the theory. It encodes the restriction to the first Gribov horizon while
keeping full compatibility with non-perturbative supersymmetric features, such as the exactly vanishing vacuum energy. 

Even though this non-perturbative framework has been constructed through the introduction of two massive parameters $\gamma, M$ which are not present in the classical action,
those new parameters are determined in a dynamical, self-consistent way via two
non-perturbative conditions: (i) the Gribov gap equation, that fixes $\gamma$ by imposing the
positivity of the Faddeev-Popov operator and eliminating a large set of Gribov copies from the
functional integral, and (ii) the vanishing of the vacuum energy, which determines the
parameter $M$ that plays the role of a supersymmetric counterpart of the Gribov parameter
$\gamma$, guaranteeing a consistent non-perturbative fermion sector. Interestingly, the
appearance of the dynamical fermionic scale $M$ has been shown to be directly related to the
formation of a gluino condensate, a well-known non-perturbative property of ${\cal N}=1$ SYM
theories. For further details, the reader is referred to \cite{Capri:2014xea}. A brief
summary of the notation adopted may also be found in the Appendix \ref{notations}.

The propagators of the theory \eqref{sgzn1} can be straightforwardly shown to be of the Gribov type. The gauge field propagator is:
\begin{equation}
\langle
A_{\mu}^a(p)A_{\nu}^b(-p)
\rangle
= \delta^{ab}\left(\delta_{\mu\nu}-\frac{p_{\mu}p_{\nu}}{p^2}\right)
\frac{p^2}{p^4+2Ng^2\gamma^4}\,,
\end{equation}
which, apart from the more complicated tensorial structure, is equivalent to the Gribov scalar propagator studied above in section \ref{confscalar}. The gauge field propagator in this Gribov-extended $\mathcal N=1$ SYM theory displays thus a confining contribution that is suppressed by an extra $1/p^4$ factor in the UV as compared to the free term.

For gluino fields we have:
\begin{eqnarray}
\langle
\bar{\lambda}_{\alpha}^a(p)
{\lambda}_{\beta}^b(-p)
\rangle
&=&
\delta^{ab}\frac{ip_{\mu}(\gamma_{\mu})_{\alpha\beta}+m(p^2)\delta_{\alpha\beta}}{p^2+m^2(p^2)}
\,,
\\
\langle \lambda^{a\rho}(p)\lambda^{b}_{\beta}(-p)\rangle 
&=& 
- \frac{\big( ip_{\mu}(\gamma_{\mu})_{\alpha\beta} + m(p^2)\delta_{\alpha\beta}\big)\delta^{ab}C^{\alpha\rho}}{p^{2} + m^{2}(p^2)}
\,,
\\
\langle \bar{\lambda}^{a}_{\alpha}(p)\bar{\lambda}^{b\tau}(-p)\rangle 
&=&
\frac{\big( ip_{\mu}(\gamma_{\mu})_{\alpha\beta} + m(p^2)\delta_{\alpha\beta}\big)\delta^{ab}C^{\beta\tau}}{p^{2} + m^{2}(p^2)}
\,,
\end{eqnarray}
where $C^{\alpha\beta}$ is the charge conjugation matrix and
\begin{equation}
m(p^2)=\frac{M^3}{p^2}\,.
\end{equation}
The presence of three two-point correlation functions involving gluino fields is a result of the lack of charge conservation for Majorana fermions. One verifies however that all of them have the form of Gribov propagators with $M$ playing an analogous role as the Gribov parameter in the gluino sector. In particular, one can easily check that the same structure observed for the Gribov fermion propagator
in the previous section (cf. Eq.\eqref{fermprop}) is found here:
\begin{eqnarray} 
\langle
\bar{\lambda}_{\alpha}^a(k)
{\lambda}_{\beta}^b(-k)
\rangle& = & \frac{i\kslash + \frac{M^3}{k^2}}{k^2 + \frac{M^6}{k^4}} 
= \frac{i\kslash }{k^2} + M^3 \Sigma_{\lambda}(k^2)\,,
\label{gluinoprop}
\end{eqnarray}
where the isolated confining contribution $\Sigma_{\lambda}$ to the gluino propagator is again highly suppressed in the UV with respect to the leading term ($\sim 1/k$):
\begin{eqnarray} 
\Sigma_{\lambda}(k^2)& = &   \frac{k^4 -i\kslash M^3  }{(k^6 + M^6)k^2}\stackrel{\rm UV}{\sim} 1/k^4
 \,.
\label{sigmag}
\end{eqnarray}

The same reasoning applied in the scalar and Yukawa theories above may be followed here in order to prove that the UV regime of the theory remains the same even after the inclusion of nonlocal confining terms in the propagators. One may compute the one-loop primitive divergences and show that the confining parameters $\gamma,M$ will not affect the UV divergent pieces, due to the high suppression observed in the Gribov-type propagators. We shall, however, use this most complicated theory analyzed in the current section to present an all-order algebraic proof of renormalizability and of the fact that the confining parameters $\gamma,M$ do not display independent renormalization.

\noindent The non-local action \eqref{sgzn1} is, however, not helpful in the algebraic renormalization procedure. Fortunately we are able to write its local form with the insertion of auxiliary fields. 

\noindent The whole action which describes our model can then be written in its local form as, 
\begin{eqnarray}
\label{fnlact1}
S &=& S_{SYM} + S_{gf} + S_{GZ'} + S_{L\tilde{G}}\nonumber \\
&&
=\int d^{4}x\; \left\{\frac{1}{4}F^{a}_{\mu \nu}F^{a}_{\mu\nu} 
+ \frac{1}{2} \bar{\lambda}^{a\alpha} (\gamma_{\mu})_{\alpha\beta} D^{ab}_{\mu}\lambda^{b\beta}
+ \frac{1}{2}\mathfrak{D}^a\mathfrak{D}^a 
+ b^{a}\partial_{\mu}A^{a}_{\mu} \right. \nonumber\\
&&
+\check{c}^{a}\left[\partial_{\mu}D^{ab}_{\mu}c^{b}
-\bar{\epsilon}^{\alpha}(\gamma_{\mu})_{\alpha\beta}\partial_{\mu}\lambda^{a\,\beta}\right]
+ \tilde{\varphi}^{ac}_{\mu}\partial_{\nu}D_{\nu}^{ab}\varphi^{bc}_{\mu} 
-\tilde{\omega}^{ac}_{\mu}\partial_{\nu}D_{\nu}^{ab}\omega^{bc}_{\mu} \nonumber \\
&&
-gf^{abc}(\partial_{\nu}\tilde{\omega}^{ad}_{\mu})(D^{bk}_{\nu}c^{k})\varphi^{cd}_{\mu}
+gf^{abc}(\partial_{\nu}\tilde{\omega}^{ad}_{\mu})(\bar{\epsilon}^\alpha(\gamma_\nu)_{\alpha\beta}\lambda^{\beta b})\varphi^{cd}_{\mu} \nonumber \\
&&
+\gamma^{2}gf^{abc}A^{a}_{\mu}(\varphi^{bc}_{\mu} + \tilde{\varphi}^{bc}_{\mu}) 
-\gamma^{4}4(N_{c}^{2}-1) 
+\hat{\zeta}^{a\alpha} (\partial^{2} - \mu^{2})\zeta^{a}_{~\alpha} \nonumber \\
&&
\left.
-\hat{\theta}^{a\alpha}(\partial^{2} - \mu^{2})\theta^{a}_{~\alpha} 
-M^{3/2}(\bar{\lambda}^{a\alpha}\theta^{a}_{~\alpha} 
+\hat{\theta}^{a\alpha}\lambda^{a}_{~\alpha})
\right\}\;.
\label{thSYM}
\end{eqnarray} 

Applying the algebraic renormalization procedure to the local action above we are able to prove that:
(i) the Gribov-extended SYM theory is renormalizable; and (ii) the massive parameters $\gamma,
M$ introduced in the infrared action do not renormalize independently, meaning that they are
consistent with dynamically generated mass scales, produced by nonperturbative interactions in
the original theory. All details of the proof were developed in the Appendix \eqref{algrenorm}.

The final results for the renormalization factors related to the confining parameters
$M,\gamma$ may be read off from the renormalization of external sources conveniently introduced
in the algebraic procedure (developed in the Appendix \eqref{algrenorm}). The renormalization of
the sources $m_{\psi}$ and $\tilde{m_{\psi}}$ give us the renormalization factor of the Gribov
parameter $\gamma^{2}$, while the renormalization of $V$ and $\hat{V}$ give us the
renormalization of $m_{\psi}^{3/2}$, when every source assumes its physical value stated at
\eqref{physval2}. We have:
\begin{eqnarray}
&&
Z_{\tilde{M}} =Z_{M} = Z^{-1/2}_{g}Z^{-1/4}_{A}\;, \nonumber \\
&&
Z_{\hat{V}} =Z_{V} = Z^{-1/2}_{\lambda}\;,
\end{eqnarray}
which clearly prove that the renormalization of the infrared parameters $m_{\psi},\gamma$ is
fixed by the renormalization factor of the original $SYM$ theory: the renormalization of the
gauge coupling, $Z_g$, the wave function renormalization of the gauge field, $Z_A$, and and the
wave function renormalization of the gluing field, $Z_{\lambda}$.

Therefore we conclude that this action is indeed a suitable nonperturbative infrared action for
${\cal N}=1$ SYM theories, reducing consistently to the ultraviolet original action. Moreover,
even in this very intricate non-Abelian gauge theory with matter fields, the good UV behavior
in the presence of confining propagators of the Gribov type shows up at all orders.

\section{Discussions about the results\label{conc5}}

Carrying on the analysis started on Chapter \ref{brstonmatter}, we have studied the UV
behavior of quantum field theory models in which the two-point correlation functions of the
elementary fields are described by confining propagators of the Gribov type. Our analysis was
not restricted to the gauge sector of the theory, but it concerns general properties in the UV
regime of any kind of fields that is said to be confined in the Gribov sense.

We could show that, order by order, the UV divergent behavior of the Feynman diagrams is not
affected by the infrared parameters of the theory. By infrared parameters we mean those
associated to the non-local term of the action, which accounts for non-perturbative
effects of the theory: $\sigma$ for the scalar field sector, and $M$ for the fermion field
sector. More precisely, we observed that contributions to Feynman diagrams stemming from the
non-local terms of the action, which are those proportional to $\sigma$ or $M$, are always
finite, being, thus, highly suppressed in the UV regime by the standard ultraviolet tree-level
propagator.

As a consequence, no new UV divergences in the infrared parameters can arise. Otherwise said,
the only UV divergences affecting the 1PI Green's functions of the theory are those present
when the infrared parameters are set to zero (and no non-perturbative effect is taken into
account). Therefore, the infrared parameters do not renormalize independently.

An all order proof was also presented, which can be checked in the Appendix \ref{algrenorm}, in the case of ${\cal N}=1$ Super Yang-Mills model within the Gribov horizon and with a \emph{horizon-like} term in the super-partner sector. We could explicitly show that both infrared parameters, $\gamma^{2}$, for the gauge sector, and $M^{3}$, for the fermion sector, do not renormalize independently, \emph{i.e.} their renormalization factor depends on the renormalization factor of fundamental fields and parameters of the original theory (out of Gribov horizon).  Namely, their renormalization factors can be read out of
\begin{eqnarray}
&&
Z_{\tilde{M}} =Z_{M} = Z^{-1/2}_{g}Z^{-1/4}_{A}\;, \nonumber \\
&&
Z_{\hat{V}} =Z_{V} = Z^{-1/2}_{\lambda}\;.
\end{eqnarray}

\chapter{The finite temperature case: the interplay between Polyakov and Gribov}
\chaptermark{The interplay between Polyakov and Gribov}
\label{Ploop1}


Until now we have dealt with Yang-Mills theories in the framework of Gribov quantization
procedure, coupled to a matter field, either being scalar, playing as the Higgs field or just
as a toy-model confined field, or being the confined quark field. We have shown that it is
possible, and perhaps reasonable, to consistently construct an effective theory for a confined
matter, inspired by the Gribov confinement mechanism in the gauge field sector.
Therefore, we have claimed that it may be possible that non-perturbative effects of the gauge
sector play some influence on the IR behavior of the matter sector.

Now, in this chapter, we are going to check in a finite temperature scenario if there exist any
interplay between gauge and matter fields in the IR regime, at all. In order to do that, we
will compute the vacuum expectation value of the Polyakov loop, implemented in an gauge
theory restricted to the Gribov horizon, using the Gribov-Zwanziger approach. 
Related computations are available using different techniques to cope with non-perturbative
propagators at finite temperature, see
e.g.~\cite{Maas:2011se,Braun:2007bx,Marhauser:2008fz,Reinhardt:2012qe,Reinhardt:2013iia,Heffner:2015zna,Reinosa:2014ooa,Reinosa:2014zta,Fischer:2009gk,Herbst:2013ufa,Bender:1996bm}.
In \cite{Zwanziger:2006sc,Lichtenegger:2008mh,Fukushima:2013xsa}, it was already
pointed out that the Gribov--Zwanziger quantization offers an interesting way to illuminate
some of the typical infrared problems for finite temperature gauge theories.

In the following section the Polyakov loop is introduced into the GZ theory via the
background field method, based on the works
\cite{Braun:2007bx,Marhauser:2008fz,Reinosa:2014ooa}. Next, section \ref{sec4} handles the
technical computation of the leading order finite temperature effective action, while in
section \ref{sec5} we discuss the gap equations, leading to our estimates for both Polyakov
loop and Gribov mass. The key finding is a deconfinement phase transition at the same
temperature at which the Gribov mass develops a cusp-like behavior. Subsequently, we also
discuss the pressure and energy anomaly. Due to a problem with the pressure in the GZ formalism
(regions of negativity), we take a preliminary look at the situation upon invoking the more
recently developed Refined Gribov--Zwanziger approach. On section \ref{RGZ-Ploop} the
refined-GZ is briefly analyzed.  We summarize in section \ref{DAR-Ploop}.

\section{The Polyakov loop and the background field formalism}
\label{Ploop}
In this section we shall investigate the confinement/deconfinement phase transition
of the $SU(2)$ gauge field theory in the presence of two static sources
of (heavy) quarks. The standard way
to achieve this goal is by probing the Polyakov loop order parameter,
\begin{eqnarray}
\mathcal{P} = \frac{1}{N}\tr \Braket{P e^{ig\int_{0}^{\beta}dt \;\;
A_{0}(t,x)}}\;,
\end{eqnarray}
with $P$ denoting path ordering, needed in the non-Abelian case to ensure
the gauge invariance of $\mathcal{P}$. This path ordering is not relevant at
one-loop order, which will considerably simplify the computations of the current work. In analytical studies of the phase transition involving the
Polyakov loop, one usually imposes the so-called ``Polyakov gauge'' on the
gauge field, in which case the time-component $A_{0}$ becomes diagonal and
independent of (imaginary) time. This means that the gauge field belongs to the Cartan
subalgebra. More details on the Polyakov gauge can be found in
\cite{Marhauser:2008fz,Fukushima:2003fw,Ratti:2005jh}. Besides the trivial
simplification of the Polyakov loop, when imposing the Polyakov gauge it
turns out that the quantity $\Braket{A_{0}}$ becomes a good alternative choice for the order
parameter instead of $\mathcal{P}$. This extra benefit can be proven by means of Jensen's
inequality for convex functions and is carefully explained in \cite{Marhauser:2008fz}, see also
\cite{Braun:2007bx,Reinhardt:2012qe,Reinhardt:2013iia,Heffner:2015zna,Reinosa:2014ooa}. For
example, for the $SU(2)$ case we have the following: if $\frac{1}{2}g\beta%
\Braket{A_{0}} = \frac{\pi}{2}$ then we are in the ``unbroken symmetry phase''
(confined or disordered phase), equivalent to $\Braket{{\cal P}} = 0$;
otherwise, if $\frac{1}{2}g\beta\Braket{A_{0}} < \frac{\pi}{2}$, we are in the
``broken symmetry phase'' (deconfined or ordered phase), equivalent to
$\Braket{{\cal P}} \neq 0$. Since $\mathcal{P}\propto e^{-F T}$ with $T$ the temperature and
$F$ the free energy of a heavy quark, it is clear that in the confinement phase, an infinite
amount of energy would be required to actually get a free quark. The broken/restored symmetry
referred to is the $\mathbb{Z}_N$ center symmetry of a pure gauge theory (no dynamical matter
in the fundamental representation).

A slightly alternative approach to access the Polyakov loop was worked out in
\cite{Reinosa:2014ooa}.
In order to probe the phase transition in a quantized non-Abelian gauge
field theory, we use, following \cite{Reinosa:2014ooa}, the Background Field Gauge (BFG)
formalism, detailed in general in e.g.~\cite{Weinberg:1996kr}. Within this framework, the
effective gauge field will be defined as the sum of a classical field $\bar{A}_{\mu}$ and a
quantum field $A_{\mu}$: $a_{\mu}(x) = a_{\mu}^{a}(x)t^{a} = \bar{A}_{\mu}+A_{\mu}
$, with $t^{a}$  the infinitesimal generators of the
$SU(N)$ symmetry group. The BFG method is a convenient approach, since the
tracking of breaking/restoration of the $\mathbb{Z}_{N}$ symmetry becomes
easier by choosing the Polyakov gauge for the background field.

Within this framework, it is convenient to define the gauge condition for the quantum field,
\begin{eqnarray}
\bar{D}_{\mu}A_{\mu} = 0\;,  \label{LDW}
\end{eqnarray}
 known as the Landau--DeWitt (LDW) gauge fixing condition, where $\bar{D}%
^{ab}_{\mu} =\delta^{ab}\partial_{\mu} - gf^{abc}\bar{A}^{c}_{\mu}$ is the
background covariant derivative. After integrating out the (gauge fixing)
auxiliary field $b^{a}$, we end up with the following Yang--Mills action,
\begin{eqnarray}
S_\text{BFG} = \int d^{d}x\; \left\{ \frac{1}{4}F^{a}_{\mu\nu}F^{a}_{\mu\nu}
- \frac{\left( \bar{D}A \right)^{2}}{2\xi} + \bar{c}^{a}\bar{D}_{\mu}^{ab}
D^{bd}_{\mu}(a)c^{d} \right\} \;.  \label{bfg}
\end{eqnarray}
Notice that, concerning the quantum field $A_{\mu}$, the condition %
\eqref{LDW} is equivalent to the Landau gauge, yet the action still has  background center symmetry. The LDW gauge is actually recovered in the limit
$\xi \to 0$, taken at the very end of each computation.

As explained for the simple Landau gauge in the previous section, the Landau
background gauge condition is also plagued by Gribov ambiguities, and the
Gribov--Zwanziger procedure is applicable also in this instance. The starting
point of our analysis is, therefore, the GZ action modified for the BFG
framework (see \cite{Zwanziger:1982na}):
\begin{multline}
S_\text{GZ+PLoop} = \int d^{d}x\; \left\{ \frac{1}{4}F^{a}_{\mu\nu}F^{a}_{%
\mu\nu} - \frac{\left( \bar{D}A \right)^{2}}{2\xi} + \bar{c}^{a}\bar{D}%
_{\mu}^{ab}D^{bd}_{\mu}(a)c^{d} + \bar{\varphi}_{\mu}^{ac} \bar{D}%
_{\nu}^{ab}D^{bd}_{\nu}(a) \varphi_\mu^{dc} \right. \\
\left. - \bar{\omega}_{\mu}^{ac} \bar{D}_{\nu}^{ab}D^{bd}_{\nu}(a)
\omega_\mu^{dc} - g\gamma ^{2} f^{abc}A_\mu^a \left( \varphi_\mu^{bc} + \bar{%
\varphi}_\mu^{bc} \right) - \gamma^{4}d(N^{2}-1) \right\}\;.  \label{gzpl}
\end{multline}
As mentioned before, with the Polyakov gauge imposed to the background
field $\bar{A}_{\mu}$, the time-component becomes diagonal and
time-independent. In other words, we have $\bar{A}_{\mu}(x) = \bar{A}%
_{0}\delta_{\mu 0}$, with $\bar{A}_{0}$ belonging to the Cartan subalgebra
of the gauge group. For instance, in the Cartan subalgebra of $SU(2)$ only
the $t^{3}$ generator is present, so that $\bar{A}^{a}_{0} = \delta^{a3}\bar{%
A}^{3}_{0}\equiv \delta^{a3}\bar A_0$. As explained in \cite{Reinosa:2014ooa}, at leading order we then simply find, using the properties of the Pauli matrices,
\begin{equation}
\mathcal{P}=\cos\frac{r}{2}\,,
\end{equation}
where we defined
\begin{equation}
  r=g\beta \bar{A}_0\,,
\end{equation}
with $\beta$ the inverse temperature. Just like before, $r=\pi$ corresponds to the confinement phase, while $0\leq r<\pi$ corresponds to deconfinement. With a slight abuse of language, we will refer to the quantity $r$ as the Polyakov loop hereafter.

Here we are limited to one-loop order, then only terms quadratic in the quantum
fields in the action \eqref{gzpl} shall be considered. One then immediately gets an action that
can be split in term coming from the two color sectors: the 3rd color direction, called Cartan
direction, which does not depend on the parameter $r$; and one coming from the $2\times 2$
block given by the 1st and 2nd color directions. This second $2\times2$ color sector is
orthogonal to the Cartan direction and does depend on $r$. The scenario can then be interpreted
as a $U(1)$ symmetric system where the vector field is coupled to a chemical potential $i rT$
and has isospins $+1$ and $-1$ related to the $2\times2$ color sector and one isospin $0$
related to the $1\times1$ color sector.

\section{The finite temperature effective action at leading order}
\sectionmark{The finite temperature effective action}
\label{sec4}

Considering only the quadratic terms of \eqref{gzpl}, the integration of the
partition function gives us the following vacuum energy at one-loop order,
\begin{equation}
\beta V\mathcal{E}_{v} ~=~ -\frac{d(N^{2}-1)}{2Ng^{2}}\lambda ^{4}+\frac{1}{2}%
(d-1)\tr\ln \frac{\mathcal{D}^{4}+\lambda ^{4}}{-\mathcal{D}^{2}}-\frac{1}{2}%
\tr\ln (-\mathcal{D}^{2})\;,  \label{vace}
\end{equation}%
according to the definition 
\begin{eqnarray}
\e^{\beta V{\cal E}_{v}} ~=~ Z\,.
\end{eqnarray}
In the vacuum energy expression \eqref{vace}, $V$ stands here, only in this Chapter, for the
spacial volume. Here, $\mathcal{D}$ is the covariant derivative in the adjoint representation
in the presence of the background $A_{0}^{3}$ field and $\lambda ^{4}=2Ng^{2}\gamma ^{4}$.
Throughout this work, it is always tacitly assumed we are working with $N=2$ colors, although
we will frequently continue to explicitly write $N$ dependence for generality. Using the usual
Matsubara formalism, we have that $\mathcal{D}^{2}=(2\pi nT+rsT)^{2}+\vec{q}^{2}$, where $n$ is
the Matsubara mode, $\vec{q}$ is the spacelike momentum component, and $s$ is the isospin,
given by $-1$, $0$, or $+1$ for the $SU(2)$ case\footnote{The $SU(3)$ case was handled in
\cite{Reinosa:2014ooa} as well (see also \cite{Serreau:2015saa}).}.

The general trace is of the form
\begin{equation}  \label{1}
\frac{1}{\beta V}\tr\ln (-\mathcal{D}^2 + m^2) = T \sum_s
\sum_{n=-\infty}^{+\infty}\int\frac{ d^{3-\epsilon}q}{(2\pi)^{3-\epsilon}}
\ln \left((2\pi nT + rsT)^2+\vec{q}^2+m^2\right)\,,
\end{equation}
which will be computed immediately below.

\subsection{The sum-integral: 2 different computations}
We want to compute the following expression:
\begin{equation}
\mathcal{I }= T \sum_{n=-\infty}^{+\infty}\int\frac{ d^{3-\epsilon}q}{%
(2\pi)^{3-\epsilon}} \ln \left((2\pi nT + rT)^2+\vec{q}^2+m^2\right) \;.
\end{equation}
One way to proceed is to start by deriving the previous expression with
respect to $m^2$. Then, one can use the well-known formula from complex
analysis
\begin{equation}  \label{3}
\sum_{n=-\infty}^{+\infty} f(n) = -\pi\sum_{z_0}\mathop{\mathcal Res}%
\limits_{z=z_0}\cot(\pi z)f(z)
\end{equation}
where the sum is over the poles $z_0$ of the function $f(z)$. Subsequently
we integrate with respect to $m^2$ (and determine the integration
constant by matching the result with the known $T=0$ case). Finally one can split off the analogous $T=0$ trace (which does not depend on the background
field) to find
\begin{equation}  \label{6}
\mathcal{I }= \int \frac{d^{4-\epsilon}q}{(2\pi)^{4-\epsilon}}\ln(q^2+m^2) +
T \int\frac{d^{3}q}{(2\pi)^{3}}\ln\left(1+e^{-2\frac{\sqrt{\vec{q}^2+m^2}}{T}%
}-2e^{-\frac{\sqrt{\vec{q}^2+m^2}}{T}}\cos r\right) \;.
\end{equation}
where the limit $\epsilon\to0$ was taken in the (convergent) second
integral. The first term in the r.h.s.~is the (divergent) zero temperature
contribution.

Another way to compute the above integral is by making use of Zeta function regularization
techniques, which are particularly useful in the computation  of the Casimir energy in various
configurations see \cite{elizalde95,bordag10}. The advantage of this second technique is that,
although it is less direct, it provides one with an easy way to analyze the high and low
temperature limits as well as the small mass limit, as we will now show. Moreover, within this
framework, the regularization procedures are often quite transparent. One starts by writing the
logarithm as $\ln x=-\lim_{s\rightarrow 0}\partial _{s}x^{-s}$, after which the integral over
the momenta can be performed:
\begin{equation}
\mathcal{I}=-T\lim_{s\rightarrow 0}\partial _{s}\left( \mu
^{2s}\sum_{n=-\infty }^{\infty }\frac{\Gamma (s-3/2)}{8\pi ^{\frac{3}{2}%
}\Gamma (s)}\left[ (2\pi nT+rT)^{2}+m^{2}\right] ^{\frac{3}{2}-s}\right) \;,
\end{equation}%
where the renormalization scale $\mu $ has been introduced to get
dimensional agreement for $s\not=0$, and where we already put $\epsilon =0$, as $s$ will function as a regulator --- i.e.~we assume $s>3/2$ and
analytically continuate to bring $s\rightarrow 0$. Using the integral
representation of the Gamma function, the previous expression can be recast
to
\begin{eqnarray}
\mathcal{I} &=&-T\lim_{s\rightarrow 0}\partial _{s}\left( \mu
^{2s}\sum_{n=-\infty }^{\infty }\frac{1}{8\pi ^{\frac{3}{2}}\Gamma (s)}%
\int_{0}^{\infty }t^{s-5/2}e^{-t\left( (2\pi nT+rT)^{2}+m^{2}\right)
}dt\right)   \notag \\
&=&-\lim_{s\rightarrow 0}\partial _{s}\left( \mu ^{2s}\frac{T^{4-2s}}{%
4^{s}\pi ^{2s-3/2}\Gamma (s)}\int_{0}^{\infty }dyy^{s-5/2}e^{-\frac{m^{2}y%
}{4\pi ^{2}T^{2}}}\sum_{n=-\infty }^{\infty }e^{-y(n+\frac{r}{2\pi }%
)^{2}}\right) \;,
\end{eqnarray}%
where the variable of integration was transformed as $y=4\pi ^{2}T^{2}t\geq0 $ in
the second line. Using the Poisson rule (valid for positive $\omega $):
\begin{equation}
\sum_{n=-\infty }^{+\infty }e^{-(n+x)^{2}\omega }=\sqrt{\frac{\pi }{\omega }}%
\left( 1+2\sum_{n=1}^{\infty }e^{-\frac{n^{2}\pi ^{2}}{\omega }}\cos {(2n\pi
x)}\right) \;,
\end{equation}%
we obtain that
\begin{multline}
\mathcal{I}=-\lim_{s\rightarrow 0}\partial _{s}\mu ^{2s}\left[ \frac{\Gamma
(s-2)T^{4-2s}}{4^{s}\pi ^{2s-2}\Gamma (s)}\left( \frac{m^{2}}{4\pi ^{2}T^{2}}%
\right) ^{2-s}+\right.  \\
\left. \frac{T^{4-2s}}{4^{s-1}\pi ^{s}\Gamma (s)}\left( \frac{m^{2}}{4\pi
^{2}T^{2}}\right) ^{1-s/2}\sum_{n=1}^{\infty }{n^{s-2}\cos {(nr)}%
K_{2-s}\left( \frac{nm}{T}\right) }\right] \;,
\end{multline}%
where $K_{\nu }(z)$ is the modified Bessel function of the second kind.
Simplifying this, we find
\begin{equation}
\mathcal{I}=\frac{m^{4}}{2(4\pi )^{2}}\left[ \ln {\left( \frac{m^{2}}{\mu
^{2}}\right) }-\frac{3}{2}\right] -\sum_{n=1}^{\infty }{\frac{m^{2}T^{2}\cos
{(nr)}}{\pi ^{2}n^{2}}K_{2}\left( \frac{nm}{T}\right) }\;,  \label{result2}
\end{equation}%
where the first term is the $T=0$ contribution, and the sum is the
finite-temperature correction. Using numerical integration and series
summation, it can be checked that both results \eqref{6} and \eqref{result2}
are indeed identical. Throughout this paper, we will mostly base ourselves
on the expression \eqref{6}. Nonetheless the Bessel series is quite useful in
obtaining the limit cases $m=0$, $T\rightarrow \infty $, and $T\rightarrow 0$
by means of the corresponding behaviour of $K_{2}(z)$. Observing that
\begin{equation}
\lim_{m\rightarrow 0}\left( -\frac{m^{2}T^{2}K_{2}\left( \frac{mn}{T}\right)
\cos (nrs)}{\pi ^{2}n^{2}}\right) =-\frac{2T^{4}\cos (nrs)}{\pi ^{2}n^{4}}\,,
\end{equation}%
we obtain
\begin{equation}
\mathcal{I}_{m=0}=-\frac{T^{4}}{\pi ^{2}}\left[ \text{Li}_{4}\left(
e^{-irs}\right) +\text{Li}_{4}\left( e^{irs}\right) \right],
\end{equation}%
where $\text{Li}_{s}(z)=\sum_{n=1}^{\infty}\frac{z^{n}}{n^{s}}$ is the polylogarithm or Jonqui\`ere's function.

Analogously,
\begin{equation*}
\lim_{T\rightarrow\infty}K_2\left( \frac{m n}{T}\right)\sim \frac{2 T^2}{m^2 n^2}-\frac{1}{2}\,,
\end{equation*}
so that
\begin{equation}
\mathcal{I}_{T\rightarrow\infty} = \frac{m^4}{2(4\pi)^2}\left[ \ln{\left(
\frac{m^2}{\mu^2}\right)}-\frac{3}{2} \right]+ \frac{T^2}{4 \pi^2 }
\left\{m^2 \left[\text{Li}_2\left(e^{-i r s}\right)+\text{Li}_2\left(e^{i r
s}\right)\right]-4 T^2 \left[\text{Li}_4\left(e^{-i r s}\right)+\text{Li}%
_4\left(e^{i r s}\right)\right]\right\}\,.
\end{equation}
Finally for $T \rightarrow 0$ we can use the asymptotic expansion of the
Bessel function \cite{as}:
\begin{equation}
K_\nu(z)\sim\sqrt{\frac{\pi}{2z}}e^{-z}\left( \sum_{k=0}^{\infty}\frac{%
a_k(\nu)}{z^k}\right),\ \ |\text{Arg}( z)|\leq\frac{3}{2}\pi\,,
\end{equation}
where $a_{k}(\nu)$ are finite factors. So, at first order ($k=0$),
\begin{eqnarray}
\mathcal{I}_{T\rightarrow0} = \frac{m^4}{2(4\pi)^2}\left[ \ln{\left( \frac{%
m^2}{\mu^2}\right)}-\frac{3}{2} \right]-\frac{m^{3/2} T^{5/2} }{2 \sqrt{2}
\pi ^{3/2}} \left[\text{Li}_{\frac{5}{2}}\left(e^{-\frac{m}{T}-i r s}\right)+%
\text{Li}_{\frac{5}{2}}\left(e^{-\frac{m}{T}+i r s}\right)\right]\,.
\end{eqnarray}

\subsection{The result for further usage}
Making use of the result \eqref{6} we may define
\begin{equation}  \label{defi}
I(m^2,r,s,T) = T \int\frac{d^{3}q}{(2\pi)^{3}}\ln\left(1+e^{-2\frac{\sqrt{%
\vec{q}^2+m^2}}{T}}-2e^{-\frac{\sqrt{\vec{q}^2+m^2}}{T}}\cos rs\right),
\end{equation}
so that the vacuum energy \eqref{vace} can be rewritten as
\begin{equation}  \label{vactwee}
\begin{aligned} {\cal E}_{v} = &- \frac{d(N^2-1)}{2Ng^2} \lambda^4 + \frac12
(d-1)(N^2-1) \tr_{T=0}\ln\frac{\partial^4+\lambda^4}{-\partial^2} - \frac12
(N^2-1) \tr_{T=0}\ln(-\partial^2) \\ &+ \sum_s \left( \frac12 (d-1)
(I(i\lambda^2,r,s,T)+I(-i\lambda^2,r,s,T)-I(0,r,s,T)) - \frac12 I(0,r,s,T)
\right) \;, \end{aligned}
\end{equation}
where $\tr_{T=0}$ denotes the trace taken at zero temperature.

\section{Minimization of the effective action, the Polyakov loop and the Gribov mass}
\sectionmark{Minimization of the effective action}
\label{sec5}

\subsection{Warming-up exercise: assuming a \texorpdfstring{$T$}{T}-independent Gribov mass \texorpdfstring{$\lambda$}{lambda}}
\label{Tonafhankelijk}

\begin{figure}[tbp]
\begin{center}
\includegraphics[width=.5\textwidth]{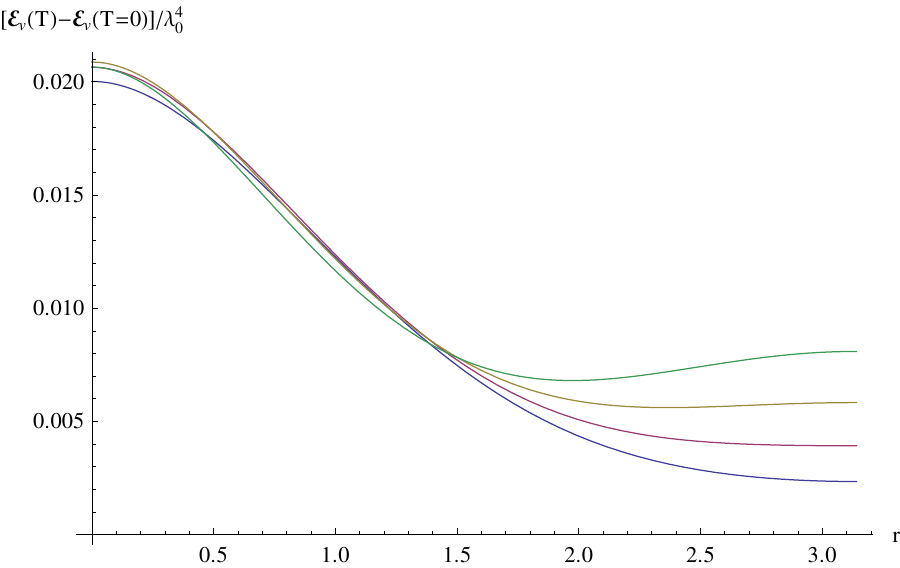}
\end{center}
\caption{The effective potential \eqref{vactwee} at the temperatures (from
below upwards at $r=\protect\pi$) 0.42, 0.44, 0.46, and 0.48 times $\protect\lambda$ as a function of $r$, with the simplifying assumption that $\protect%
\lambda$ maintains its zero-temperature value $\lambda_0$ throughout. It can be seen that
the minimum of the potential moves away from $r=\protect\pi$ in between $T=%
\unit{0.44}{\protect\gamma}$ and $\unit{0.46}{\protect\gamma}$. }
\label{energieplot}
\end{figure}

As a first simpler case, let us simplify matters slightly by assuming that the temperature does
not influence the Gribov parameter $\lambda$. This means that $\lambda$ will be supposed to
assume its zero-temperature value, which we will call $\lambda_{0}$, given by the solution of
the gap equation \eqref{hc2} (or horizon condition) at zero temperature,
\begin{equation}
\left\langle H(A)   \right\rangle = 4V \left(  N^{2}-1\right) \;.
\end{equation}
In this case, only
the terms with the function $I$ really matter in \eqref{vactwee}, since the other terms do not
explicitly depend on the Polyakov line $r$. Plotting this part of the potential (see
\figurename\ \ref{energieplot}), one finds by visual inspection that a second-order phase
transition occurs from the minimum with $r=\pi$ to a minimum with $r\not=\pi$. The transition
can be identified by the condition
\begin{equation}  
\label{dvdr} 
\left. \frac{d^2}{dr^2}
\mathcal{E}_{v} \right|_{r=\pi} = 0 \;.
\end{equation}
Using the fact that
\begin{equation}
\frac{\partial^2I}{\partial r^2}(m^2,r=\pi,s,T) = -2T \int\frac{d^{3}q}{%
(2\pi)^{3}}\frac{e^{-\frac{\sqrt{\vec{q}^2+m^2}}{T}}}{\left(1+e^{-\frac{%
\sqrt{\vec{q}^2+m^2}}T}\right)^2}
\end{equation}
when $s=\pm1$ and zero when $s=0$, the equation \eqref{dvdr} can be
straightforwardly solved numerically for the critical temperature. We find
\begin{equation}  \label{Tcrit}
	T_\text{crit} = \unit{0.45}{\lambda}_{0} \;.
\end{equation}

\subsection{The \texorpdfstring{$T$}{T}-dependence of the Gribov mass \texorpdfstring{$\lambda$}{lambda}}

\label{gammavanT} Let us now investigate what happens to the Gribov
parameter $\lambda$ when the temperature is nonzero. Taking the derivative of
the effective potential \eqref{vactwee} with respect to $\lambda^2$ and
dividing by $d(N^2-1)\lambda^2/Ng^2$ (as we are not interested in the solution $\lambda^2=0$) yields the gap equation for general number of colors $N$:
\begin{equation}  \label{gapeen}
1 = \frac12 \frac{d-1}d Ng^2 \tr \frac1{\partial^4+\lambda^4} + \frac12
\frac{d-1}d \frac{Ng^2}{N^2-1} \frac i{\lambda^2} \sum_s \left(\frac{%
\partial I}{\partial m^2}(i\lambda^2,r,s,T) - \frac{\partial I}{\partial m^2}%
(-i\lambda^2,r,s,T)\right) \;,
\end{equation}
where the notation $\partial I/\partial m^2$ denotes the derivative of $I$
with respect to its first argument (written $m^2$ in \eqref{defi}). If we
now define $\lambda_0$ to be the solution to the gap equation at $T=0$:
\begin{equation}
1 = \frac12 \frac{d-1}d Ng^2 \tr \frac1{\partial^4+\lambda_0^4} \;,
\label{t0gapeq}
\end{equation}
then we can subtract this equation from the general gap equation %
\eqref{gapeen}. After dividing through $(d-1)Ng^2/2d$ and setting $d=4$ and $%
N=2$, the result is
\begin{equation}
\int \frac{d^4q}{(2\pi)^4} \left(\frac1{q^4+\lambda^4}-\frac1{q^4+%
\lambda_0^4}\right) + \frac i{3\lambda^2} \sum_s \left(\frac{\partial I}{%
\partial m^2}(i\lambda^2,r,s,T) - \frac{\partial I}{\partial m^2}%
(-i\lambda^2,r,s,T)\right) = 0 \;,
\end{equation}
where now all integrations are convergent. This equation can be easily
solved numerically to yield $\lambda$ as a function of temperature $T$ and
background $r$, in units $\lambda_0$. This is shown in \figurename\ \ref%
{gammaplot}.

\begin{figure}[tbp]
\begin{center}
\includegraphics[width=.5\textwidth]{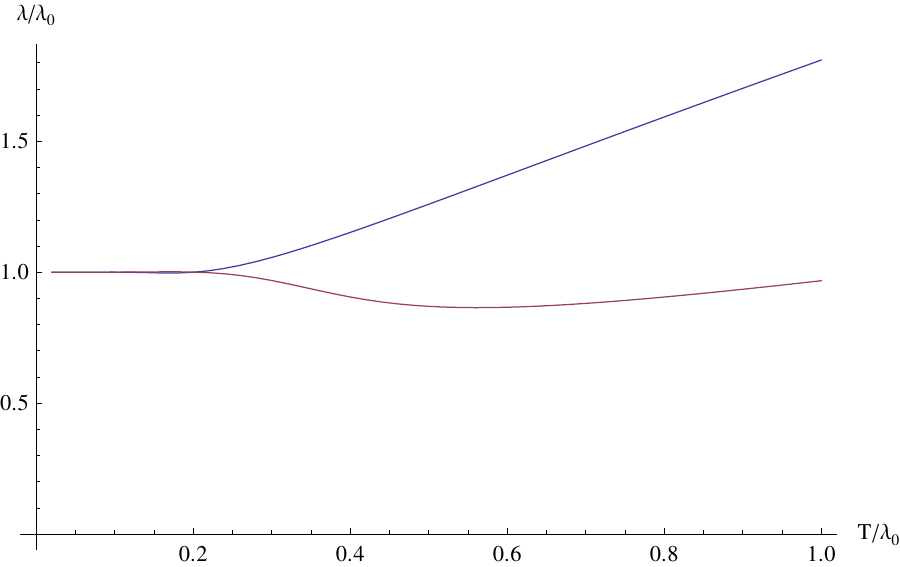}
\end{center}
\caption{The Gribov parameter $\protect\lambda$ as a function of the
temperature $T$ at $r$ equals to zero (upper line) and $\protect\pi$ (lower
line), in units of the zero-temperature Gribov parameter $\protect\lambda_0$.
}
\label{gammaplot}
\end{figure}

\subsection{Absolute minimum of the effective action}

As $\lambda$ does not change much when including its dependence on
temperature and background, the transition is still second order and its
temperature is, therefore, still given by the condition \eqref{dvdr}. Now,
however, the potential depends explicitely on $r$, but also implicitely due
to the presence of the $r$-dependent $\lambda$. We therefore have
\begin{equation}
\left. \frac{d^2}{dr^2} \mathcal{E}_{v} \right|_{r=\pi} = \left. \frac{%
\partial^2\mathcal{E}_{v}}{\partial r^2} + 2 \frac{d\lambda}{dr} \frac{%
\partial^2\mathcal{E}_{v}}{\partial r\partial\lambda} + \frac{d^2\lambda}{%
dr^2} \frac{\partial \mathcal{E}_{v}}{\partial\lambda} + \left(\frac{d\lambda%
}{dr}\right)^2 \frac{\partial^2\mathcal{E}_{v}}{\partial\lambda^2}%
\right|_{\lambda=\lambda(r),r=\pi} \;.
\end{equation}
Now, $d\lambda/dr|_{r=\pi} = 0$ due to the symmetry at that point.
Furthermore, as we are considering $\lambda\neq0$, $\partial \mathcal{E}_{v}/\partial\lambda = 0$ is the gap
equation and is solved by $\lambda(r)$. Therefore, we find for the condition
of the transition:
\begin{equation}  \label{dvdrpartial}
\left. \frac{\partial^2\mathcal{E}_{v}}{\partial r^2} (r,\lambda,T) \right|_{r=\pi} = 0 \;,
\end{equation}
where the derivative is taken with respect to the explicit $r$ only.

We already solved equation \eqref{dvdrpartial} in the subsection \ref{Tonafhankelijk}, giving
\eqref{Tcrit}:
\begin{equation}
T = 0.45 \lambda(r,T) \;.
\end{equation}
As we computed $\lambda$ as a function of $r$ and $T$ in the subsection \ref{gammavanT}
already, it is again straightforward to solve this equation to give the eventual critical
temperature:
\begin{equation}
T_\text{crit} = \unit{0.40}{\lambda_0} \;,
\end{equation}
as expected only slightly different from the simplified estimate \eqref{Tcrit} found before. 

\subsection{The \texorpdfstring{$T$}{T}-dependence of the Polyakov loop \texorpdfstring{$r$}{r} and the equation of
state}

\subsubsection{Deconfinement transition and its imprint on the Gribov mass}

\begin{figure}[h]
\begin{center}
\includegraphics[width=.5\textwidth]{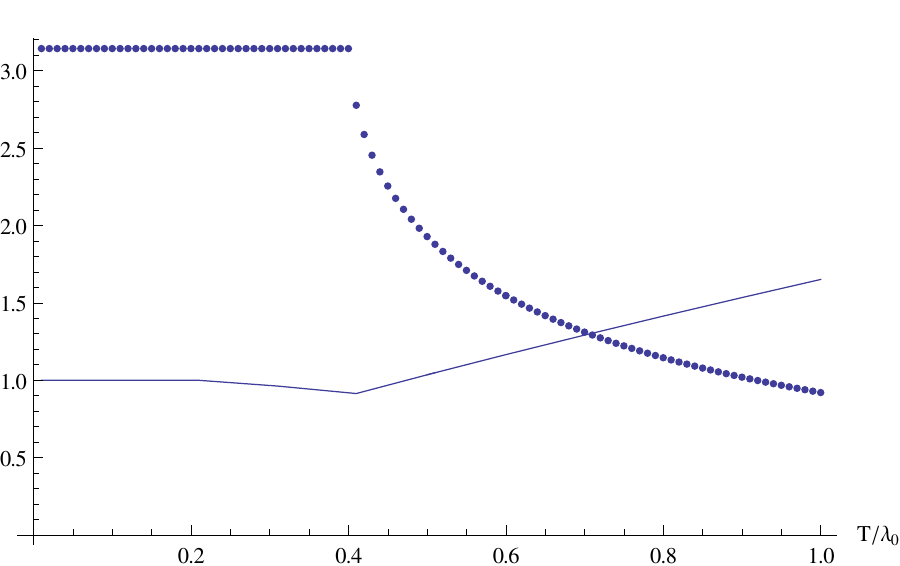}
\end{center}
\caption{The dotted line curve represents $r(T)$, while the continuous line is $\protect\lambda(T)$. At $T\approx 0.40\protect\lambda_{0}$, both curves clearly have a discontinuous derivative.}
\label{randlambda}
\end{figure}

Let us now investigate the temperature dependence of $r$. The physical value of the background field $r$ is found by minimizing the vacuum energy:
\begin{eqnarray}
\frac{d}{dr} {\cal E}_{v} =0\;.
\end{eqnarray}
From the vacuum energy  \eqref{vactwee} we have
\begin{eqnarray}
\frac{\partial\mathcal{E}_{v}}{\partial r} = (d-1) \left[ \frac{\partial
I}{\partial r}(i\gamma^2,r,T) + \frac{\partial I}{\partial r}(-i\gamma^2,r,T)
- \frac{d}{(d-1)} \frac{\partial I}{\partial r}(0,r,T) \right] = 0\;.
\label{rgapeq}
\end{eqnarray}
The expression \eqref{rgapeq} was obtained after  summation over the
possible values of $s$. Furthermore, we used the fact that $%
I(m^{2},r,+1,T)=I(m^{2},r,-1,T)$ and that $s=0$ accounts for terms
independent of $r$, which are cancelled by the derivation w.r.t. $r$. From %
\eqref{defi} one can get, whenever $s=\pm1$:
\begin{eqnarray}
\frac{\partial I(m^{2},r,T)}{\partial r} = T \int\frac{d^{3}q}{(2\pi)^{3}}
\frac{2e^{-\frac{\sqrt{\vec{q}^2+m^2}}{T}}\sin r}{\left(1+e^{-2\frac{\sqrt{%
\vec{q}^2+m^2}}{T}}-2e^{-\frac{\sqrt{\vec{q}^2+m^2}}{T}}\cos r \right)}\;.
\end{eqnarray}
Since \eqref{rgapeq} is finite, we can numerically obtain $r$ as a function of
temperature. From the dotted curve in Figure \ref{randlambda} one can easily see that, for $T > T_{\text{crit}} \approx 0.40 \lambda_{0}$, we have $%
r \neq \pi$, pointing to a deconfined phase,  confirming
the computations of the previous section. In the same figure, $\lambda(T)$ is plotted in a continuous line. We observe very clearly
that the Gribov mass $\lambda(T)$ develops a cusp-like behaviour exactly at
the critical temperature $T =T_{\text{crit}}$.

\subsubsection{Equation of state}

\begin{figure}[h]
\begin{center}
\includegraphics[width=.45\textwidth]{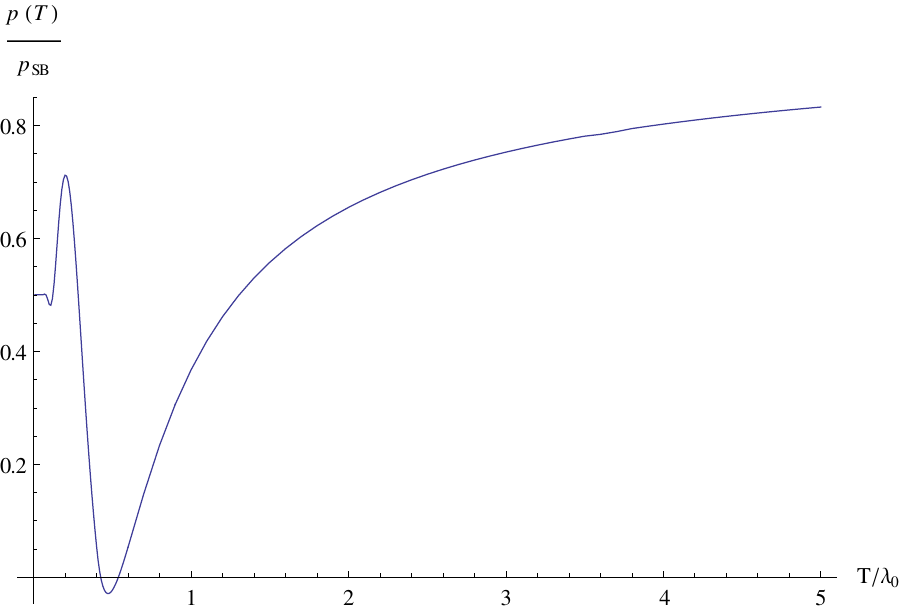} \hspace{10mm} %
\includegraphics[width=.45\textwidth]{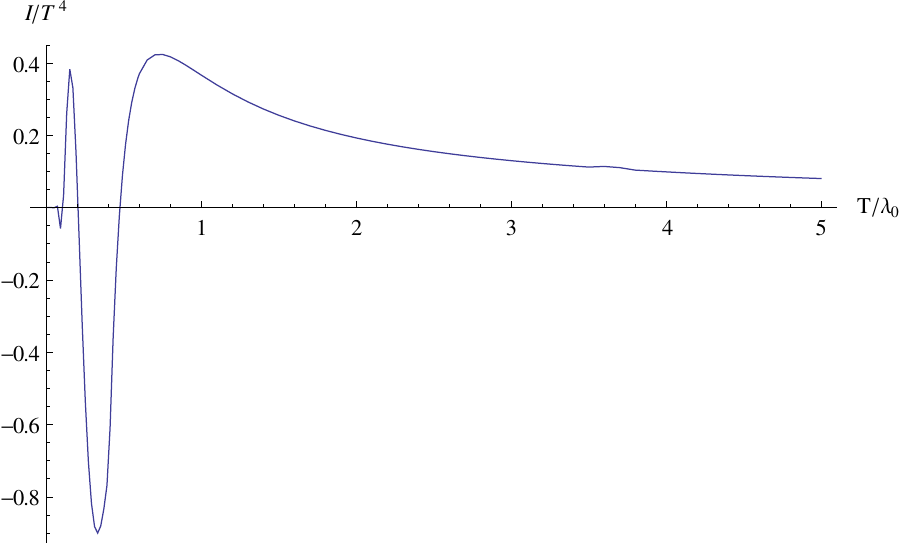}
\end{center}
\caption{Left: GZ pressure (relative to the Stefan-Boltzmann limit pressure $%
\sim T^{4}$). Right: GZ trace anomaly.}
\label{PTraceAnom}
\end{figure}

Following \cite{Philipsen:2012nu}, we can also extract an estimate for the
(density) pressure $p$ and the interaction measure $I/T^{4}$, shown in
Figure \ref{PTraceAnom} (left and right  respectively). As usual the
(density) pressure is defined as
\begin{eqnarray}
p = \frac{1}{\beta V}\ln Z_{GZ}\;,
\end{eqnarray}
which is related to the free energy by $p = -\mathcal{E}_{v}$. Here the plot of the pressure is given relative to the Stefan--Boltzmann limit pressure: $p_{SB} = \kappa T^{4}$, where $\kappa = (N^2 -1)\pi T^{4}/45$ is the Stefan--Boltzmann constant accounting for all degrees of freedom of the system at high temperature. We subtract the zero-temperature value, such that the pressure becomes zero at zero temperature: $p(T) = - [\mathcal{E}_{v}(T) - \mathcal{E}_{v}(T=0)]$. Namely, after using the $\MSbar$ renormalization prescription and choosing the renormalization parameter $\bar{\mu}$ so that the zero temperature gap equation is satisfied,
\begin{eqnarray}
{\bar{\mu}}^2 = \lambda^{2}_{0}e^{-\left(  \frac{5}{6} - \frac{32\pi^2}{3g^{2}} \right)}\;,
\end{eqnarray}
we have the following expression for the pressure (in units of $\lambda_{0}^{4}$),
\begin{eqnarray}
-\frac{p(T)}{\lambda_{0}^{4}} &=& 3 \left[  I(i\lambda'^{2},r,T') + I(-i\lambda'^{2},r,T') - \frac{4}{3}I(0,r,T')\right]
\nonumber \\
&+&
\frac{3}{2} \left[ I(i\lambda'^{2},0,T') + I(-i\lambda'^{2},0,T') - \frac{4}{3}I(0,0,T')\right]
\nonumber \\
&-&
\frac{9\lambda'^{4}}{32\pi^2 } \left(  \ln \lambda'^{2}  - \frac12 \right) - \frac{9}{64\pi^{2}}
\;. \label{vcen1}
\end{eqnarray}
In \eqref{vcen1} prime quantities stand for quantities in units of $\lambda_{0}$, while $\lambda$ and $\lambda_{0}$ satisfy their gap equation. The last term of \eqref{vcen1} accounts for the zero temperature subtraction, so that $p(0) = 0$, according to the definition of $I(m^2, r,T)$ in \eqref{defi}. Note that the coupling constant does not explicitly appear in \eqref{vcen1} and that $\lambda_{0}$ stands for the Gribov parameter at $T=0$.

The interaction measure $I$ is defined as the trace anomaly in units of $T^{4}$, and $I$ is nothing less than the trace of the of the stress-energy
tensor, given by
\begin{eqnarray}
\theta_{\mu\nu} = (p + \epsilon)u_{\mu}u_{\nu} - p\eta_{\mu\nu}\;,
\end{eqnarray}
with $\epsilon$ being the internal energy density, which is defined as $\epsilon = {\cal E}_{v} + Ts$ (with $s$ the entropy density), $u = (1,0,0,0)$ and $\eta_{\mu\nu}$ the (Euclidean) metric of the space-time. Given the thermodynamic definitions of each quantity (energy, pressure and entropy), we obtain
\begin{eqnarray}
I = \theta_{\mu\mu} = T^{5}\frac{\partial}{\partial T}\left(\frac{p}{T^{4}}\right)\;.
\end{eqnarray}
Both quantities display a behavior similar to that presented in \cite{Fukushima:2013xsa} (but note that in they plot the temperature in units of the critical temperature ($T_{c}$ in their notation), while we use units $\lambda_{0}$). Besides this, and the fact that we included the effect of Polyakov loop on the Gribov parameter, in \cite{Fukushima:2013xsa} a lattice-inspired effective coupling was introduced at finite temperature while we used the exact one-loop perturbative expression, which is consistent with the order of all the computations made here.

However, we notice that at temperatures relatively close to our $T_{c}$, the
pressure becomes negative. This is clearly an unphysical feature, possibly
related to some missing essential physics. For higher temperatures, the
situation is fine and the pressure moreover displays a behaviour similar to what is seen
in lattice simulations for the nonperturbative pressure (see \cite{Borsanyi:2012ve} for the
$SU(3)$ case). A similar problem is present in one
of the plots of [Fig.~4] of \cite{Fukushima:2013xsa}, although no comment is made about it. Another strange feature is the oscillating behaviour of both
pressure and interaction measure at low temperatures. Something similar was already observed in
\cite{Benic:2012ec} where a quark model was employed with complex conjugate quark mass. It is
well-known that the gluon propagator develops two complex conjugate masses in Gribov--Zwanziger
quantization, see e.g.~\cite{Dudal:2010cd,Dudal:2013wja,Baulieu:2009ha,Cucchieri:2011ig} for
some more details, so we confirm the findings of \cite{Benic:2012ec} that, at least at leading
order, the thermodynamic quantities develop an oscillatory behaviour. We expect this
oscillatory behaviour would in principle also be present in \cite{Fukushima:2013xsa} if the
pressure and interaction energy were to be computed at lower temperatures than shown there. In
any case, the presence of complex masses and their consequences gives us a warning that a
certain care is needed when using GZ dynamics, also at the level of spectral properties as done
in \cite{Su:2014rma,Florkowski:2015rua}, see also \cite{Baulieu:2008fy,Dudal:2010wn}.

These peculiarities justify giving an outline in the next Section of the behaviours of the pressure and interaction measure in an improved formalism, such as in the Refined Gribov-Zwanziger one.

\section{Outlook to the Refined Gribov--Zwanziger formalism\label{RGZ-Ploop}}
\sectionmark{Outlook to the Refined GZ formalism}

The previous results can be slightly generalized to the case of the Refined
Gribov--Zwanziger (RGZ) formalism studied in \cite%
{Dudal:2007cw,Dudal:2008sp,Dudal:2011gd,Gracey:2010cg,Thelan:2014mza}. In
this refined case, additional nonperturbative vacuum condensates such as $%
\langle A_\mu^2\rangle$ and $\langle\bar\varphi_\mu^{ab}\varphi_\mu^{ab}%
\rangle$ are to be introduced. The corresponding mass dimension two
operators get a nonzero vacuum expectation value (thereby further lowering
the vacuum energy) and thus influence the form of the propagator and
effective action computation. The predictions for the RGZ propagators, see
also \cite{Dudal:2010tf,Dudal:2012zx,Oliveira:2012eh}, are in fine agreement
with ruling $T=0$ lattice data, see e.g.~also
\cite{Bogolubsky:2007ud,Cucchieri:2007md,Sternbeck:2007ug,Maas:2008ri,Oliveira:2008uf,Cucchieri:2008fc,Cucchieri:2007rg,Cucchieri:2008qm,Bogolubsky:2009dc}.
This is in contrast with the original GZ predictions, such that it could happen that the finite
temperature version of RGZ is also better suited to describe the phase transition and/or
thermodynamical properties of
the pure gauge theory.

Due to the more complex nature of the RGZ effective action (more vacuum
condensates), we will relegate a detailed (variational) analysis of their
finite temperature counterparts\footnote{From
\cite{Chernodub:2008kf,Dudal:2009tq,Vercauteren:2010rk}, the nontrivial response of the $d=2$
condensate $\braket{A^2}$ to temperature already became clear.} to future work, as this will
require
new tools. Here, we only wish to present a first estimate of the
deconfinement critical temperature $T_c$ using as input the $T=0$ RGZ gluon
propagator where the nonperturbative mass parameters are fitted to lattice
data for the same propagator. More precisely, we use \cite{Cucchieri:2011ig}
\begin{equation}
\Delta_{\mu\nu}^{ab}(p) = \delta^{ab} \frac{p^2+M^2+\rho_1}{%
p^4+p^2(M^2+m^2+\rho_1) + m^2(M^2+\rho_1)+\lambda^4} \left(\delta_{\mu\nu} -
\frac{p_\mu p_\nu}{p^2}\right) \;.
\label{rgzgluonprop}
\end{equation}
where we omitted the global normalization factor $Z$ which drops out from
our leading order computation\footnote{This $Z$ is related to the choice of a MOM renormalization scheme, the kind
of scheme that can also be applied to lattice Green functions, in contrast
with the $\overline{\mbox{MS}}$ scheme.}. In this expression, we have that
\begin{equation}
\langle A_\mu^aA_\mu^a\rangle \to -m^2 \;, \qquad
\langle\bar\varphi_\mu^{ab}\varphi_\mu^{ab}\rangle \to M^2 \;, \qquad
\frac12 \langle \varphi_\mu^{ab}\varphi_\mu^{ab} +
\bar\varphi_\mu^{ab}\bar\varphi_\mu^{ab} \rangle \to \rho_1 \;.
\end{equation}
The free energy associated to the RGZ framework can be obtained by following the same steps as in section \ref{sec4}, leading to
\begin{eqnarray}
{\cal E}_{v}(T) &=& (d-1)\left[ I(r_{+}^{2}, r,T) + I(r_{-}^{2}, r,T) - I(N^{2}, r,T) - \frac{1}{d-1}I(0, r,T) \right]
\nonumber \\
&&
+\frac{(d-1)}{2}\left[ I(r_{+}^{2}, 0,T) + I(r_{-}^{2}, 0,T) - I(N^{2}, 0,T) - \frac{1}{d-1}I(0, 0,T) \right]
\nonumber \\
&&
+\int \frac{d^{d}p}{ (2\pi)^{d} }\,\, \ln \left( \frac{ p^{4} + (m^{2}+ N^{2})p^{2} + (m^{2}N^{2} + \lambda^{4}) }{p^{2}+N^{2}}  \right)
-  \frac{3\lambda^{4}d}{4g^{2}}\;,
\label{vacfreeenergy}
\end{eqnarray}
with $r_{\pm}^{2}$ standing for minus the roots of the denominator of the gluon propagator \eqref{rgzgluonprop}, $N^{2} = M^{2} + \rho_{1}$, and $I(m^{2},r,T)$  given by \eqref{defi}. Explicitly, the roots are
\begin{eqnarray}
r_{\pm}^{2} = \frac{ (m^{2}+N^{2}) \pm \sqrt{ (m^{2}+N^{2})^{2} - 4(m^{2}N^{2}+\lambda^{4}) } }{2}\;.
\end{eqnarray}
The (central) condensate values were extracted from \cite{Cucchieri:2011ig}:
\begin{subequations}\label{latticec}
\begin{align}
N^2 = M^2+\rho_1 &= \unit{2.51}{\giga\electronvolt\squared} \;, \\
m^2 &= \unit{-1.92}{\giga\electronvolt\squared} \;, \\
\lambda^4 &= \unit{5.3}{\power{\giga\electronvolt}4} \;.
\end{align}
\end{subequations}
Once again the vacuum energy will be minimized with respect to the Polyakov loop expectation
value $r$. For the analysis of thermodynamic quantities, only contributions coming from terms
proportional to $I(m^{2},r,T)$ will be needed. Therefore, we will always consider the
difference ${\cal E}_{v}(T) - {\cal E}_{v}(T=0)$. Since in the present  (RGZ) prescription the
condensates are given by the zero temperature lattice results \eqref{latticec} instead of
satisfying gap equations, the divergent contributions to the free energy are subtracted, and no
specific choice of renormalization scheme is needed. Furthermore, explicit dependence on the
coupling constant seems to drop out of the one-loop expression, such that no renormalization
scale has to be chosen. Following the steps taken in Section \ref{Tonafhankelijk}, we find a
second order phase transition at the temperature:
\begin{equation}
T_\text{crit} = \unit{0.25}{\giga\electronvolt} \;,
\end{equation}
which is not that far from the value of the $SU(2)$ deconfinement temperature  found on the lattice: $T_c\approx\unit{0.295}{\giga\electronvolt}$, as
quoted in \cite{Cucchieri:2007ta,Fingberg:1992ju}.

In future work, it would in particular be interesting to find out whether
---upon using the RGZ formalism--- the Gribov mass and/or RGZ condensates
directly feel the deconfinement transition, similar to the cusp we
discovered in the Gribov parameter following the exploratory restricted
analysis of this paper. This might also allow to shed further light on the
ongoing discussion of whether the deconfinement transition should be felt at
the level of the correlation functions, in particular the electric screening
mass associated with the longitudinal gluon propagator \cite{Cucchieri:2012nx,Cucchieri:2007ta,Cucchieri:2014nya,Silva:2013maa}.

\begin{figure}[h]
\begin{center}
\includegraphics[width=.45\textwidth]{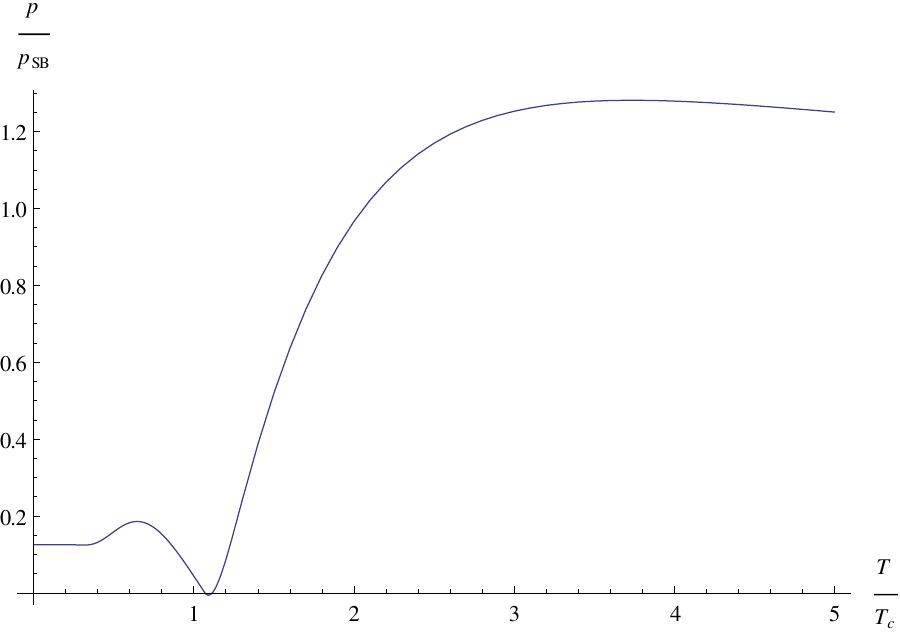} \hspace{10mm}
\includegraphics[width=.45\textwidth]{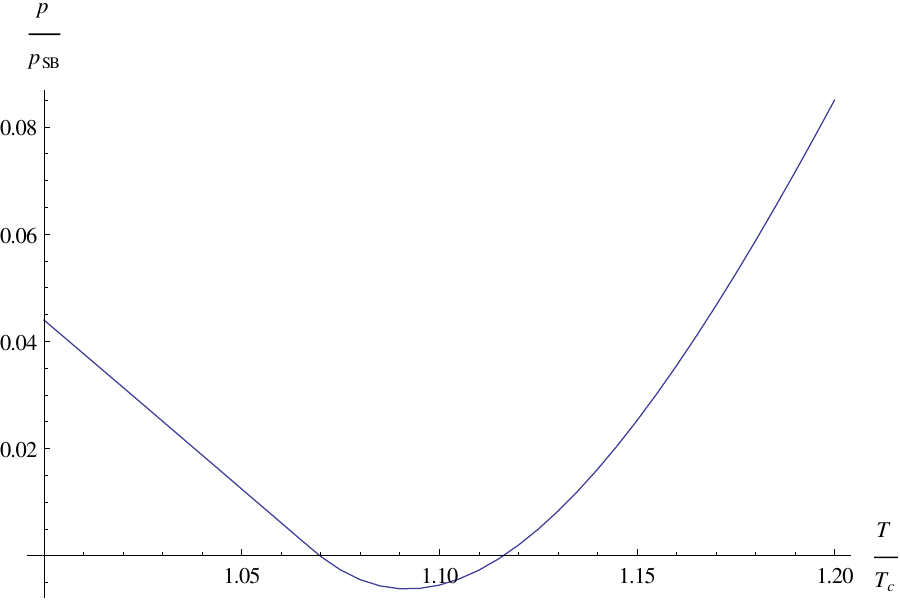}
\end{center}
\caption{Right and left plots refer to the RGZ pressure in terms of $T/T_{c}$ and in units of $T^{4}$. In the left plot, a wide temperature range of is shown. In the right plot, a zoom is made for temperatures around $1.10\; T_{c}$ to show the existence of negative pressure.
}
\label{rgzP}
\end{figure}

\begin{figure}[h]
\begin{center}
\includegraphics[width=.45\textwidth]{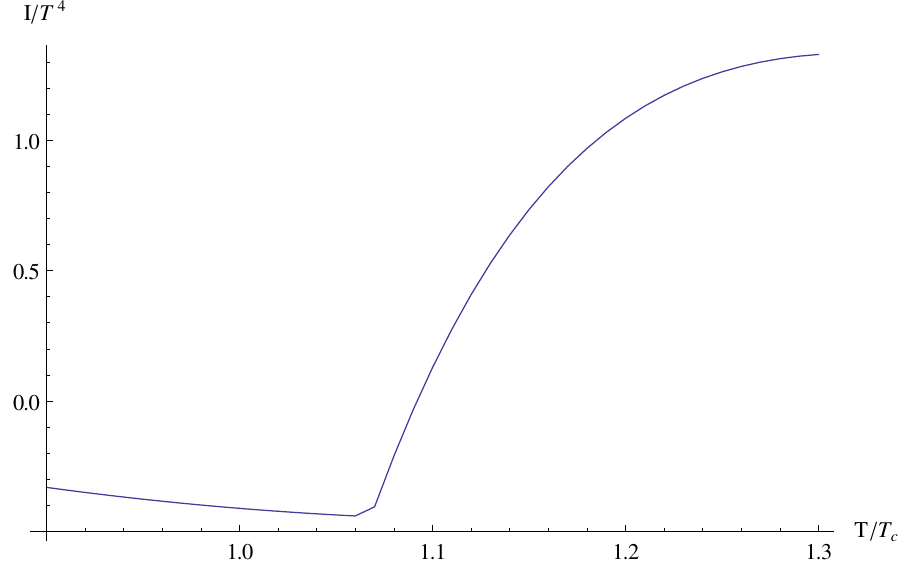}
\end{center}
\caption{The RGZ interaction measure $I/T^{4}$ in units $T/T_{c}$.}
\label{rgzTraceAnom}
\end{figure}

Let us also consider the pressure and interaction measure once more. The results are shown in
\figurename\ \ref{rgzP} and \figurename\ \ref{rgzTraceAnom}, respectively. The oscillating
behaviour at low temperature persists at leading order, while a small region of negative
pressure is still present --- see the right plot of \figurename\ \ref{rgzP}. These findings are
similar to \cite{Fukushima:2012qa} (low temperature results are not shown there), where two
sets of finite temperature RGZ fits to the $SU(3)$ lattice data were used
\cite{Aouane:2011fv,Aouane:2012bk}, in contrast with our usage of zero temperature $SU(2)$
data. In any case, a more involved analysis of the RGZ finite temperature dynamics is needed to
make firmer statements. As already mentioned before, there is also the possibility that
important low temperature physics is missing, as for instance the proposal of
\cite{Fukushima:2012qa} related to the possible effect of light electric glueballs near the
deconfinement phase transition \cite{Ishii:2001zq,Hatta:2003ga}. Obviously, these effects are
absent in the current treatment (or in most other treatments in fact).

\section{Discussions about the results \label{DAR-Ploop}}

In this chapter we studied the Gribov--Zwanziger (GZ) action for $SU(2)$ gauge theories with
the Polyakov loop coupled to it via the background field formalism. Doing
so, we were able to compute in a simultaneous fashion the finite temperature
value of the Polyakov loop and Gribov mass to the leading one-loop
approximation. The latter dynamical scale enters the theory as a result of
the restriction of the domain of gauge field integration to avoid
(infinitesimally connected) Gribov copies. Our main result is that we found
clear evidence of a second order deconfinement phase transition at finite
temperature, an occurrence accompanied by a cusp in the Gribov mass, which
thus directly feels the transition. It is perhaps worthwhile to stress here
that at temperatures above $T_{c}$, the Gribov mass is nonzero, indicating
that the gluon propagator still violates positivity and as such it rather
describes a quasi- than a \textquotedblleft free\textquotedblright\
observable particle, see also \cite{Maas:2011se,Haas:2013hpa} for more on this.

We also presented the pressure and trace anomaly, indicating there is a
problem at temperatures around the critical value when using the original GZ
formulation. We ended with a first look at the changes a full-fledged
analysis with the Refined Gribov--Zwanziger (RGZ) formalism might afflict, given
that the latter provides an adequate description of zero temperature gauge
dynamics, in contrast to the GZ predictions. This will be studied further in
upcoming work. Note that, even not considering finite temperature corrections to the
condensates in the RGZ formalism, the region of negative pressure is considerably smaller than
the region found with the GZ formalism.

A further result, interesting from the methodological point of view, is that it shows
explicitly that finite-temperature computations (such as the computation of the vacuum
expectation value of the Polyakov loop) are very suitable to be analyzed using analytical
Casimir-like techniques. The interesting issue of Casimir-style computations at finite
temperatures is that, although they can be more involved, they provide one
with easy tools to analyze the high and low temperature limits as well as
the small mass limit. Moreover, within the Casimir framework, the
regularization procedures are often quite transparent. Indeed, in the
present paper, we have shown that the computation of the vacuum expectation value of the Polyakov
loop is very similar to the computation of the Casimir energy between two
plates. We believe that this point of view can be useful in different
contexts as well.

\chapter{Final words
\label{finalwords}}

The present thesis was devoted to the study of aspects of the Gribov problem in Euclidean
Yang-Mills theories coupled to matter field.  We presented some evidences that point to the
existence of a possible interplay between the gauge sector and the
matter sector, in regimes of sufficiently low energy (known as the infrared (IR) regime). Our
work is analytic, based on the quantization procedure of Gribov where (infinitesimal) gauge
fixing ambiguities are taken into account by getting rid of nonzero modes of the Faddeev-Popov
operator. This framework was briefly introduced in the second chapter, where some fundamental
concepts were discussed, and some important quantities were derived. We presented that, to get
rid of such gauge ambiguities a nonlocal term must be added to the Faddeev-Popov action, which
leads to two important consequences: the soft BRST breaking and a deeply modified gauge
field propagator, whose expression displays complex conjugate poles. Such behaviour
of the gauge field may be interpreted as a sign of confinement, providing us an alternative
tool to analytically investigate gauge confinement.

Our study of Yang-Mills theories coupled to matter field started with the analysis of
Yang-Mills models coupled to the Higgs field, in Euclidean space-time with dimensionality $d=3$
and $d=4$. Two different representations of the scalar field was considered, the fundamental
one, which is an example of a nontrivial representation, and the adjoint representation, which
is an example of a trivial one. The scalar Higgs field was considered to be frozen in its
vacuum configuration. Our analysis concern the direct observation of the
propagator of the gauge field: if it presents or not complex conjugate poles, or negative
residues. We use to say that the gauge propagator is of the Gribov-type when it presents
\emph{cc} poles, and the Gribov's alternative confinement criterion applies. We could generally
notice that the representation of the scalar field is of great importance to the analysis.

The point we would like to emphasise here is that our work shares remarkable similarities with
the seminal paper of Fradkin \& Shenker \cite{Fradkin:1978dv} and others lattice works
\cite{Nadkarni:1989na,Hart:1996ac}, despite the existence of fundamental differences between
them. The authors of \cite{Fradkin:1978dv} made use of the lattice formalism to investigate the
structure of gauge theories coupled to the Higgs field. In particular, the Wilson loop was
measured, as a suitable order parameter of (static quark) confinement, in mainly two different
scenarios, for the scalar field in its fundamental configuration, and for the scalar field in
its adjoint representation. In the other side, our work was made in the continuum space-time,
by means of the Gribov effect model, to probe for gauge field confinement in Yang-Mills + Higgs
theories.

Therefore, when we say that both works share some similarities, we do not mean we are measuring
the same thing, or else, that we are obtaining the same results. But rather, we mean that the
structure of the confinement spectrum of the gauge field, in the light of Gribov's approach, is
quite similar to the structure of (static) quark confinement, \`a la Wilson loop. Namely, for
the Higgs field in the fundamental representation we could find: 1)
the existence of two distinct regimes, the Higgs-like and the confined-like, corresponding to
weak and strong coupling regions, respectively; 2) these two detected regimes, Higgs- and
confined-like, were found to be continuously connected, in the sense that the parameters of the
theory are allowed to continuously vary from one region to another, without leading to any
discontinuity or singularity of the vacuum energy or the two-point Green function. The
similarity with Fradkin \& Shenker indeed exists, although we could not properly talk about the
existence of an \emph{analyticity region} in our work. Besides being a perturbative work, we
could also find out a region in our model where perturbation theory is not trustworthy anymore.
Such unreliable region lies in between the Higgs- and confined-regions and prevents us from
proving the existence of an analyticity region.

Similarities are still present in the adjoint Higgs field case: 1) in both works, the
connection between the two distinct regimes is not a smooth connection anymore. In our work we
could find a point of discontinuity in the vacuum energy, although our perturbative
computations are not reliable at precisely this point. Perhaps, this feature may be a sign, in
the gauge sector, of the phase transition associated with the breaking of the center symmetry,
since this kind of phase transition is only possible for the matter field in the adjoint
representation (or in its absence);
 2) we could also find the existence of a third regime, besides the confined- and Higgs-like
ones. We could detect a kind of $U(1)$ confined-like regime, where the third component of the
gauge field has a propagator of the Gribov-type and the off-diagonal sector is massive.
Interestingly, something similar to this has been already detected on lattice studies of the
three-dimensional Georgi-Glashow model \cite{Nadkarni:1989na,Hart:1996ac}.

Finally, we may conclude that, by means of the Gribov's approach to the quantization of the
gauge field, leading to an alternative criterion of gauge field confinement, the structure of
the Yang-Mills + Higgs field's spectrum shares some resemblances with what is find out by works
on the lattice, where order parameters of (static quark) confinement is measured, such as the
Wilson loop. Thus, we obtained our first indicative sign that IR features of the gauge sector
may be reflected, in some sense, in the IR behaviour of the matter sector.

Subsequently, we proposed an effective model to the matter sector, inspired by the
Gribov-Zwanziger structure of the gauge sector that leads to the confinement interpretation of
the gauge field, where a kind of horizon-function is consistently plugged to the matter field.
By consistently we mean that we could show that such implementation does not lead to any new UV
divergences, but only to those already present in the standard, noneffective, Yang-Mills +
matter theory. Such UV safety was order by order analysed in the fifth chapter, by means of a
careful analysis of the Feynman diagrams. An all order renormalizability proof has been
provided in the Appendix, together with the example of the ${\cal N}=1$ supersymmetric
Yang-Mills model within the Gribov horizon.

The matter field was considered in its adjoint representation, so that we could show that it is
possible to construct an operator ${\cal R}^{ai}_{\;\;F}$ for matter field,
\begin{eqnarray}
{\cal R}^{ai}_{\;\;F}(x)  &=&  g \int d^4z\;  ({\cal M}^{-1})^{ab} (x,z)   \;(T^b)^{ij}
\;F^{j}(z)  
\,,
\end{eqnarray}
in analogy with the gauge field restricted to the first Gribov region.
Furthermore, we could also show that the correlation function $\langle{\cal R}_{\;\;F}{\cal
R}_{\;\;F}\rangle$ is nonvanishing and, from the available lattice data, seems to behave like
the Boson-ghost propagator in the IR regime, \eqref{RRgluon}, namely
\begin{eqnarray}
\langle \tilde {\cal R}^{ai}_{\;\;F}(k)\tilde {\cal R}^{bj}_{\;\;F}(-k)\rangle&\stackrel{k\to
0}{\sim}&\frac{1}{k^4} 
\,.
\end{eqnarray}
After constructing such effective nonlocal model to the matter sector we could show that, the
nonvanishing of $\langle{\cal R}_{\;\;F}{\cal R}_{\;\;F}\rangle$ indicates the soft breaking
of the BRST symmetry in the matter sector, since the \emph{vev} of ${\cal R}_{\;\;F}{\cal
R}_{\;\;F}$ can be written in terms the \emph{vev} of a BRST exact local operator. Therefore,
the correlation function $\langle{\cal R}_{\;\;F}{\cal R}_{\;\;F}\rangle$ could be regarded as
a direct signature for BRST breaking, being accessible both analytically as well as through
numerical lattice simulations.

Another important outcome of this effective model is that, by fitting our effective matter
propagator to the most recent lattice data we could find that, in this scenario, the matter
field is deprived of an asymptotic physical particle interpretation, since its propagator
displays positivity violation, so not satisfying every reality condition of
Osterwalder-Schrader (just as the gauge field). In this sense the adjoint scalar
propagators  consistently represent confined degrees of freedom, that do not exhibit a physical
propagating pole. We could also qualitatively show that our results are renormalization scheme
independent. A more quantitative analysis would require further simulations with improved
statistics and even larger lattices.

Subsequently, we studied the Gribov-Zwanziger (GZ) action for $SU(2)$ gauge theories
with the Polyakov loop coupled to it via the background field formalism. Doing so, we were able
to compute simultaneously the finite temperature value of the Polyakov loop and of the
Gribov mass parameter, up to the leading order. Within this formalism we could confirm the
existence of a second order deconfinement phase transition of static quarks. Besides, we could
also observe that the GZ mass parameter evidently feels the effects of quark confinement: 
such a mass parameter develops a cusp-like behaviour precisely at the critical temperature of
quark confinement, probed by the Polyakov loop parameter. It is perhaps worthwhile to stress
here that at temperatures above $T_{c}$, the Gribov mass is nonzero, indicating that the gluon
propagator still violates positivity and as such it rather describes a quasi- than a
\textquotedblleft free\textquotedblright\ observable particle. It would means that the gauge
sector is, indeed, sensible to IR effects of the matter sector. In a sense, it may corroborate
our feelings that the IR structure of the gauge sector may be transferred to the matter sector.

Besides that, we could also find that our model is plagued by the existence of an instability
region in the vicinity of the critical temperature. To investigate such region we computed the
pressure and the trace anomaly, so that we could find out a region of negative pressure. It is
worthwhile to emphasise that those results were obtained in the GZ formalism. The RGZ formalism
was adopted with the status of a first computation, while a full finite-temperature approach of
the RGZ in the presence of the Polyakov loop is currently being carried out. The main outcome
of the RGZ formalism is that the instability region is considerably smaller, possibly because
the RGZ framework provides an adequate description of the zero-temperature gauge dynamics. One
should also keep in mind that we are dealing with a perturbative effective theory, so that
higher loop order computations, within the RGZ framework, should be carried out in order to
make any reliable assertion about the existence of instability regions.

Finally, we close this thesis stating that a lot of work still has to be done in the direction
of a deeper understanding of nonperturbative effects of QCD. Precisely, the concept of gauge
confinement is not as clear as the one of quark confinement, regarding that a definite
understanding of confinement is lacking, at all. A step towards the reconciliation of Gribov's
mechanism and BRST breaking has been made, so that the possibility to define physical states
even in the (R)GZ framework still exists \cite{Capri:2015ixa,Capri:2015nzw,Capri:2016aqq}.
Concomitantly, the same construction of a nonlocal effective model of the matter field that
still keeps the action invariant under a nonperturbative BRST transformation has been worked
out \cite{Capri:2016aqq}. Equally, there are currently efforts of carrying on investigations 
on finite temperature Yang-Mills theory, within the RGZ framework, in order to better
understand our earlier results. It is becoming clear for us that a nonlocal horizon-like term
in the matter sector sufficiently describes some IR features of this sector. However,
fundamental arguments that would justify the existence of that horizon-like term in
matter are still lacking. We hope that further work would point us to the right answer.

\begin{appendix}
\appendix

\chapter{General effective action with soft BRST symmetry breaking: algebraic renormalization}
\chaptermark{General effective action with soft BRST symmetry breaking}
\label{ARscalaraction}

In order to prove the renormalizability of the action $S_{ loc}$, eq.\eqref{actlc}, we proceed as in   \cite{Zwanziger:1988jt,Zwanziger:1989mf,Zwanziger:1992qr,Dudal:2007cw,Dudal:2008sp,Dudal:2011gd} and we embed the theory into an extended action $\Sigma$ enjoying exact BRST symmetry, given by
\begin{eqnarray}
\label{fullact}
\Sigma &=& \int d^4x\; \Biggl\{
\frac{1}{4}F^a_{\mu\nu} F^a_{\mu\nu}
+ b^{a}\partial_{\mu}A^{a}_{\mu}
+ \bar{c}^{a}\partial_{\mu}D^{ab}_{\mu}c^{b}
+ \frac{1}{2}(D^{ab}_{\mu}\phi^{b})^{2} + \frac{m^2_{\phi}}{2} \phi^a \phi^a
+ \frac{\lambda}{4!}(\phi^{a}\phi^{a})^{2}
+ \bar{\varphi}^{ac}_{\nu}\partial_{\mu}D^{ab}_{\mu}\varphi^{bc}_{\nu}
\nonumber \\
&&
- \bar{\omega}^{ac}_{\nu}\partial_{\mu}D^{ab}_{\mu}\omega^{bc}_{\nu}
- gf^{abc}(\partial_{\mu}\bar{\omega}^{ae}_{\nu})(D^{bd}_{\mu}c^{d})\varphi^{ce}_\nu
-{N}^{ac}_{\mu\nu}\,D^{ab}_{\mu}\bar{\omega}^{bc}_{\nu}
-{M}^{ae}_{\mu\nu}\Bigl[D^{ab}_{\mu}\bar{\varphi}^{be}_{\nu}
-gf^{abc}(D^{bd}_{\mu}c^{d})\bar{\omega}^{ce}_{\nu}
\Bigr]
\nonumber \\
&&
- \bar{M}^{ac}_{\mu\nu}\,D^{ab}_{\mu}\varphi^{bc}_{\nu}
+ \bar{N}^{ae}_{\mu\nu}\Bigl[D^{ab}_{\mu}\omega^{be}_{\nu}
- gf^{abc}(D^{bd}_{\mu}c^{d})\varphi^{ce}_{\nu}
\Bigr]
- \bar{M}^{ac}_{\mu\nu}{M}^{ac}_{\mu\nu}
+ \bar{N}^{ac}_{\mu\nu}{N}^{ac}_{\mu\nu}
+ \tilde{\eta}^{ac}(\partial_{\mu}D_{\mu}^{ab})\eta^{bc}
\nonumber \\
&&
- \tilde{\theta}^{ac}(\partial_{\mu}D_{\mu}^{ab})\theta^{bc}
- gf^{abc}(\partial_{\mu}\tilde{\theta}^{ae})(D^{bd}_{\mu}c^{d})\eta^{ce}
+ gf^{abc}\tilde{V}^{ad}\phi^{b}\eta^{cd}
+ gf^{abc}V^{ad}\left( 
- gf^{bde}\phi^{d}c^{e}\tilde{\theta}^{cd}
+ \phi^{b}\tilde{\eta}^{cd}
\right)
\nonumber \\
&&
+ \rho\left(\tilde{V}^{ab}V^{ab} - \tilde{U}^{ab}U^{ab}\right)
+ gf^{abc}\tilde{U}^{al}\left( gf^{bde}\phi^{d}c^{e}\eta^{cl}
- \phi^{b}\theta^{cl}\right)
+ gf^{abc}U^{ad}\phi^{b}\tilde{\theta}^{cd}
- K^{a}_{\mu}D^{ab}_{\mu}c^{b}
+ \frac{g}{2}f^{abc} L^{a}c^{b}c^{c}
\nonumber \\
&&
- gf^{abc} F^{a}\phi^{b}c^{c}
\Biggr\}\;,
\end{eqnarray}
where $\left( M^{ab}_{\mu\nu}, \bar{M}^{ab}_{\mu\nu}, N^{ab}_{\mu\nu}, \bar{N}^{ab}_{\mu\nu}, V^{abc}, \tilde{V}^{abc}, U^{abc}, \tilde{U}^{abc} \right)$ are external sources. 
The original local action $S_{ loc}$, \eqref{actlc}, can be re-obtained from the extended action $\Sigma$  by letting the external fields to assume their  physical values namely
\begin{eqnarray}
\label{physval1}
&&
M^{ab}_{\mu\nu}\Big{|}_{phys}=\bar{M}^{ab}_{\mu\nu}\Big{|}_{phys}=
\gamma^{2}\delta^{ab}\delta_{\mu\nu}
\;;\nonumber \\
&&
V^{ab}\Big{|}_{phys}=\tilde{V}^{ab}\Big{|}_{phys}=
\sigma^{2}\delta^{ab}
\;;\nonumber \\
&&
N^{ab}_{\mu\nu}\Big{|}_{phys}=\bar{N}^{ab}_{\mu\nu}\Big{|}_{phys}=
U^{ab}\Big{|}_{phys}=\tilde{U}^{ab}\Big{|}_{phys}=0
\;. \nonumber\\
&& K^{a}_\mu = L^a = F^a =0 \;, 
\end{eqnarray}
so that 
\begin{equation} 
\Sigma \Big|_{phys}  = S_{ loc}  + V \rho \;\sigma^4 g^2 N(N^2-1)  \;, \label{limit} 
\end{equation}
where the parameter $\rho$ has been introduced in order to take into account possible divergences in the vacuum energy associated to the term $\sigma^4$. This term stems from the source term $\rho\tilde{V}^{abc}V^{abc}$, which  is allowed by power counting. In the physical limit the vertex $\phi c \tilde{\theta}$ remains non-vanishing. Though, it is harmless, due to the absence of mixed propagators $\langle c\; \tilde{\theta}\rangle $ and $\langle {\bar c}\; \theta\rangle $. \\\\It is easy to check that the extended action $\Sigma$ enjoys exact BRST invariance, {\it i.e.} 
\begin{equation} 
s \Sigma = 0  \;, \label{sinv}
\end{equation}
where
\begin{eqnarray}
\label{brst2}
sA^{a}_{\mu} &=& - D^{ab}_{\mu}c^{b}\;,\nonumber \\
s\phi^{a}& = & -gf^{abc}\phi^{b}c^{c} \;,    \nonumber \\
s c^{a} &=& \frac{1}{2}gf^{abc}c^{b}c^{c} \;, \nonumber \\
s{\bar c}^{a} &=& b^{a}\;, \qquad \; \; 
sb^{a} = 0 \;, \nonumber \\
s{\bar \omega}^{ab}_\mu & = & {\bar \varphi}^{ab}_\mu \;, \qquad  s {\bar \varphi}^{ab}_\mu =0\;, \nonumber \\
s { \varphi}^{ab}_\mu&  = & {\omega}^{ab}_\mu  \;, \qquad s {\omega}^{ab}_\mu = 0 \;,  \nonumber \\
s\tilde{\theta}^{ab} & =&  \tilde{\eta}^{ab}\;, \qquad s\tilde{\eta}^{ab} =0\;, \nonumber \\
s\eta^{ab}& = & \theta^{ab}\;, \qquad s\theta^{ab}=0\;, 
\end{eqnarray}
and 
\begin{eqnarray}
&&
sM^{ab}_{\mu\nu} = N^{ab}_{\mu\nu}\;; \qquad sN^{ab}_{\mu\nu} = 0\;; \nonumber \\
&&
s\bar{N}^{ab}_{\mu\nu} = \bar{M}^{ab}_{\mu\nu}\;; \qquad s\bar{M}^{ab}_{\mu\nu} = 0\;; \nonumber \\
&&
s\tilde{U}^{ab}=\tilde{V}^{ab}\,,\qquad s\tilde{V}^{ab}=0\;; \nonumber \\
&&
sV^{ab}=U^{ab}\,,\qquad sU^{ab}=0\;; \nonumber \\
&&
sK^{a}=sL^{a}=sF^{a}=0\;.
\end{eqnarray}
As noticed in \cite{Zwanziger:1988jt,Zwanziger:1989mf,Zwanziger:1992qr,Dudal:2007cw,Dudal:2008sp,Dudal:2011gd}, it is useful introducing a multi-index notation for the localizing auxiliary fields $({\bar \varphi}^{ab}_\mu, { \varphi}^{ab}_\mu, {\bar \omega}^{ab}_\mu, {\bar \omega}^{ab}_\mu) = ({\bar \varphi}^{a}_i, { \varphi}^{a}_i, {\bar \omega}^{a}_i, {\bar \omega}^{a}_i) $ where the multi-index $i=(b,\mu)$ runs from $1$ to $4(N^2-1)$. The important reason in order to introduce the multi-index notation is related to the existence of a global symmetry $U(4(N^2-1))$ in the index $i$, which plays an important role in the proof of the algebraic renormalization. Analogously, one can introduce a second index $I$ for the localizing fields of the matter scalar sector $(\tilde{\eta}^{ab},{\eta}^{ab},\tilde{\theta}^{ab},{\theta}^{ab})= (\tilde{\eta}^{aI},{\eta}^{aI},\tilde{\theta}^{aI},{\theta}^{aI})$, where $I=1,..,(N^2-1)$. Again, the introduction of the index $I$ is related to the existence of a second global symmetry $U(N^2-1)$. In the multi-index notation, the action \eqref{fullact} reads 
\begin{eqnarray}
\label{fullactm}
\Sigma &=& \int d^4x\; \Biggl\{
\frac{1}{4}F^a_{\mu\nu} F^a_{\mu\nu}
+ b^{a}\partial_{\mu}A^{a}_{\mu}
+ \bar{c}^{a}\partial_{\mu}D^{ab}_{\mu}c^{b}
+ \frac{1}{2}(D^{ab}_{\mu}\phi^{b})^{2} + \frac{m^2_{\phi}}{2} \phi^a \phi^a 
+ \frac{\lambda}{4!}(\phi^{a}\phi^{a})^{2}
+ \bar{\varphi}^{a}_{i}\partial_{\mu}D^{ab}_{\mu}\varphi^{b}_{i}
\nonumber \\
&&
- \bar{\omega}^{a}_{i}\partial_{\mu}D^{ab}_{\mu}\omega^{b}_{i}
- gf^{abc}(\partial_{\mu}\bar{\omega}^{a}_{i})(D^{bd}_{\mu}c^{d})\varphi^{c}_{i}
-{N}^{a}_{\mu{i}}\,D^{ab}_{\mu}\bar{\omega}^{b}_{i}
-{M}^{a}_{\mu{i}}\Bigl[D^{ab}_{\mu}\bar{\varphi}^{b}_{i}
-gf^{abc}(D^{bd}_{\mu}c^{d})\bar{\omega}^{c}_{i}
\Bigr]
\nonumber \\
&&
- \bar{M}^{a}_{\mu{i}}\,D^{ab}_{\mu}\varphi^{b}_{i}
+ \bar{N}^{a}_{\mu{i}}\Bigl[D^{ab}_{\mu}\omega^{b}_{i}
- gf^{abc}(D^{bd}_{\mu}c^{d})\varphi^{c}_{i}
\Bigr]
- \bar{M}^{a}_{\mu{i}}{M}^{a}_{\mu{i}}
+ \bar{N}^{a}_{\mu{i}}{N}^{a}_{\mu{i}}
+ \tilde{\eta}^{aI}(\partial_{\mu}D_{\mu}^{ab})\eta^{bI}
\nonumber \\
&&
- \tilde{\theta}^{aI}(\partial_{\mu}D_{\mu}^{ab})\theta^{bI}
- gf^{abc}(\partial_{\mu}\tilde{\theta}^{aI})(D^{bd}_{\mu}c^{d})\eta^{cI}
+ gf^{abc}\tilde{V}^{aI}\phi^{b}\eta^{cI}
+ gf^{abc}V^{aI}\left( 
- gf^{bde}\phi^{d}c^{e}\tilde{\theta}^{cI}
+ \phi^{b}\tilde{\eta}^{cI}
\right)
\nonumber \\
&&
+ \rho\left(\tilde{V}^{aI}V^{aI} - \tilde{U}^{aI}U^{aI}\right)
+ gf^{abc}\tilde{U}^{aI}\left( gf^{bde}\phi^{d}c^{e}\eta^{cI}
- \phi^{b}\theta^{cI}\right)
+ gf^{abc}U^{aI}\phi^{b}\tilde{\theta}^{cI}
- K^{a}_{\mu}D^{ab}_{\mu}c^{b} \nonumber \\ &&
+ \frac{g}{2}f^{abc} L^{a}c^{b}c^{c}
- gf^{abc} F^{a}\phi^{b}c^{c}
\Biggr\}\;,
\end{eqnarray}
We are now ready to write down the large set of Ward identities fulfilled by the action \eqref{fullactm}. These are given by: 

\noindent {\bf $\bullet$ The Slavnov-Taylor identity}:
\begin{equation} 
S(\Sigma) = 0  \;, \label{st}
\end{equation} 
where 
\begin{eqnarray}
S(\Sigma) &=& \int d^{4}x \Biggl\{
\frac{\delta \Sigma}{\delta K^{a}_{\mu}}\frac{\delta \Sigma}{\delta A^{a}_{\mu}}
+ \frac{\delta \Sigma}{\delta F^{a}}\frac{\delta \Sigma}{\delta \phi^{a}}
+ \frac{\delta \Sigma}{\delta L^{a}}\frac{\delta \Sigma}{\delta c^{a}}
+ b^{a}\frac{\delta \Sigma}{\delta \bar{c}^{a}}
+ \omega^{a}_{i}\frac{\delta \Sigma}{\delta \varphi^{a}_{i}}
+ \bar{\varphi}^{a}_{i}\frac{\delta \Sigma}{\delta \bar{\omega}^{a}_{i}}
\nonumber \\
&&
+ \tilde{\eta}^{aI}\frac{\delta \Sigma}{\delta \tilde{\theta}^{aI}}
+ \theta^{aI}\frac{\delta \Sigma}{\delta \eta^{aI}}
+ N^{a}_{\mu i}\frac{\delta \Sigma}{\delta M^{a}_{\mu i}}
+ \bar{M}^{a}_{\mu i}\frac{\delta \Sigma}{\delta \bar{N}^{a}_{\mu i}}
+ \tilde{V}^{aI}\frac{\delta \Sigma}{\delta \tilde{U}^{aI}}
+ U^{aI}\frac{\delta \Sigma}{\delta V^{aI}}
\Biggr\}  \;.     \label{stop}
\end{eqnarray}
For future convenience, let us also introduce the so-called linearized Slavnov-Taylor operator ${\cal B}_{\Sigma}$, given by 
\begin{eqnarray}
{\cal B}_{\Sigma} &=& \int d^{4}x \Biggl\{
\frac{\delta \Sigma}{\delta K^{a}_{\mu}}\frac{\delta }{\delta A^{a}_{\mu}}
+ \frac{\delta \Sigma}{\delta A^{a}_{\mu}}\frac{\delta }{\delta K^{a}_{\mu}}
+ \frac{\delta \Sigma}{\delta F^{a}}\frac{\delta }{\delta \phi^{a}}
+ \frac{\delta \Sigma}{\delta \phi^{a}}\frac{\delta }{\delta F^{a}}
+ \frac{\delta \Sigma}{\delta L^{a}}\frac{\delta }{\delta c^{a}}
+ \frac{\delta \Sigma}{\delta c^{a}}\frac{\delta }{\delta L^{a}}
+ b^{a}\frac{\delta }{\delta \bar{c}^{a}}
\nonumber \\
&&
+ \omega^{a}_{i}\frac{\delta }{\delta \varphi^{a}_{i}}
+ \bar{\varphi}^{a}_{i}\frac{\delta }{\delta \bar{\omega}^{a}_{i}}
+ \tilde{\eta}^{aI}\frac{\delta }{\delta \tilde{\theta}^{aI}}
+ \theta^{aI}\frac{\delta }{\delta \eta^{aI}}
+ N^{a}_{\mu i}\frac{\delta }{\delta M^{a}_{\mu i}}
+ \bar{M}^{a}_{\mu i}\frac{\delta }{\delta \bar{N}^{a}_{\mu i}}
+ \tilde{V}^{aI}\frac{\delta }{\delta \tilde{U}^{aI}}
+ U^{aI}\frac{\delta }{\delta V^{aI}}
\Biggr\}\;.  \nonumber \\    \label{lst}
\end{eqnarray}
The operator ${\cal B}_{\Sigma}$ enjoys the important property of being nilpotent
\begin{equation}
{\cal B}_{\Sigma} {\cal B}_{\Sigma} = 0 \;.    \label{nlst}
\end{equation}
{\bf $\bullet$  The gauge-fixing and anti-ghost equations}:
\begin{equation}
\frac{\delta\Sigma}{\delta b^{a}}=\partial_{\mu}A^{a}_{\mu}\,,\qquad
\frac{\delta\Sigma}{\delta\bar{c}^{a}}+\partial_{\mu}\frac{\delta\Sigma}{\delta K^{a}_{\mu}}=0\,.
\label{GFandAntiGhost2}
\end{equation}
\noindent {\bf $\bullet$  The linearly broken Ward identities}:
\begin{eqnarray}
&&\frac{\delta \Sigma}{\delta \bar{\varphi}^{a}_{i}} + \partial_{\mu}\frac{\delta\Sigma}{\delta \bar{M}^{a}_{\mu i}}  = 0\,,
\\
&&\frac{\delta\Sigma}{\delta\omega^{a}_{i}} + \partial_{\mu}\frac{\delta\Sigma}{\delta N^{a}_{\mu i}} - gf^{abc} \frac{\delta \Sigma}{\delta b^{c}} \bar{\omega}^{b}_{i} = 0\,,
\\
&&
\frac{\delta \Sigma}{\delta \bar{\omega}^{a}_{i}} 
+ \partial_{\mu}\frac{\delta\Sigma}{\delta\bar{N}^{a}_{\mu i}} 
- gf^{abc} M^{b}_{\mu i}\frac{\delta\Sigma}{\delta K^{c}_{\mu}} = 0\,,
\\
&&
\frac{\delta\Sigma}{\delta\varphi^{a}_{i}} 
+ \partial_{\mu}\frac{\delta\Sigma}{\delta M^{a}_{\mu i}} 
- gf^{abc}\left( \frac{\delta \Sigma}{\delta b^{c}}\bar{\varphi}^{b}_{i} 
+ \frac{\delta \Sigma}{\delta \bar{c}^{b}} \bar{\omega}^{c}_{i} 
- \bar{N}^{c}_{\mu i} \frac{\delta \Sigma}{\delta K^{b}_{\mu}} \right) = 0 \,,
\\
&&
\int d^{4}x\; \left[ c^{a}\frac{\delta }{\delta \omega^{a}_{i}} + \bar{\omega}^{a}_{i}\frac{\delta }{\delta\bar{c}^{a}} + \bar{N}^{a}_{\mu i}\frac{\delta }{\delta K^{a}_{\mu}} \right]\Sigma  = 0  \;,
\\
&&
\int d^{4}x\; \left[ c^{a}\frac{\delta }{\delta \theta^{aI}} + \tilde{\theta}^{aI}\frac{\delta }{\delta\bar{c}^{a}} - \tilde{U}^{aI}\frac{\delta }{\delta F^{a}} \right]\Sigma  = 0  \;,
\\
&&
\int d^{4}x\; \left[ \frac{\delta }{\delta \eta^{bI}} - gf^{abc}\tilde{U}^{aI}\frac{\delta}{\delta F^{c}} - gf^{abe}\left(\tilde{\eta}^{aI}\frac{\delta}{\delta b^{e}}-\tilde{\theta}^{aI}\frac{\delta}{\delta \bar{c}^{e}}\right) \right]\Sigma  = \int d^{4}x\; gf^{abc}V^{aI}\phi^{c}  \;,
\\
&&
\int d^{4}x\; \left[ \frac{\delta }{\delta \theta^{bI}} - gf^{abe}\tilde{\theta}^{aI}\frac{\delta}{\delta b^{e}} \right]\Sigma  = -\int d^{4}x\; gf^{abc}\tilde{U}^{aI}\phi^{c}  \;,
\\
&&
\int d^{4}x\; \left[ \frac{\delta}{\delta \tilde{\theta}^{aI}} - gf^{abc}V^{cI}\frac{\delta}{\delta F^{b}}\right] \Sigma= \int d^{4}x gf^{abc}U^{cI}\phi^{b}\;,
\\
&&
\int d^{4}x\; \frac{\delta \Sigma}{\delta \tilde{\eta}^{bI}} =-  \int d^{4}x\; gf^{abc}V^{aI}\phi^{c}  \;.
\end{eqnarray}
\noindent {\bf $\bullet$ The ghost equation}:
\begin{equation}
{\cal G}^{a}(\Sigma) = \Delta^{a}_{class}\;,
\end{equation}
where
\begin{eqnarray}
{\cal G}^{a} &=& \int d^{4}x \left[ \frac{\delta}{\delta c^{a}} + gf^{abc}\left( \bar{c}^{b}\frac{\delta}{\delta b^{c}} + \bar{\omega}^{b}_{i}\frac{\delta}{\delta \varphi^{c}_{i}} + \varphi^{b}_{i}\frac{\delta}{\delta\omega^{c}_{i}} + M^{b}_{\mu i}\frac{\delta}{\delta N^{c}_{\mu i}} + \bar{N}^{b}_{\mu i}\frac{\delta}{\delta \bar{M}^{c}_{\mu i}} +\tilde{\theta}^{bI}\frac{\delta}{\delta \tilde{\eta}^{cI}}
\right.
\right.
\nonumber \\
&&
\phantom{\int d^{4}x\,}  
\left.
\left.
\eta^{bI}\frac{\delta}{\delta \theta^{cI}} + \tilde{U}^{bI}\frac{\delta}{\delta \tilde{V}^{cI}} + V^{bI}\frac{\delta}{\delta U^{cI}}
\right)\right]
\end{eqnarray}
and
\begin{equation}
\Delta^{a}_{class} = \int d^{4}x gf^{abc}\left( K^{b}_{\mu}A^{c}_{\mu} - L^{b}c^{c} + F^{b}\phi^{c} \right)\;.
\end{equation}

\noindent {\bf $\bullet$ The global symmetry $U(f=4(N^{2}-1))$}:
\begin{eqnarray}
 \mathcal{L}_{ij}(\Sigma) &=&  \int d^{4}x \left[ 
 \varphi^{c}_{i}\frac{\delta}{\delta\varphi^{c}_{j}}
- \bar{\varphi}^{c}_{i}\frac{\delta }{\delta \bar{\varphi}^{c}_{j}}
+ \omega^{c}_{i}\frac{\delta}{\delta\omega^{c}_{j}}    
- \bar{\omega}^{c}_{i}\frac{\delta }{\delta \bar{\omega}^{c}_{j}}
+ M^{c}_{\mu i}\frac{\delta}{\delta{M}^{c}_{\mu j}}  
- \bar{M}^{a}_{\mu i}\frac{\delta }{\delta \bar{M}^{a}_{\mu j}}
 \right. 
\nonumber \\
&&
\left. 
\phantom{\int d^{4}x\,}  
+ N^{a}_{\mu i}\frac{\delta}{\delta{N}^{a}_{\mu j}}  
- \bar{N}^{a}_{\mu i}\frac{\delta }{\delta \bar{N}^{a}_{\mu j}}
\right]\Sigma = 0\;.
\end{eqnarray}

\noindent {\bf $\bullet$  The global symmetry $U(f'=(N^{2}-1))$}:
\begin{eqnarray}
{\cal L}^{IJ}(\Sigma) &=& \int d^{4}x\, \left[
\theta^{bI} \frac{\delta}{\delta \theta^{bJ}} 
- \tilde{\theta}^{bI} \frac{\delta}{\delta \tilde{\theta}^{bJ}}
+ \eta^{bI} \frac{\delta}{\delta \eta^{bJ}}
- \tilde{\eta}^{bI} \frac{\delta}{\delta \tilde{\eta}^{bJ}}
+ V^{aI} \frac{\delta}{\delta V^{aJ}}
- \tilde{V}^{aI} \frac{\delta}{\delta \tilde{V}^{aJ}}
\right.
\nonumber \\
&&
\left.
\phantom{\int d^{4}x\,}  
+ U^{aI} \frac{\delta}{\delta U^{aJ}}
- \tilde{U}^{aI} \frac{\delta}{\delta \tilde{U}^{aJ}}
\right]\Sigma = 0\;.
\end{eqnarray}
Let us also dispslay below the quantum numbers of all fields and sources 

\noindent {\bf $\bullet$ Table of quantum numbers} (``B" is for bosonic fields and ``F" is for fermionic fields) :
\begin{center}
\begin{tabular}{l|c|c|c|c|c|c|c|c|c|c|c|c|c|}
&$A$&$\phi$&$c$&$\bar{c}$&$b$&$\varphi$&$\bar\varphi$&$\omega$&$\bar\omega$&$\eta$&$\tilde{\eta}$&$\theta$&$\tilde{\theta}$\cr
\hline\hline
$\phantom{\Bigl|}\!\!$Dim
&1&$1$&0&2&2&1&1&1&1&1&1&1&1\cr
\hline
$\phantom{\Bigl|}\!\!$Ghost\#
&0&0&1&$-1$&$0$&0&0&1&$-1$&0&0&1&$-1$\cr
\hline
$\phantom{\Bigl|}\!\!$Charge-$q_f$
&0&0&0&0&0&1&$-1$&1&$-1$&0&0&0&0\cr
\hline
$\phantom{\Bigl|}\!\!$Charge-$q_{f'}$
&0&0&0&0&0&0&0&$0$&0&$1$&$-1$&1&$-1$\cr
\hline
$\phantom{\Bigl|}\!\!$Nature
&B&B&F&F&B&B&B&F&F&B&B&F&F
\end{tabular}
\end{center}

\begin{center}
\begin{tabular}{l|c|c|c|c|c|c|c|c|c|c|c|}
&$M$&$\bar{M}$&$N$&$\bar{N}$&$U$&$\tilde{U}$&$V$&$\tilde{V}$&$K$&$L$&$F$\cr
\hline\hline
$\phantom{\Bigl|}\!\!$Dim
&2&2&2&2&2&2&2&2&3&4&3\cr
\hline
$\phantom{\Bigl|}\!\!$Ghost\#
&0&0&1&$-1$&$1$&$-1$&0&0&$-1$&$-2$&$-1$\cr
\hline
$\phantom{\Bigl|}\!\!$Charge-$q_f$
&1&$-1$&1&$-1$&0&0&0&0&0&0&0\cr
\hline
$\phantom{\Bigl|}\!\!$Charge-$q_{f'}$
&0&0&0&0&1&$-1$&1&$-1$&0&$0$&0\cr
\hline
$\phantom{\Bigl|}\!\!$Nature
&B&B&F&F&F&F&B&B&F&B&F
\end{tabular}
\end{center}

\section{Algebraic characterisation of the invariant counter-term and renormalizability}
\sectionmark{Algebraic characterisation of the counter-term}

In order to determine the most general invariant counter-term which can be freely added to each order of perturbation theory, we follow the Algebraic Renormalization framework  \cite{Piguet:1995er} and perturb  the complete action $\Sigma$ by adding an integrated local polynomial in the fields and sources with dimension bounded by four and vanishing ghost number, $\Sigma_{ct}$, and we require that the perturbed action, $(\Sigma + \varepsilon \Sigma_{ct})$, where $\varepsilon$ is an infinitesimal expansion parameter, obeys the same Ward identities fulfilled by $\Sigma$ to the first order in the parameter $\varepsilon$. Therefore, in the  case of the Slavnov-Taylor identity  \eqref{st}, we have
\begin{equation}
S\left( \Sigma + \varepsilon\Sigma_{ct} \right) = 0 + {\cal O}(\varepsilon^{2})\;,   \label{pertb1}
\end{equation}
which leads to
\begin{equation}
{\cal B}_{\Sigma}\left( \Sigma_{ct} \right) = 0\;, \label{perturblin}
\end{equation}
implying that $\Sigma_{ct}$ belongs to the cohomology of the linearized Slavnov-Taylor operator
in the sector of the local integrated polynomials of dimension bounded by four. From the
general results on the cohomology of Yang-Mills theories, see \cite{Piguet:1995er},  the
counter-term $\Sigma_{ct}$ can be parametrized as follows 
\begin{equation}
	\Sigma_{ct} = a_{0}S_{\text{YM}}+a_{1}\frac{\lambda}{4!}(\phi^{a}\phi^{a})^{2}+a_{2}\frac{m_{\phi}^{2}}{2}\phi^{a}\phi^{a} + {\cal B}_{\Sigma}(\Delta^{-1})\;, \label{ctp}
\end{equation}
where $a_{0}, a_{1}, a_{2}$ are free arbitrary coefficients and  $\Delta^{-1}$ is an integrated polynomial in the fields and sources with dimension bounded by $4$ and with ghost number $-1$. The most general expression  for $\Delta^{-1}$  is given by 
\begin{eqnarray}
\Delta^{-1} &=& \int d^{4}x\; \left\{ a_{3}(\partial_{\mu}\bar{c}^{a} + K^{a}_{\mu})A^{a}_{\mu}
+ a_{4}L^{a}c^{a}
+ a_{5}\phi^{a}F^{a}
+ a_{6}\partial_{\mu}\varphi^{a}_{i}\partial_{\mu}\bar{\omega}^{a}_{i}
+ a_{7}\partial_{\mu}\eta^{aI}\partial_{\mu}\tilde{\theta}^{aI}
\right.
\nonumber \\
&&
\phantom{\int d^{4}x\;}
+ a_{8}\partial_{\mu}\bar{\omega}^{a}_{i}M^{a}_{\mu i}
+ a_{9}\bar{N}^{a}_{\mu i}\partial_{\mu}\varphi^{a}_{i}
+ a_{10}M^{a}_{\mu i}\bar{N}^{a}_{\mu i}
+ a_{11}V^{aI}\tilde{U}^{aI}
+ a_{12}m_{\phi}^{2}\varphi^{a}_{i}\bar{\omega}^{a}_{i}
\nonumber \\
&&
\phantom{\int d^{4}x\;}
+ a_{13}m_{\phi}^{2}\eta^{aI}\tilde{\theta}^{aI}
+ a_{14}gf^{abc}V^{aI}\phi^{b}\tilde{\theta}^{cI}
+ a_{15}gf^{abc}\tilde{U}^{aI}\phi^{b}\eta^{cI}
\nonumber \\
&&
\phantom{\int d^{4}x\;}
+ a_{16}gf^{abc}\partial_{\mu}A^{a}_{\mu}\varphi^{b}_{i}\bar{\omega}^{c}_{i}
+ a_{17}gf^{abc}A^{a}_{\mu}\partial_{\mu}\varphi^{b}_{i}\bar{\omega}^{c}_{i}
+ a_{18}gf^{abc}A^{a}_{\mu}\varphi^{b}_{i}\partial_{\mu}\bar{\omega}^{c}_{i}
\nonumber \\
&&
\phantom{\int d^{4}x\;}
+ a_{19}gf^{abc}A^{a}_{\mu}M^{b}_{\mu i}\bar{\omega}^{c}_{i}
+ a_{20}gf^{abc}A^{a}_{\mu}\bar{N}^{b}_{\mu i}\varphi^{c}_{i}
+ a_{21}gf^{abc}\partial_{\mu}A^{a}_{\mu}\eta^{bI}\tilde{\theta}^{cI}
\nonumber \\
&&
\phantom{\int d^{4}x\;}
+ a_{22}gf^{abc}A^{a}_{\mu}\partial_{\mu}\eta^{bI}\tilde{\theta}^{cI}
+ a_{23}gf^{abc}A^{a}_{\mu}\eta^{bI}\partial_{\mu}\tilde{\theta}^{cI}
\nonumber \\
&&
\phantom{\int d^{4}x\;}
+\mathbb{C}^{abcd}_{1}\phi^{a}\phi^{b}\varphi^{c}_{i}\bar{\omega}^{d}_{i}
+ \mathbb{C}^{abcd}_2\phi^{a}\phi^{b}\eta^{cI}\tilde{\theta}^{dI}
+ \mathbb{C}^{abcdIJKL}_3\eta^{aI}\tilde{\theta}^{bJ}\theta^{cK}\tilde{\theta}^{dL}
\nonumber \\
&&
\phantom{\int d^{4}x\;}
+ \mathbb{C}^{abcdIJKL}_4\eta^{aI}\tilde{\theta}^{bJ}\eta^{cK}\tilde{\eta}^{dL}
+ \mathbb{C}^{abcd}_5\varphi^{a}_{i}\bar{\varphi}^{b}_{i}\eta^{cI}\tilde{\theta}^{dI}
+ \mathbb{C}^{abcd}_6\omega^{a}_{i}\bar{\omega}^{b}_{i}\eta^{cI}\tilde{\theta}^{dI}
\nonumber \\
&&
\phantom{\int d^{4}x\;}
+ \mathbb{C}^{abcd}_7\varphi^{a}_{i}\bar{\omega}^{b}_{i}\theta^{cI}\tilde{\theta}^{dI}
+\mathbb{C}^{abcd}_8\varphi^{a}_{i}\bar{\omega}^{b}_{i}\eta^{cI}\tilde{\eta}^{dI}
+ \mathbb{C}^{abcdijkl}_9\varphi^{a}_{i}\bar{\omega}^{b}_{j}\varphi^{c}_{k}\bar{\varphi}^{d}_{l}
\nonumber \\
&&
\phantom{\int d^{4}x\;}
+ \left. \mathbb{C}^{abcdijkl}_{10}\varphi^{a}_{i}\bar{\omega}^{b}_{j}\omega^{c}_{k}\bar{\omega}^{d}_{l}
\right\}\;,
\end{eqnarray}
where $\left( \mathbb{C}^{abcd}_{1}, \mathbb{C}^{abcd}_2, \mathbb{C}^{abcdIJKL}_3, \mathbb{C}^{abcdIJKL}_4, \mathbb{C}^{abcd}_5, \mathbb{C}^{abcd}_6, \mathbb{C}^{abcd}_7, \mathbb{C}^{abcd}_8,  \mathbb{C}^{abcdijkl}_9, \mathbb{C}^{abcdijkl}_{10} \right)$ are arbitrary coefficients. After imposition of all other Ward identities it turns out that  the non-vanishing parameters which remain at the end of a lengthy algebraic analysis are:  
\begin{equation}
a_{3} = a_{6} = a_{7} = a_{8} = a_{9} = a_{10} = a_{17} = -a_{18} = a_{19} = a_{22} \neq 0   \;, \label{p1}
\end{equation}
as well as 
\begin{equation}
-a_{5} = a_{16} = a_{17}\neq 0 \;,  \qquad  \qquad a_{11}\neq 0 \;.   \label{p2}   
\end{equation}
Therefore, for the final expression of the invariant  counter-term one finds
\begin{eqnarray}
\label{fullct}
\Sigma_{ct} &=& \int d^{4}x \left\{ a_{0} F^{a}_{\mu\nu}F^{a}_{\mu\nu}
+ a_{1}\frac{\lambda}{4!}(\phi^{a}\phi^{a})^{2}
+ a_{2}\frac{m_{\phi}^{2}}{2}\phi^{a}\phi^{a}
+ a_{3}\left[ \frac{\delta S_{YM}}{\delta A^{a}_{\mu}}A^{a}_{\mu} 
+ \partial_{\mu}\bar{c}^{a}\partial_{\mu}c^{a}
\right.
\right.
\nonumber \\
&&
\phantom{\int d^{4}x}
+ K^{a}_{\mu}\partial_{\mu}c^{a}
-\bar{\varphi}^{a}_{i}\partial^{2}\varphi^{a}_{i}
+ \bar{\omega}^{a}_{i}\partial^{2}\omega^{a}_{i}
-\tilde{\eta}^{aI}\partial^{2}\eta^{aI}
+ \tilde{\theta}^{aI}\partial^{2}\theta^{aI}
- \bar{\varphi}^{a}_{i}\partial_{\mu}M^{a}_{\mu i}
\nonumber \\
&&
\phantom{\int d^{4}x}
+ N^{a}_{\mu i}\partial_{\mu}\bar{\omega}^{a}_{i}
+ \bar{M}^{a}_{\mu i}\partial_{\mu}\varphi^{a}_{i}
- \omega^{a}_{i}\partial_{\mu}\bar{N}^{a}_{\mu i}
- \bar{N}^{a}_{\mu i}N^{a}_{\mu i}
+ \bar{M}^{a}_{\mu i}M^{a}_{\mu i}
\nonumber \\
&&
\phantom{\int d^{4}x}
\left.
+ gf^{abc}\left(
- \partial_{\mu}c^{a}\varphi^{b}_{i}\partial_{\mu}\bar{\omega}^{c}_{i}
- \partial_{\mu}c^{a}\bar{N}^{b}_{\mu i}\varphi^{c}_{i}
+ \partial_{\mu}c^{a}M^{a}_{\mu i}\bar{\omega}^{c}_{i}
- \partial_{\mu}c^{a}\eta^{bI}\partial_{\mu}\tilde{\theta}^{cI}
\right)
\right]
\nonumber \\
&&
\phantom{\int d^{4}x}
+ a_{5}\left[gf^{abc}F^{a}\phi^{b}c^{c}
+ D^{ab}_{\mu}\phi^{b}D^{ac}_{\mu}\phi^{c}
+ m_{\phi}^{2}\phi^{a}\phi^{a}
+ \frac{\lambda}{3!}(\phi^{a}\phi^{a})^{2}
\right]
\nonumber \\
&&
\phantom{\int d^{4}x}
\left.
+ a_{11}\left(\tilde{V}^{aI}V^{aI} - \tilde{U}^{aI}U^{aI}\right)
\right\}\;.
\end{eqnarray}
It remains now to check that the counter-term $\Sigma_{ct}$ can be reabsorbed into the initial action $\Sigma$, through a redefinition of the fields, sources and parameters, according to 
\begin{equation}
\label{renorm}
\Sigma(F,S,\xi)+\varepsilon \Sigma_{ct}(F,S,\xi) = \Sigma(F_{0},S_{0},\xi_{0}) + {\cal O}(\varepsilon^{2})\;,
\end{equation}
with
\begin{equation}
F_{0} = Z^{1/2}_{F}F\;, \qquad S_{0} = Z_{S}S \qquad \text{and} \qquad \xi_{0} = Z_{\xi}\xi\;, 
\end{equation}
where $\{F\}$ stands for all fields, $\{S\}$ for all sources and $\{xi\}$ for all parameters, {\it i.e.} $\xi=g,m_{\phi}, \lambda, \rho$. \\\\Therefore, by direct application of \eqref{renorm} we get
\begin{eqnarray}
&&
Z^{1/2}_{A} = 1+ \varepsilon\left(\frac{a_{0}}{2} + a_{3}\right) \\
&&
Z^{1/2}_{\phi}= 1 +\varepsilon a_{5}\\
&&
Z^{1/2}_{b} = Z^{-1/2}_{A} \\
&&
Z^{1/2}_{\bar{c}} =Z^{1/2}_{c} =Z^{-1/2}_{g} Z^{-1/4}_{A} \\
&&
Z^{1/2}_{\bar{\varphi}} =Z^{1/2}_{\varphi} =Z^{-1/2}_{g} Z^{-1/4}_{A}\\
&&
Z^{1/2}_{\bar{\omega}} =Z^{-1}_{g} \\
&&
Z^{1/2}_{\omega} =Z^{-1/2}_{A}\\
&&
Z^{1/2}_{\theta} = Z^{-1/2}_{A}\\
&&
Z^{1/2}_{\bar{\theta}} = Z^{-1}_{g}\\
&&
Z^{1/2}_{\eta} = Z^{1/2}_{\bar{\eta}} = Z^{-1/2}_{g}Z^{-1/4}_{A}\\
&&
Z_{N} =Z^{-1/2}_{A} \\
&&
Z^{1/2}_{\bar{N}} =Z^{-1}_{g} \\
&&
Z_{M} =Z_{\bar{M}} =Z^{-1/2}_{g}Z^{-1/4}_{A} \\
&&
Z_{V} = Z_{\bar{V}} = Z^{-1/2}_{\phi} Z^{1/2}_{g}Z^{1/4}_{A}\\
&&
Z_{U} = Z^{-1/2}_{\phi} \\
&&
Z_{\bar{U}} = Z^{-1}_{g}Z^{1/2}_{A}Z^{-1/2}_{\phi} \\
&&
Z_{K} =Z^{1/2}_{\bar{c}}\\
&&
Z_{F} = Z^{-1}_{\phi}Z^{1/4}_{A}Z^{-1/2}_{g}\;.
\end{eqnarray}
and
\begin{eqnarray}
&&
Z_{g} = 1-\varepsilon\frac{a_{0}}{2}\\
&&
Z_{m_{\phi}} = 1+\varepsilon a_{2}\\
&&
Z_{\lambda} = 1+\varepsilon a_{1}\\
&&
Z_{\rho} = (1+\varepsilon a_{11})Z^{-1}_{g}Z^{1/2}_{A}Z^{-1}_{\phi}\;.
\end{eqnarray}
These equations show that the invariant  counter-term  $\Sigma_{ct}$, eq.\eqref{fullct}, can be reabsorbed into the initial action $\Sigma$ through a multiplecative redefinition of the fields, sources and parameters. This concludes the algebraic proof of the all order renormalizability of $\Sigma$.

\chapter{Notations, conventions and identities of SUSY theories in Euclidean space-time}
\chaptermark{SUSY notations, conventions and identities}
\label{notations}


\noindent
\textbf{Units}: $\hbar=c=1$.

\noindent
\textbf{Euclidean metric}: $\delta_{\mu\nu}=diag(+,+,+,+)$.

\noindent
\textbf{Wick rotations:} $X_0\rightarrow -iX_4\Rightarrow\partial_0\rightarrow+i\partial_4$, $A_0\rightarrow+iA_4$

\noindent
\textbf{Gamma matrices:}
\[\gamma_4=\left( \begin{array}{cc}
0 & \mathbb{1} \\
\mathbb{1} & 0 \end{array} \right),~~
\gamma_k=-i\left( \begin{array}{cc} 0 & \sigma_k \\ -\sigma_k & 0 \end{array} \right)\]

\noindent
\textbf{Pauli matrices}:
\[\sigma_4=\left( \begin{array}{cc}
1 & 0 \\
0 & 1 \end{array} \right),~~
\sigma_1=\left( \begin{array}{cc} 0 & 1\\ 1 & 0 \end{array} \right),~~
\sigma_2=\left( \begin{array}{cc} 0 & -i\\ i & 0 \end{array} \right),~~
\sigma_3=\left( \begin{array}{cc} 1 & 0 \\ 0 & -1 \end{array} \right)\]

\noindent
The Gamma matrices obey the following properties:
\begin{eqnarray}
 \gamma_\mu=\gamma_\mu^\dagger\\
  \{\gamma_\mu,\gamma_\nu\}&=&2\delta_{\mu\nu}
\end{eqnarray}

\noindent
We also define the $\gamma_5$ matrix as:
\[\gamma_5=\gamma_4\gamma_1\gamma_2\gamma_3=\left(\begin{array}{cc}
                                                   \mathbb{1} & 0\\
                                                   0 & -\mathbb{1}\\
                                                  \end{array}\right)\]

\noindent
with the following properties:
\begin{equation}
 \{\gamma_5,\gamma_\mu\}=0,~~(\gamma_5)^2=\mathbb{1},~~\gamma_5^\dagger=\gamma_5
\end{equation}

\noindent
The charge conjugation matrix is:
\begin{equation}
\label{Cmtrx}
\mathcal{C}=\gamma_4\gamma_2=i\left(\begin{array}{cc} \sigma_2 & 0 \\ 0 & -\sigma_2\\ \end{array}\right)
\end{equation}

\noindent
with the following properties:
\begin{equation}
 \mathcal{C}^{-1}=-\mathcal{C}=\mathcal{C}^T,~~\mathcal{C}^{-1}\gamma_\mu\mathcal{C}=-
\gamma_\mu^T
\end{equation}

\noindent
The $\sigma^{\mu\nu}$ tensor is defined as
\begin{equation}
  (\sigma_{\mu\nu})^{~\beta}_\alpha\equiv\frac{1}{2}[\gamma_\mu,\gamma_\nu]^{~\beta}_\alpha
\end{equation}
and has the property $\sigma_{\mu\nu}^\dagger=-\sigma_{\mu\nu}$.

\noindent
\textbf{Majorana fermions:}

\noindent
The Majorana condition reads:
\begin{equation}
\label{mjconj1}
 \lambda^\mathcal{C}=\lambda=\mathcal{C}\bar{\lambda}^T~~
\Longleftrightarrow~~\bar{\lambda}=\lambda^T\mathcal{C}\;,
\end{equation}
leading to the following relations
\begin{equation}
\bar{\lambda}\gamma_{\mu}\epsilon = \bar{\epsilon}\gamma_{\mu}\lambda \qquad \text{and} \qquad \bar{\lambda}\gamma_{\mu}\gamma_{5}\epsilon = - \bar{\epsilon}\gamma_{\mu}\gamma_{5}\lambda\;.
\end{equation}

\noindent
\textbf{Fierz identity (in Euclidean space-time):}
\begin{eqnarray}
\epsilon_{1}\bar{\epsilon}_{2} &=&  \frac{1}{4}(\bar{\epsilon}_{2}\epsilon_{1})\mathbb{1} 
+ \frac{1}{4}(\bar{\epsilon}_{2}\gamma_{5}\epsilon_{1})\gamma_{5}
+ \frac{1}{4}(\bar{\epsilon}_{2}\gamma_{\mu}\epsilon_{1})\gamma_{\mu}
- \frac{1}{4}(\bar{\epsilon}_{2}\gamma_{\mu}\gamma_{5}\epsilon_{1})\gamma_{\mu}\gamma_{5}\nonumber \\
&&
- \frac{1}{8}(\bar{\epsilon}_{2}\sigma_{\mu\nu}\epsilon_{1})\sigma_{\mu\nu}    \;.
\end{eqnarray}


\noindent {\bf Indices notations}:
\begin{center}
\begin{tabular}{ll}
$\bullet$&The Lorentz indices: $\mu,\nu,\rho,\sigma,\lambda\in\{1,2,3,4\}$\,;$\phantom{\Bigl|}$\\
$\bullet$&The Spinor indices: $\alpha,\beta,\gamma,\delta,\eta\in\{1,2,3,4\}$\,;$\phantom{\Bigl|}$\\
$\bullet$&The $SU(N)$ group indices: $a,b,c,d,e\in\{1,\dots,N^{2}-1\}$\,;$\phantom{\Bigl|}$\\
$\bullet$&The multi-index $(a,\mu)$: $i,j,k,l\in\{1,\dots,f=4(N^{2}-1)\}$\,;$\phantom{\Bigl|}$\\
$\bullet$&The multi-index $(a,\alpha)$: $I,J,K,L\in\{1,\dots,f'=4(N^{2}-1)\}$\,.$\phantom{\Bigl|}$\\
\end{tabular}
\end{center}

\noindent {\bf Table of quantum numbers} (``C" is for commutating and ``A" is for anti-commutating) :
\begin{center}
\begin{tabular}{l|c|c|c|c|c|c|c|c|c|c|c|c|c|c|c|c}
&$A$&$\lambda$&$\mathfrak{D}$&$c$&$\check{c}$&$b$&$\varphi$&$\tilde\varphi$&$\omega$&$\tilde\omega$&$\zeta$&$\hat{\zeta}$&$\theta$&$\hat{\theta}$
&$\epsilon$&$\bar{\epsilon}$\cr
\hline\hline
$\phantom{\Bigl|}\!\!$Dim
&1&$\frac{3}{2}$&2&1&1&2&1&1&2&0&0&2&1&1&$\frac{1}{2}$&$\frac{1}{2}$\cr
\hline
$\phantom{\Bigl|}\!\!$Ghost\#
&0&0&0&1&$-1$&0&0&0&1&$-1$&$-1$&1&0&0&1&1\cr
\hline
$\phantom{\Bigl|}\!\!$Charge-$q_f$
&0&0&0&0&0&0&1&$-1$&1&$-1$&0&0&0&0&0&0\cr
\hline
$\phantom{\Bigl|}\!\!$Charge-$q_{f'}$
&0&0&0&0&0&0&0&$0$&0&$0$&1&$-1$&1&$-1$&0&0\cr
\hline
$\phantom{\Bigl|}\!\!$Nature
&C&A&C&A&A&C&C&C&A&A&C&C&A&A&C&C
\end{tabular}
\end{center}

\begin{center}
\begin{tabular}{l|c|c|c|c|c|c|c|c|c|c|c|c|c|c|c|c}
&$U$&$\hat{U}$&$V$&$\hat{V}$&$M$&$\tilde{M}$&$N$&$\tilde{N}$&$K$&$\Omega$&$L$&$\Lambda$&$T$&$J$&$Y$&$X$\cr
\hline\hline
$\phantom{\Bigl|}\!\!$Dim
&$\frac{1}{2}$&$\frac{1}{2}$&$\frac{3}{2}$&$\frac{3}{2}$&2&2&3&1&2&3&2&3&1&2&$\frac{3}{2}$&$\frac{5}{2}$\cr
\hline
$\phantom{\Bigl|}\!\!$Ghost\#
&$-1$&$-1$&0&0&$0$&0&1&$-1$&$-1$&$0$&$-2$&$-1$&$-1$&0&$-1$&0\cr
\hline
$\phantom{\Bigl|}\!\!$Charge-$q_f$
&0&0&0&0&1&$-1$&1&$-1$&0&$0$&0&0&0&0&0&0\cr
\hline
$\phantom{\Bigl|}\!\!$Charge-$q_{f'}$
&1&$-1$&1&$-1$&0&0&0&$0$&0&$0$&0&$0$&0&$0$&0&0\cr
\hline
$\phantom{\Bigl|}\!\!$Nature
&A&A&C&C&C&C&A&A&A&C&C&A&A&C&C&A
\end{tabular}
\end{center}

\chapter{Algebraic renormalization of $\mathcal N=1$ Super Yang--Mills in Wess--Zumino gauge within the Gribov--Zwanziger approach}
\chaptermark{Algebraic renormalization: $\mathcal N=1$ SYM in the GZ approach}
\label{algrenorm}

In order to apply the algebraic renormalization procedure to study the renormalizability of the Gribov-extended SYM theory, we must first write down a local action and the associated set of symmetries and Ward identities. We shall proceed by first analyzing the theory with a Gribov gauge sector and then including the confining gluino term to obtain the final action to be studied.

\noindent The $\mathcal N=1$ Euclidean super Yang--Mills action with Majorana fermions in the superfield components and within the Wess--Zummino gauge, without a matter field, is
\begin{equation}
\label{SYM1}
S_\text{SYM} = \int d^{4}x \left[ \frac{1}{4}F^{a}_{\mu \nu}F^{a}_{\mu\nu} 
+ \frac{1}{2} \bar{\lambda}^{a\alpha} (\gamma_{\mu})_{\alpha\beta} D^{ab}_{\mu}\lambda^{b\beta}
+ \frac{1}{2}\mathfrak{D}^a\mathfrak{D}^a\right]\;,
\end{equation}
where $\lambda_{\alpha}$ is a four component Majorana spinor.

\noindent This action is left invariant under the the usual $SUSY$ transformations $\epsilon^{\alpha}\delta_{\alpha} = \epsilon^{\alpha}\delta_{\alpha} + \bar{\epsilon}^{\dot{\alpha}}\bar{\delta}_{\dot{\alpha}}$, where the right-hand-side is in terms of the Weyl spinors, however, to avoid the infinite chain of new generators we also include the $BRST$ transformation, $s$, so that the full transformation $Q=s+\epsilon^{\alpha}\delta_{\alpha}$ applied to the superfield components gives
\begin{eqnarray}
\label{susytransf1}
&&
QA^{a}_{\mu} = - D^{ab}_{\mu}c^{b} 
+\bar{\epsilon}^\alpha(\gamma_\mu)_{\alpha\beta}\lambda^{a\beta}\;,\nonumber \\
&&
Q\lambda^{a\alpha} = gf^{abc}c^{b}\lambda^{c\alpha}
- \frac{1}{2}(\sigma_{\mu\nu})^{\alpha\beta}\epsilon_{\beta} F_{\mu\nu}^{a}
+ (\gamma_{5})^{\alpha\beta}\epsilon_{\beta} \mathfrak{D}^a\;, \nonumber \\
&&
Q\mathfrak{D}^a = gf^{abc}c^{b}\mathfrak{D}^c 
+ \bar{\epsilon}^{\alpha}(\gamma_{\mu})_{\alpha\beta}(\gamma_{5})^{\beta\eta}D_{\mu}^{ab}\lambda^{b}_{\eta} \;, \\
&&
Qc^{a} = \frac{1}{2}gf^{abc}c^{b}c^{c} 
- \bar{\epsilon}^{\alpha}(\gamma_{\mu})_{\alpha\beta}\epsilon^{\beta} A^{a}_{\mu}\;, \nonumber \\
&&
Q\bar{c}^{a} = b^{a}\;, \nonumber \\
&&
Qb^{a} = \nabla\bar{c}^{a}\;, \nonumber\\
&&
Q^{2}=\nabla\;,
\end{eqnarray}

where we define the translation operator 
\begin{equation}
\label{top1}
\nabla := \bar{\epsilon}^{\alpha}(\gamma_{\mu})_{\alpha\beta}\epsilon^{\beta} \partial_{\mu}\,.
\end{equation}
With the above transformations we can show that the starting action \eqref{SYM1} is $Q$ invariant or have a {\it super-BRST} symmetry ({\it S-BRST}).

\noindent In order to fix the gauge freedom of the Yang--Mills field, let us chose the  Landau gauge, with which we ensure the existence of the Gribov copies,
\begin{equation}
S_\text{gf} = Q\int d^{4}x (\check{c}^{a}\partial_{\mu}A^{a}_{\mu})\;,
\end{equation}
so that, according to \eqref{susytransf1}, we have
\begin{equation}
S_\text{gf} = \int d^{4}x \left[ \check{c}^{a}\partial_{\mu}D^{ab}_{\mu}c^{b} + b^{a}\partial_{\mu}A^{a}_{\mu}   - \check{c}^{a}\bar{\epsilon}^{\alpha}(\gamma_{\mu})_{\alpha\beta}\partial_{\mu}\lambda^{a\,\beta} \right]\;.
\end{equation}
Therefore, the Yang--Mills action in the Landau gauge can be written as
\begin{eqnarray}
\label{act2}
S' &=& S_{SYM} + S_{gf} \nonumber \\
&=&
\int d^{4}x \left\{\frac{1}{4}F^{a}_{\mu \nu}F^{a}_{\mu\nu} 
+ \frac{1}{2}\bar{\lambda}^{a\alpha}(\gamma_{\mu})_{\alpha \beta}D^{ab}_{\mu} \lambda^{b\beta}
+ \frac{1}{2}\mathfrak{D}^{2} \right. \nonumber \\
&&
\left.
+ b^{a}\partial_{\mu}A^{a}_{\mu}
+\check{c}^{a}\left[\partial_{\mu}D^{ab}_{\mu}c^{b} - \bar{\epsilon}^{\alpha}(\gamma_{\mu})_{\alpha \beta}\partial_{\mu}\lambda^{a\beta} \right]\right\}.
\label{gfaction1}
\end{eqnarray}

\noindent One can easily check the invariance of the $S'$ action under the set of transformations \eqref{susytransf1}, as well as the nilpotency of $Q$ under spacetime integration. In other words, making use of some relations given at the Appendix, one can show that 
\begin{equation}
QS'=0
\end{equation}
 and that
\begin{equation}
 Q^{2} \equiv\nabla=\bar{\epsilon}^{\alpha}(\gamma_{\mu})_{\alpha\beta}\epsilon^{\beta} \partial_{\mu}\;
\end{equation}
for all fields.

\noindent The Gribov problem, inherent to the Yang--Mills fields, will be considered here in a kind of {\it extended} localized Gribov--Zwanziger (GZ) action, namely
\begin{equation}
S_\text{GZ'} = Q\int d^{4}x \left[\tilde{\omega}^{ac}_{\mu}\partial_{\nu}D_{\nu}^{ab}\varphi^{bc}_{\mu} \right] + \int d^{4}x \left[\gamma^{2}gf^{abc}A^{b}_{\mu}(\varphi^{ac}_{\mu}+\tilde{\varphi}^{ac}_{\mu})  - \gamma^{4}4(N_{c}^{2}-1)\right]\;,
\end{equation}
\noindent
where, making use of the set of transformations \eqref{susytransf1}, one can easily see that
\begin{eqnarray}
\label{gzac}
S_\text{GZ'} &=& \int d^{4}x \left[ \tilde{\varphi}^{ac}_{\mu}\partial_{\nu}D_{\nu}^{ab}\varphi^{bc}_{\mu} - \tilde{\omega}^{ac}_{\mu}\partial_{\nu}D_{\nu}^{ab}\omega^{bc}_{\mu} - gf^{abc}(\partial_{\nu}\tilde{\omega}^{ad}_{\mu})(D^{bk}_{\nu}c^{k})\varphi^{cd}_{\mu} \right. \nonumber \\
&+& \left. gf^{abc}(\partial_{\nu}\tilde{\omega}^{ad}_{\mu})(\bar{\epsilon}^\alpha(\gamma_\nu)_{\alpha\beta}\lambda^{\beta b})\varphi^{cd}_{\mu}\right] + \gamma^{2} \int d^{4}x \left[ gf^{abc}A^{a}_{\mu}(\varphi^{bc}_{\mu} + \tilde{\varphi}^{bc}_{\mu}) - \gamma^{2}4(N_{c}^{2}-1)\right]\;.
\end{eqnarray}
The above $GZ'$ action is said to be {\it extended} as it is related to the $BRST$ algebra, as it must be, and also to the $SUSY$ algebra in addition.

\noindent In order to keep the supersymmetric algebra as well as the BRST one, the localizing fields $(\omega,\tilde{\omega})$ and $(\varphi,\tilde{\varphi})$ must transform under $Q$ as 
\begin{eqnarray}
\label{lcf}
Q\varphi^{ac}_{\mu} &=& \omega^{ac}_{\mu} \nonumber \\
Q\omega^{ac}_{\mu} &=& \nabla\varphi^{ac}_{\mu} \nonumber \\
Q\tilde{\omega}^{ac}_{\mu} &=& \tilde{\varphi}^{ac}_{\mu} \nonumber \\
Q\tilde{\varphi}^{ac}_{\mu} &=& \nabla\tilde{\omega}^{ac}_{\mu}\;, \nonumber \\
\end{eqnarray}

\noindent With this term added to $S'$ an explicit super-BRST breaking is observed, which comes from the term proportional to $\gamma$ in \eqref{gzac}, and can be easily checked making use of \eqref{lcf}. As we saw that $QS'=0$, we just need to verify the invariance of the Gribov--Zwanziger action, which is said to be
\begin{eqnarray}
 QS&=&Q^{2}\int d^{4}x \left[\tilde{\omega}^{ac}_{\mu}\partial_{\nu}D_{\nu}^{ab}\varphi^{bc}_{\mu} \right]+Q\int d^{4}x \left[\gamma^{2}gf^{abc}A^{b}_{\mu}(\varphi^{ac}_{\mu}+\tilde{\varphi}^{ac}_{\mu})\right]\nonumber\\
&=&\int d^4x \Big(\gamma^2gf^{abc}\left[(QA_\mu^b)(\varphi^{ac}_\mu+\tilde{\varphi}^{ac}_{\mu})+A_\mu^b(Q\varphi^{ac}_\mu+
Q\tilde{\varphi}^{ac}_{\mu})\right]\Big)\nonumber\\
&=& \gamma^2 \int d^4x gf^{abc}\Big[\Big(- D^{ab}_{\mu}c^{b} -\bar{\epsilon}^\alpha(\gamma_\mu)_{\alpha\beta}\lambda^{\beta a}\Big)(\varphi^{ac}_\mu+\tilde{\varphi}^{ac}_{\mu})+\nonumber\\
&+&A_\mu^b\Big(\omega^{ac}_{\mu} -2\epsilon^{\alpha}(\gamma_{\mu})_{\alpha\beta}\bar{\epsilon}^\beta\partial_{\mu}\tilde{\omega}^{ac}_{\mu}\Big)
\Big]\;.
\end{eqnarray}
Summarizing, we have
\begin{equation}
QS = \gamma^{2} \Delta\;,
\end{equation}
with $\Delta$ being a dimension two integrated polynomial of the fields, known as composite operator, characterizing then a soft breaking.
We may now proceed by including the confining term in the gluino sector,
\begin{equation}
S_{\tilde{G}}=- \frac{1}{2}\int d^{4}x \left( \bar{\lambda}^{a\alpha}\frac{M^{3}\delta_{\alpha\beta}}{(\partial^{2}-\mu^{2})}\lambda^{a\beta}\right)\;,
\end{equation}
leading to the following non-local action,
\begin{equation}
S = S_\text{SYM} + S_\text{gf} + S_\text{GZ'} - \frac{1}{2}\int d^{4}x \left( \bar{\lambda}^{a\alpha}\frac{M^{3}\delta_{\alpha\beta}}{(\partial^{2}-\mu^{2})}\lambda^{a\beta}\right)\;,
\label{glnlcal}
\end{equation}
with $M$ and $\mu$ having dimension of mass. That subscribed $\tilde{G}$ is not a index, it stands for the {\it Gluino} field.
One can write this term in a local form with the insertion of two more fields. Namely,
\begin{equation}
\label{actlocal}
S_{L\tilde{G}} = \int d^{4}x \left[ \hat{\zeta}^{a\alpha} (\partial^{2} - \mu^{2})\zeta^{a}_{~\alpha} - \hat{\theta}^{a\alpha}(\partial^{2} - \mu^{2})\theta^{a}_{~\alpha} - M^{3/2}(\bar{\lambda}^{a\alpha}\theta^{a}_{~\alpha} + \hat{\theta}^{a\alpha}\lambda^{a}_{~\alpha})\right]\;.
\end{equation}
The non-local version can be easily recovered after integration over the new Dirac spinor fields. The localizing fields $\hat{\zeta}^{a\alpha}$, $\zeta^{a\alpha}$ are bosonic while $\hat{\theta}^{a\alpha}$ and $\theta^{a\alpha}$ are fermionic and they form doublets under $Q$ transformation, $(\hat{\theta}^{a\alpha},\, \hat{\zeta}^{a\alpha})$ and $(\theta^{a\alpha},\,\zeta^{a\alpha})$.

\noindent Equivalently to the gluon sector, the $Q$-symmetry is also broken by the gluino term with the breaking coming from the term proportional to $M^{3/2}$. According to the $Q$ transformation of the localizing fields $(\hat{\theta}^{a\alpha},\, \hat{\zeta}^{a\alpha})$ and $(\theta^{a\alpha},\,\zeta^{a\alpha})$, namely
\begin{eqnarray}
\label{loctrans}
Q\hat{\theta}^{a}_{\alpha} &=& \hat{\zeta}^{a}_{\alpha} \;; \nonumber \\
Q\hat{\zeta}^{a}_{\alpha} &=& \nabla\hat{\theta}^{a}_{\alpha} \;; \nonumber \\
Q\zeta^{a}_{\alpha} &=& \theta^{a}_{\alpha} \;; \nonumber \\
Q\theta^{a}_{\alpha} &=& \nabla\zeta^{a}_{\alpha}\;,
\end{eqnarray}
it is not hard to see that the breaking is soft and thus can be restored by insertions of external fields. The whole action which describes our model can then be written in its local form as, 
\begin{eqnarray}
\label{fnlact}
S &=& S_{SYM} + S_{gf} + S_{GZ'} + S_{L\tilde{G}}\nonumber \\
&&
=\int d^{4}x\; \left\{\frac{1}{4}F^{a}_{\mu \nu}F^{a}_{\mu\nu} 
+ \frac{1}{2} \bar{\lambda}^{a\alpha} (\gamma_{\mu})_{\alpha\beta} D^{ab}_{\mu}\lambda^{b\beta}
+ \frac{1}{2}\mathfrak{D}^a\mathfrak{D}^a 
+ b^{a}\partial_{\mu}A^{a}_{\mu} \right. \nonumber\\
&&
+\check{c}^{a}\left[\partial_{\mu}D^{ab}_{\mu}c^{b}
-\bar{\epsilon}^{\alpha}(\gamma_{\mu})_{\alpha\beta}\partial_{\mu}\lambda^{a\,\beta}\right]
+ \tilde{\varphi}^{ac}_{\mu}\partial_{\nu}D_{\nu}^{ab}\varphi^{bc}_{\mu} 
-\tilde{\omega}^{ac}_{\mu}\partial_{\nu}D_{\nu}^{ab}\omega^{bc}_{\mu} \nonumber \\
&&
-gf^{abc}(\partial_{\nu}\tilde{\omega}^{ad}_{\mu})(D^{bk}_{\nu}c^{k})\varphi^{cd}_{\mu}
+gf^{abc}(\partial_{\nu}\tilde{\omega}^{ad}_{\mu})(\bar{\epsilon}^\alpha(\gamma_\nu)_{\alpha\beta}\lambda^{\beta b})\varphi^{cd}_{\mu} \nonumber \\
&&
+\gamma^{2}gf^{abc}A^{a}_{\mu}(\varphi^{bc}_{\mu} + \tilde{\varphi}^{bc}_{\mu}) 
-\gamma^{4}4(N_{c}^{2}-1) 
+\hat{\zeta}^{a\alpha} (\partial^{2} - \mu^{2})\zeta^{a}_{~\alpha} \nonumber \\
&&
\left.
-\hat{\theta}^{a\alpha}(\partial^{2} - \mu^{2})\theta^{a}_{~\alpha} 
-M^{3/2}(\bar{\lambda}^{a\alpha}\theta^{a}_{~\alpha} 
+\hat{\theta}^{a\alpha}\lambda^{a}_{~\alpha})
\right\}\;.
\end{eqnarray}
It is straightforward to see that the action \eqref{actlocal} is not $Q$ invariant, with the $Q$ transformation of the localizing fields given by \eqref{loctrans}.
More preciselly there is a softy symmetry breaking which comes from the GZ term and from the non-local proposed term in the gluino sector. Hence, in order to prove the renormalizability with the algebraic renormalization approach we have to include external sources to restore the $Q$ invariance of the local action ensuring thus the Slavnov-Taylor identity. These source will be included as doublets, namely
\begin{equation}
(\hat{V}^{ab\,\alpha\beta},\,V^{ab\,\alpha\beta}); ~~~ (\hat{U}^{ab\,\alpha\beta},\,U^{ab\,\alpha\beta}); ~~~ (\tilde{M}^{ab}_{\mu\nu},\,M^{ab}_{\mu\nu}); ~~~ (\tilde{N}^{ab}_{\mu\nu},\,N^{ab}_{\mu\nu})\;,
\end{equation}
whose $Q$ transformation of each field is given by
\begin{eqnarray}
QU^{ab\,\alpha\beta} &=& V^{ab\,\alpha\beta}\;; \nonumber \\
QV^{ab\,\alpha\beta} &=& \nabla U^{ab\,\alpha\beta}\;; \nonumber \\
Q\hat{U}^{ab\,\alpha\beta} &=& \hat{V}^{ab\,\alpha\beta}\; ; \nonumber \\
Q \hat{V}^{ab\,\alpha\beta} &=& \nabla\hat{U}^{ab\,\alpha\beta}\;; \nonumber \\
{}\nonumber \\
QM^{ab}_{\mu\nu} &=& N^{ab}_{\mu\nu}\;; \nonumber \\
QN^{ab}_{\mu\nu} &=& \nabla M^{ab}_{\mu\nu}\;; \nonumber \\
Q\tilde{N}^{ab}_{\mu\nu} &=& \tilde{M}^{ab}_{\mu\nu}\;; \nonumber \\
Q\tilde{M}^{ab}_{\mu\nu} &=& \nabla \tilde{N}^{ab}_{\mu\nu}\;. \nonumber
\end{eqnarray}
Thus, the local action invariant under the $Q$ transformation is
\begin{eqnarray}
\label{ac1}
S &=& \int d^{4}x \left[ \frac{1}{4}F^{a}_{\mu \nu}F^{a}_{\mu\nu} 
+ \bar{\lambda}^{a\,\alpha}(\gamma_{\mu})_{\alpha\beta}D^{ab}_{\mu}\lambda^{b\,\beta}
+ \frac{1}{2}\mathfrak{D}^{2} + \check{c}^{a}\partial_{\mu}D^{ab}_{\mu}c^{b} 
+ b^{a}\partial_{\mu}A^{a}_{\mu} \right. 
\nonumber \\
&&
\left. 
+ \check{c}^{a}\bar{\epsilon}^{\alpha}(\gamma_{\mu})_{\alpha\beta}\partial_{\mu}\lambda^{a\,\beta}
+\tilde{\varphi}^{ac}_{\mu}\partial_{\nu}D_{\nu}^{ab}\varphi^{bc}_{\mu} 
- \tilde{\omega}^{ac}_{\mu}\partial_{\nu}D_{\nu}^{ab}\omega^{bc}_{\mu} 
- gf^{abc}(\partial_{\nu}\tilde{\omega}^{ad}_{\mu})(D^{bk}_{\nu}c^{k})\varphi^{cd}_{\mu} \right. 
\nonumber \\
&&
+ gf^{abc}(\partial_{\nu}\tilde{\omega}^{ad}_{\mu})(\bar{\epsilon}^\alpha(\gamma_\nu)_{\alpha\beta}\lambda^{\beta b})\varphi^{cd}_{\mu} 
- N^{ab}_{\mu\nu}D^{ac}_{\mu}\tilde{\omega}^{cb}_{\nu} 
- M^{ab}_{\mu\nu}D^{ac}_{\mu}\tilde{\varphi}^{cb}_{\nu} 
+ gf^{adc}M^{ab}_{\mu\nu}D^{dl}_{\mu}c^{l}\tilde{\omega}^{cb}_{\nu} 
\nonumber \\
&&
- gf^{adc}\bar{\epsilon}^{\alpha}(\gamma_{\mu})_{\alpha\beta}M^{ab}_{\mu\nu}\lambda^{d\beta}\tilde{\omega}^{cb}_{\nu} 
- \tilde{M}^{ab}_{\mu\nu}D^{ac}_{\mu}\varphi^{cb}_{\nu} 
+ \tilde{N}^{ab}_{\mu\nu}D^{ac}_{\mu}\omega^{cb}_{\nu} 
- gf^{adc}\tilde{N}^{ab}_{\mu\nu}D^{dl}_{\mu}c^{l}\varphi^{cb}_{\nu} 
\nonumber \\
&&
+ gf^{adc}\bar{\epsilon}^{\alpha}(\gamma_{\mu})_{\alpha\beta}\tilde{N}^{ab}_{\mu\nu}\lambda^{d\beta}\varphi^{cb}_{\nu} 
- \tilde{M}^{ab}_{\mu\nu}M^{ab}_{\mu\nu} + \tilde{N}^{ab}_{\mu\nu}N^{ab}_{\mu\nu} 
+ \hat{\zeta}^{a\alpha}(\partial^{2}-\mu^{2})\zeta^{a}_{\alpha} 
- \hat{\theta}^{a\alpha}(\partial^{2}-\mu^{2})\theta^{a}_{\alpha} 
\nonumber \\
&&
- \hat{V}^{ab\,\alpha\beta}\bar{\lambda}^{a}_{\alpha}\theta^{b}_{\beta} 
- gf^{adc}\hat{U}^{ab\,\alpha\beta}c^{d}\bar{\lambda}^{c}_{\alpha}\theta^{b}_{\beta} 
+ \frac{1}{2}\hat{U}^{ab\,\alpha\beta}\bar{\epsilon}^{\gamma}(\sigma_{\mu\nu})_{\gamma\alpha}F^{a}_{\mu\nu}\theta^{b}_{\beta} 
- \hat{U}^{ab\,\alpha\beta}\bar{\epsilon}^{\gamma}(\gamma_{5})_{\gamma\alpha}\mathfrak{D}^{a}\theta^{b}_{\beta}
\nonumber \\
&&
+ \epsilon^{\gamma}(\gamma_{\mu})_{\gamma\eta}\bar{\epsilon}^{\eta}\hat{U}^{ab\,\alpha\beta}\bar{\lambda}^{a}_{\alpha}\partial_{\mu}\zeta^{b}_{\beta} 
- V^{ab\alpha\beta} \hat{\theta}^{b}_{\beta}\lambda^{a}_{\alpha}
- U^{ab\alpha\beta}\hat{\zeta}^{b}_{\beta}\lambda^{a}_{\alpha}
+ gf^{adc}U^{ab\alpha\beta}\hat{\theta}^{b}_{\beta}c^{d}\lambda^{c}_{\alpha}
\nonumber \\
&&
\left. 
- \frac{1}{2}U^{ab\,\alpha\beta}\hat{\theta}^{b}_{\beta} (\sigma_{\mu\nu})_{\alpha\gamma}\epsilon^{\gamma}F^{a}_{\mu\nu} 
+ U^{ab\,\alpha\beta}\hat{\theta}^{b}_{\beta}(\gamma_{5})_{\alpha\gamma}\epsilon^{\gamma}\mathfrak{D}^{a} \right]
\;.
\end{eqnarray}
This external sources must assume a physical value in the future so that one must fall back to the original local and explicitly broken action. Due to the behavior $Q^{2} = \epsilon^{\alpha}(\gamma_{\mu})_{\alpha\beta} \bar{\epsilon}^{\beta} \partial_{\mu}$ over all fields, as given by the set of equations \eqref{susytransf1}, this action is left $Q$ invariant, for a given boundary condition, {\it i.e.}
\begin{equation}
QS = 0\;.
\end{equation}
One should also put sources coupled to the non-linear transformations, as $QA^{a}_{\mu}$, $Q\lambda^{a\beta}$, $QD^{a}$ and $Qc^{a}$, in order to take a well defined vacuum expectation value of these quantities. Thus, let us add the following doublets of sources,
\begin{equation}
\left\{\begin{matrix}QK^{a}_{\mu}=\Omega^{a}_{\mu}\phantom{\Bigl|}\cr
Q\Omega^{a}_{\mu}=\nabla K^{a}_{\mu}\phantom{\Bigl|}\end{matrix}\right.\,,\qquad
\left\{\begin{matrix}QL^{a}=\Lambda^{a}\phantom{\Bigl|}\cr
Q\Lambda^{a}=\nabla L^{a}\phantom{\Bigl|}\end{matrix}\right.\,,\qquad
\left\{\begin{matrix}QT^{a}=J^{a}\phantom{\Bigl|}\cr
QJ^{a}=\nabla T^{a}\phantom{\Bigl|}\end{matrix}\right.\,,\qquad
\left\{\begin{matrix}QY^{a\alpha}=X^{a\alpha}\phantom{\Bigl|}\cr
QX^{a\alpha}=\nabla Y^{a\alpha}\phantom{\Bigl|}\;.\end{matrix}\right.
\end{equation}
The action which must be added to the action \eqref{ac1} is
\begin{equation}
\int d^{4}x \left[ -Q(K^{a}_{\mu} A^{a}_{\mu}) + Q(L^{a}c^{a}) - Q(T^{a}D^{a}) + Q(Y^{a\alpha}\lambda^{a}_{\alpha}) \right]\;.
\end{equation}

\noindent Therefore the action in its full glory is\footnote{For the index notation and about the quantum number of each field take a look at the notations in the Appendix \ref{notations}.}
\begin{eqnarray}
\label{fullact2}
\Sigma &=& \int d^{4}x \biggl\{ 
\frac{1}{4}F^{a}_{\mu \nu}F^{a}_{\mu\nu} 
+ \frac{1}{2}\bar{\lambda}^{a\,\alpha} (\gamma_{\mu})_{\alpha\beta}\,D^{ab}_{\mu}\lambda^{b\,\beta}
+ \frac{1}{2}\mathfrak{D}^{a}\mathfrak{D}^{a}
+ b^{a}\partial_{\mu}A^{a}_{\mu}
+ \check{c}^{a}\Bigl[\partial_{\mu}D^{ab}_{\mu}c^{b}
- \bar{\epsilon}^{\alpha}(\gamma_{\mu})_{\alpha\beta}\partial_{\mu}\lambda^{a\,\beta}\Bigr]\nonumber \\
&&
+\tilde{\varphi}^{a}_{i}\partial_{\mu}D_{\mu}^{ab}\varphi^{b}_{i}
-\tilde{\omega}^{a}_{i}\partial_{\mu}D_{\mu}^{ab}\omega^{b}_{i}
-gf^{abc}(\partial_{\mu}\tilde{\omega}^{a}_{i})\Bigl[(D^{bd}_{\mu}c^{d})
-\bar{\epsilon}^\alpha(\gamma_\mu)_{\alpha\beta}\lambda^{b\beta}\Bigl]
\varphi^{c}_{i}\nonumber\\
&&
-{N}^{a}_{\mu{i}}\,D^{ab}_{\mu}\tilde{\omega}^{b}_{i}
-{M}^{a}_{\mu{i}}\Bigl[D^{ab}_{\mu}\tilde{\varphi}^{b}_{i}
-gf^{abc}(D^{bd}_{\mu}c^{d})\tilde{\omega}^{c}_{i}
+gf^{abc}\bar{\epsilon}^{\alpha}(\gamma_{\mu})_{\alpha\beta}\lambda^{b\beta}\tilde{\omega}^{c}_{i}\Bigr]\nonumber \\
&&
-\tilde{M}^{a}_{\mu{i}}\,D^{ab}_{\mu}\varphi^{b}_{i}
+\tilde{N}^{a}_{\mu{i}}\Bigl[D^{ab}_{\mu}\omega^{b}_{i}
- gf^{abc}(D^{bd}_{\mu}c^{d})\varphi^{c}_{i}
+ gf^{abc}\bar{\epsilon}^{\alpha}(\gamma_{\mu})_{\alpha\beta}\lambda^{b\beta}\varphi^{c}_{i}\Bigr]\nonumber \\
&&
- \tilde{M}^{a}_{\mu{i}}{M}^{a}_{\mu{i}}
+ \tilde{N}^{a}_{\mu{i}}{N}^{a}_{\mu{i}}
+ \hat{\zeta}^{I}(\partial^{2}-\mu^{2})\zeta_{I}
- \hat{\theta}^{I}(\partial^{2}-\mu^{2})\theta_{I}
+ \hat{V}^{Ia\alpha}\,\bar{\lambda}^{a}_{\alpha}\theta_{I}\nonumber \\
&&
- \hat{U}^{Ia\alpha}\Bigl[gf^{abc}c^{b}\bar{\lambda}^{c}_{\alpha}\theta_{I}
-\bar{\lambda}^{a}_{\alpha}\nabla\zeta_{I}
-\frac{1}{2}\bar{\epsilon}^{\gamma}(\sigma_{\mu\nu})_{\gamma\alpha}F^{a}_{\mu\nu}\theta_{I} 
+\bar{\epsilon}^{\gamma}(\gamma_{5})_{\gamma\alpha}\mathfrak{D}^{a}\theta_{I}\Bigr]\nonumber \\
&&
+ V^{Ia\alpha}\,\hat{\theta}_{I}\lambda^{a}_{\alpha}
+ U^{Ia\alpha}\Bigl[-\hat{\zeta}_{I}\lambda^{a}_{\alpha}
+ gf^{abc}\hat{\theta}_{I}c^{b}\lambda^{c}_{\alpha}
- \frac{1}{2}\hat{\theta}_{I}(\sigma_{\mu\nu})_{\alpha\gamma}\epsilon^{\gamma}F^{a}_{\mu\nu}  
+ \hat{\theta}_{I}(\gamma_{5})_{\alpha\gamma}\epsilon^{\gamma}\mathfrak{D}^{a}\Bigr] \nonumber \\
&&
- \Omega^{a}_{\mu} A^{a}_{\mu}
- K^{a}_{\mu}\Bigl[D^{ab}_{\mu}c^{b} - \bar{\epsilon}^\alpha(\gamma_\mu)_{\alpha\beta}\lambda^{a\beta}\Bigr]
+ \Lambda^{a}c^{a}
+ L^{a}\Bigl[ \frac{g}{2}f^{abc}c^{b}c^{c}
- \bar{\epsilon}^\alpha(\gamma_{\mu})_{\alpha\beta}\epsilon^\beta A^{a}_{\mu}\Bigr]\nonumber \\
&&
- J^{a}\mathfrak{D}^{a}
+ T^{a}\Bigl[gf^{abc}c^{b}\mathfrak{D}^{c}
+ \bar{\epsilon}^{\alpha}(\gamma_{\mu})_{\alpha \beta}(\gamma_{5})^{\beta\eta}D_{\mu}^{ab}\lambda^{b}_{\eta}\Bigr]\nonumber\\
&&+ X^{a\alpha}\lambda^{a}_{\alpha}
+Y^{a\alpha}\Bigl[ gf^{abc}c^{b}\lambda^{c}_{\alpha} - \frac{1}{2}(\sigma_{\mu\nu})_{\alpha\beta} F_{\mu\nu}^{a}\epsilon^{\beta}
+ (\gamma_{5})_{\alpha\beta}\epsilon^{\beta} \mathfrak{D}^a \Bigr]
- X^{a\alpha}(\gamma_{5})_{\alpha\beta}\varepsilon^{\beta}T^{a}
\nonumber \\
&&
- \hat{V}^{Ia\alpha}\bar{\varepsilon}^{\beta}(\gamma_{5})_{\beta\alpha}\theta_{I}T^{a}
+ \hat{U}^{Ia\alpha}\bar{\varepsilon}^{\beta}(\gamma_{5})_{\beta\alpha}\nabla\zeta_{I} T^{a}
- V^{Ia\alpha}(\gamma_{5})_{\alpha\beta}\varepsilon^{\beta}\hat{\theta}_{I}T^{a}
+ U^{Ia\alpha}(\gamma_{5})_{\alpha\beta}\varepsilon^{\beta}\hat{\zeta}_{I}T^{a}
\nonumber \\
&&
- J^{a}  \left(
Y^{a\alpha}(\gamma_{5})_{\alpha \beta}\varepsilon^{\beta}
+ \hat{U}^{Ia}_{\;\;\alpha}C^{\beta\alpha}\theta_{I}(\gamma_{5})_{\beta\eta}\varepsilon^{\eta}
+ U^{Ia\alpha}\hat{\theta}_{I}(\gamma_{5})_{\alpha\beta}\varepsilon^{\beta}
\right)
\biggr\}\;.
\end{eqnarray}
In the above action the quadratic terms in the source $-Q(Y\gamma_{5}\varepsilon T)$, $-Q(\hat{U}\bar{\varepsilon}\gamma_{5}\theta T)$ and $-Q(U\gamma_{5}\varepsilon \hat{\theta}T)$ were introduced by hand without really change the physical content of the model. These terms are needed in order to account for new terms which appear in the renormalized action due to the matrix renormalization method.

\noindent When all the external fields reach its respective physical value we fall back to the original broken local action \eqref{fnlact}. Namely, its physical values are
\begin{eqnarray}
\label{physval2}
&&
M^{ab}_{\mu\nu}\Big{|}_{phy}=\tilde{M}^{ab}_{\mu\nu}\Big{|}_{phy}=
\gamma^{2}\delta^{ab}\delta_{\mu\nu}
\;;\nonumber \\
&&
V^{ab\alpha\beta}\Big{|}_{phy}=\hat{V}^{ab\alpha\beta}\Big{|}_{phy}=
-M^{3/2}\delta^{ab}\delta^{\alpha\beta}
\;;\nonumber \\
&&
N^{ab}_{\mu\nu}\Big{|}_{phy}=\tilde{N}^{ab}_{\mu\nu}\Big{|}_{phy}=
U^{ab\alpha\beta}\Big{|}_{phy}=\hat{U}^{ab\alpha\beta}\Big{|}_{phy}=0
\;.
\end{eqnarray}

\section{Symmetry content of the model}

\subsection*{Ward identities}
\begin{itemize}
{\item The Slavnov-Taylor identity:}

\begin{eqnarray}
\mathcal{S}(\Sigma) &=& \int d^{4}x \biggl\{\biggl(\frac{\delta \Sigma}{\delta A^{a}_{\mu}}
+ \Omega^{a}_{\mu}\biggr)\frac{\delta \Sigma}{\delta K^{a}_{\mu}}
+ \biggl(\frac{\delta \Sigma}{\delta \lambda^{a\alpha}}
+ X^{a\alpha}\biggr)\frac{\delta \Sigma}{\delta Y^{a\alpha}}
+\biggl(\frac{\delta \Sigma}{\delta c^{a}}
+ \Lambda^{a}\biggr)\frac{\delta \Sigma}{\delta L^{a}} \nonumber \\
&&
+ \biggl(\frac{\delta \Sigma}{\delta \mathfrak{D}^{a}}
+ {J}^{a}\biggr)\frac{\delta \Sigma}{\delta T^{a}}
+ b^{a}\frac{\delta \Sigma}{\delta \check{c}^{a}}
+ \omega^{a}_{i}\frac{\delta \Sigma}{\delta \varphi^{a}_{i}}
+ \tilde{\varphi}^{a}_{i}\frac{\delta \Sigma}{\delta \tilde{\omega}^{a}_{i}}
+ \hat{\zeta}^{I}\frac{\delta \Sigma}{\delta \hat{\theta}^{I}}
+ \theta^{I}\frac{\delta \Sigma}{\delta \zeta^{I}}\nonumber \\
&&
+{V}^{Ia\alpha}\frac{\delta \Sigma}{\delta {U}^{Ia\alpha}}
+ \hat{V}^{Ia\alpha}\frac{\delta \Sigma}{\delta \hat{U}^{Ia\alpha}}
+ {N}^{a}_{\mu i}\frac{\delta \Sigma}{\delta{M}^{a}_{\mu i}}
+ \tilde{M}^{a}_{\mu i}\frac{\delta \Sigma}{\delta \tilde{N}^{a}_{\mu i}} \nonumber  \\
&&
+(\nabla{U}^{Ia\alpha})\frac{\delta \Sigma}{\delta{V}^{Ia\alpha}}
+(\nabla\hat{U}^{Ia\alpha})\frac{\delta \Sigma}{\delta \hat{V}^{Ia\alpha}}
+(\nabla{M}^{a}_{\mu i})\frac{\delta \Sigma}{\delta {N}^{a}_{\mu i}} \nonumber \\
&&
+(\nabla\tilde{N}^{a}_{\mu i})\frac{\delta \Sigma}{\delta \tilde{M}^{a}_{\mu i}}
+(\nabla K^{a}_{\mu})\frac{\delta \Sigma}{\delta \Omega^{a}_{\mu}}
+(\nabla Y^{a\alpha})\frac{\delta \Sigma}{\delta {X}^{a\alpha}}
+(\nabla T^{a})\frac{\delta \Sigma}{\delta {J}^{a}}
+(\nabla L^{a})\frac{\delta \Sigma}{\delta {\Lambda}^{a}}\nonumber  \\
&&
+(\nabla \check{c}^{a})\frac{\delta \Sigma}{\delta b^{a}}
+(\nabla\varphi^{a}_{i})\frac{\delta \Sigma}{\delta \omega^{a}_{i}}
+(\nabla\tilde{\omega}^{a}_{i})\frac{\delta \Sigma}{\delta \tilde{\varphi}^{a}_{i}}
+(\nabla\hat{\theta}^{I})\frac{\delta \Sigma}{\delta \hat{\zeta}^{I}}
+(\nabla\zeta^{I})\frac{\delta \Sigma}{\delta \theta^{I}}  \biggr\}\nonumber\\
&=& 0\,.
\label{ST1}
\end{eqnarray}

\item{The linearized Slavnov-Taylor operator:}
\begin{eqnarray}
{\cal B}_{\Sigma} &=& \int d^{4}x \biggl\{
  \biggl(\frac{\delta \Sigma}{\delta A^{a}_{\mu}}
+ \Omega^{a}_{\mu} \biggr) \frac{\delta }{\delta K^{a}_{\mu}}
+ \frac{\delta \Sigma}{\delta K^{a}_{\mu}}\frac{\delta }{\delta A^{a}_{\mu}}
+ \biggl(\frac{\delta \Sigma}{\delta \lambda^{a\alpha}}
+ X^{a\alpha}\biggr)\frac{\delta }{\delta Y^{a\alpha}}
+ \frac{\delta \Sigma}{\delta Y^{a\alpha}}\frac{\delta }{\delta \lambda^{a\alpha}}
+\biggl(\frac{\delta \Sigma}{\delta c^{a}}
+ \Lambda^{a}\biggr)\frac{\delta }{\delta L^{a}} 
\nonumber \\
&&
+ \frac{\delta \Sigma}{\delta L^{a}}\frac{\delta }{\delta c^{a}}
+ \biggl(\frac{\delta \Sigma}{\delta \mathfrak{D}^{a}}
+ {J}^{a}\biggr)\frac{\delta }{\delta T^{a}}
+ \frac{\delta \Sigma}{\delta T^{a}}\frac{\delta }{\delta \mathfrak{D}^{a}}
+ b^{a}\frac{\delta \Sigma}{\delta \check{c}^{a}}
+ \omega^{a}_{i}\frac{\delta \Sigma}{\delta \varphi^{a}_{i}}
+ \tilde{\varphi}^{a}_{i}\frac{\delta \Sigma}{\delta \tilde{\omega}^{a}_{i}}
+ \hat{\zeta}^{I}\frac{\delta \Sigma}{\delta \hat{\theta}^{I}}
+ \theta^{I}\frac{\delta \Sigma}{\delta \zeta^{I}}
\nonumber \\
&&
+{V}^{Ia\alpha}\frac{\delta \Sigma}{\delta {U}^{Ia\alpha}}
+ \hat{V}^{Ia\alpha}\frac{\delta \Sigma}{\delta \hat{U}^{Ia\alpha}}
+ {N}^{a}_{\mu i}\frac{\delta \Sigma}{\delta{M}^{a}_{\mu i}}
+ \tilde{M}^{a}_{\mu i}\frac{\delta \Sigma}{\delta \tilde{N}^{a}_{\mu i}} 
+(\nabla{U}^{Ia\alpha})\frac{\delta \Sigma}{\delta{V}^{Ia\alpha}}
+(\nabla\hat{U}^{Ia\alpha})\frac{\delta \Sigma}{\delta \hat{V}^{Ia\alpha}}
\nonumber  \\
&&
+(\nabla{M}^{a}_{\mu i})\frac{\delta \Sigma}{\delta {N}^{a}_{\mu i}} 
+(\nabla\tilde{N}^{a}_{\mu i})\frac{\delta \Sigma}{\delta \tilde{M}^{a}_{\mu i}}
+(\nabla K^{a}_{\mu})\frac{\delta \Sigma}{\delta \Omega^{a}_{\mu}}
+(\nabla Y^{a\alpha})\frac{\delta \Sigma}{\delta {X}^{a\alpha}}
+(\nabla T^{a})\frac{\delta \Sigma}{\delta {J}^{a}}
+(\nabla L^{a})\frac{\delta \Sigma}{\delta {\Lambda}^{a}}\nonumber  \\
&&
+(\nabla \check{c}^{a})\frac{\delta \Sigma}{\delta b^{a}}
+(\nabla\varphi^{a}_{i})\frac{\delta \Sigma}{\delta \omega^{a}_{i}}
+(\nabla\tilde{\omega}^{a}_{i})\frac{\delta \Sigma}{\delta \tilde{\varphi}^{a}_{i}}
+(\nabla\hat{\theta}^{I})\frac{\delta \Sigma}{\delta \hat{\zeta}^{I}}
+(\nabla\zeta^{I})\frac{\delta \Sigma}{\delta \theta^{I}}  \biggr\} \;.
\label{ST2}
\end{eqnarray}

\item{The gauge-fixing and anti-ghost equations:}
\begin{equation}
\frac{\delta\Sigma}{\delta b^{a}}=\partial_{\mu}A^{a}_{\mu}\,,\qquad
\frac{\delta\Sigma}{\delta\check{c}^{a}}+\partial_{\mu}\frac{\delta\Sigma}{\delta K^{a}_{\mu}}=0\,.
\label{GFandAntiGhost1}
\end{equation}

\item{The equations of motion of the auxiliary fields:}

\begin{eqnarray}
&&\frac{\delta \Sigma}{\delta \tilde{\varphi}^{a}_{i}} + \partial_{\mu}\frac{\delta\Sigma}{\delta \tilde{M}^{a}_{\mu i}} - gf^{abc}M^{b}_{\mu i}\frac{\delta\Sigma}{\delta\Omega^{c}_{\mu}} = 0 \,,
\\
&&\frac{\delta\Sigma}{\delta\omega^{a}_{i}} + \partial_{\mu}\frac{\delta\Sigma}{\delta N^{a}_{\mu i}} - gf^{abc} \left( \frac{\delta \Sigma}{\delta b^{c}} \tilde{\omega}^{b}_{i} + \frac{\delta \Sigma}{\delta \Omega^{c}_{\mu}}\tilde{N}^{b}_{\mu i} \right) = 0 \,,
\\
&&
\frac{\delta \Sigma}{\delta \tilde{\omega}^{a}_{i}} 
+ \partial_{\mu}\frac{\delta\Sigma}{\delta\tilde{N}^{a}_{\mu i}} 
- gf^{abc}\left( M^{b}_{\mu i}\frac{\delta\Sigma}{\delta K^{c}_{\mu}} 
- N^{b}_{\mu i} \frac{\delta\Sigma}{\delta\Omega^{c}_{\mu}} \right)=
0 \,,
\\
&&
\frac{\delta\Sigma}{\delta\varphi^{a}_{i}} 
+ \partial_{\mu}\frac{\delta\Sigma}{\delta M^{a}_{\mu i}} 
- gf^{abc}\left( \frac{\delta \Sigma}{\delta b^{c}}\tilde{\varphi}^{b}_{i} 
+ \frac{\delta \Sigma}{\delta \Omega^{c}_{\mu}}\tilde{M}^{b}_{\mu i} 
+ \frac{\delta \Sigma}{\delta \check{c}^{b}} \tilde{\omega}^{c}_{i} 
- \tilde{N}^{c}_{\mu i} \frac{\delta \Sigma}{\delta K^{b}_{\mu}} \right) = 0 \,,
\\
&&
\frac{\delta\Sigma}{\delta\zeta_{I}}=
(\partial^{2}-\mu^{2})\hat{\zeta}^{I}
- \nabla(\hat{U}^{Ia\alpha}\,\bar\lambda^{a}_{\alpha})\,,
\\
&&\frac{\delta\Sigma}{\delta\hat\zeta_{I}}=(\partial^{2}-\mu^{2}){\zeta}^{I}
-{U}^{Ia\alpha}\,\lambda^{a}_{\alpha}\,,
\\
&&\frac{\delta\Sigma}{\delta\hat\theta_{I}}
-{U}^{Ia\alpha}\,\frac{\delta\Sigma}{\delta Y^{a\alpha}}=-(\partial^{2}-\mu^{2}){\theta}^{I}
+{V}^{Ia\alpha}\,\lambda^{a}_{\alpha}\,,
\\
&&\frac{\delta\Sigma}{\delta\theta_{I}}
-\left(\frac{\delta\Sigma}{\delta Y^{a}}\right)^{T}_{\beta}\mathcal{C}^{\beta}_{\;\;\alpha}\;{\hat{U}}^{Ia\alpha}=(\partial^{2}-\mu^{2}){\hat\theta}^{I}
-\bar\lambda^{a}_{\;\;\alpha}\;{\hat{V}}^{Ia\alpha}\,,
\\
&&\frac{\delta\Sigma}{\delta{\mathfrak{D}^{a}}}=-\mathfrak{D}^{a}- J^{a} + gf^{abc}c^{b}T^{c}-Y^{a\alpha}(\gamma_{5})_{\alpha\beta}\,\varepsilon^{\beta}
+\hat{U}^{Ia\alpha}\,\bar{\varepsilon}_{\beta}(\gamma_{5})^{\beta\alpha}\,\theta_{I}
-U^{Ia\alpha}\,\hat\theta_{I}\,(\gamma_{5})^{\alpha\beta}\,\varepsilon_{\beta}\,.
\end{eqnarray}

\item{The equations of motion of the external BRST sources:}
\begin{equation}
\frac{\delta\Sigma}{\delta{\Omega^{a}_{\mu}}}=A^{a}_{\mu}\,,\qquad
\frac{\delta\Sigma}{\delta{\Lambda^{a}}}=c^{a}\,,\qquad
\frac{\delta\Sigma}{\delta{J^{a}}}=-\mathfrak{D}^{a}\,,\qquad
\frac{\delta\Sigma}{\delta{X^{a\alpha}}}=\lambda^{a}_{\alpha}\,.
\end{equation}

\item{The $U(f=4(N^2-1))$ invariance and the multi-index $i\equiv(a,\mu)$:}
\begin{eqnarray}
 \mathcal{L}^{ab}_{\mu\nu}(\Sigma) &=&- \int d^{4}x \left( 
\tilde{\varphi}^{ca}_{\mu}\frac{\delta \Sigma}{\delta \tilde{\varphi}^{cb}_{\nu}}
 - \varphi^{cb}_{\nu}\frac{\delta\Sigma}{\delta\varphi^{ca}_{\mu}} 
+\tilde{\omega}^{ca}_{\mu}\frac{\delta \Sigma}{\delta \tilde{\omega}^{cb}_{\nu}}
 - \omega^{cb}_{\nu}\frac{\delta\Sigma}{\delta\omega^{ca}_{\mu}}   
+\tilde{M}^{ca}_{\sigma\mu}\frac{\delta \Sigma}{\delta \tilde{M}^{cb}_{\sigma\nu}}
 - M^{cb}_{\sigma\nu}\frac{\delta\Sigma}{\delta{M}^{ca}_{\sigma\mu}}  
 \right. \nonumber \\
&&\phantom{\int d^{4}x\,}  \left. 
+\tilde{N}^{ca}_{\sigma\mu}\frac{\delta \Sigma}{\delta \tilde{N}^{cb}_{\sigma\nu}}
 - N^{cb}_{\sigma\nu}\frac{\delta\Sigma}{\delta{N}^{ca}_{\sigma\mu}}  
\right) = 0\;.
\end{eqnarray}

\noindent These fields has a $q$ charge and this relation defines a $(c,\mu)\!:~ i,j,k,l$ 
multi-index.

\item{The $U(f'=4(N^2-1))$ invariance and the multi-index $I\equiv(a,\alpha)$:}
\begin{eqnarray}
\mathcal{L}^{ab\phantom{\alpha}\beta}_{\phantom{ab}\alpha}(\Sigma)&=&\int d^{4}x\,\biggl(
\zeta^{a}_{\alpha}\frac{\delta\Sigma}{\delta\zeta^{b}_{\beta}}
-\bar\zeta^{b}_{\beta}\frac{\delta\Sigma}{\delta\bar\zeta^{a\alpha}}
+\theta^{a}_{\alpha}\frac{\delta\Sigma}{\delta\theta^{b}_{\beta}}
-\bar\theta^{b}_{\beta}\frac{\delta\Sigma}{\delta\bar\theta^{a\alpha}}
+\widetilde{\mathcal{V}}^{ca}_{\phantom{ca}\gamma\alpha}
\frac{\delta\Sigma}{\delta\widetilde{\mathcal{V}}^{cb\phantom{\gamma}\beta}_{\phantom{cb}\gamma}}
-{\mathcal{V}}^{cb\phantom{\gamma}\beta}_{\phantom{cb}\gamma}
\frac{\delta\Sigma}{\delta{\mathcal{V}}^{ca\phantom{\gamma}\alpha}_{\phantom{ca}\gamma}}
\nonumber\\
&&\phantom{\int d^{4}x\,}
+\widetilde{\mathcal{U}}^{ca}_{\phantom{ca}\gamma\alpha}
\frac{\delta\Sigma}{\delta\widetilde{\mathcal{U}}^{cb\phantom{\gamma}\beta}_{\phantom{cb}\gamma}}
-{\mathcal{U}}^{cb\phantom{\gamma}\beta}_{\phantom{cb}\gamma}
\frac{\delta\Sigma}{\delta{\mathcal{U}}^{ca\phantom{\gamma}\alpha}_{\phantom{ca}\gamma}}\biggr)\nonumber\\
&=&0\,.
\end{eqnarray}

\noindent These fields has a $q'$ charge and this relation define a $(a,\alpha)\!:~ I,J,K,L$ 
multi-index.

\item{The ghost equation:}
\begin{equation}
G^{a}(\Sigma)=\Delta^{a}_{\mathrm{class}}\,,
\end{equation}
where
\begin{eqnarray}
G^{a}&:=&\int d^{4}x\,\biggl[\frac{\delta}{\delta{c}^{a}}+gf^{abc}\biggl(\check{c}^{b}\frac{\delta}{\delta{b}^{c}}
+\varphi^{b}_{i}\frac{\delta}{\delta\omega^{c}_{i}}
+\tilde\omega^{b}_{i}\frac{\delta}{\delta\tilde\varphi^{c}_{i}}
+\tilde{N}^{b}_{\mu i}\frac{\delta}{\delta\tilde{M}^{c}_{\mu i}}
+M^{b}_{\mu i}\frac{\delta}{\delta N^{c}_{\mu i}}\nonumber\\
&&\phantom{\int d^{4}x\,}+\hat{U}^{Ib\alpha}\frac{\delta}{\delta\hat{V}^{Ic\alpha}}
-U^{Ib\alpha}\frac{\delta}{\delta V^{Ic\alpha}}\biggr)\biggr]\,,
\end{eqnarray}
and
\begin{equation}
\Delta^{a}_{\mathrm{class}}=\int d^{4}x\,\left[gf^{abc}\left(K^{b}_{\mu}A^{c}_{\mu}
-L^{b}c^{c}+T^{b}\mathfrak{D}^{a}
-Y^{b\alpha}\lambda^{c}_{\alpha}\right)-\Lambda^{a}\right]\,.
\end{equation}

\item The equation of the source $T^{a}$:
\begin{equation}
\frac{\delta\Sigma}{\delta{T^{a}}}
+\frac{\delta\Sigma}{\delta{\lambda^{a}_{\alpha}}}(\gamma_{5})_{\alpha\beta}\,\varepsilon^{\beta}
+gf^{abc} c^{b}\frac{\delta\Sigma}{\delta{\mathfrak{D}^{c}}}
+gf^{abc}T^{b}\frac{\delta\Sigma}{\delta{L^{c}}}=\tilde\Delta^{a}_{\mathrm{class}}\,,
\end{equation}
where
\begin{eqnarray}
\tilde\Delta^{a}_{\mathrm{class}} &=&
3gf^{abc}\bar{\epsilon}^{\alpha}(\gamma_{\mu})_{\alpha\beta}\epsilon^{\beta}T^{b}A^{c}_{\mu} 
+ \nabla T^{b}
-gf^{abc}c^{b}J^{c}
\nonumber \\
&&
-\epsilon^{\beta}(\gamma_{5})_{\beta\alpha}X^{a\alpha}
-\bar{\epsilon}^{\alpha}(\gamma_{\mu})_{\alpha\eta}(\gamma_{5})^{\eta\beta}\epsilon_{\beta} \left( \partial_{\mu}\bar{c}^{a} + K^{a}_{\mu} \right) \;.
\end{eqnarray}

\noindent This equation can also be obtained from the commutation relation between the linearized Slavnov-Taylor operator and $\delta/\delta{\mathfrak{D}}^{a}$.


\end{itemize}

\subsection*{Discrete symmetries}

\noindent Let the $\gamma$ matrices change as $\gamma_{\mu}\to e^{n\pi i}\,\gamma_{\mu}$, with $n$ positive integer. In this case $\gamma_5$, $\mathcal{C}$, $\sigma_{\mu\nu}$ and $\{\gamma_{\mu},\gamma_{\nu}\}$ remain unchanged and the action $\Sigma$ is left invariant under the following sets of transformations,
\begin{equation}
\begin{tabular}{c|c|c}
First set&Second set&Third set\\
$\gamma_{\mu} \to e^{n\pi i} \gamma_{\mu}$&$\gamma_{\mu} \to e^{n\pi i}\gamma_{\mu}$&$\gamma_{\mu} \to e^{n\pi i}\gamma_{\mu}$\\
$\lambda \to e^{-\frac{n\pi i}{2}} \lambda$&$\lambda \to e^{\frac{n\pi i}{2}}\lambda$&$\lambda \to e^{-\frac{3}{2}n\pi i}\lambda$\\
$\bar{\lambda} \to e^{-\frac{n\pi i}{2}} \bar{\lambda}$&$\bar{\lambda} \to e^{\frac{n\pi i}{2}}\bar{\lambda}$&$\bar{\lambda} \to e^{\frac{1}{2}n\pi i}\bar{\lambda}$\\
$\bar{\epsilon} \to e^{-\frac{n\pi i}{2}} \bar{\epsilon}$&$\bar{\epsilon} \to e^{\frac{n\pi i}{2}} \bar{\epsilon}$&$\bar{\epsilon} \to 
e^{-\frac{3}{2}n\pi i} \bar{\epsilon}$\\
$\epsilon \to e^{-\frac{n\pi i}{2}} \epsilon$&$\epsilon \to e^{\frac{n\pi i}{2}}\epsilon$&$\epsilon \to e^{-\frac{3}{2}n\pi i}\epsilon$\\
$\theta \to e^{\frac{n\pi i}{2}m} \theta$&$\theta \to e^{\frac{n\pi i}{2}} \theta$&$\theta \to e^{-\frac{3}{2}n\pi i} \theta$\\
$\hat{\theta} \to e^{-\frac{n\pi i}{2}m} \hat{\theta}$&$\hat{\theta} \to e^{\frac{3 n\pi i}{2}} \hat{\theta}$&$\hat{\theta} \to 
e^{-\frac{1}{2} n\pi i} \hat{\theta}$\\
$\zeta \to e^{\frac{n\pi i}{2}m} \zeta$&$\zeta \to e^{\frac{n\pi i}{2}} \zeta$&$\zeta \to e^{-\frac{3}{2}n\pi i} \zeta$\\
$\hat{\zeta} \to e^{-\frac{n\pi i}{2}m} \hat{\zeta}$&$\hat{\zeta} \to e^{\frac{3n\pi i}{2}} \hat{\zeta}$&$\hat{\zeta} \to 
e^{\frac{3}{2}n\pi i} \hat{\zeta}$\\
$\hat{V} \to e^{(1-m)\frac{n\pi i}{2}} \hat{V}$&$\hat{V} \to e^{n\pi i} \hat{V}$&$\hat{V} \to e^{n\pi i} \hat{V}$\\
$V \to e^{(1+m)\frac{n\pi i}{2}} V $&$V \to V $&$V \to V$\\
$\hat{U} \to e^{(1-m)\frac{n\pi i}{2}} \hat{U}$&$\hat{U} \to e^{n\pi i} \hat{U}$&$\hat{U} \to e^{n\pi i} \hat{U}$\\
$U \to e^{(1+m)\frac{n\pi i}{2}} U$&$U \to  U$&$U \to U$\\
$Y \to e^{\frac{n\pi i}{2}} Y$&$Y \to e^{\frac{3n\pi i}{2}} Y$&$Y \to e^{-\frac{1}{2}n\pi i} Y$\\
$X \to e^{\frac{n\pi i}{2}} X$&$X \to e^{\frac{3n\pi i}{2}} X$&$X \to e^{-\frac{1}{2}n\pi i} X$
\end{tabular}
\end{equation}
where $m\in \Re$ in the first set. Unmentioned fields are known to be transformed in itself.

\noindent It is important to note the existence of particular cases where $n=1$ and when $n$ is even.

Now, let $x_{4}\to-x_{4}$ (the same is possible for $x_{2}\to-x_{2}$). In this case we can transform the $\gamma$ matrices as
\begin{equation}
\gamma_{4}\to-\gamma_{4}\,,\qquad\gamma_{k}\to\gamma_{k},\qquad k=1,2,3\,.
\end{equation}
Notice that the anti-commutation relation $\{\gamma_{\mu},\gamma_{\nu}\}=2\delta_{\mu\nu}$ remains unchanged by the transformations above, but
\begin{equation}
\gamma_{5}\to-\gamma_{5}\,,\qquad
\mathcal{C}\to-\mathcal{C}\,,\qquad
\sigma_{4k}\to-\sigma_{4k}\,,\qquad
\sigma_{kl}\to\sigma_{kl}\,,\qquad
k,l=1,2,3\,. 
\end{equation}
In this case we have two sets of transformations that let the action $\Sigma$ invariant:
\begin{equation}
\begin{tabular}{c|c}
First set&Second set\\
\hline
$A_{4}\to-A_{4}$&$A_{4}\to-A_{4}$\\
$\lambda\to+i\lambda$&$\lambda\to+i\lambda$\\
$\bar\lambda\to-i\bar\lambda$&$\bar\lambda\to-i\bar\lambda$\\
$\epsilon\to+i\epsilon$&$\epsilon\to+i\epsilon$\\
$\bar\epsilon\to-i\bar\epsilon$&$\bar\epsilon\to-i\bar\epsilon$\\
$\tilde{M}^{ab}_{4\nu}\to-\tilde{M}^{ab}_{4\nu}$&$\tilde{M}^{ab}_{4\nu}\to-\tilde{M}^{ab}_{4\nu}$\\
$M^{ab}_{4\nu}\to -M^{ab}_{4\nu}$&$M^{ab}_{4\nu}\to-M^{ab}_{4\nu}$\\
$\tilde{N}^{ab}_{4\nu}\to-\tilde{N}^{ab}_{4\nu}$&$\tilde{N}^{ab}_{4\nu}\to-\tilde{N}^{ab}_{4\nu}$\\
$N^{ab}_{4\nu}\to-N^{ab}_{4\nu}$&$N^{ab}_{4\nu}\to-N^{ab}_{4\nu}$\\
$K^{a}_{4} \to - K^{a}_{4}$&$K^{a}_{4} \to - K^{a}_{4}$\\
$\Omega^{a}_{4} \to - \Omega^{a}_{4}$&$\Omega^{a}_{4} \to - \Omega^{a}_{4}$\\
$\mathfrak{D}\to-\mathfrak{D}$&$\mathfrak{D}\to-\mathfrak{D}$\\
${T}\to-{T}$&${T}\to-{T}$\\
$J\to-J$&$J\to-J$\\
$Y\to-iY$&$Y\to-iY$\\
$X\to-iX$&$X\to-iX$\\
$\hat{V}\to+i\hat{V}$&$\theta\to+i\theta$\\
$V\to-iV$&$\hat\theta\to-i\hat\theta$\\
$\hat{U}\to+i\hat{U}$&$\zeta\to+i\zeta$\\
$U\to-iU$&$\hat\zeta\to-i\hat\zeta$
\end{tabular}
\end{equation}
Finally, let $x_1\to-x_1$ (or $x_3\to-x_3$). In this case we have:
\begin{equation}
\gamma_{1}\to-\gamma_{1}\,,\qquad\gamma_{k}\to\gamma_{k}\,,\qquad k=2,3,4\,.
\end{equation}
Also in this case the anti-commutation relation between the $\gamma$ matrices remains unchanged, but
\begin{equation}
\gamma_{5}\to-\gamma_{5}\,,\qquad
\mathcal{C}\to\mathcal{C}\,,\qquad
\sigma_{1k}\to-\sigma_{1k}\,,\qquad
\sigma_{kl}\to\sigma_{kl}\,,\qquad
k,l=2,3,4\,.
\end{equation}
One can show in this case that the action $\Sigma$ is then invariant by the following set of transformations:
\begin{equation}
\begin{tabular}{c}
$A_1\to-A_{1}$\\
$\mathfrak{D}\to-\mathfrak{D}$\\
$T\to-T$\\
$J\to-J$\\
$\tilde{M}^{ab}_{1\nu}\to-\tilde{M}^{ab}_{1\nu}$\\
$M^{ab}_{1\nu}\to-M^{ab}_{1\nu}$\\
$\tilde{N}^{ab}_{1\nu}\to-\tilde{N}^{ab}_{1\nu}$\\
$N^{ab}_{1\nu}\to-N^{ab}_{1\nu}$\\
$K^{a}_{1} \to - K^{a}_{1}$\\
$\Omega^{a}_{1} \to - \Omega^{a}_{1}$
\end{tabular}
\end{equation}

\section{Determining the counter-term}

\begin{equation}
\label{count}
\Sigma_{count}=a_{0}\,S_{\mathrm{SYM}}+\mathcal{B}_{\Sigma}\Delta^{(-1)}\,.
\end{equation}
Here, $\Delta^{(-1)}$ is an integrated polynomial in the fields and in the sources of dimension $3$, ghost number $-1$,  and $q_{f}=q_{f'}=0$. Taking into account some symmetries, the most general $\Delta^{(-1)}$ with $39$ terms is given by\footnote{The most general $\Delta^{(-1)}$, with $350$ terms, was found by M. Capri}

\begin{eqnarray}
\Delta^{-1} &=& \int d^{4}x \left\{ 
  a_{1}(\partial_{\mu}\check{c}^{a}+K^{a}_{\mu})A^{a}_{\mu}
+ a_{2}c^{a}L^{a}
+ a_{3}(\partial_{\mu}\tilde{\omega}^{a}_{i})M^{a}_{\mu i}
+ a_{4}gf^{abc}A^{c}_{\mu}M^{a}_{\mu i}\tilde{\omega}^{b}_{i}
+ a_{5}(\partial_{\mu}\tilde{\omega}^{a}_{i})\partial_{\mu}\varphi^{a}_{i}
\right. \nonumber \\
&&
+ a_{6}gf^{abc}A^{c}_{\mu}\tilde{\omega}^{a}_{i}\partial_{\mu}\varphi^{b}_{i}
+ a_{7}gf^{abc}A^{c}_{\mu}(\partial_{\mu}\tilde{\omega}^{a}_{i})\varphi^{b}_{i}
+ a_{8}\kappa\tilde{M}^{a}_{\mu i}M^{a}_{\mu i}
+ a_{9}\tilde{N}^{a}_{\mu i}\partial_{\mu}\varphi^{a}_{i}
+ a_{10}gf^{abc}A^{c}_{\mu}\tilde{N}^{a}_{\mu i}\varphi^{b}_{i}
\nonumber \\
&&
+ a_{12}gf^{abc}T^{a}T^{b}c^{c}
+ a_{13}\varUpsilon^{a\alpha}\lambda^{a}_{\alpha}
+ a_{19}\varUpsilon^{a\alpha}(\gamma_{5})_{\alpha\beta}\varepsilon^{\beta}T^{a}
+ a_{22}J^{a}T^{a}
+ a_{23}\mathfrak{D}^{a}T^{a}
\nonumber \\
&&
+ a_{35}gf^{abc}U^{Ib\alpha}(\gamma_{5})_{\alpha\beta}\hat{U}_{I}^{c\gamma}
\bar{\varepsilon}^{\beta}\varepsilon_{\beta}T^{a}
+ a_{36}gf^{abc}U^{Ib\alpha}\hat{U}_{I}^{c\beta}(\gamma_{5})_{\beta\eta}
\bar{\varepsilon}_{\alpha}\epsilon^{\eta}T^{a}
\nonumber \\
&&
+ a_{41}gf^{abc}U^{Ib\alpha}(\gamma_{5})_{\alpha\beta}\hat{U}_{I}^{c\gamma}
\varepsilon^{\beta}\bar{\varepsilon}_{\gamma}T^{a}
+ a_{42}gf^{abc}U^{Ib\alpha}\hat{U}_{I}^{c\beta}(\gamma_{5})_{\beta\eta}
\varepsilon_{\alpha}\bar{\varepsilon}^{\eta}T^{a}
\nonumber \\
&&
+ a_{47}gf^{abc}U^{Ib\alpha}\hat{U}_{I\;\;\alpha}^{\;\;c}\bar{\varepsilon}^{\beta}
(\gamma_{5})_{\beta\eta}\varepsilon^{\eta}T^{a}
+ a_{48}gf^{abc}U^{Ib\alpha}
(\gamma_{5})_{\alpha\beta}\hat{U}_{I}^{c\beta}\bar{\varepsilon}^{\eta}
\varepsilon_{\eta}T^{a}
\nonumber \\
&&
+ a_{50}gf^{abc}U^{Ib\alpha}(\gamma_{5})_{\alpha\beta}C^{\beta\gamma}
\hat{U}_{I\;\;\gamma}^{\;\;c}\varepsilon^{\eta}\varepsilon_{\eta}
+ a_{51}gf^{abc}U^{Ib\alpha}C_{\alpha\beta}\hat{U}_{I}^{\;c\beta}\varepsilon^{\eta}(\gamma_{5})_{\eta\delta}\varepsilon^{\delta}T^{a}
\nonumber \\
&&
+ a_{68}g^{2}f^{abe}f^{ecd}\tilde{\omega}^{a}_{i}\varphi^{b}_{i}\tilde{\varphi}^{c}_{j}\varphi^{d}_{j}
+ a_{69}g^{2}f^{abe}f^{ecd}\tilde{\omega}^{a}_{i}\varphi^{b}_{j}\tilde{\varphi}^{c}_{i}\varphi^{d}_{j}
+ a_{70}g^{2}f^{abe}f^{ecd}\tilde{\omega}^{a}_{i}\varphi^{b}_{j}\tilde{\omega}^{c}_{i}\omega^{d}_{j}
\nonumber \\
&&
+ a_{71}g^{2}f^{abe}f^{ecd}\tilde{\omega}^{a}_{i}\varphi^{b}_{j}\tilde{\omega}^{c}_{i}\omega^{d}_{j}
+ a_{72}\mu^{2}\tilde{\omega}^{a}_{i}\varphi^{a}_{i}
+ a_{76}g^{3}f^{abn}f^{lcd}f^{nle}\tilde{\omega}^{a}_{i}\varphi^{b}_{i}\tilde{\omega}^{c}_{j}\varphi^{d}_{j}c^{e}
\nonumber \\
&&
+ a_{77}g^{3}f^{abn}f^{lcd}f^{nle}\tilde{\omega}^{a}_{i}\varphi^{b}_{j}\tilde{\omega}^{c}_{i}\varphi^{d}_{j}c^{e}
+ a_{78}gf^{abc}(Y^{a\alpha}-\hat{U}^{Ia}_{\beta}C^{\alpha\beta}\theta_{I} - U^{Ia\alpha}\hat{\theta}_{I})\varepsilon_{\alpha}\tilde{\omega}^{b}_{i}\varphi^{c}_{i}
\nonumber \\
&&
+ a_{118}g^{2}f^{abe}f^{ecd}U^{Ia\alpha}\hat{U}_{I}^{\;\;b\beta}\bar{\varepsilon}_{\alpha}\varepsilon_{\beta}\tilde{\omega}^{c}_{i}\varphi^{d}_{i}
+ a_{119}g^{2}f^{abe}f^{ecd}U^{Ia\alpha}\hat{U}_{I}^{\;\;b\beta}\varepsilon_{\alpha}\bar{\varepsilon}_{\beta}\tilde{\omega}^{c}_{i}\varphi^{d}_{i}
\nonumber \\
&&
+ a_{122}g^{2}f^{abe}f^{ecd}U^{Ia\alpha}\hat{U}_{I\;\alpha}^{\;\;b}\bar{\varepsilon}^{\beta}\varepsilon_{\beta}\tilde{\omega}^{c}_{i}\varphi^{d}_{i}
+ a_{123}g^{2}f^{abe}f^{ecd}U^{Ia\alpha}C_{\alpha\beta}\hat{U}_{I}^{\;\;b\beta}\varepsilon^{\eta}\varepsilon_{\eta}\tilde{\omega}^{c}_{i}\varphi^{d}_{i}
\nonumber \\
&&
+ a_{130}g^{2}f^{abe}f^{ecd}U^{Ia\alpha}(\gamma_{5})_{\alpha\beta}\hat{U}_{I}^{\;\;b\gamma}(\gamma_{5})_{\gamma\eta}\bar{\varepsilon}^{\beta}\varepsilon^{\eta}\tilde{\omega}^{c}_{i}\varphi^{d}_{i}
\nonumber \\
&&
+ a_{136}g^{2}f^{abe}f^{ecd}U^{Ia\alpha}(\gamma_{5})_{\alpha\beta}\hat{U}_{I}^{\;\;b\gamma}(\gamma_{5})_{\gamma\eta}\varepsilon^{\beta}\bar{\varepsilon}^{\eta}\tilde{\omega}^{c}_{i}\varphi^{d}_{i}
\nonumber \\
&&
+ a_{142}g^{2}f^{abe}f^{ecd}U^{Ia\alpha}(\gamma_{5})_{\alpha\beta}\hat{U}_{I}^{\;\;b\beta}\bar{\varepsilon}^{\eta}(\gamma_{5})_{\eta\delta}\varepsilon^{\delta}\tilde{\omega}^{c}_{i}\varphi^{d}_{i}
\nonumber \\
&&
\left.
+ a_{145}g^{2}f^{abe}f^{ecd}U^{Ia\alpha}(\gamma_{5})_{\alpha\beta}C^{\beta\gamma}\hat{U}_{I\;\gamma}^{\;\;b}\varepsilon^{\eta}(\gamma_{5})_{\eta\delta}\varepsilon^{\delta}\tilde{\omega}^{c}_{i}\varphi^{d}_{i}\right\}\;.
\end{eqnarray}

\noindent After applying the stated Ward Identities we end up with only three parameters, which are
\begin{equation}
a_{23}=-\frac{a_0}{2}, \qquad a_{19}=\frac{a_0}{2} - a_{13} \qquad \text{and} \qquad -a_{4}=a_{3}=a_{5}=a_{7}=a_{8}=a_{9}=a_{10}=a_{1}\;.
\end{equation}
All the others are null. Note that we can write all the non-null parameters as $a_{0}$, $a_{1}$ and $a_{13}$.\\
This result is in full agreement with the more simple case where the Gribov ambiguity was not took into account, and can be checked just by turning off all the auxiliary fields, or even taking $\gamma^{2}=M^{3/2}=0$.

\noindent Therefore, for the exact part of the counter-term, which is obtained applying the linearized Slavnov-Taylor over $\Delta^{(-1)}$, we get
\begin{eqnarray}
\label{extct}
\mathcal{B}_{\Sigma}(\Delta^{(-1)}) &=& \int d^{4}x\;\left\{
  \frac{a_{0}}{2}\left[
  \left(\frac{\delta\Sigma}{\delta\mathfrak{D}^{a}}+J^{a}\right)
  \left( Y^{a\alpha}(\gamma_{5})_{\alpha\eta}\varepsilon^{\eta}
- \hat{U}^{Ia}_{\;\;\beta}C^{\alpha\beta}(\gamma_{5})_{\alpha\eta}\varepsilon^{\eta}\theta_{I}
- U^{Ia\alpha}(\gamma_{5})_{\alpha\eta}\varepsilon^{\eta}\hat{\theta}_{I}
- \mathfrak{D}^{a}
  \right)
  \right.
  \right.
\nonumber \\
&&
+ \left(\frac{\delta\Sigma}{\delta\lambda^{a}_{\alpha}}+X^{a\alpha}
- \hat{V}^{Ia}_{\;\;\beta}C^{\alpha\beta}\theta_{I}
+ \hat{U}^{Ia}_{\;\;\beta}C^{\alpha\beta}(\nabla\zeta_{I})
- V^{Ia\alpha}\hat{\theta}_{I}
+ U^{Ia\alpha}\hat{\zeta}_{I} \right)
  (\gamma_{5})_{\alpha\eta}\varepsilon^{\eta}T^{a}
\nonumber \\
&&
\left.
- \frac{\delta\Sigma}{\delta T^{a}}T^{a}
\right]
+ a_{1}\left[ \left(\frac{\delta\Sigma}{\delta A^{a}_{\mu}}+\Omega^{a}_{\mu}\right)A^{a}_{\mu}
+ \frac{\delta\Sigma}{\delta K^{a}_{\mu}}(\partial_{\mu}\check{c}^{a}+K^{a}_{\mu})
- b^{a}\partial_{\mu}A^{a}_{\mu}
- \tilde{\varphi}^{a}_{i}\partial_{\mu}\partial_{\mu}\varphi^{a}_{i}
  \right.
\nonumber \\
&&
+ \tilde{\omega}^{a}_{i}\partial_{\mu}\partial_{\mu}\omega^{a}_{i}
+ \tilde{M}^{a}_{\mu i}\partial_{\mu}\varphi^{a}_{i}
- \tilde{N}^{a}_{\mu i}\partial_{\mu}\omega^{a}_{i}
+ \kappa M^{a}_{\mu i}\tilde{M}^{a}_{\mu i}
+ \kappa N^{a}_{\mu i}\tilde{N}^{a}_{\mu i}
-\tilde{\varphi}^{a}_{i}\partial_{\mu}M^{a}_{\mu i}
+ N^{a}_{\mu i}\partial_{\mu}\tilde{\omega}^{a}_{i}
\nonumber \\
&&
+ gf^{abc}A^{c}_{\mu}
  \left(
- (\partial_{\mu}\tilde{\varphi}^{a}_{i})\varphi^{b}_{i}
+ (\partial_{\mu}\tilde{\omega}^{a}_{i})\omega^{b}_{i}
- \tilde{M}^{a}_{\mu i}\varphi^{b}_{i}
+ \tilde{N}^{a}_{\mu i}\omega^{b}_{i}
-  N^{a}_{\mu i}\tilde{\omega}^{b}_{i}
- M^{a}_{\mu i}\tilde{\varphi}^{b}_{i}
\right)
\nonumber \\
&&
\left.
- gf^{abc}\frac{\delta\Sigma}{\delta K^{c}_{\mu}}
\left( 
 (\partial_{\mu}\tilde{\omega}^{a}_{i})\varphi^{b}_{i}
+ M^{a}_{\mu i}\tilde{\omega}^{b}_{i}
+ \tilde{N}^{a}_{\mu i}\varphi^{b}_{i}
\right)
\right]
+ a_{13}\left[ 
\left(\frac{\delta\Sigma}{\delta\lambda^{a}_{\alpha}}+X^{a\alpha}
- \hat{V}^{Ia}_{\;\;\beta}C^{\alpha\beta}\theta_{I}
\right.
\right.
\nonumber \\
&&
\left.
+ \hat{U}^{Ia}_{\;\;\beta}C^{\alpha\beta}(\nabla\zeta_{I})
- V^{Ia\alpha}\hat{\theta}_{I}
+ U^{Ia\alpha}\hat{\zeta}_{I}
\right)
\left(
  \lambda^{a}_{\;\alpha}
- (\gamma_{5})_{\alpha\eta}\varepsilon^{\eta}T^{a}
\right)
\nonumber \\
&&
\left.
\left.
+ \left(
  Y^{a\alpha}
- \hat{U}^{Ia}_{\;\;\beta}C^{\alpha\beta}\theta_{I}
- U^{Ia\alpha}\hat{\theta}_{I}
\right)
\left(
\frac{\delta\Sigma}{\delta Y^{a\alpha}}
- (\gamma_{5})_{\alpha\eta}\varepsilon^{\eta}
\left( \frac{\delta\Sigma}{\delta\mathfrak{D}^{a}} + J^{a} \right)
\right)
\right]
\right\}\;.
\end{eqnarray}

\noindent One sees that $\Sigma_{count}$ contains three arbitrary coefficients, $a_0, a_1,
a_{13}$, which will identify the renormalization factors of all fields, sources and coupling
constant. To complete the analysis of the algebraic renormalization of the model, we need to
show that the counter-term $\Sigma_{count}$ can be reabsorbed into the starting action $\Sigma$ through a redefinition of the fields and parameters $\{\phi \}$, of the sources $\{ S \}$ and coupling constant $g$, namely 
\begin{equation}
\label{ration1}
\Sigma(\phi,S,g) + \omega \Sigma_{count}(\phi,S,g)  = \Sigma(\phi_0,S_0,g_0) + O(\omega^2) \;, 
\end{equation}
where $(\phi_0, S_0, g_0)$ stand for the so-called bare fields, sources and coupling constant:
\begin{equation}
\label{renormfs1}
\phi_{0}=Z^{1/2}_{\phi}\,\phi  \qquad\;,   \qquad
S_{0}=Z_{S}\,S\,,  \qquad g_0 = Z_g g   \;, 
\end{equation}
and the renormalization factors  $Z$ can be written as
\begin{equation}
Z^{1/2}_{\phi}=(1+\omega\,z_\phi)^{1/2}=1+\omega \frac{z_{\phi}}{2}+O(\omega^{2})\,,\qquad
Z_{S}=1+\omega\,z_S\,, \qquad Z_g = 1 +\omega z_g \;.  
\end{equation}
Moreover, in the present case, a little care has to be taken with the potential mixing of quantities which have the same quantum numbers. In fact, from equation \eqref{extct} one can easily notice  that the field $\lambda^{a\alpha}$ and the combination $\gamma_{5}\epsilon T^{a}$ have the same dimension and quantum numbers as well as the field $\mathfrak{D}^{a}$ and the combination $\left(Y^{a}-\hat{U}^{Ia}C\theta_{I}-U^{Ia}\hat{\theta}_{I}\right)\gamma_{5}\epsilon$, as it can be checked from Table 1.   As a consequence, these quantities can mix at the quantum level, a well known property of renormalization theory. This feature can be properly taken into account by writing the renormalization of the fields $\lambda$ and $\mathfrak{D}$ in matrix form, {\it i.e.}  
\begin{equation}
\label{lrenorm1}
\lambda^{a\alpha}_{0}=Z^{1/2}_{\lambda}\,\lambda^{a\alpha}+\omega\, z_{1}\,T^{a}(\gamma_{5})^{\alpha\beta}\varepsilon_{\beta}
\end{equation}
and
\begin{equation}
\label{drenorm1}
\mathfrak{D}^{a}_{0}=Z^{1/2}_{\mathfrak{D}}\,\mathfrak{D}^{a}+\omega\,\left( z_{2}\,Y^{a\alpha}(\gamma_{5})_{\alpha\beta}\varepsilon^{\beta} + z_{3}\hat{U}^{Ia}_{\;\;\beta}C^{\alpha\beta}\theta_{I}(\gamma_{5})_{\alpha\eta}\varepsilon^{\eta} + z_{4}U^{Ia}_{\;\;\beta}\hat{\theta}_{I}(\gamma_{5})_{\alpha\eta}\varepsilon^{\eta}  \right)\,,
\end{equation}
while the remaining fields, sources and parameters still obey \eqref{renormfs1}.

\noindent By just applying the definition of the bare fields as well as the counter-term \eqref{count}, with \eqref{SYM1} and \eqref{extct}, in eq. \eqref{ration1} we can find the respective renormalization parameters as follows,
\begin{eqnarray}
Z^{1/2}_{A} = 1 + \omega\left(\frac{a_{0}}{2}+a_{1}\right)\;, \nonumber \\
Z^{1/2}_{\lambda} = 1+ \omega \left(\frac{a_{0}}{2}-a_{13}\right)\;, \nonumber \\
Z_{g}=1-\omega\frac{a_{0}}{2}\;.
\end{eqnarray}

\begin{eqnarray}
&&
Z_{\mathfrak{D}}^{1/2}=1\;, \nonumber \\
&&
Z^{1/2}_{\bar{\varphi}} = Z^{1/2}_{\varphi} = Z^{1/2}_{c}=Z^{1/2}_{\check{c}} = Z_{K} = Z^{-1/2}_{g}Z^{-1/4}_{A}\;, \nonumber \\
&&
Z^{1/2}_{\bar{\omega}} = Z^{-1}_{g}\;, \nonumber \\
&&
Z^{1/2}_{\omega} = Z^{-1/2}_{A} \nonumber \\
&&
Z^{1/2}_{\theta}=Z^{1/2}_{\hat{\theta}} = 1\;, \nonumber \\
&&
Z^{1/2}_{\zeta}=Z^{-1/2}_{\hat{\zeta}} = Z^{1/2}_{g}Z^{-1/4}_{A}
\end{eqnarray}

\noindent The renormalization of $M$ and $\tilde{M}$ give us the renormalization factor of the Gribov parameter $\gamma^{2}$, while the renormalization of $V$ and $\hat{V}$ give us the renormalization of $M^{3/2}$, when every source assume its physical value stated at \eqref{physval2}.
 Namely,
\begin{eqnarray}
&&
Z_{\tilde{M}} =Z_{M} = Z^{-1/2}_{g}Z^{-1/4}_{A}\;, \nonumber \\
&&
Z_{\hat{V}} =Z_{V} = Z^{-1/2}_{\lambda}\;.
\end{eqnarray}

\noindent The other sources renormalize as
\begin{eqnarray}
&&
Z_{N} = Z^{-1/2}_{A}\;, \nonumber \\
&&
Z_{\bar{N}} = Z^{-1}_{g}\;, \nonumber \\
&&
Z_{\hat{U}} = Z^{-1/2}_{g}Z^{1/4}_{A}Z^{-1/2}_{\lambda}\;, \nonumber \\
&&
Z_{U} = Z^{-1/2}_{g}Z^{1/4}_{A}Z^{-1/2}_{\lambda}\;, \nonumber \\
&&
Z_{Y} = Z^{-1/2}_{g}Z^{1/4}_{A}Z^{-1/2}_{\lambda}\;, \nonumber \\
&&
Z_{L} = Z^{1/2}_{A}\;, \nonumber \\
&&
Z_{\Omega} = Z^{-1/2}_{A}\;, \nonumber \\
&&
Z_{T} = Z^{-1/2}_{g}Z^{1/4}_{A}\;, \nonumber \\
&&
Z_{\Lambda} = Z^{1/2}_{g}Z^{1/4}_{A}\;, \nonumber \\
&&
Z_{X} = Z^{-1/2}_{\lambda}\;, \nonumber \\
&&
Z_{J} = 1
\end{eqnarray}

\noindent The renormalization parameter of the SUSY parameter $\varepsilon$
\begin{equation}
Z_{\varepsilon}=Z^{1/2}_{g}Z^{-1/4}_{A}\;,
\end{equation}

\begin{eqnarray}
-z_{1} = z_{2} =  z_{3} =  z_{4} = \frac{a_{0}}{2} - a_{13}\;, \nonumber
\end{eqnarray}

\end{appendix}


\bibliographystyle{unsrt}

\end{document}